\documentclass[11pt,a4paper,twoside,final]{report}
\usepackage{psfig,epsfig,graphicx}
\usepackage{amssymb,amsmath,bbm,rotating,subfigure,here,citesort}
\newcommand{\befig}{\begin{figure}}
\newcommand{\efig}{\end{figure}}
\newcommand{\betab}{\begin{table}}
\newcommand{\etab}{\end{table}}
\newcommand{\barray}{\begin{array}}
\newcommand{\earray}{\end{array}}
\newcommand{\be}{\begin{equation}}
\newcommand{\ee}{\end{equation}}
\newcommand{\bea}{\begin{eqnarray}}
\newcommand{\eea}{\end{eqnarray}}
\newcommand{\benn}{\begin{displaymath}}
\newcommand{\eenn}{\end{displaymath}}
\newcommand{\beann}{\begin{eqnarray*}}
\newcommand{\eeann}{\end{eqnarray*}}
\newcommand{\inv}{\frac{1}}
\newcommand{\gtsim}{\gtrsim}
\newcommand{\ltsim}{\lesssim}
\newcommand{\im}{\mbox{Im}}    
\newcommand{\Order}{{\cal O}}   
\newcommand{\fm}{\mbox{fm}}
\newcommand{\mb}{\mbox{mb}}
\newcommand{\nb}{\mbox{nb}}
\newcommand{\MeV}{\mbox{MeV}}
\newcommand{\GeV}{\mbox{GeV}}
\newcommand{\TeV}{\mbox{TeV}}
\newcommand{\alphaS}{\alpha_s}
\newcommand{\alphaEM}{\alpha}

\newcommand{\SVM}{SVM}
\newcommand{\pert}{P}
\newcommand{\nprt}{N\!P}
\newcommand{\glueball}{G\!B}
\newcommand{\WW}{Wegner-Wilson}
\newcommand{\pot}{\mbox{\scriptsize pot}}
\newcommand{\self}{\mbox{\scriptsize self}}
\newcommand{\actiondensity}{\mathit{s}}
\newcommand{\pbar}{{\bar{p}}}
\newcommand{\qbar}{{\bar{q}}}

\newcommand{\Reggeon}{I\!\!R}
\newcommand{\Pomeron}{I\!\!P}
\newcommand{\Lagrangian}{{\cal L}}
\newcommand{\QCD}{\mathrm{QCD}}
\newcommand{\G}{{\cal G}}       
\newcommand{\GG}{\hat{\cal{G}}} 
\newcommand{\Identity}{{1\!\rm l}}

\newcommand{\impactT}{{\cal T}}
\newcommand{\Pc}{{\cal P}}      
\newcommand{\Ps}{{\cal P}_S}    
\newcommand{\Tr}{\mbox{Tr}}             
\newcommand{\rTr}{\Tr_{r}}        
\newcommand{\ronexrtwoTr}{\Tr_{r_1 \otimes r_2}}  
\newcommand{\nrTr}{\tilde{\Tr}_{r}}             
\newcommand{\nroneTr}{\tilde{\Tr}_{r_1}}              
\newcommand{\nrtwoTr}{\tilde{\Tr}_{r_2}}              
\newcommand{\nronexrtwoTr}{\tilde{\Tr}_{r_1 \otimes r_2}}
\newcommand{\roneIdentity}{\Identity_{r_1}}
\newcommand{\rtwoIdentity}{\Identity_{r_2}}
\newcommand{\ronexrtwoIdentity}{\Identity_{r_1 \otimes r_2}}
\newcommand{\Projector}{\mbox{P}} 
\newcommand{\tensor}{t_{r_1 \otimes r_2}} 
\newcommand{\fundamental}{\mbox{\scriptsize N}_c}
\newcommand{\adjoint}{\mbox{\scriptsize N}_c^2\!-\!1}
\newcommand{\Fundamental}{\mbox{N}_c}
\newcommand{\Adjoint}{\mbox{N}_c^2\!-\!1}
\begin{document}
%
%
\pagenumbering{roman}
%
%
\begin{titlepage}
\pagestyle{empty}
%
%
%
\begin{center}
  \Huge \bf \sc
  From Static Potentials\\
  to\\
  High-Energy Scattering
\end{center}
\vfill
\begin{center}
  {
  \Large Dissertation\\}
  \vspace{0.3cm}
  {\large
  submitted to the\\
  Combined Faculties for the Natural Sciences and for Mathematics\\
  of the Ruperto--Carola University of Heidelberg, Germany\\
  for the degree of\\ 
  Doctor of Natural Sciences}
\end{center}
\vfill
\begin{center}
\begin{large}
  presented by\\
  \vspace{0.5cm}
  {\LARGE\sc Frank Daniel Steffen}\\
  \vspace{0.5cm}
    born in Wattenscheid\\
\end{large}
\end{center}
\vfill
\begin{center}
\begin{large}
\begin{tabular}{ll}
Referees: & Prof.~Dr.~Hans G\"unter Dosch\\
& Prof.~Dr.~Bogdan~Povh\\
\end{tabular}

Oral examination: \, February 14, 2003 \, \hphantom{.}

\end{large}
\end{center}
\newpage
%
%
\enlargethispage{2cm}
\vspace*{-2cm}
%
\begin{center}
  {\bf \large Von Statischen Potenzialen zur Hochenergiestreuung}\\[.2cm]
  {\bf Zusammenfassung}
\end{center}
{\small
  Wir entwickeln ein Loop-Loop Korrelations Modell zur einheitlichen
  Beschreibung von sta\-tischen Farbdipol Potenzialen, einschliessenden
  QCD Strings, und hadronischen Hochenergiereaktionen mit besonderer
  Ber\"ucksichtigung von S\"attigungseffekten, die die $S$-Matrix
  Unitarit\"at bei extrem hohen Energien manifestieren.
  Das Modell verbindet st\"orungs\-theoretischen Gluonaustausch mit
  dem nicht-st\"orungstheoretischen Modell des Stochastischen Vakuums,
  das den Einschluss von Farbladungen durch Flussschlauchbildung der
  Farbfelder beschreibt.
  Wir berechnen Farbfeldverteilungen statischer Farbdipole in
  verschiedenen $SU(N_c)$ Darstellungen und finden Casimir
  Skalierungsverhalten in \"Ubereinstimmung mit aktuellen Gitter-QCD
  Ergebnissen.
  Wir untersuchen die im einschliessenden String gespeicherte Energie
  und zeigen mit Niederenergie\-theoremen die Konsistenz mit dem
  statischen Quark-Antiquark Potenzial.
  Wir ver\-allgemeinern Meggiolaros Analytische Fortsetzung von
  Parton-Parton auf Dipol-Dipol Streuung und er\-halten einen
  Euklidischen Zugang zur Hochenergiestreuung, der prinzipiell
  erlaubt, Streu\-matrix\-elemente in Gitter-QCD zu berechnen.
  Mit dem Euklidischen Loop-Loop Korrelations Modell berechnen wir in
  diesen Zugang Dipol-Dipol Streuung bei hohen Energien.
  Das Ergebnis bildet zusammen mit einer universellen
  Energieabh\"angigkeit und reaktionsspezifischen Wellenfunktionen die
  Grundlage f\"ur eine einheitliche Beschreibung von pp, $\pi$p, Kp,
  $\gamma^*$p und $\gamma\gamma$ Reaktionen in guter \"Ubereinstimmung
  mit experimentellen Daten f\"ur Wirkungsquerschnitte,
  Steigungsparameter und Strukturfunktionen.
  Die erhaltenen Stossparameterprofile f\"ur pp und $\gamma^*_{L}$p
  Reaktionen und die stossparameterabh\"angige Gluonverteilung des
  Protons $xG(x,Q^2,|\vec{b}_{\!\perp}|)$ zeigen S\"attigung bei
  extrem hohen Energien in \"Ubereinstimmung mit Unitarit\"atsgrenzen.
}
%
%
\begin{center}
  {\bf \large From Static Potentials to High-Energy Scattering}\\[.2cm]
  {\bf  Abstract}
\end{center}
{\small
  We develop a loop-loop correlation model for a unified description
  of static color dipole potentials, confining QCD strings, and
  hadronic high-energy reactions with special emphasis on saturation
  effects manifesting $S$-matrix unitarity at ultra-high energies.
  The model combines perturbative gluon exchange with the
  non-perturbative stochastic vacuum model which describes color
  confinement via flux-tube formation of color fields.
  We compute the chromo-field distributions of static color dipoles in
  various $SU(N_c)$ representations and find Casimir scaling in
  agreement with recent lattice QCD results.
  We investigate the energy stored in the confining string and use
  low-energy theorems to show consistency with the static
  quark-antiquark potential.
  We generalize Meggiolaro's analytic continuation from parton-parton
  to dipole-dipole scattering and obtain a Euclidean approach to
  high-energy scattering that allows us in principle to calculate
  $S$-matrix elements in lattice QCD.
  In this approach we compute high-energy dipole-dipole scattering
  with the Euclidean loop-loop correlation model.
  Together with a universal energy dependence and reaction-specific
  wave functions, the result forms the basis for a unified description
  of pp, $\pi$p, Kp, $\gamma^*$p, and $\gamma\gamma$ reactions in good
  agreement with experimental data for cross sections, slope
  parameters, and structure functions.
  The obtained impact parameter profiles for pp and $\gamma^*_{L}$p
  reactions and the impact parameter dependent gluon distribution of
  the proton $xG(x,Q^2,|\vec{b}_{\!\perp}|)$ show saturation at
  ultra-high energies in accordance with unitarity constraints.
} \vfill
\cleardoublepage
%
%
%
\end{titlepage}
%


%
%
\tableofcontents
%
%
%
%
%
\clearpage{\pagestyle{empty} \cleardoublepage} 
\setcounter{page}{1}
\pagenumbering{arabic}
\pagestyle{plain}
%
\makeatletter
\@addtoreset{equation}{chapter}
\makeatother
\renewcommand{\theequation}{\thechapter.\arabic{equation}}
%
%
\chapter{Introduction}
\label{Sec_Introduction}


According to our present understanding, {\em quantum chromodynamics}
(QCD) is the theory of strong interactions and thus describes the
diversity of strong interaction phenomena~\cite{Fritzsch:pi+X}. QCD is
the gauge field theory defined by the non-Abelian color gauge group
$SU(N_c)$ with $N_c=3$ colors~\cite{Bar:2001qk} and the presence of a
certain number of quark fields $\psi_q$ ($q = u,d,s,...$) in the
fundamental representation. The Lagrange density of QCD has an
extremely simple form
\be
        \Lagrangian_{\QCD}(x) = 
        -\frac{1}{4}\G^a_{\mu\nu}(x)\G^{a\,\mu\nu}(x)
        +\sum_q \bar{\psi}_q(x)
        \left\{i\gamma^{\mu}(\partial_{\mu}+ig\G^a_{\mu}t^a_{\fundamental})-m_q\right\}
        \psi_q(x)
\label{Eq_L_QCD}
\ee
and involves only very few parameters, i.e.\ the gauge coupling $g$
and the current quark masses $m_q$. The first term describes pure
gauge theory of the massless gluon potentials $\G^a_{\mu}(x)$
($a=1,...,N_c^2-1$) in terms of the gluon field strengths
\be
        \G^a_{\mu\nu}(x)
        = \partial_\mu \G_\nu^a(x) 
        - \partial_\nu \G_\mu^a(x)
        - g f^{abc} \G^b_\mu(x)\G_\nu^c(x)
        \ ,
\label{Eq_Gluon_Field_Strength}
\ee
where $f^{abc}$ are the structure constants of the gauge group
$SU(N_c)$. The second term is a sum over the contributions of the
quarks with flavor $q=u,d,s,...$ and involves the $SU(N_c)$ generators
$t^a_r$ in the fundamental representation $r=\Fundamental$.


The simple Lagrangian~(\ref{Eq_L_QCD}) brings along a very rich
structure. Due to vacuum polarization, the effective coupling depends
on the distance scale, or equivalently the (inverse) energy scale, at
which it is measured: As long as there are no more than $N_f=16$ quark
flavors, the renormalization group tells us that the effective
coupling becomes small at short distances and thus that QCD is an
asymptotically free theory~\cite{Politzer:fx+X}. Indeed, high-energy
deep inelastic scattering experiments reveal that quarks behave as
free particles at short distances. Accordingly, perturbation theory is
applicable in this regime and allows reliable analytic calculations,
for example, of the total cross section for electron-positron
annihilation into hadrons~\cite{Pascual:zb,Ellis:qj}. At distance
scales of order of the proton size ($\approx 1\,\fm$), the effective
coupling becomes large so that perturbation theory breaks down.


Genuinely non-perturbative methods are needed to describe the physics
of hadrons and low-energy interactions~\cite{Dosch:wj}. Indeed, an
analytic derivation of color confinement -- the phenomenon that quarks
and gluons cannot be observed as isolated particles -- from the QCD
Lagrangian is still missing and among the ultimate goals in
theoretical physics.  Further fundamental strong interaction phenomena
at low energy are spontaneous chiral symmetry breaking and dynamical
mass generation, i.e.\ the hadron spectrum and the origin of the
hadron mass, which are inherently non-perturbative phenomena as well
not yet proven analytically from the QCD Lagrangian.


A challenge at high energy is the description and understanding of
hadronic high-energy scattering~\cite{Hebecker:2001ex}. For small
momentum transfers, the effective QCD coupling is again too large for
a reliable perturbative treatment and non-perturbative methods are
needed. In particular, it is a key issue to unravel the effects of
confinement and topologically non-trivial gauge field configurations
(such as instantons) on such reactions~\cite{Dosch:JHW2002,Ringwald:2002iy}.


The most interesting phenomenon in hadronic high-energy scattering is
the rise of the total cross sections with increasing c.m.\ energy
$\sqrt{s}$: While the rise is slow in hadronic reactions of {\em
  large} particles such as protons, pions, kaons, or real
photons~\cite{Hagiwara:fs}, it is steep if only one {\em small}
particle is involved such as an incoming virtual
photon~\cite{Adloff:1997mf,Adloff:2001qj,Breitweg:1997hz,Breitweg:2000yn,Adloff:2001qk}
or an outgoing charmonium~\cite{Breitweg:1997rg,Adloff:2000vm}. This
energy behavior is best displayed in the proton structure function
$F_2(x,Q^2)$ that is equivalent to the total $\gamma^* p$ cross
section $\sigma_{\gamma^*p}^{tot}(s,Q^2)$. With increasing photon
virtuality $Q^2$, the increase of $F_2(x,Q^2)$ towards small Bjorken
$x=Q^2/s$ becomes significantly stronger. Together with the steep rise
of the gluon distribution in the proton $xG(x,Q^2)$ with decreasing
$x$, the rise of the structure function $F_2(x,Q^2)$ towards small
$x$~\cite{Adloff:1997mf,Adloff:2001qj,Breitweg:1997hz,Breitweg:2000yn,Adloff:2001qk}
is one of the most exciting results of the HERA experiments.


In high-energy collisions of stable hadrons, the rise of the total
cross sections is limited: The Froissart bound, derived from very
general principles such as unitarity and analyticity of the
$S$-matrix, allows at most a logarithmic energy dependence of the
cross sections at asymptotic energies~\cite{Froissart:1961ux}.
Analogously, the rise of $\sigma_{\gamma^* p}^{tot}(s,Q^2)$ is
expected to slow down. The microscopic picture behind this slow-down
is the concept of gluon saturation: Since the gluon density in the
proton becomes large at high energies $\sqrt{s}$ (small $x$), gluon
fusion processes are expected to tame the growth of $\sigma_{\gamma^*
  p}^{tot}(s,Q^2)$.  It is a key issue to determine the energy at
which these processes become significant.


Lattice QCD is the principal theoretical tool to study
non-perturbative aspects of the QCD Lagrangian from first principles.
Numerical simulations of QCD on Euclidean lattices give strong
evidence for color confinement and spontaneous chiral symmetry
breaking and describe dynamical mass generation from the QCD
Lagrangian~\cite{Rothe:1997kp,Kronfeld:2002pi,Muller-Preussker:nz}.
However, since lattice QCD is limited to the Euclidean formulation of
QCD, it cannot be applied in Minkowski space-time to simulate
high-energy reactions in which particles are inherently moving near
the light-cone. Furthermore, it is hard to understand from simulations
the important QCD mechanisms that lead, for example, to color
confinement. Here (phenomenological) models that allow analytic
calculations are important.


In this thesis we develop a {\em loop-loop correlation model} (LLCM)
for a unified description of static quark-antiquark potentials and
confining QCD strings in Euclidean space-time~\cite{Shoshi:2002rd} and
hadronic high-energy reactions in Minkowski
space-time~\cite{Shoshi:2002in,Shoshi:2002fq} with special emphasis on
saturation effects manifesting $S$-matrix unitarity at ultra-high
energies~\cite{Shoshi:2002ri,Shoshi:2002mt}. The model is constructed
in Euclidean space-time. It combines perturbative gluon exchange with
the {\em stochastic vacuum model} (SVM) of Dosch and
Simonov~\cite{Dosch:1987sk+X} which leads to confinement via a string
of color fields~\cite{Rueter:1994cn,Dosch:1995fz,Shoshi:2002rd}. For
applications to high-energy scattering, the LLCM can be analytically
continued to Minkowski space-time~\cite{Shoshi:2002in} following the
procedure introduced for applications of the SVM to high-energy
reactions~\cite{Kramer:1990tr,Dosch:1994ym,Dosch:RioLecture}. In this
way the LLCM allows us to investigate manifestations of the confining
QCD string in unintegrated gluon distributions of hadrons and
photons~\cite{Shoshi:2002fq}. We present an alternative Euclidean
approach to high-energy scattering that shows how one can access
high-energy scattering in lattice simulations of QCD. Indeed, this
approach allows us to compute $S$-matrix elements for dipole-dipole
scattering in the Euclidean LLCM and confirms the analytic
continuation of the model to Minkowski
space-time~\cite{Shoshi:2002rd}. The applications of the LLCM to
high-energy scattering are based on the functional integral approach
developed for parton-parton
scattering~\cite{Nachtmann:1991ua+X,Nachtmann:ed.kt} and extended to
gauge-invariant dipole-dipole
scattering~\cite{Kramer:1990tr,Dosch:1994ym,Dosch:RioLecture}; see
also Chap.~8 of~\cite{Donnachie:en}. Together with a universal energy
dependence introduced phenomenologically, the functional integral
approach to dipole-dipole scattering is the key to the presented
unified description of hadron-hadron, photon-hadron, and photon-photon
reactions.


The central element in our approach is the gauge-invariant
Wegner-Wilson loop~\cite{Wegner:1971qt,Wilson:1974sk}: The considered
physical quantities are obtained from the vacuum expectation value
(VEV) of one {\WW} loop, $\langle W_{r}[C] \rangle$, and the
correlation of two {\WW} loops, $\langle W_{r_1}[C_1] W_{r_2}[C_2]
\rangle$. Here $r_{(i)}$ indicates the $SU(N_c)$ representation of the
{\WW} loops which we keep as general as possible. In phenomenological
applications, the propagation of (anti-)quarks requires the
fundamental representation, $r=\Fundamental$, and the propagation of
gluons the adjoint representation, $r=\Adjoint$.  We express $\langle
W_{r}[C] \rangle$ and $\langle W_{r_1}[C_1] W_{r_2}[C_2] \rangle$ in
terms of the gauge-invariant bilocal gluon field strength correlator
integrated over minimal surfaces by using the non-Abelian Stokes
theorem and a matrix cumulant expansion in the Gaussian approximation.
We decompose the gluon field strength correlator into a perturbative
and a non-perturbative component. Here the SVM is used for the
non-perturbative low-frequency background field~\cite{Dosch:1987sk+X}
and perturbative gluon exchange for the additional high-frequency
contributions. This combination allows us to describe long and short
distance correlations in agreement with lattice calculations of the
gluon field strength
correlator~\cite{DiGiacomo:1992df+X,Meggiolaro:1999yn}. Moreover, it
leads to a static quark-antiquark potential with color Coulomb
behavior for small source separations and confining linear rise for
large source separations. We calculate the static quark-antiquark
potential with the LLCM parameters adjusted in fits to high-energy
scattering data~\cite{Shoshi:2002in} and find good agreement with
lattice data. We thus have one model that describes both static
hadronic properties and high-energy reactions of hadrons and photons
in good agreement with experimental and lattice QCD data.


We apply the LLCM to compute the {\em chromo-electric fields}
generated by a static color dipole in the fundamental and adjoint
representation of $SU(N_c)$. The non-perturbative SVM component
describes the formation of a color flux tube that confines the two
color sources in the
dipole~\cite{Rueter:1994cn,Dosch:1995fz,Shoshi:2002rd} while the
perturbative component leads to color Coulomb fields. We find {\em
  Casimir scaling} for both the perturbative and non-perturbative
contributions to the chromo-electric fields in agreement with lattice
data and our results for the static dipole potential, which is the
potential of a static quark-antiquark pair in case of the fundamental
representation and the potential of a gluino pair in case of the
adjoint representation. The mean squared radius of the confining QCD
string is calculated as a function of the dipole size.  Transverse and
longitudinal energy density profiles are provided to study the
interplay between perturbative and non-perturbative physics for
different dipole sizes. The transition from perturbative to string
behavior is found at source separations of about $0.5\,\fm$ in
agreement with the recent results of L\"uscher and
Weisz~\cite{Luscher:2002qv}.


The {\em low-energy theorems}, known in lattice QCD as Michael sum
rules~\cite{Michael:1986yi}, relate the energy and action stored in
the chromo-fields of a static color dipole to the corresponding ground
state energy. The Michael sum rules, however, are incomplete in their
original form~\cite{Michael:1986yi}. We present the complete energy
and action sum rules~\cite{Rothe:1995hu+X,Michael:1995pv,Green:1996be}
in continuum theory taking into account the contributions to the
action sum rule found in~\cite{Dosch:1995fz} and the trace anomaly
contribution to the energy sum rule found in~\cite{Rothe:1995hu+X}.
Using these corrected low-energy theorems, we compare the energy and
action stored in the confining string with the confining part of the
static quark-antiquark potential. This allows us to confirm
consistency of the model results and to determine the values of the
$\beta$\,-\,function and the strong coupling $\alphaS$ at the
renormalization scale at which the non-perturbative SVM component is
working. Earlier investigations along these lines have been incomplete
since only the contribution from the traceless part of the
energy-momentum tensor has been considered in the energy sum rule.


To study the effect of the confining QCD string examined in Euclidean
space-time on high-energy reactions in Minkowski space-time, an {\em
  analytic continuation} from Euclidean to Minkowski space-time is
needed. For investigations of high-energy reactions in our model
constructed in Euclidean space-time, the gauge-invariant bilocal gluon
field strength correlator can be analytically continued from Euclidean
to Minkowski space-time.  As mentioned above, this analytic
continuation has been introduced by Dosch and collaborators for
applications of the SVM to high-energy
reactions~\cite{Kramer:1990tr,Dosch:1994ym,Dosch:RioLecture} and is
used also in our Minkowskian applications of the
LLCM~\cite{Shoshi:2002in,Shoshi:2002ri,Shoshi:2002fq,Shoshi:2002mt}.
Recently, an alternative analytic continuation for parton-parton
scattering has been established in the perturbative context by
Meggiolaro~\cite{Meggiolaro:1996hf+X}. This analytic continuation has
already been used to access high-energy scattering from the
supergravity side of the AdS/CFT correspondence~\cite{Janik:2000zk+X},
which requires a positive definite metric in the definition of the
minimal surface~\cite{Rho:1999jm}, and to examine the effect of
instantons in high-energy scattering~\cite{Shuryak:2000df+X}.


In this thesis we generalize Meggiolaro's analytic
continuation~\cite{Meggiolaro:1996hf+X} from parton-parton to
gauge-invariant {\em dipole-dipole scattering} such that $S$-matrix
elements for high-energy reactions can be computed from configurations
of {\WW} loops in Euclidean space-time and with {\em Euclidean}
functional integrals. This evidently shows how one can access
high-energy reactions directly in lattice QCD. First attempts in this
direction have already been carried out but only very few signals
could be extracted, while most of the data was dominated by
noise~\cite{DiGiacomo:2002PC}. We apply this approach to compute the
scattering of dipoles at high-energy in the Euclidean LLCM: We recover
exactly the result derived with the analytic continuation of the gluon
field strength correlator~\cite{Shoshi:2002in}. This confirms the
analytic continuation used in all earlier applications of the SVM to
high-energy
scattering~\cite{Kramer:1990tr,Dosch:1994ym,Dosch:RioLecture,Rueter:1996yb,Dosch:1997ss,Dosch:1998nw,Rueter:1998qy,Kulzinger:1999hw,Rueter:1998up,D'Alesio:1999sf,Berger:1999gu,Donnachie:2000kp,Donnachie:2001wt,Dosch:2001jg,Kulzinger:2002iu}
including the Minkowskian applications of the
LLCM~\cite{Shoshi:2002in,Shoshi:2002ri,Shoshi:2002fq,Shoshi:2002mt}.
Here we use the obtained $S$-matrix element $S_{DD}$ as the basis for
our unified description of hadronic high-energy reactions with special
emphasis on saturation effects in hadronic cross sections and gluon
saturation~\cite{Shoshi:2002in,Shoshi:2002ri,Shoshi:2002mt}.
Moreover, we have used the obtained $S$-matrix element to investigate
manifestations of the confining string in high-energy reactions of
hadrons and photons~\cite{Shoshi:2002fq}. In particular, we have found
that the string can be represented as an integral over stringless
dipoles with a given dipole number density. This {\em decomposition of
  the confining string into dipoles} gives insights into the
microscopic structure of the model. It allows us to calculate
unintegrated gluon distributions of hadrons and photons from
dipole-hadron and dipole-photon cross sections via
$|\vec{k}_{\!\perp}|$ factorization.  Our result shows explicitly that
non-perturbative physics dominates the unintegrated gluon
distributions at small transverse momenta
$|\vec{k}_{\!\perp}|$~\cite{Shoshi:2002fq}.


Aiming at a unified description of hadron-hadron, photon-hadron, and
photon-photon reactions, we follow the {\em functional integral
  approach} to high-energy scattering in the eikonal
approximation~\cite{Nachtmann:1991ua+X,Nachtmann:ed.kt,Kramer:1990tr,Dosch:1994ym,Nachtmann:ed.kt}
(cf.\ also Chap.~8 of~\cite{Donnachie:en}) in which $S$-matrix
elements factorize into the universal $S$-matrix element for elastic
high-energy dipole-dipole scattering $S_{DD}$ and reaction-specific
light-cone wave functions. The color dipoles -- described by
light-like {\WW} loops -- are given by the quark and antiquark in the
meson or photon and in a simplified picture by a quark and diquark in
the baryon. Consequently, hadrons and photons are described as color
dipoles with size and orientation determined by appropriate light-cone
wave functions~\cite{Dosch:1994ym,Nachtmann:ed.kt}.


We introduce a phenomenological {\em energy dependence} into the
univeral $S$-matrix element for dipole-dipole scattering $S_{DD}$ in
order to describe simultaneously the energy behavior in hadron-hadron,
photon-hadron, and photon-photon reactions involving real and virtual
photons as well. Motivated by the two-pomeron picture of Donnachie and
Landshoff~\cite{Donnachie:1998gm+X}, we ascribe to our
non-perturbative (soft) and perturbative (hard) component a weak and
strong energy dependence, respectively. Including {\em multiple
  gluonic interactions}, we obtain $S$-matrix elements with a
universal energy dependence that respects unitarity constraints in
impact parameter space.


To study saturation effects that manifest $S$-matrix unitarity, we
consider the scattering amplitudes in impact parameter space, where
$S$-matrix unitarity imposes rigid limits on the impact parameter
profiles such as the {\em black disc limit}. We confirm explicitly
that our model respects this unitarity constraint for dipole-dipole
scattering, which is the underlying process of each considered
reaction in our approach. We calculate the {\em impact parameter
  profiles} for proton-proton and longitudinal photon-proton
scattering. The profile functions describe the blackness or opacity of
the interacting particles and give an intuitive geometrical picture
for the energy dependence of the cross sections. At ultra-high
energies, the hadron opacity saturates at the black disc limit which
tames the growth of the hadronic cross sections in agreement with the
Froissart bound~\cite{Froissart:1961ux}. We estimate the {\em impact
  parameter dependent gluon distribution} of the proton
$xG(x,Q^2,|\vec{b}_{\perp}|)$ from the profile function for
longitudinal photon-proton scattering and find gluon saturation at
small Bjorken $x$ that tames the steep rise of the integrated gluon
distribution $xG(x,Q^2)$ towards small~$x$. These saturation effects
manifest $S$-matrix unitarity in hadronic collisions and should be
observable in future cosmic ray and accelerator experiments at
ultra-high energies. The c.m.\ energies and Bjorken $x$ at which
saturation sets in are determined and LHC and THERA predictions are
given.


With the intuitive geometrical picture gained in impact parameter
space, we turn to experimental observables to analyse the energy
dependence of the cross sections and to localize saturation effects.
We compare the LLCM results with experimental data and provide
predictions for future cosmic ray and accelerator experiments. Total
cross sections $\sigma^{tot}$, the structure function of the proton
$F_2$, slope parameters $B$, differential elastic cross sections
$d\sigma^{el}/dt$, elastic cross sections $\sigma^{el}$, and the
ratios $\sigma^{el}/\sigma^{tot}$ and $\sigma^{tot}/B$ are considered
for proton-proton, pion-proton, kaon-proton, photon-proton, and
photon-photon reactions involving real and virtual photons as well.


The outline of this thesis is as follows: In Chap.~\ref{Sec_The_Model}
the loop-loop correlation model is developed in its Euclidean version
and the general computations of $\langle W_{r}[C] \rangle$ and
$\langle W_{r_1}[C_1] W_{r_2}[C_2] \rangle$ are presented. Based on
these evaluations, we compute in Chap.~\ref{Sec_Static_Sources}
potentials and chromo-field distributions of static color dipoles with
emphasis on Casimir scaling and the interplay between perturbative
color Coulomb behavior and non-per\-tur\-ba\-tive formation of the
confining QCD string. Moreover, low-energy theorems are discussed and
used to show consistency of the model results and to determine the
values of $\beta$ and $\alphaS$ at the renormalization scale at which
the non-perturbative SVM component is working. In
Chap.~\ref{Sec_DD_Scattering} the Euclidean approach to high-energy
scattering is presented and applied to compute high-energy
dipole-dipole scattering in our Euclidean model. The additional
ingredients for the unified description of hadronic high-energy
scattering, i.e.\ hadron and photon wave functions and the
phenomenological universal energy dependence, are introduced in
Chap.~\ref{Sec_High-Energy_Scattering}. Going to impact parameter
space in Chap.~\ref{Sec_Impact_Parameter}, we confirm the unitarity
condition in our model, study the impact parameter profiles for
proton-proton and photon-proton scattering, and discuss the impact
parameter dependent gluon distribution of the proton
$xG(x,Q^2,|\vec{b}_{\perp}|)$ and gluon saturation. Finally, in
Chap.~\ref{Sec_Comparison_Data} we present the phenomenological
performance of the LLCM and provide predictions for saturation effects
in experimental observables. In the Appendices we give explicit
parametrizations of the loops and the minimal surfaces and provide the
detailed computations for the results in the main text.



%
\cleardoublepage
%
\chapter{The Loop-Loop Correlation Model}
\label{Sec_The_Model}

In this chapter the vacuum expectation value (VEV) of one {\WW} loop
and the correlation of two {\WW} loops are computed for arbitrary loop
geometries within a Gaussian approximation in the gluon field
strengths. The results are applied in the following chapters. We
describe our model for the QCD vacuum in which the stochastic vacuum
model (SVM) of Dosch and Simonov~\cite{Dosch:1987sk+X} is used for the
non-perturbative low-frequency background field (long-distance
correlations) and perturbative gluon exchange for the additional
high-frequency contributions (short-distance correlations). In this
and the next chapter we work in Euclidean space-time as indicated by
exclusively subscript Dirac indices and space-time variables written
in capital letters.

\section{Vacuum Expectation Value of a {\WW} Loop}
\label{Sec_<W[C]>}

A crucial quantity in gauge theories is the Wegner-Wilson loop
operator~\cite{Wegner:1971qt,Wilson:1974sk}
\be
        W_r[C] = 
        \nrTr\,\Pc
        \exp\!\left[
        -i g \oint_{\scriptsize C} dZ_{\mu}\,\G_{\mu}^a(Z)\,t_r^a 
        \right]      
        \ .
\label{Eq_WW_loop}        
\ee
Concentrating on $SU(N_c)$ {\WW} loops, where $N_c$ is the number of
colors, the subscript $r$ indicates a representation of $SU(N_c)$,
$\nrTr(...) = \rTr(...)/\rTr \Identity_r$ is the normalized
trace in the corresponding color space with unit element
$\Identity_r$, $g$ is the strong coupling, and $\G_{\mu}(Z) =
\G_{\mu}^a(Z) t_r^a$ is the gluon potential with the $SU(N_c)$ group
generators in the corresponding representation, $t_r^a$, that demand
the path ordering indicated by $\Pc$ on the closed path $C$ in
space-time. A distinguishing theoretical feature of the {\WW} loop is
its invariance under local gauge transformations in color space.
Therefore, it is the basic object in lattice gauge
theories~\cite{Wegner:1971qt,Wilson:1974sk,Rothe:1997kp,Bali:2001gf,Kronfeld:2002pi}
and has been considered as the fundamental building block for a gauge
theory in terms of gauge invariant variables~\cite{Migdal:1983gj}.
Phenomenologically, the {\WW} loop represents the phase factor
associated to the propagation of a very massive or very fast color
source in the representation $r$ of the gauge group $SU(N_c)$.

To compute the expectation value of a {\WW} loop~(\ref{Eq_WW_loop}) in
the QCD vacuum
\be
        \Big\langle W_r[C] \Big\rangle_G
        = \Big\langle
        \nrTr\,\Pc
        \exp\!\left[-i\,g \oint_{\scriptsize C} dZ_{\mu}\,\G_{\mu}^a(Z)\,t_r^a \right]      
        \Big\rangle_G
        \ ,
\label{Eq_<W[C]>}
\ee
we transform the line integral over the loop $C$ into an integral over
the surface $S$ with $\partial S = C$ by applying the {\em non-Abelian
  Stokes' theorem}~\cite{Arefeva:dp+X}
\be
        \Big\langle W_r[C] \Big\rangle_G
        = \Big\langle
        \nrTr\,\Ps
          \exp \left[-i\,\frac{g}{2} 
                \int_{S} \! d\sigma_{\mu\nu}(Z) 
                \G^a_{\mu\nu}(O,Z;C_{ZO})\,t_r^a 
          \right] 
        \Big\rangle_G
        \ ,
\label{Eq_Non-Abelian_Stokes_<W[C]>}
\ee
where $\Ps$ indicates surface ordering and $O$ is an arbitrary
reference point on the surface $S$. In
Eq.~(\ref{Eq_Non-Abelian_Stokes_<W[C]>}) the gluon field strength
tensor, $\G_{\mu\nu}(Z) = \G_{\mu\nu}^{a}(Z)\,t^{a}$, is parallel
transported to the reference point $O$ along the path $C_{ZO}$
\be
        \G_{\mu\nu}(O,Z;C_{ZO}) 
        = \Phi(O,Z;C_{ZO})^{-1} \G_{\mu\nu}(Z) \Phi(O,Z;C_{ZO})
\label{Eq_gluon_field_strength_tensor}
\ee
with the QCD Schwinger string
\be
        \Phi(O,Z;C_{ZO}) 
        = \Pc \exp 
        \left[-i\,g \int_{C_{ZO}} \!\!dZ_{\mu}\G^a_{\mu}(Z)\,t_r^a \right] 
        \ .
\label{Eq_parallel_transport}
\ee
The QCD vacuum expectation value $\langle \ldots \rangle_G$ represents
functional integrals in which the functional integration over the
fermion fields has already been carried out as indicated by the
subscript $G$~\cite{Nachtmann:ed.kt}.  The model we use for the QCD
vacuum works in the {\em quenched approximation} that does not allow
string breaking through dynamical quark-antiquark production.

Due to the linearity of the functional
integral, $\langle \nrTr \ldots \rangle = \nrTr \langle \ldots
\rangle$, we can write
\be
        \Big\langle W_r[C] \Big\rangle_G
        = \nrTr 
        \Big\langle
        \Ps \exp \left[-i\,\frac{g}{2} 
                \int_{S} \! d\sigma_{\mu\nu}(Z) 
                \G^a_{\mu\nu}(O,Z;C_{ZO})\,t_r^a 
          \right] 
        \Big\rangle_G
        \ .
\label{Eq_Tr_<W[C]>}
\ee
For the evaluation of~(\ref{Eq_Tr_<W[C]>}), a {\em matrix cumulant
  expansion} is used as explained in~\cite{Nachtmann:ed.kt}
(cf.~also~\cite{VAN_KAMPEN_1974_1976+X})
\bea
        && \Big\langle 
                \Ps \, \exp 
                \left[-i\,\frac{g}{2} 
                \int_{S} \! d\sigma(Z) \G(O,Z;C_{ZO})
                \right] 
           \Big\rangle_G \nonumber \\
        && 
        = \exp[\,\,\sum_{n=1}^{\infty}\frac{1}{n !}(-i\,\frac{g}{2})^n
        \int d\sigma(X_1)\cdots d\sigma(X_n)\,K_n(X_1,\cdots,X_n)]
        \ ,
\label{Eq_matrix_cumulant_expansion}
\eea
where space-time indices are suppressed to simplify notation. The
cumulants $K_n$ consist of expectation values of {\em ordered}
products of the non-commuting matrices $\G(O,Z;C_{ZO})$. The leading
matrix cumulants are
\bea
        K_1(X)       
        & = & \langle \G(O,X;C_X) \rangle_G, 
\label{Eq_K_1_matrix_cumulant}\\
        K_2(X_1,X_2) 
        & = & \langle \Ps
        [\G(O,X_1;C_{X_1})\G(O,X_2;C_{X_2})]\rangle_G 
        \nonumber\\
        &   & - \frac{1}{2}
        \left(\langle \G(O,X_1;C_{X_1})\rangle_G 
        \langle \G(O,X_2;C_{X_2})\rangle_G 
        + (1 \leftrightarrow 2)\,\right) \ .
\label{Eq_K_2_matrix_cumulant}
\eea
Since the vacuum does not prefer a specific color direction, $K_1$
vanishes and $K_2$ becomes
\be
        K_2(X_1,X_2) 
        = \langle\Ps [\G(O,X_1;C_{X_1})\G(O,X_2;C_{X_2})]\rangle_G
        \ .
\label{Eq_K_2_matrix_cumulant<-no_color_direction_preferred}
\ee
Now, we approximate the functional integral associated with the
expectation values $\langle \ldots \rangle_G$ as a {\em Gaussian
  integral} in the gluon field strength. Consequently, the cumulants
factorize into two-point field correlators such that all higher
cumulants $K_n$ with $n>2$ vanish\footnote{We are going to use the
  cumulant expansion in the Gaussian approximation also for
  perturbative gluon exchange.  Here certainly the higher cumulants
  are non-zero.} and $\langle W_r[C] \rangle_G$ can be expressed in
terms of $K_2$
\bea
&& \!\!\!\!\!\!\!\!\!\!\!
        \Big\langle W_r[C] \Big\rangle_G = 
        \nrTr 
        \exp\!\left[-\frac{g^2}{8} \!
          \int_{S} \! d\sigma_{\mu\nu}(X_1) \!
          \int_{S} \! d\sigma_{\rho\sigma}(X_2) 
        \right.
        \nonumber \\
&& \!\!\!\!\!\!\!\!\!\!\!
        \hphantom{\Big\langle W_r[C] \Big\rangle_G = \nrTr \exp}
        \left.
          \Big\langle \Ps 
          [\G^a_{\mu\nu}(O,X_1;C_{X_1 O})\,t_r^a\,\,
          \G^b_{\rho\sigma}(O,X_2;C_{X_2 O})\,t_r^b] 
          \Big\rangle_G 
        \right]
\label{Eq_matrix_cumulant_expansion_<W[C]>}
\eea
Due to the color neutrality of the vacuum, the gauge-invariant bilocal
gluon field strength correlator contains a $\delta$-function in
color space,
\be
        \Big\langle
        \frac{g^2}{4\pi^2}
        \left[\G^a_{\mu\nu}(O,X_1;C_{X_1 O})
        \G^b_{\rho\sigma}(O,X_2;C_{X_2 O})\right]
        \Big\rangle_G
        =: \inv{4}\delta^{ab} 
        F_{\mu\nu\rho\sigma}(X_1,X_2,O;C_{X_1 O},C_{X_2 O}) 
\label{Eq_Ansatz}
\ee
which makes the surface ordering $\Ps$
in~(\ref{Eq_matrix_cumulant_expansion_<W[C]>}) irrelevant. The tensor
$F_{\mu\nu\rho\sigma}$ will be specified in
Sec.~\ref{Sec_QCD_Components}.  With~(\ref{Eq_Ansatz}) and the
quadratic Casimir operator $C_2(r)$,
\be
        t_r^a\,t_r^a = t_r^2 = C_2(r)\,\Identity_r
        \ ,
\label{Eq_quadratic_Casimir_operator}
\ee
Eq.~(\ref{Eq_matrix_cumulant_expansion_<W[C]>}) reads
\be
        \Big\langle W_r[C] \Big\rangle_G
        = \nrTr 
        \exp\left[
        - \frac{C_2(r)}{2}\,\chi_{SS}\,\Identity_r
        \right] 
        = \exp \left[-\frac{C_2(r)}{2}\,\chi_{SS}\right] 
        \ ,
\label{Eq_final_result_<W[C]>}
\ee
where
\be
        \chi_{SS}
        :=  \frac{\pi^2}{4} 
        \int_{S} \! d\sigma_{\mu\nu}(X_1) 
        \int_{S} \! d\sigma_{\rho\sigma}(X_2)
        F_{\mu\nu\rho\sigma}(X_1,X_2,O;C_{X_1 O},C_{X_2 O}) 
        \ .
\label{Eq_chi_SS}        
\ee
In this rather general result~(\ref{Eq_final_result_<W[C]>}) obtained
directly from the color neutrality of the vacuum and the Gaussian
approximation in the gluon field strengths, the more detailed aspects
of the QCD vacuum and the geometry of the considered {\WW} loop are
encoded in the function $\chi_{SS}$ which is computed in
Appendix~\ref{Sec_Chi_Computation} for the rectangular loop shown in
Fig.~\ref{Fig_ONE_WWL}.

In explicit computations we use for $S$ the {\em minimal surface},
which is the planar surface spanned by the loop, $C = \partial S$,
that leads to Wilson's area law~\cite{Dosch:1987sk+X}. The minimal
surface is represented in the upcoming figures by the shaded areas
(cf.\ Figs.~\ref{Fig_ONE_WWL} and~\ref{Fig_OneLoop_MinimalSurface}).
Of course, the results should not dependent on the surface choice. In
our model this will be fulfilled for the perturbative and
non-perturbative non-confining component but not for the
non-perturbative confining component in $F_{\mu\nu\rho\sigma}$
(specified in Sec.~\ref{Sec_QCD_Components}) due to the Gaussian
approximation and the associated truncation of the cumulant expansion.
Nevertheless, since our results for the VEV of a rectangular {\WW}
loop lead to a static quark-antiquark potential that is in good
agreement with lattice data (see Sec.~\ref{Sec_Static_Potential}), we
are led to conclude that the choice of the minimal surface is required
by the Gaussian approximation in the gluon field strengths. The
minimal surface is also favored by other complementary approaches such
as the strong coupling expansion in lattice QCD, where plaquettes
cover the minimal surface, or large-$N_c$ investigations, where the
planar gluon diagrams dominate in the large-$N_c$ limit. Within
bosonic string theory, our minimal surface represents the world-sheet
of the {\em rigid} string: Our model does not describe fluctuations or
excitations of the string and thus cannot reproduce the L\"uscher term
which has recently been confirmed with unprecedented precision by
L\"uscher and Weisz~\cite{Luscher:2002qv}.

\section{The Loop-Loop Correlation Function}
\label{Sec_<W[C_1]W[C_2]>}

The computation of the {\em loop-loop correlation function} $\langle
W_{r_1}[C_{1}] W_{r_2}[C_{2}] \rangle_G$ starts again with the
application of the non-Abelian Stokes' theorem~\cite{Arefeva:dp+X}
that allows us to transform the line integrals over the loops
$C_{1,2}$ into integrals over surfaces $S_{1,2}$ with $\partial
S_{1,2} = C_{1,2}$
\bea
        &&
        \Big\langle W_{r_1}[C_{1}] W_{r_2}[C_{2}] \Big\rangle_G 
        = \Big\langle 
        \nroneTr\,\Ps
          \exp \left[-i\,\frac{g}{2} 
                \int_{S_1} \! d\sigma_{\mu\nu}(X_1) 
                \G^a_{\mu\nu}(O_1,X_1;C_{X_1 O_1})\,t_{r_1}^a 
          \right] 
        \nonumber \\
        &&
        \quad\quad\quad\quad\quad
        \times\,\nrtwoTr\,\Ps
          \exp \left[-i\,\frac{g}{2} 
                \int_{S_2} \! d\sigma_{\rho\sigma}(X_2) 
                \G^b_{\rho\sigma}(O_2,X_2;C_{X_2 O_2})\,t_{r_2}^b 
          \right] 
        \Big\rangle_G
\label{Eq_Non-Abelian_Stokes_<W[C1]W[C2]>}
\eea
where $O_{1}$ and $O_{2}$ are the reference points on the surfaces
$S_{1}$ and $S_{2}$, respectively, that enter through the non-Abelian
Stokes' theorem. In order to ensure gauge invariance in our model, the
gluon field strengths associated with the loops must be compared at
{\em one} reference point $O$. Due to this physical constraint, the
surfaces $S_{1}$ and $S_{2}$ are required to touch at a common
reference point $O_{1} = O_{2} = O$.

To treat the product of the two traces
in~(\ref{Eq_Non-Abelian_Stokes_<W[C1]W[C2]>}), we transfer the
approach of Berger and Nachtmann~\cite{Berger:1999gu}, cf.\ 
also~\cite{Shoshi:2002in}, to Euclidean space-time. Accordingly, the
product of the two traces respectively over $SU(N_c)$ matrices in the
$r_1$ and $r_2$ representation, $\nroneTr(...)\,\nrtwoTr(...)$,
is interpreted as one trace
$\nronexrtwoTr(...):=\ronexrtwoTr(...)/\ronexrtwoTr(\ronexrtwoIdentity)$
that acts in the tensor product space built from the $r_1$ and $r_2$
representations
\bea
        \Big\langle W_{\!r_1}[C_{1}] W_{\!r_2}[C_{2}] \Big\rangle_G 
        \!\!\!\! & \!\!\! = \!\!\! &\!
        \Big\langle 
        \nronexrtwoTr
          \!\left\{\!\!
            \Big[\Ps \exp\!\big[\!-\!i\frac{g}{2} \!
                \int_{S_{1}} \!\!\!\! d\sigma_{\mu\nu}(X_{1}) 
                \G^a_{\mu\nu}(O,X_{1};C_{X_{1} O})\,t_{r_1}^a \big] 
                \,\otimes\,\Identity_{r_2}\Big]
          \right.
        \nonumber \\
        &&\!\!\!\!\!\!\!\!\!\!\!\!\!
        \times
          \left. 
          \Big[\Identity_{r_1}\,\otimes\,
          \Ps \exp\! \big[\!-\!i\frac{g}{2} \!
                \int_{S_{2}} \!\!\!\! d\sigma_{\rho\sigma}(X_{2}) 
                \G^b_{\rho\sigma}(O,X_{2};C_{X_{2} O})\,t_{r_2}^b \big]
          \Big]
          \!\!\right\}\!
        \Big\rangle_G 
\label{Eq_trace_trick_<W[C1]W[C2]>}
\eea
With the identities
\bea
        \exp\left(\,t_{r_1}^a\,\right) \,\otimes\, \Identity_{r_2} 
        & = & \exp\left(\,t_{r_1}^a \,\otimes\, \Identity_{r_2}\,\right) \\
        \Identity_{r_1} \,\otimes\, \exp\left(\,t_{r_2}^a\,\right) 
        & = & \exp\left(\,\Identity_{r_1} \,\otimes\, t_{r_2}^a\,\right)
\label{Eq_exp(t^a)_times_1_identities}
\eea
the tensor products can be shifted into the exponents. Using the
matrix multiplication relations in the tensor product space
\bea
        \big( t_{r_1}^a \,\otimes\, \Identity_{r_2} \big)
        \big( t_{r_1}^b \,\otimes\, \Identity_{r_2} \big) 
        & = & t_{r_1}^a t_{r_1}^b \,\otimes\, \Identity_{r_2}
        \nonumber
        \\
        \big( t_{r_1}^a \,\otimes\, \Identity_{r_2} \big)
        \big( \Identity_{r_1} \,\otimes\, t_{r_2}^b \big) 
        & = & t_{r_1}^a \,\otimes\, t_{r_2}^b
\label{Eq_matrix_multiplication_in_tensor_product_space}
\eea
and the vanishing of the commutator
\be
        \left[t_{r_1}^a \otimes \Identity_{r_2}, 
        \Identity_{r_1} \otimes t_{r_2}^b\right] 
        = 0
        \ ,
\label{Eq_[t_x_1,1_x_t]}
\ee
the two exponentials in (\ref{Eq_trace_trick_<W[C1]W[C2]>}) commute
and can be written as one exponential
\be
        \Big\langle W[C_{1}] W[C_{2}] \Big\rangle_G =
        \Big\langle 
        \nronexrtwoTr\,\Ps \exp\!
            \left[-i\,\frac{g}{2} 
                \int_{S} \! d\sigma_{\mu\nu}(X) 
                \GG_{\mu\nu}(O,X;C_{XO}) 
            \right]
        \Big\rangle_G    
\label{Eq_<W[C1]W[C2]>_analogous_to_<W[C]>}
\ee
with the following gluon field strength tensor acting in the tensor product space
\be
        \GG_{\mu\nu}(O,X;C_{XO})
        := \left\{ \begin{array}{cc}
            \G_{\mu\nu}^a(O,X;C_{XO})
                \big( t_{r_1}^a \,\otimes\, \Identity_{r_2} \big)
            & \mbox{for $\,\,\, X\,\,\, \in \,\,\, S_1$} \\
            \G_{\mu\nu}^a(O,X;C_{XO})
                \big( \Identity_{r_1} \,\otimes\, t_{r_2}^a \big)
            & \mbox{for $\,\,\, X\,\,\, \in \,\,\, S_2$}
        \end{array}\right.
\label{Eq_GG} 
\ee
In Eq.~(\ref{Eq_<W[C1]W[C2]>_analogous_to_<W[C]>}) the surface
integrals over $S_1$ and $S_2$ are written as one integral over the
combined surface $S = S_1 + S_2$ so that the left-hand side (lhs)
of~(\ref{Eq_<W[C1]W[C2]>_analogous_to_<W[C]>}) becomes very similar to
the lhs of~(\ref{Eq_Non-Abelian_Stokes_<W[C]>}). This allows us to
proceed analogously to the computation of $\langle W_r[C] \rangle_G$
in the previous section. After exploiting the linearity of the
functional integral, the matrix cumulant expansion is applied, which
holds for $\GG_{\mu\nu}(O,X;C_{XO})$ as well. Then, with the
color neutrality of the vacuum and by imposing the Gaussian
approximation now in the color components of the gluon field strength
tensor, only the $n=2$ term of the matrix cumulant expansion survives,
which leads to
\bea
&& \!\!\!\!\!\!\!\!\!\!\!
        \Big\langle W_{r_1}[C_{1}] W_{r_2}[C_{2}] \Big\rangle_G 
\label{Eq_matrix_cumulant_expansion_<W[C1]W[C2]>}\\
&& \!\!\!\!\!\!\!\!\!\!\!
        = \nronexrtwoTr
        \exp\!\left[\!-\frac{g^2}{8} \!\!
          \int_{S} \!\! d\sigma_{\mu\nu}(X_1) \!
          \int_{S} \!\! d\sigma_{\rho\sigma}(X_2) 
          \Big\langle\!\Ps 
          [\GG_{\mu\nu}(O,X_1;C_{X_1 O}) 
          \GG_{\rho\sigma}(O,X_2;C_{X_2 O})] 
          \!\Big\rangle_{\!\!G} 
        \right]
\nonumber  
\eea

Note that the Gaussian approximation on the level of the color
components of the gluon field strength tensor (component
factorization) differs from the one on the level of the gluon field
strength tensor (matrix factorization) used to compute $\langle
W_{r}[C] \rangle$ in the original version of the
SVM~\cite{Dosch:1987sk+X}. Nevertheless, with the additional ordering
rule~\cite{Rueter:1994cn} explained in detail in Sec.~2.4
of~\cite{Dosch:2000va}, a modified component factorization is obtained
that leads to the same area law as the matrix factorization.

Using definition~(\ref{Eq_GG}) and
relations~(\ref{Eq_matrix_multiplication_in_tensor_product_space}), we
now redivide the exponent
in~(\ref{Eq_matrix_cumulant_expansion_<W[C1]W[C2]>}) into integrals
of the ordinary parallel transported gluon field strengths over the
separate surfaces $S_{1}$ and $S_{2}$
\bea
        && \Big\langle W_{r_1}[C_{1}] W_{r_2}[C_{2}] \Big\rangle_G = 
        \nronexrtwoTr
        \exp \Bigg[             
\label{Eq_exponent_decomposition_<W[C1]W[C2]>}
\\ &&\hspace{-0.8cm}
          -\frac{g^2}{8} \!\!
          \int_{S_1} \!\!\! d\sigma_{\mu\nu}(X_1) \!\! 
          \int_{S_2} \!\!\! d\sigma_{\rho\sigma}(X_2)\, 
          \Ps \!\left [ \Big\langle\! 
          \G^a_{\mu\nu}(O,X_1;C_{X_1 O}) 
          \G^b_{\rho\sigma}(O,X_2;C_{X_2 O})\!\Big\rangle_{\!\!G}
                \big(t_{r_1}^a  \,\otimes\, t_{r_2}^b\big) \right ] 
          \nonumber \\
        &&\hspace{-0.8cm}
          -\frac{g^2}{8} \!\!
          \int_{S_2} \!\!\! d\sigma_{\mu\nu}(X_1) \!\! 
          \int_{S_1} \!\!\! d\sigma_{\rho\sigma}(X_2)\, 
          \Ps \!\left [ \Big\langle\! 
            \G^a_{\mu\nu}(O,X_1;C_{X_1 O}) 
            \G^b_{\rho\sigma}(O,X_2;C_{X_2 O})\!\Big\rangle_{\!\!G}
                \big(t_{r_1}^a  \,\otimes\, t_{r_2}^b\big) \right ] 
            \nonumber \\
        &&\hspace{-0.8cm}
          -\frac{g^2}{8} \!\!
          \int_{S_1} \!\!\! d\sigma_{\mu\nu}(X_1) \!\! 
          \int_{S_1} \!\!\! d\sigma_{\rho\sigma}(X_2)\, 
          \Ps \!\left [ \Big\langle \!
            \G^a_{\mu\nu}(O,X_1;C_{X_1 O}) 
            \G^b_{\rho\sigma}(O,X_2;C_{X_2 O})\!\Big\rangle_{\!\!G}
                \big(t_{r_1}^a t_{r_1}^b \,\otimes\, \Identity_{r_2}\big)\right ] 
        \nonumber \\
        &&\hspace{-0.8cm}
        \left.
          -\frac{g^2}{8}\!\! 
          \int_{S_2} \!\!\! d\sigma_{\mu\nu}(X_1) \!\! 
          \int_{S_2} \!\!\! d\sigma_{\rho\sigma}(X_2)\, 
           \Ps \!\left [ \Big\langle\! 
             \G^a_{\mu\nu}(O,X_1;C_{X_1 O}) 
             \G^b_{\rho\sigma}(O,X_2;C_{X_2 O})\!\Big\rangle_{\!\!G}
                \big(\Identity_{r_1} \,\otimes\, t_{r_2}^a t_{r_2}^b\big) \right ]
              \!\right]
        \nonumber
\eea
Here the surface ordering $\Ps$ is again irrelevant due to the
color neutrality of the vacuum~(\ref{Eq_Ansatz}), and
(\ref{Eq_exponent_decomposition_<W[C1]W[C2]>}) becomes
\bea
        && \Big\langle W_{r_1}[C_{1}] W_{r_2}[C_{2}] \Big\rangle_G 
        = \nronexrtwoTr
        \exp\!\Bigg[
                - \frac{\chi_{S_1 S_2}+\chi_{S_2 S_1}}{2}\,
                \big(t_{r_1}^a \,\otimes\, t_{r_2}^a\big) 
        \nonumber \\
        && \quad\quad\quad\quad  
        - \,\frac{\chi_{S_1 S_1}}{2}
                \big(t_{r_1}^a t_{r_1}^a\,\otimes\,\rtwoIdentity\big) 
            - \frac{\chi_{S_2 S_2}}{2}
                \big(\roneIdentity\,\otimes\,t_{r_2}^a t_{r_2}^a\big) 
              \Bigg]
\label{Eq_eikonal_functions_<W[C1]W[C2]>}
\eea
with
\be
        \chi_{S_i S_j}
        := \frac{\pi^2}{4} 
        \int_{S_i} \! d\sigma_{\mu\nu}(X_1) 
        \int_{S_j} \! d\sigma_{\rho\sigma}(X_2)
        F_{\mu\nu\rho\sigma}(X_1,X_2,O;C_{X_1 O},C_{X_2 O}) 
        \ .
\label{Eq_chi_Si_Sj}        
\ee
The symmetries in the tensor structure of $F_{\mu\nu\rho\sigma}$ --
see (\ref{Eq_F_decomposition}), (\ref{Eq_PGE_Ansatz_F}), and
(\ref{Eq_MSV_Ansatz_F}) -- lead to $\chi_{S_1 S_2} = \chi_{S_2 S_1}$.
With the quadratic Casimir
operator~(\ref{Eq_quadratic_Casimir_operator}) our final Euclidean
result for general $SU(N_c)$ representations $r_1$ and $r_2$
becomes\footnote{Note that the Euclidean $\chi_{S_i S_i} \neq 0$ in
  contrast to $\chi_{S_i S_i} = 0$ for Minkowskian light-like loops
  $C_i$ considered in the original version of the Berger-Nachtmann
  approach~\cite{Berger:1999gu,Shoshi:2002in}.}
\bea
        &&\Big\langle W_{r_1}[C_{1}] W_{r_2}[C_{2}] \Big\rangle_{\!G} 
\label{Eq_final_general_Euclidean_result_<W[C1]W[C2]>}\\
        && = \nronexrtwoTr
        \exp\!\Bigg[
                - \chi_{S_1 S_2}\,
                \big(t_{r_1}^a \,\otimes\, t_{r_2}^a\big) 
                - \Big(\frac{C_2(r_1)}{2}\,\chi_{S_1 S_1} + \frac{C_2(r_2)}{2}\,\chi_{S_2 S_2}\Big)\,
                \ronexrtwoIdentity
        \Bigg]
        \nonumber
\eea
where $\ronexrtwoIdentity = \roneIdentity\otimes\rtwoIdentity$. After
specifying the representations $r_1$ and $r_2$, the tensor product
$\tensor:=t_{r_1}^a \,\otimes\, t_{r_2}^a$ can be expressed as a sum
of projection operators $\Projector_i$ with the property $\Projector_i
\,\tensor = \lambda_i \,\Projector_i$ (no sum over $i$),
\be
        \tensor = \sum_i \lambda_i\,\Projector_i
        \quad\quad \mbox{with} \quad\quad 
        \lambda_i = 
        \frac{\nronexrtwoTr\big(\Projector_i \,\tensor\big)}
        {\nronexrtwoTr \big(\Projector_i\big)}
        \ ,
\label{Eq_tensor_decomposition}
\ee
which corresponds to the decomposition of the tensor product space
into irreducible representations.

For two {\WW} loops in the {\em fundamental representation} of
$SU(N_c)$, $r_1 = r_2 = \Fundamental$, that could describe the
trajectories of two quark-antiquark pairs, the
decomposition~(\ref{Eq_tensor_decomposition}) is trivial
\be
        t_{\fundamental}^a \,\otimes\, t_{\fundamental}^a
        = \frac{N_c-1}{2N_c} \Projector_s - \frac{N_c+1}{2N_c} \Projector_a
        \ ,
\label{Eq_projector_tFa_x_tFa_relation}
\ee
with the projection operators
\bea
        &&
        (\Projector_s)_{(\alpha_1 \alpha_2)( \beta_1 \beta_2)} =
        \frac{1}{2}
        (\delta_{\alpha_1 \beta_1} \delta_{\alpha_2 \beta_2} 
        +\delta_{\alpha_1 \beta_2} \delta_{\alpha_2 \beta_1})
        \\
        &&
        (\Projector_a)_{(\alpha_1 \alpha_2)( \beta_1 \beta_2)} =
        \frac{1}{2}
        (\delta_{\alpha_1 \beta_1} \delta_{\alpha_2 \beta_2} 
        -\delta_{\alpha_1 \beta_2} \delta_{\alpha_2 \beta_1})
\label{Eq_projectors}
\eea
that decompose the direct product space of two fundamental $SU(N_c)$
representations into the irreducible representations
\be
        \mbox{N}_c \,\otimes\, \mbox{N}_c 
        = (\mbox{N}_c + 1)\mbox{N}_c/2 \,\oplus\, \overline{\mbox{N}_c(\mbox{N}_c - 1)/2}
        \ .
\label{Eq_tensor_product_fundamental_decomposition}
\ee
With
$\Tr_{\fundamental\otimes\fundamental}\,\Identity_{\fundamental\otimes\fundamental}
= N_c^2$ and the projector properties
\be
        \Projector^2_{s,a} = \Projector_{s,a}
        \ , \,\,\, 
        \Tr_{\fundamental\otimes\fundamental} \,\Projector_s = (N_c + 1)N_c/2
        \quad \mbox{and} \quad
        \Tr_{\fundamental\otimes\fundamental} \,\Projector_a = (N_c - 1)N_c/2
\label{Eq_projector_properties_fundamental}
\ee 
we find for the loop-loop correlation function with both loops in the
fundamental $SU(N_c)$ representation
\bea
        && \Big\langle W_{\fundamental}[C_{1}] W_{\fundamental}[C_{2}] \Big\rangle_G 
        = \exp\!\left[- \frac{C_2(\!\Fundamental\!)}{2}\Big(\chi_{S_1 S_1} + \chi_{S_2 S_2}\Big)\right]
\label{Eq_final_Euclidean_result_<W[C1]W[C2]>_fundamental}\\
        && \quad\quad\quad\quad
        \times
        \Bigg(\frac{N_c+1}{2N_c}\exp\!\left[-\frac{N_c-1}{2N_c}\chi_{S_1 S_2}\right]
        + \frac{N_c-1}{2N_c}\exp\!\left[ \frac{N_c+1}{2N_c}\chi_{S_1 S_2}\right]\Bigg)
        \nonumber
\eea
where
\be
        C_2(\!\Fundamental\!) = \frac{N_c^2-1}{2N_c}
        \ .
\label{Eq_Casimir_fundamental}
\ee

For one {\WW} loop in the {\em fundamental} and one in the {\em adjoint
  representation} of $SU(N_c)$, $r_1 = \Fundamental$ and $r_2 =
\Adjoint$, which is needed in Sec.~\ref{Sec_Flux_Tube} to investigate
the chromo-field distributions around color sources in the adjoint
representation, the decomposition~(\ref{Eq_tensor_decomposition})
reads
\be
        t_{\fundamental}^a \,\otimes\, t_{\adjoint}^a
        \,\,=\,\, 
        -\,\frac{N_c}{2}\,\Projector_1
        \,+\, \inv{2}\,\Projector_2 
        \,-\, \inv{2}\,\Projector_3 
\label{Eq_projector_tFa_x_tAa_relation}
\ee
with the projection operators\footnote{The explicit form of the
  projection operators $\Projector_1$, $\Projector_2$, and
  $\Projector_3$ can be found in~\cite{Cvitanovic_1984} but note that
  we use the Gell-Mann (conventional) normalization of the gluons. The
  eigenvalues, $\lambda_i$, of the projection operators
  in~(\ref{Eq_projector_tFa_x_tAa_relation}) can be evaluated
  conveniently with the computer program
  ``Colour''~\cite{Hakkinen:1996bb}.}  $\Projector_1$, $\Projector_2$,
and $\Projector_3$ that decompose the direct product space of one
fundamental and one adjoint representation of $SU(N_c)$ into the
irreducible representations
\be
        \Fundamental\,\otimes\,\Adjoint
        \,\,=\,\,
        \mbox{N}_c
        \,\,\oplus\,\,\inv{2}\,\mbox{N}_c(\mbox{N}_c-1)(\mbox{N}_c+2)
        \,\,\oplus\,\,\inv{2}\,\mbox{N}_c(\mbox{N}_c+1)(\mbox{N}_c-2)
\label{Eq_tensor_product_f_x_a_decomposition}
\ee
which reduces for $N_c = 3$ to the well-known $SU(3)$ decomposition
\be
        3\,\otimes\,8 
        = 3\,\oplus\,15\,\oplus\,6
        \ .
\label{Eq_tensor_product_f_x_a_SU(3)_decomposition}
\ee
With
$\Tr_{\fundamental\otimes\adjoint}\,\Identity_{\fundamental\otimes\adjoint}
= N_c(N_c^2 - 1)$ and projector properties analogous
to~(\ref{Eq_projector_properties_fundamental}), we obtain the
loop-loop correlation function for one loop in the fundamental and one
loop in the adjoint representation of $SU(N_c)$
\bea
        && \!\!\!\!\!\!\!\!
        \Big\langle W_{\fundamental}[C_{1}] W_{\adjoint}[C_{2}] \Big\rangle_G 
        = \exp\!\left[- \Big(\frac{C_2(\!\Fundamental\!)}{2}\,\chi_{S_1 S_1} 
        + \frac{C_2(\!\Adjoint\!)}{2}\,\chi_{S_2 S_2}\Big)\right]
\label{Eq_final_Euclidean_result_<Wf[C1]Wa[C2]>}\\
        && \!\!\!\!\!\!\!\!
        \times\,
        \Bigg(\!\inv{N_c^2\!-\!1}\exp\!\Big[\frac{N_c}{2}\chi_{S_1 S_2}\Big]
        +\frac{N_c\!+\!2}{2(N_c\!+\!1)}\exp\!\Big[\!-\inv{2}\chi_{S_1 S_2}\Big]
        +\frac{N_c\!-\!2}{2(N_c\!-\!1)}\exp\!\Big[\inv{2}\chi_{S_1 S_2}\Big]
        \!\Bigg)
\nonumber
\eea
where
\be
        C_2(\!\Adjoint\!) = N_c
        \ .
\label{Eq_Casimir_adjoint}
\ee

Note that our expressions for the loop-loop correlation
function~(\ref{Eq_final_general_Euclidean_result_<W[C1]W[C2]>}) and,
more specifically,
(\ref{Eq_final_Euclidean_result_<W[C1]W[C2]>_fundamental})
and~(\ref{Eq_final_Euclidean_result_<Wf[C1]Wa[C2]>}), are rather
general results -- as our result for the VEV of one \WW\ 
loop~(\ref{Eq_final_result_<W[C]>}) -- obtained directly from the
color neutrality of the vacuum and the Gaussian approximation in the
gluon field strengths. The loop geometries, which characterize the
problem under investigation, are again encoded in the functions
$\chi_{S_i S_j}$, where also more detailed aspects of the QCD vacuum
enter in terms of $F_{\mu\nu\rho\sigma}$, i.e.\ the gauge-invariant
bilocal gluon field strength correlator~(\ref{Eq_Ansatz}).

For the explicit computations of $\chi_{S_1 S_2}$ presented in
Appendix~\ref{Sec_Chi_Computation}, one has to specify surfaces
$S_{1,2}$ with the restriction $\partial S_{1,2} = C_{1,2}$ according
to the non-Abelian Stokes' theorem. As illustrated in
Figs.~\ref{Fig_PW_arrangement} and~\ref{Fig_tilted_loops}, we choose
for $S_{1,2}$ the {\em minimal surfaces} that are built from the
planar areas spanned by the corresponding loops $C_{1,2}$ and the
infinitesimally thin tube which connects the two surfaces $S_1$ and
$S_2$. This is in line with our surface choice in applications of the
LLCM to high-energy
reactions~\cite{Shoshi:2002in,Shoshi:2002ri,Shoshi:2002fq} illustrated
in Fig.~\ref{Fig_loop_loop_scattering_surfaces}. The thin tube allows
us to compare the gluon field strengths in surface $S_1$ with the
gluon field strengths in surface $S_2$.

Due to the Gaussian approximation and the associated truncation of the
cumulant expansion, the non-perturbative confining contribution (see
Sec.~\ref{Sec_QCD_Components}) to the loop-loop correlation function
depends on the surface choice.  Consequently, our results for the
chromo-field distributions of color dipoles obtained with the minimal
surfaces (see Sec.~\ref{Sec_Flux_Tube}) differ from the ones obtained
with the pyramid mantle choice for the surfaces~\cite{Rueter:1994cn}
even if the same parameters are used. With low-energy theorems we show
in Sec.~\ref{Sec_Low_Energy_Theorems} that the minimal surfaces are
actually required to ensure the consistency of our results for the VEV
of one loop, $\langle W_{r}[C] \rangle$, and the loop-loop correlation
function, $\langle W_{r_1}[C_1] W_{r_2}[C_2] \rangle$.

In applications of the model to high-energy
scattering~\cite{Shoshi:2002in,Shoshi:2002fq,Shoshi:2002ri,Shoshi:2002mt}
the surfaces are interpreted as the world-sheets of the confining
strings in line with the picture obtained for the static dipole
potential from the VEV of one loop. The minimal surfaces are the most
natural choice to examine the scattering of two rigid strings without
any fluctuations or excitations. Our model does unfortunately not
choose the surface dynamically and, thus, cannot describe string flips
between two non-perturbative color dipoles.  Recently, new
developments towards a dynamical surface choice and a theory for the
dynamics of the confining strings have been
reported~\cite{Shevchenko:2002xi}.

\section[Perturbative and Non-Perturbative QCD Components]{\!\!Perturbative and Non-Perturbative QCD Components}
\label{Sec_QCD_Components}

We decompose the gauge-invariant bilocal gluon field strength
correlator~(\ref{Eq_Ansatz}) into a perturbative ($\pert$) and
non-perturbative ($\nprt$) component
\be
        F_{\mu\nu\rho\sigma} 
        = F_{\mu\nu\rho\sigma}^{\pert} + F_{\mu\nu\rho\sigma}^{\nprt} 
        \ ,
\label{Eq_F_decomposition}
\ee
where $F_{\mu\nu\rho\sigma}^{\nprt}$ gives the low-frequency
background field contribution modeled by the non-perturbative {\em
  stochastic vacuum model} (SVM)~\cite{Dosch:1987sk+X} and
$F_{\mu\nu\rho\sigma}^{\pert}$ the additional high-frequency
contribution described by perturbative gluon exchange. This
combination allows us to describe long and short distance correlations
in agreement with lattice calculations of the gluon field strength
correlator~\cite{DiGiacomo:1992df+X,Meggiolaro:1999yn}. Moreover, this
two component ansatz leads to the static quark-antiquark potential
with color Coulomb behavior for small and confining linear rise for
large source separations in good agreement with lattice data as shown
in Sec.~\ref{Sec_Static_Potential}. Note that besides our two
component ansatz an ongoing effort to reconcile the non-perturbative
SVM with perturbative gluon exchange has led to complementary
methods~\cite{Simonov:kt,Shevchenko:1998ej,Shevchenko:2002xi}.

We compute the perturbative correlator $\!F_{\!\!\mu\nu\rho\sigma}^{\pert}\!$
from the gluon propagator in Feynman-'t~Hooft gauge
\be
        \Big\langle  \G^a_{\mu}(X_1)\G^b_{\nu}(X_2) \Big\rangle
        = \int
        \frac{d^4K}{(2\pi)^4}
        \,\frac{-i\delta^{ab}\delta_{\mu\nu}}{K^2-m_G^2}
        \, e^{-iK(X_1-X_2)}
        \ ,
\label{Eq_massive_gluon_propagator}
\ee
where we introduce an {\em effective gluon mass} of $m_G = m_{\rho} =
0.77\,\GeV$ to limit the range of the perturbative interaction in the
infrared (IR) region. This value is, of course, important for the
interplay between the perturbative and non-perturbative component
which comes out reasonable as illustrated in
Sec.~\ref{Sec_Static_Potential} for the static quark-antiquark
potential. Moreover, it gives the ``perturbative glueball'' ($GB$)
generated by our perturbative component a reasonable finite mass of
$M_{\glueball}^{\pert} = 2m_G = 1.54\,\GeV$.

In leading order in the strong coupling $g$, the resulting
bilocal gluon field strength correlator is gauge-invariant already
without the parallel transport to a common reference point so that
$F_{\mu\nu\rho\sigma}^{\pert}$ depends only on the difference $Z= X_1
- X_2$
\bea
        F_{\mu\nu\rho\sigma}^{\pert}(Z)
        \!\!&=&\!\!\frac{g^2}{\pi^2}\, \inv{2}\Bigl[
                       \frac{\partial}{\partial Z_\nu}
                         \left(Z_\sigma \delta_{\mu\rho}
                         -Z_\rho \delta_{\mu\sigma}\right)
                       +\frac{\partial}{\partial Z_\mu}
                         \left(Z_\rho \delta_{\nu\sigma}
                         -Z_\sigma \delta_{\nu\rho}\right)\Bigr]\,
              D_{\pert}(Z^2)
\label{Eq_PGE_Ansatz_F}\\ 
        \!\!& &\!\! 
        \hspace{-2cm}
        = \,\,
        -\,\frac{g^2}{\pi^2}\!
                \int \!\!\frac{d^4K}{(2\pi)^4} \,e^{-iKZ}\,\Bigl[
                K_\nu K_\sigma \delta_{\mu\rho}  - K_\nu K_\rho   \delta_{\mu\sigma}
              + K_\mu K_\rho  \delta_{\nu\sigma} - K_\mu K_\sigma \delta_{\nu\rho} \Bigr]\,
           \tilde{D}_{\pert}^{\prime}(K^2)
        \nonumber
\eea
with the perturbative correlation function
\bea
        D_{\pert}(Z^2)
        & = & \frac{m_G^2}{2\,\pi^2 Z^2}\,K_2(m_G\,|Z|)
\label{Eq_Dp(z,mg)}\\
        \tilde{D}_{\pert}^{\prime}(K^2)
        &:= & 
        \frac{d}{dK^2}\int d^4Z\,e^{iKZ}\,D_{\pert}(Z^2)
        \,\, =\,\, 
        -\,\inv{K^2 + m_G^2}
         \ .
\label{Eq_D'p(K,mg)}
\eea

The perturbative gluon field strength correlator has also been
considered at next-to-leading order, where the dependence of the
correlator on both the renormalization scale and the renormalization
scheme becomes explicit and an additional tensor structure arises
together with a path dependence of the
correlator~\cite{Eidemuller:1997bb}. However, cancellations of
contributions from this additional tensor structure have been
shown~\cite{Shevchenko:1998ej}. We refer to Sec.~3.3
of~\cite{Dosch:2000va} for a more detailed discussion of this issue.

We describe the perturbative correlations in our phenomenological
applications only with the leading tensor
structure~(\ref{Eq_PGE_Ansatz_F}) and take into account radiative
corrections by replacing the constant coupling $g^2$ with the running
coupling
\be
        g^2(Z^2)
        = 4 \pi \alphaS(Z^2)
        = \frac{48 \pi^2}
        {(33-2 N_f) 
        \ln\left[
                (Z^{-2} + M^2)/\Lambda_{QCD}^2
        \right]}
\label{Eq_g2(z_perp)}
\ee
in the final step of the computation of the $\chi$-function, where the
Euclidean distance $|Z|$ over which the correlation occurs provides
the renormalization scale. In Eq.~(\ref{Eq_g2(z_perp)}) $N_f$ denotes
the number of dynamical quark flavors, which is set to $N_f = 0$ in
agreement with the quenched approximation, $\Lambda_{QCD} =
0.25\;\GeV$, and $M$ allows us to freeze $g^2$ for
$|Z|\rightarrow\infty$. Relying on low-energy theorems, we freeze the
running coupling at the value $g^2 = 10.2$ ($\equiv \alphaS = 0.81$),
i.e.\ $M = 0.488\,\GeV$, at which our non-perturbative results for the
confining potential and the total flux tube energy of a static
quark-antiquark pair coincide (see
Sec.~\ref{Sec_Low_Energy_Theorems}).

The tensor structure~(\ref{Eq_PGE_Ansatz_F}) together with the
perturbative correlation function~(\ref{Eq_Dp(z,mg)})
or~(\ref{Eq_D'p(K,mg)}) leads to the color Yukawa potential (which
reduces for $m_G = 0$ to the color Coulomb potential) as shown in
Sec.~\ref{Sec_Static_Potential}. The perturbative contribution
thus dominates the full potential at small quark-antiquark
separations.

If the path connecting the points $X_1$ and $X_2$ is a straight line,
the non-perturbative correlator $F_{\mu\nu\rho\sigma}^{\nprt}$ depends
also only on the difference $Z=X_1-X_2$. Then, the most general form
of the correlator that respects translational, $O(4)$, and parity
invariance reads~\cite{Dosch:1987sk+X}
\bea
        F_{\mu\nu\rho\sigma}^{\nprt}(Z) 
        & = & F_{\mu\nu\rho\sigma}^{\nprt\,c}(Z) +  
               F_{\mu\nu\rho\sigma}^{\nprt\,nc}(Z)
\label{Eq_MSV_Ansatz_F}\\
        &  = & \inv{3(N_c^2-1)}\,G_2\, \Bigl\{
        \kappa\, \left(\delta_{\mu\rho}\delta_{\nu\sigma}
          -\delta_{\mu\sigma}\delta_{\nu\rho}\right) \,
        D(Z^2)                                
        \nonumber\\
        &   &  
        +\,(1-\kappa)\,\inv{2}\Bigl[
        \frac{\partial}{\partial Z_\nu}
        \left(Z_\sigma \delta_{\mu\rho}
          -Z_\rho \delta_{\mu\sigma}\right)
        +\frac{\partial}{\partial Z_\mu}
        \left(Z_\rho \delta_{\nu\sigma}
          -Z_\sigma \delta_{\nu\rho}\right)\Bigr]\,
        D_1(Z^2) \Bigr\}
        \nonumber\\
        &  = & \inv{3(N_c^2-1)}\,G_2 
        \int \frac{d^4K}{(2\pi)^4} \,e^{-iKZ}\,\Bigl\{
        \kappa \,\left(\delta_{\mu\rho}\delta_{\nu\sigma}
          -\delta_{\mu\sigma}\delta_{\nu\rho}\right)\, 
        \tilde{D}(K^2)                                
        \nonumber\\
        &   &  -\,(1-\kappa)\,\Bigl[
        K_\nu K_\sigma \delta_{\mu\rho}   - K_\nu K_\rho   \delta_{\mu\sigma}
        + K_\mu K_\rho  \delta_{\nu\sigma} - K_\mu K_\sigma \delta_{\nu\rho} \Bigr]\,
               \tilde{D}_{1}^{\prime}(K^2) \Bigr\} 
        \nonumber \ ,
\eea
where
\be
        \tilde{D}_{1}^{\prime}(K^2) 
        := \frac{d}{dK^2} \int d^4Z\, D_{1}(Z^2)\, e^{iKZ} 
        \ .
\label{Eq_D1_prime}
\ee
In all previous applications of the SVM, this form depending only on
$Z=X_1-X_2$ has been used. New lattice results on the path dependence
of the correlator show a dominance of the shortest
path~\cite{DiGiacomo:2002mq}. This result is effectively incorporated
in the model since the straight paths dominate in the average over all
paths.

The non-perturbative correlator~(\ref{Eq_MSV_Ansatz_F}) involves the
gluon condensate \cite{Shifman:1979bx+X} $G_2 := \langle
\frac{g^2}{4\pi^2} \G^a_{\mu\nu}(0) \G^a_{\mu\nu}(0) \rangle$, the
parameter $\kappa$ that determines the non-Abelian character of the
correlator, and the correlation length $a$ that enters through the
non-perturbative correlation functions $D$ and $D_1$.

We adopt for our calculations a simple {\em exponential correlation
  function}
\be
        D(Z^2) = D_1(Z^2) = \exp(-|Z|/a)
        \ ,
\label{Eq_SVM_correlation_functions}
\ee
which is motivated by lattice QCD measurements of the gluon field
strength correlator~\cite{DiGiacomo:1992df+X,Meggiolaro:1999yn}. This
correlation function stays positive for all Euclidean distances $|Z|$
and thus is compatible with a spectral representation of the
correlation function~\cite{Dosch:1998th}. This means a conceptual
improvement since the correlation function that has been used in
several earlier applications of the
SVM~\cite{Dosch:1994ym,Rueter:1994cn,Dosch:1995fz,Rueter:1996yb,Dosch:1997ss,Dosch:1998nw,Rueter:1998qy,Kulzinger:1999hw,Rueter:1998up,D'Alesio:1999sf,Berger:1999gu,Dosch:2001jg,Kulzinger:2002iu}
becomes negative at large distances.

With the exponential correlation
function~(\ref{Eq_SVM_correlation_functions}) the lattice data of the
gluon field strength correlator down to distances of $0.4\,\fm$ give
the following values for the parameters of the non-perturbative
correlator~\cite{Meggiolaro:1999yn}: $G_2 = 0.173\,\GeV^4$, $\kappa =
0.746$, and $a = 0.219\,\fm$. We have optimized these parameters in
our fit to high-energy scattering data~\cite{Shoshi:2002in} presented
in Chap.~\ref{Sec_Comparison_Data} (see also
Sec.~\ref{Sec_Model_Parameters}):
\be
        a =  0.302\,\fm, \quad 
        \kappa = 0.74, \quad 
        G_2 = 0.074\,\GeV^4
        \ .
\label{Eq_MSV_scattering_fit_parameter_results}
\ee
We use these optimized
parameters~(\ref{Eq_MSV_scattering_fit_parameter_results}) throughout
this work. They lead to a static quark-antiquark potential that is in
good agreement with lattice data and, in particular, give a QCD string
tension of $\sigma_3 = 0.22\,\GeV^2 \equiv 1.12 \,\GeV/\fm$ as shown
in Sec.~\ref{Sec_Static_Potential}. This value is consistent with
hadron spectroscopy~\cite{Kwong:1987mj}, Regge
theory~\cite{Goddard:1973qh+X}, and lattice QCD
investigations~\cite{Bali:2001gf}. Moreover, the non-perturbative
component with $a=0.302\,\fm$ generates a ``non-perturbative
glueball'' with a mass of $M_{\glueball}^{\nprt} = 2/a = 1.31\,\GeV$
which is smaller than $M_{\glueball}^{\pert}=1.54\,\GeV$ and thus
governs the long-range correlations as expected. We thus have one
model that describes both static hadronic properties and high-energy
reactions of hadrons and photons in good agreement with experimental
and lattice QCD data.

Finally, let us emphasize that the non-perturbative
correlator~(\ref{Eq_MSV_Ansatz_F}) is a sum of the two different
tensor structures, $F_{\mu\nu\rho\sigma}^{\nprt\,nc}$ and
$F_{\mu\nu\rho\sigma}^{\nprt\,c}$, with characteristic behavior: The
tensor structure $F_{\mu\nu\rho\sigma}^{\nprt\,nc}$ is characteristic
for Abelian gauge theories, exhibits the same tensor structure as the
perturbative correlator~(\ref{Eq_PGE_Ansatz_F}) and does not lead to
confinement~\cite{Dosch:1987sk+X}, i.e.\ it gives an exponentially
vanishing static color dipole potential at large dipole sizes as shown
explictly in Sec.~\ref{Sec_Static_Potential}. In contrast, the tensor
structure $F_{\mu\nu\rho\sigma}^{\nprt\,c}$ can only occur in
non-Abelian gauge theories and Abelian gauge-theories with monopoles.
It leads in the case of $\kappa \neq 0$ to
confinement~\cite{Dosch:1987sk+X}, i.e.\ to the confining linear
increase of the static potential at large dipole sizes as demonstrated
in Sec.~\ref{Sec_Static_Potential}. Therefore, we call the tensor
structure multiplied by $(1-\kappa)$ non-confining ($nc$) and the one
multiplied by $\kappa$ confining ($c$).


%
\cleardoublepage
%
\chapter{Static Color Dipoles and Confining QCD Strings}
\label{Sec_Static_Sources}

In this chapter we apply the loop-loop correlation model to compute
the QCD potential and the chromo-field distributions of static color
dipoles in the fundamental and adjoint representation of $SU(N_c)$.
Special emphasis is on Casimir scaling behavior and the interplay
between perturbative Coulomb behavior and non-perturbative formation
of the confining QCD string. Moreover, low-energy theorems are
discussed that relate the energy and action stored in the
chromo-fields to the static quark-antiquark potential.  These energy
and action sum rules allow us to show consistency of the model results
and to determine the values of $\beta$ and $\alphaS$ at the
renormalization scale at which the non-perturbative SVM component is
working.

\section{The Static Color Dipole Potential}
\label{Sec_Static_Potential}

\begin{figure}[b!]
  \centerline{\epsfig{figure=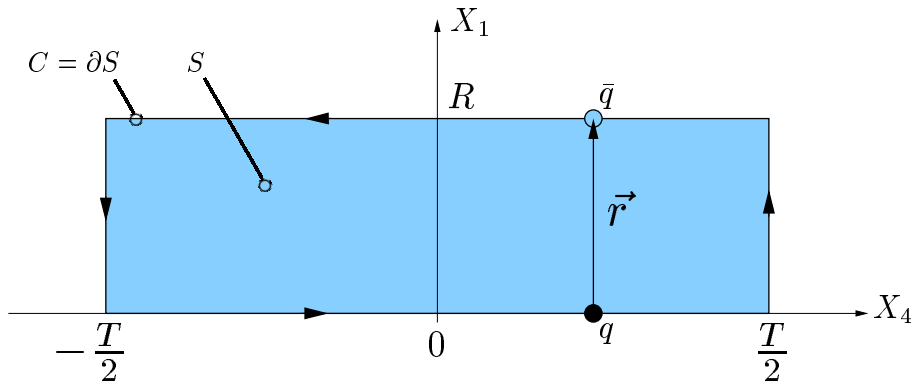,width=12.cm}}
\caption{\small 
  A static color dipole of size $R$ in the fundamental representation.
  The rectangular path $C$ of spatial extension $R$ and temporal
  extension $T$ indicates the world-line of the dipole described the
  {\WW} loop $W_{\fundamental}[C]$. The shaded area bounded by the
  loop $C=\partial S$ represents the minimal surface $S$ used to
  compute the static dipole potential.}
\label{Fig_ONE_WWL}
\end{figure}

The static color dipole -- two static color sources separated by a
distance $R$ in a net color singlet state -- is described by a {\WW}
loop $W_r[C]$ with a rectangular path $C$ of spatial extension $R$ and
temporal extension $T\to\infty$ where $r$ indicates the $SU(N_c)$
representation of the considered sources.  Figure~\ref{Fig_ONE_WWL}
illustrates a static color dipole in the fundamental representation
$r=\Fundamental$.  The potential of the static color dipole is
obtained from the VEV of the corresponding Wegner-Wilson
loop~\cite{Wilson:1974sk,Brown:1979ya}
\be
        V_r(R) 
        = - \lim_{T \to \infty} \inv{T} 
        \ln \langle W_r[C] \rangle_{\pot}
        \ ,
\label{Eq_static_potential}
\ee
where ``pot'' indicates the subtraction of the self-energy of the
color sources. The static quark-antiquark potential $V_{\fundamental}$
is obtained from a loop in the fundamental rep\-re\-sen\-ta\-tion
($r\!=\!\Fundamental$) and the potential of a static gluino pair
$V_{\adjoint}$ from a loop in the adjoint representation
($r\!=\!\Adjoint$).

With our result for $\langle W_r[C] \rangle$,
(\ref{Eq_final_result_<W[C]>}), obtained with the Gaussian
approximation in the gluon field strength, the static potential reads
\be
        V_r(R) = \frac{C_2(r)}{2}\,\lim_{T \to \infty} \inv{T}\,\chi_{SS\,\pot}
        \ , 
\label{Eq_Vr(R)_Gaussian_approximation}
\ee
with the self-energy subtracted, i.e.\ $\chi_{SS\,\pot} := \chi_{SS} -
\chi_{SS\,\self}$ (see Appendix~\ref{Sec_Chi_Computation}). According
to the structure of the gluon field strength correlator,
(\ref{Eq_Ansatz}) and~(\ref{Eq_F_decomposition}), there are
perturbative ($\pert$) and non-perturbative ($\nprt$) contributions to
the static potential
\be
        V_r(R) = \frac{C_2(r)}{2}\,\lim_{T \to \infty} \inv{T}\,
                 \left\{\chi_{SS\,\pot}^{\pert}
                   +\left(\chi_{SS\,\pot}^{\nprt\,\,nc} +
                     \chi_{SS\,\pot}^{\nprt\,\,c} \right) \right\}
        \ ,
\label{Eq_Vr(R)_P+NP}
\ee
where the explicit form of the $\chi$\,-\,functions is given
in~(\ref{Eq_chi_SS_NP_c_T->infty_V_E}),
(\ref{Eq_chi_SS_NP_nc_T->infty_pot_E}),
and~(\ref{Eq_chi_SS_P_T->infty_pot_E}).

The perturbative contribution to the static potential describes the
{\em color Yukawa potential} (which reduces to the {\em color Coulomb
  potential}~\cite{Kogut:1979wt} for $m_G=0$)
\be
        V_r^{\pert}(R) 
        = - C_2(r)\,\frac{g^2(R)}{4 \pi R} \exp[-m_G R] 
        \ .
\label{Eq_Vr(R)_color-Yukawa}
\ee
Here we have used the result for $\chi_{SS\,\pot}^{\pert}$ given
in~(\ref{Eq_chi_SS_P_T->infty_pot_E}) and the perturbative correlation
function
\be
        D^{\prime\,(3)}_{\pert}(\vec{Z}^2)
        := \int \frac{d^4K}{(2\pi)^3}\,e^{iKZ}\,
        \tilde{D}^{\prime\,(3)}_{\pert}(K^2)\,\delta(K_4)
         = -\,\frac{\exp[-\,m_G\,|\vec{Z}|]}{4\pi|\vec{Z}|}
\label{Eq_D'(3)p(z,mg)}
\ee
which is obtained from the massive gluon
propagator~(\ref{Eq_massive_gluon_propagator}). As shown below, the
perturbative contribution dominates the static potential for
small dipoles sizes $R$. 

The non-perturbative contributions to the static potential, the {\em
  non-confining} component ($nc$) and the {\em confining} component
($c$), read
\bea
        V_r^{\nprt\,\,nc}(R)
        & = & 
        C_2(r)\,\,
        \frac{\pi^2 G_2 (1-\kappa)}{3(N_c^2-1)}\,\,
        D_1^{\prime\,(3)}(R^2)
\label{Eq_Vr(R)_NP_nc}\\
        V_r^{\nprt\,\,c}(R) 
        & = & 
        C_2(r)\,\,
        \frac{\pi^2 G_2 \kappa}{3(N_c^2-1)}\,\,
        \int_0^R \!\! d\rho\,
        (R-\rho)\,
        D^{(3)}(\rho^2)
        \ ,
\label{Eq_Vr(R)_NP_c}
\eea
where we have used the results for $\chi_{SS\,\pot}^{\nprt\,\,nc}$ and
$\chi_{SS\,\pot}^{\nprt\,\,c}=\chi_{SS}^{\nprt\,\,c}$ given
respectively in~(\ref{Eq_chi_SS_NP_nc_T->infty_pot_E})
and~(\ref{Eq_chi_SS_NP_c_T->infty_V_E}) obtained with the minimal
surface, i.e.\ the planar surface bounded by the loop as indicated by
the shaded area in Fig.~\ref{Fig_ONE_WWL}. With the exponential
correlation function~(\ref{Eq_SVM_correlation_functions}), the
correlation functions in~(\ref{Eq_Vr(R)_NP_nc})
and~(\ref{Eq_Vr(R)_NP_c}) read
\bea
        D^{\prime\,(3)}_1(\vec{Z}^2)
        &\!\!:=\!\!&
        \int \frac{d^4K}{(2\pi)^3}\,e^{iKZ}\,
        \tilde{D}^{\prime\,(3)}_1(K^2)\,\delta(K_4)
        \,\,=\,\, -\,a\,|\vec{Z}|^2\,K_2[|\vec{Z}|/a]
        \ ,
\label{Eq_D'(3)np_nc(z,a)}\\
        D^{(3)}(\vec{Z}^2)      
        &\!\!:=\!\! & \int \frac{d^4K}{(2\pi)^3}\,e^{iKZ}\,\tilde{D}(K^2)\,\delta(K_4)
        \,\,=\,\, 2\,|\vec{Z}|\,K_1[|\vec{Z}|/a]
        \ .
\label{Eq_D(3)np_c(z,a)}
\eea
For large dipole sizes, $R \gtsim 0.5\ \fm$, the non-confining
contribution~(\ref{Eq_Vr(R)_NP_nc}) vanishes exponentially while the
confining contribution~(\ref{Eq_Vr(R)_NP_c}) -- as anticipated --
leads to {\em confinement}~\cite{Dosch:1987sk+X}, i.e.\ the confining
linear increase,
\be
        V_r^{\nprt\,\,c}(R)\Big|_{R\,\gtsim\,0.5\,\mbox{\scriptsize fm}} 
       = \sigma_r R  + \mbox{const.} \ .
\label{Eq_Vr(R)_NP_c_linear}
\ee
Thus, the QCD {\em string tension} is given by the confining SVM
component~\cite{Dosch:1987sk+X}: For a color dipole in the $SU(N_c)$
representation $r$, it reads
\be
        \sigma_r 
        = C_2(r)\,\,\frac{\pi^3 G_2 \kappa}{48} 
          \int_0^\infty dZ^2 D(Z^2) 
        = C_2(r)\,\,\frac{\pi^3 \kappa G_2 a^2}{24}
        \ ,
\label{Eq_string_tension}
\ee
where the exponential correlation
function~(\ref{Eq_SVM_correlation_functions}) is used in the final
step.  Since the string tension can be computed from first principles
within lattice QCD~\cite{Bali:2001gf},
relation~(\ref{Eq_string_tension}) puts an important constraint on the
three parameters of the non-perturbative QCD vacuum $a$, $G_2$, and
$\kappa$. With the values for $a$, $G_2$, and $\kappa$ given
in~(\ref{Eq_MSV_scattering_fit_parameter_results}), that are used
throughout this work, one obtains for the string tension of the
$SU(3)$ quark-antiquark potential ($r=3$) a reasonable value of
\be
        \sigma_3 
        = 0.22\,\GeV^2 \equiv 1.12 \,\GeV/\fm
        \ .
\label{Eq_sting_tension_from_exp_correlation}
\ee 

\begin{figure}[t!]
  \centerline{\epsfig{figure=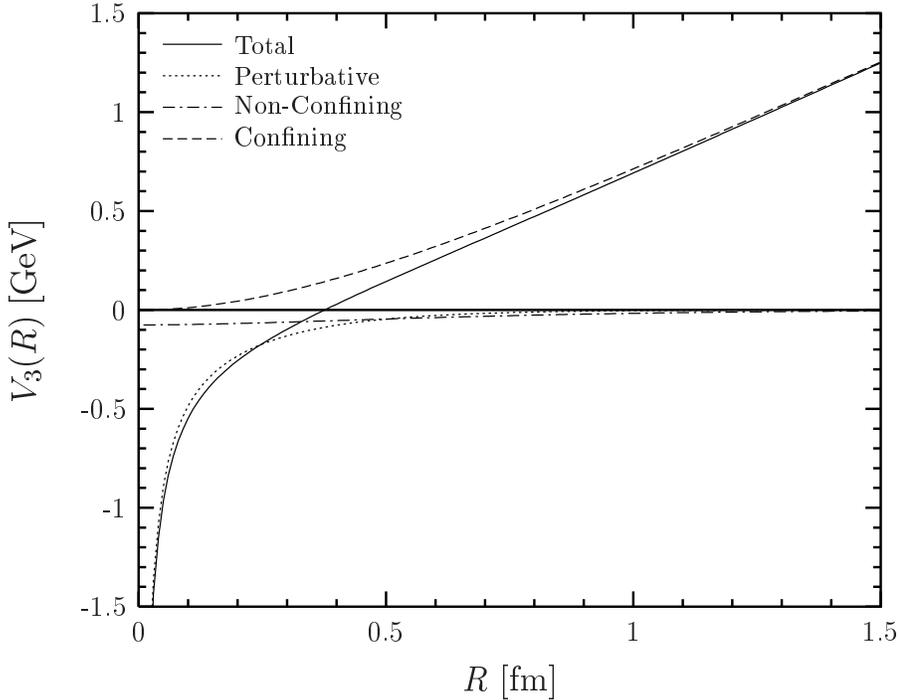,width=12.cm}}
\caption{\small 
  The static $SU(N_c = 3)$ quark-antiquark potential
  $V_{\fundamental}(R) = V_3(R)$ as a function of the quark-antiquark
  separation $R$. The solid, dotted, and dashed lines indicate the
  full static potential and its perturbative and non-perturbative
  contributions, respectively. For small quark-antiquark separations,
  $R \ltsim 0.5\,\fm$, the perturbative contribution dominates and
  gives rise to the well-known color Coulomb behavior at small
  distances. For medium and large quark-antiquark separations, $R
  \gtsim 0.5\,\fm$, the non-perturbative contribution dominates and
  leads to the confining linear rise of the static potential. As our
  model is working in the quenched approximation, string breaking
  cannot be described, which is expected to stop the linear increase
  for $R\,\gtsim\,1\,\fm$~\cite{Laermann:1998gm,Bali:2001gf}.}
\label{Fig_Static_Quark-Antiquark_Potential_Components}
\end{figure}

The static $SU(N_c = 3)$ quark-antiquark potential
$V_{\fundamental}(R) = V_3(R)$ is shown as a function of the
quark-antiquark separation $R$ in
Fig.~\ref{Fig_Static_Quark-Antiquark_Potential_Components},
where the solid, dotted, and dashed lines indicate the full static
potential and its perturbative and non-perturbative contributions,
respectively.  For small quark-antiquark separations $R \ltsim
0.5\,\fm$, the perturbative contribution dominates giving rise to the
well-known color Coulomb behavior. For medium and large
quark-antiquark separations $R \gtsim 0.5\,\fm$, the non-perturbative
contribution dominates and leads to the confining linear rise of the
static potential. The transition from perturbative to string behavior
takes place at source separations of about $0.5\,\fm$ in agreement
with the recent results of L\"uscher and Weisz~\cite{Luscher:2002qv}.
This supports our value for the gluon mass $m_G=m_{\rho}=0.77\,\GeV$
which is only important around $R\approx 0.4\,\fm$, i.e.\ for the
interplay between perturbative and non-perturbative physics. For
$R\ltsim 0.3\,\fm$ and $R\gtsim 0.5\,\fm$, the effect of the gluon
mass, introduced as an IR regulator in our perturbative component, is
negligible. String breaking is expected to stop the linear increase
for $R\,\gtsim\,1\,\fm$ where lattice investigations show deviations
from the linear rise in full QCD~\cite{Laermann:1998gm,Bali:2001gf}.
As our model is working in the quenched approximation, string breaking
through dynamical quark-antiquark production is excluded.

\begin{figure}[t!]
\centerline{\epsfig{figure=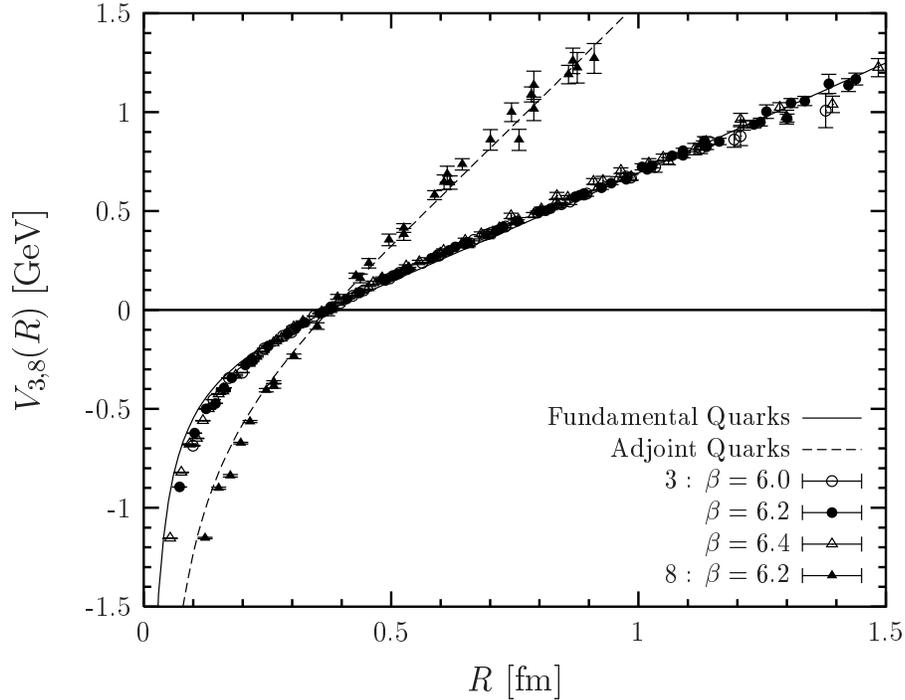,width=12.cm}}
\caption{\small 
  The static $SU(N_c = 3)$ potential of color dipoles in the
  fundamental representation $V_3(R)$ (solid line) and adjoint
  representation $V_8(R)$ (dashed line) as a function of the dipole
  size $R$ in comparison to $SU(3)$ lattice data for $\beta = 6.0$,
  6.2, and 6.4~\cite{Bali:2000un,Bali:2001gf}. The model results are
  in good agreement with the lattice data. This particularly holds for
  the obtained Casimir scaling behavior.}
\label{Fig_Static_Quark-Antiquark_Potential_F_vs_A}
\end{figure}

As can be seen from~(\ref{Eq_Vr(R)_Gaussian_approximation}), the
static potential shows {\em Casimir scaling} which emerges in our
approach as a trivial consequence of the Gaussian approximation used
to truncate the cumulant
expansion~(\ref{Eq_matrix_cumulant_expansion}). Indeed, the Casimir
scaling hypothesis~\cite{Ambjorn:1984dp} has been verified to high
accuracy for $SU(3)$ on the lattice~\cite{Deldar:1999vi,Bali:2000un}
(see also Fig.~\ref{Fig_Static_Quark-Antiquark_Potential_F_vs_A}).
These lattice results have been interpreted as a strong hint towards
Gaussian dominance in the QCD vacuum and thus as evidence for a strong
suppression of higher cumulant
contributions~\cite{Shevchenko:2000du,Shevchenko:2001ij}. In contrast
to our model, the instanton model can neither describe Casimir
scaling~\cite{Shevchenko:2001ij} nor the linear rise of the confining
potential~\cite{Chen:1999ct}.

Figure~\ref{Fig_Static_Quark-Antiquark_Potential_F_vs_A}
shows the static $SU(N_c = 3)$ potential for fundamental sources
$V_{\fundamental}(R) = V_3(R)$ (solid line) and adjoint sources
$V_{\adjoint}(R) = V_8(R)$ (dashed line) as a function of the dipole
size $R$ in comparison to $SU(3)$ lattice
data~\cite{Bali:2000un,Bali:2001gf}.  The model results are in good
agreement with the lattice data. In particular, the obtained Casimir
scaling behavior is strongly supported by $SU(3)$ lattice
data~\cite{Deldar:1999vi,Bali:2000un}.  This, however, points also to
a shortcoming of our model: From
Eq.~(\ref{Eq_Vr(R)_Gaussian_approximation}) and
Fig.~\ref{Fig_Static_Quark-Antiquark_Potential_F_vs_A} it is clear
that {\em string breaking} is neither described for fundamental nor
for adjoint dipoles in our model which indicates that not only
dynamical fermions (quenched approximation) are missing but also some
gluon dynamics.

\section{Chromo-Field Distributions of Color Dipoles}
\label{Sec_Flux_Tube}

\begin{figure}[t]
\centerline{\epsfig{figure=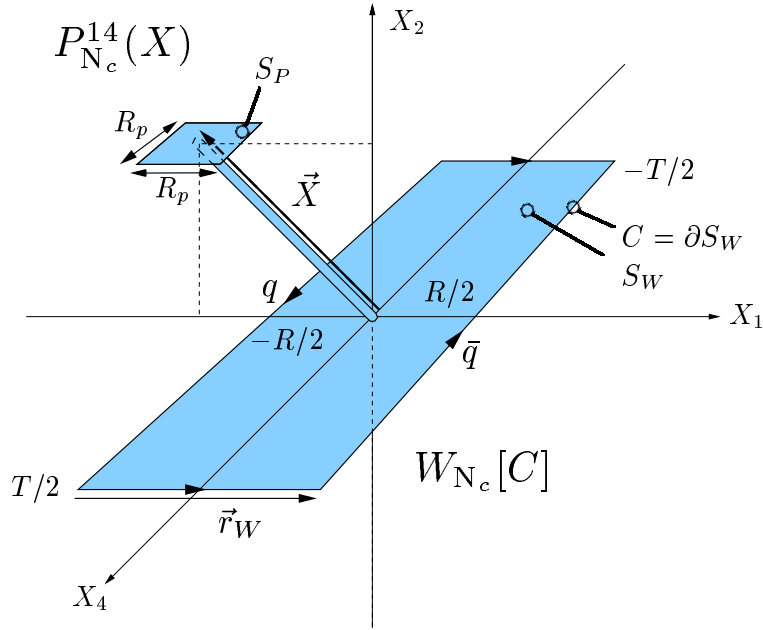,width=10.cm}}
\caption{\small
  The plaquette-loop geometry needed to compute the squared
  chromo-electric field $E_{\parallel}^2(X)$ generated by a static
  color dipole in the fundamental $SU(N_c)$ representation
  ($r=\Fundamental$).  The rectangular path $C$ indicates the
  world-line of the static dipole described the {\WW} loop
  $W_{\fundamental}[C]$. The square with side length $R_P$ illustrates
  the plaquette $P_{\fundamental}^{14}(X)$.  The shaded areas
  represent the minimal surfaces used in our computation of the
  chromo-field distributions.  The thin tube allows us to compare the
  gluon field strengths in surface $S_P$ with the gluon field
  strengths in surface $S_W$.}
\label{Fig_PW_arrangement}
\end{figure}

As already explained in Sec.~\ref{Sec_Static_Potential}, the static
color dipole -- two static color sources separated by a distance $R$
in a net color singlet state -- is described by a {\WW} loop $W_r[C]$
with a rectangular path $C$ of spatial extension $R$ and temporal
extension $T\to\infty$ (cf.\ Fig.~\ref{Fig_ONE_WWL}) where $r$
indicates the $SU(N_c)$ representation of the considered sources.  A
second small quadratic loop or plaquette in the fundamental
representation placed at the space-time point $X$ with side length
$R_P\to 0$ and oriented along the $\alpha\beta$-axes
\be
        P_{\fundamental}^{\alpha \beta}(X) 
        = \tilde{\Tr}_{\fundamental}
        \exp\!\!\left[
        -i g \oint_{C_P}\!\!\!dZ_{\mu} \G_{\mu}^a(Z) t_{\fundamental}^a 
        \right] 
        = 1 
        - R_P^4\frac{g^2}{4N_c}\G_{\alpha\beta}^a(X)\G_{\alpha\beta}^a(X) 
        + \Order(R_P^6)
\label{Eq_plaquette}
\ee
is needed -- as a ``Hall probe'' -- to calculate the chromo-field
distributions at the space-time point $X$ caused by the static
sources~\cite{Fukugita:1983du,Flower:gs}
\bea
        \Delta G_{r\,\alpha \beta}^2(X) 
        & := &
        \Big\langle 
        \frac{g^2}{4\pi^2}\G_{\alpha\beta}^a(X)\G_{\alpha\beta}^a(X)
        \Big\rangle_{W_r[C]}
        - 
        \Big\langle 
        \frac{g^2}{4\pi^2}\G_{\alpha\beta}^a(X)\G_{\alpha\beta}^a(X)
        \Big\rangle_{\mbox{\scriptsize vac}}
\label{Eq_DeltaG2_definition}\\
        & = &
        -\,\lim_{R_P \to 0}\inv{R_P^4} \frac{N_c}{\pi^2} 
        \left[
        \frac
        {\langle W_r[C] P_{\fundamental}^{\alpha \beta}(X) \rangle}
        {\langle W_r[C] \rangle}
        - \langle P_{\fundamental}^{\alpha \beta}(X) \rangle
        \right]
\label{Eq_DeltaG2_formula}
\eea
with {\em no} summation over $\alpha$ and $\beta$
in~(\ref{Eq_plaquette}), (\ref{Eq_DeltaG2_definition}),
and~(\ref{Eq_DeltaG2_formula}). In
definition~(\ref{Eq_DeltaG2_definition})
$\langle\ldots\rangle_{W_r[C]}$ indicates the VEV in the presence of
the static color dipole while $\langle\ldots\rangle_{\mbox{\scriptsize
    vac}}$ indicates the VEV in the absence of any color sources.
Depending on the plaquette orientation indicated by $\alpha$ and
$\beta$, one obtains from~(\ref{Eq_DeltaG2_formula}) the squared
components of the chromo-electric and chromo-magnetic field at the
space-time point $X$
\be
        \Delta G_{r\,\alpha \beta}^2(X) 
        = \frac{g^2}{4\pi^2}
        \left( \barray{cccc}
        0       & B_z^2 & B_y^2 & E_x^2 \\
        B_z^2   & 0     & B_x^2 & E_y^2 \\
        B_y^2   & B_x^2 & 0     & E_z^2 \\
        E_x^2   & E_y^2 & E_z^2 & 0
        \earray \right)(X)
        \ , 
\label{Eq_chromo-electromagnetic_fields}        
\ee
i.e.\ space-time plaquettes ($\alpha\beta=i4$) measure chromo-electric
fields and space-space plaquettes ($\alpha\beta=ij$) chromo-magnetic
fields. As shown in Fig.~\ref{Fig_PW_arrangement}, we place the static
color sources on the $X_1$-axis at $(X_1 = \pm R/2,0,0,X_4)$ and use
the following notation plausible from symmetry arguments
\be
        E_{\parallel}^2 = E_x^2
        \ ,\quad
        E_{\perp}^2 = E_y^2 = E_z^2
        \ ,\quad
        B_{\parallel}^2 = B_x^2
        \ ,\quad
        B_{\perp}^2 = B_y^2 = B_z^2
        \ .
\label{Eq_E_B_para_perp}
\ee
Figure~\ref{Fig_PW_arrangement} illustrates also the plaquette
$P_{\fundamental}^{14}(X)$ at $X = (X_1, X_2,0,0)$ needed to compute
$E_{\parallel}^2(X)$. Due to symmetry arguments, the complete
information on the chromo-field distributions is obtained from
plaquettes in ``transverse'' space $\mbox{$X = (X_1, X_2,0,0)$}$ with
four different orientations, $\alpha\beta = 14,\,24,\,13,\,23$,
cf.~(\ref{Eq_E_B_para_perp}).

The {\em energy} and {\em action density distributions} around a
static color dipole in the $SU(N_c)$ representation $r$ are given by
the squared chromo-field distributions
\bea
        \varepsilon_r(X) 
        & = & 
        \inv{2}\left(-\vec{E}^2(X)+\vec{B}^2(X)\right) 
\label{Eq_energy_density}\\
        \actiondensity_r(X)
        & = &
        -\inv{2}\left(\vec{E}^2(X)+\vec{B}^2(X)\right)
\label{Eq_action_density}
\eea
with signs according to Euclidean space-time conventions. Low-energy
theorems that relate the energy and action stored in the chromo-fields
of the static color dipole to the corresponding ground state energy
are discussed in the next section.

For the chromo-field distributions of a static color dipole in the
{\em fundamental} representation of $SU(N_c)$, i.e.\ a static
quark-antiquark pair, we obtain with our results for the VEV of one
loop~(\ref{Eq_final_result_<W[C]>}) and the correlation of two loops
in the fundamental
representation~(\ref{Eq_final_Euclidean_result_<W[C1]W[C2]>_fundamental})
\bea
        &&\Delta G_{\fundamental\,\alpha\beta}^2(X) =
        -\lim_{R_P \to 0}\inv{R_P^4}\frac{N_c}{\pi^2} 
        \exp\left[-\frac{C_2(\!\Fundamental\!)}{2}\,\chi_{S_P S_P}\right]
\label{Eq_chromo_fields_F_1}\\
        &&\hspace*{2cm}\times 
        \Bigg(\frac{N_c+1}{2N_c}\exp\!\left[-\frac{N_c-1}{2N_c}\chi_{S_P S_W}\right]
        + \frac{N_c-1}{2N_c}\exp\!\left[ \frac{N_c+1}{2N_c}\chi_{S_P S_W}\right] - 1\Bigg)
\nonumber
\eea
where $\chi_{S_i S_j}$ is defined in~(\ref{Eq_chi_Si_Sj}). The
subscripts $P$ and $W$ indicate surface integrations to be performed
over the surfaces spanned by the plaquette and the Wegner-Wilson-loop,
respectively. Choosing the surfaces -- as illustrated by the shaded
areas in Fig.~\ref{Fig_PW_arrangement} -- to be the minimal surfaces
connected by an infinitesimal thin tube (which gives no contribution
to the integrals) it is clear that $\chi_{S_P S_P} \propto R_P^4$ and
$\chi_{S_P S_W} \propto R_P^2$. Being interested in the chromo-fields
at the space-time point $X$, the extension of the quadratic plaquette
is taken to be infinitesimally small, $R_P \rightarrow 0$, so that one
can expand the exponential functions and keep only the term of lowest
order in $R_P$
\be
        \Delta G_{\fundamental\,\alpha\beta}^2(X) = 
        -\,C_2(\!\Fundamental\!)\,\lim_{R_P \to 0}\inv{R_P^4}\,\inv{4\pi^2}\,\chi_{S_P S_W}^2
        \ .
\label{Eq_chromo_fields_F_final_result}
\ee
This result -- obtained with the matrix cumulant expansion in a very
straightforward way -- agrees exactly with the result derived
in~\cite{Rueter:1994cn} with the expansion method. Indeed, the
expansion method agrees for small $\chi$-functions with the matrix
cumulant expansion (Berger-Nachtmann approach) used in this work but
breaks down for large $\chi$-functions, where the matrix cumulant
expansion is still applicable.

The chromo-field distributions of a static color dipole in the {\em
  adjoint} representation of $SU(N_c)$, i.e.\ a static gluino pair,
are computed analogously. Using our
result~(\ref{Eq_final_Euclidean_result_<Wf[C1]Wa[C2]>}) for the
correlation of one loop in the fundamental representation (plaquette)
with one loop in the adjoint representation (static sources), one
obtains
\bea
        \!\!\!\!&&\!\!\!\!\!\!\!\!
        \Delta G_{\adjoint\,\alpha\beta}^2(X) =
        -\lim_{R_P \to 0}\inv{R_P^4}\frac{N_c}{\pi^2} 
        \exp\left[-\frac{C_2(\!\Fundamental\!)}{2}\chi_{S_P S_P}\right]\,
        \Bigg(\!\inv{N_c^2\!-\!1}\,\exp\!\Big[\frac{N_c}{2}\,\chi_{S_P S_W}\Big]
\nonumber\\
        && \hskip 1.5cm
        +\,\frac{N_c\!+\!2}{2(N_c\!+\!1)}\exp\!\Big[\!-\inv{2}\,\chi_{S_P S_W}\Big]
        +\frac{N_c\!-\!2}{2(N_c\!-\!1)}\exp\!\Big[\inv{2}\,\chi_{S_P S_W}\Big]
        -1\!\Bigg)
\label{Eq_chromo_fields_A_1}
\eea
which reduces -- as explained for sources in the fundamental
representation -- to
\be
        \Delta G_{\adjoint\,\alpha\beta}^2(X) = 
        -\,C_2(\!\Adjoint\!)\,\lim_{R_P \to 0}\inv{R_P^4}\,\inv{4\pi^2}\,\chi_{S_P S_W}^2
        \ .
\label{Eq_chromo_fields_A_final_result}
\ee
Thus, the squared chromo-electric fields of an adjoint dipole differ
from those of a fundamental dipole only in the eigenvalue of the
corresponding quadratic Casimir operator $C_2(r)$. In fact, {\em
  Casimir scaling} of the chromo-field distributions holds for dipoles
in any representation $r$ of $SU(N_c)$ in our model. As can be seen
with the low-energy theorems discussed below, this is in line with the
Casimir scaling of the static dipole potential found in the previous
section. Besides lattice investigations that show Casimir scaling of
the static dipole potential to high accuracy in
$SU(3)$~\cite{Deldar:1999vi,Bali:2000un}, Casimir scaling of the
chromo-field distributions has been considered on the lattice as well
but only for $SU(2)$~\cite{Trottier:1995fx}. Here only slight
deviations from the Casimir scaling hypothesis have been found that
were interpreted as hints towards adjoint quark screening.

In our model the shape of the field distributions around the color
dipole is identical for all $SU(N_c)$ representations $r$ and given by
$\chi_{S_P S_W}^2$. This again illustrates the shortcoming of our
model discussed in the previous section. Working in the quenched
approximation, one expects a difference between fundamental and
adjoint dipoles: {\em string breaking} cannot occur in fundamental
dipoles as dynamical quark-antiquark production is excluded but should
be present for adjoint dipoles because of gluonic vacuum polarization.
Comparing~(\ref{Eq_chromo_fields_F_final_result})
with~(\ref{Eq_chromo_fields_A_final_result}) it is clear that this
difference is not described in our model. In fact, as shown in
Sec.~\ref{Sec_Static_Potential}, string breaking is neither described
for fundamental nor for adjoint dipoles. Interestingly, even on the
lattice there has been no striking evidence for adjoint quark
screening in quenched QCD~\cite{Kallio:2000jc}. It is even conjectured
that the {\WW} loop operator is not suited to studies of string
breaking~\cite{Gusken:1997sa+X}.

In the LLCM there are perturbative ($\pert$) and non-perturbative
($\nprt$) contributions to the chromo-electric fields according to the
structure of the gluon field strength correlator, (\ref{Eq_Ansatz})
and~(\ref{Eq_F_decomposition}),
\bea
        \Delta G_{r\,\alpha\beta}^2(X) 
        & = & 
        -C_2(r)\,\lim_{R_P \to 0}\inv{R_P^4}\,\inv{\pi^2}\,
\label{Eq_chromo_fields_F_no_interference}\\
        && \times\left\{
          \left(\chi_{S_P S_W}^{\pert}(X)\right)_{\alpha\beta}^2 
        + \left[
          \left(\chi_{S_P S_W}^{\nprt\,nc}(X)\right)_{\alpha\beta}
          +  \left(\chi_{S_P S_W}^{\nprt\,c}(X)\right)_{\alpha\beta}
        \right]^2
      \right\}
\nonumber
\eea
where we have demanded the non-interference of perturbative and
non-perturbative correlations in line with the Minkowskian
applications of our
model~\cite{Shoshi:2002in,Shoshi:2002ri,Shoshi:2002fq,Shoshi:2002mt}.
In the following we give only the final results of the
$\chi$\,-\,functions for the minimal surfaces shown in
Fig.~\ref{Fig_PW_arrangement}.  Details on their derivation can be
found in Appendix~\ref{Sec_Chi_Computation}.

The {\em perturbative contribution} ($P$) described by massive gluon
exchange leads, of course, to the well-known {\em color Yukawa field}
that reduces to the {\em color Coulomb field} for $m_g=0$. It
contributes only to the chromo-electric fields, $E_{\parallel}^2 =
E_x^2$ ($\alpha\beta=14$) and $E_{\perp}^2 = E_y^2 = E_z^2$
($\alpha\beta=24$), and reads explicitly for $X = (X_1, X_2, 0, 0)$
\bea
\!\!\!\!\!\!\!\!\!\!\!\!
        \left(\chi_{S_P S_W}^{\pert}(X)\right)_{14}
        &\!\!=\!\!& -\,\frac{R_P^2}{2}\!\int_{-\infty}^{\infty}\!\!\!d\tau
        \left\{
        (X_1 - R/2)\,
        g^2(Z_{1A}^2)\,D_\pert(Z_{1A}^2)
        \right.
\nonumber\\
        && \hphantom{-\,\frac{R_P^2}{2}\!\int_{-\infty}^{\infty}\!\!\!d\tau\Big(}
        \left.       
        - \,(X_1 + R/2)\,
        g^2(Z_{1C}^2)\,D_\pert(Z_{1C}^2)
        \right\}
\label{Eq_Chi_PW_p_14}\\
\!\!\!\!\!\!\!\!\!\!\!\!
        \left(\chi_{S_P S_W}^{\pert}(X)\right)_{24}
        &\!\!=\!\!& -\,\frac{R_P^2}{2}\!\int_{-\infty}^{\infty}\!\!\!d\tau\,X_2
        \left\{
        g^2(Z_{1A}^2)\,D_\pert(Z_{1A}^2)
        - g^2(Z_{1C}^2)\,D_\pert(Z_{1C}^2)
        \right\}
\label{Eq_Chi_PW_p_24}
\eea
with the perturbative correlation function~(\ref{Eq_Dp(z,mg)}), the
running coupling~(\ref{Eq_g2(z_perp)}), and
\be
        Z_{1A}^2 = \left(X_1\!-\!\frac{R}{2}\right)^2+X_2^2+\tau^2
        \quad \mbox{and} \quad
        Z_{1C}^2 = \left(X_1\!+\!\frac{R}{2}\right)^2+X_2^2+\tau^2
        \ .
\label{Eq_|Z1A|_|Z1C|}
\ee

The {\em non-confining non-perturbative contribution} ($\nprt\,nc$) has
the same structure as the perturbative contribution -- as expected
from the identical tensor structure -- but differs, of course, in the
prefactors and the correlation function, $D_1 \neq D_p$. Its
contributions to the chromo-electric fields $E_{\parallel}^2 = E_x^2$
($\alpha\beta=14$) and $E_{\perp}^2 = E_y^2 = E_z^2$
($\alpha\beta=24$) read for $X = (X_1, X_2, 0, 0)$
\bea
        \left(\chi_{S_P S_W}^{\nprt\,\,nc}(X)\right)_{14}
        &=&
        -\,\frac{R_P^2 \pi^2 G_2 (1\!-\!\kappa)}{6\,(N_c^2\!-\!1)}
        \!\int_{-\infty}^{\infty}\!\!\!d\tau
        \Big\{
        (X_1 - R/2)\,D_1(Z_{1A}^2)
\nonumber\\
        && 
        \hphantom{-\,\frac{R_P^2 \pi^2 G_2}{6\,(N_c^2\!-\!1)}}
        - \,(X_1 + R/2)\,D_1(Z_{1C}^2)
        \Big\}
\label{Eq_Chi_PW_np_nc_14}\\
        \left(\chi_{S_P S_W}^{\nprt\,\,nc}(X)\right)_{24}
        &=&
        -\,\frac{R_P^2 \pi^2 G_2 (1\!-\!\kappa)}{6\,(N_c^2\!-\!1)}
        \!\int_{-\infty}^{\infty}\!\!\!d\tau\,X_2
        \Big\{D_1(Z_{1A}^2)-D_1(Z_{1C}^2)\Big\}
\label{Eq_Chi_PW_np_nc_24}
\eea
with the exponential correlation
function~(\ref{Eq_SVM_correlation_functions}) and $Z_{1A}^2$ and
$Z_{1C}^2$ as given in~(\ref{Eq_|Z1A|_|Z1C|}).

The {\em confining non-perturbative contribution} ($\nprt\,c$) has a
different structure that leads to confinement and flux-tube formation.
It gives only contributions to the chromo-electric field
$E_{\parallel}^2 = E_x^2$ ($\alpha\beta=14$) that read for $X = (X_1,
X_2, 0, 0)$
\bea
        \left(\chi_{S_P S_W}^{\nprt\,\,c}(X)\right)_{14}
        & = & 
        R_P^2 R 
        \frac{\pi^2 G_2 \kappa}{3\,(N_c^2\!-\!1)}
        \int_{0}^{1} d\rho\,
        D^{(3)}(\vec{Z_{\perp}}^2)
        \ ,
\label{Eq_Chi_PW_np_c_14}
\eea
with the correlation function given in~(\ref{Eq_D(3)np_c(z,a)}) as
derived from the exponential correlation
function~(\ref{Eq_SVM_correlation_functions}), and
\be
        \vec{Z}_{\perp}^2 = [X_1+(1/2-\rho)R]^2+X_2^2
        \ .
\label{Eq_|Z|}
\ee

In our model there are no contributions to the {\em chromo-magnetic
  fields}, i.e.\ the static color charges do not affect the magnetic
background field
\be
        B_{\parallel}^2 = B_x^2 = 0
        \quad \mbox{and} \quad
        B_{\perp}^2 = B_y^2 = B_z^2 = 0
        \ ,
\label{Eq_B^2=0}
\ee
which can be seen from the corresponding plaquette-loop geometries as
pointed out in Appendix~\ref{Sec_Chi_Computation}. Thus, the energy
and action densities are identical in our approach and completely
determined by the squared chromo-electric fields
\be
        \varepsilon_r(X) 
        \,\,=\,\,\actiondensity_r(X)
        \,\,=\,\, -\inv{2}\,\vec{E}^2(X)
        \ .
\label{Eq_energy=action_density}
\ee
This picture is in agreement with other effective theories of
confinement such as the `t~Hooft-Mandelstam
picture~\cite{Mandelstam:1976pi+X} or dual QCD~\cite{Baker:bc} and,
indeed, a relation between the dual Abelian Higgs model and the SVM
has been established~\cite{Baker:1998jw}. In contrast, lattice
investigations work at scales at which the chromo-electric and
chromo-magnetic fields are of similar
magnitude~\cite{Bali:1994de,Green:1996be}. Using low-energy theorems,
we will see in the next section, that the vanishing of the
chromo-magnetic fields determines the value of the $\beta$-function at
the renormalization scale at which the non-perturbative component of
our model is working.

\begin{figure}[h]
\centerline{
\epsfig{figure=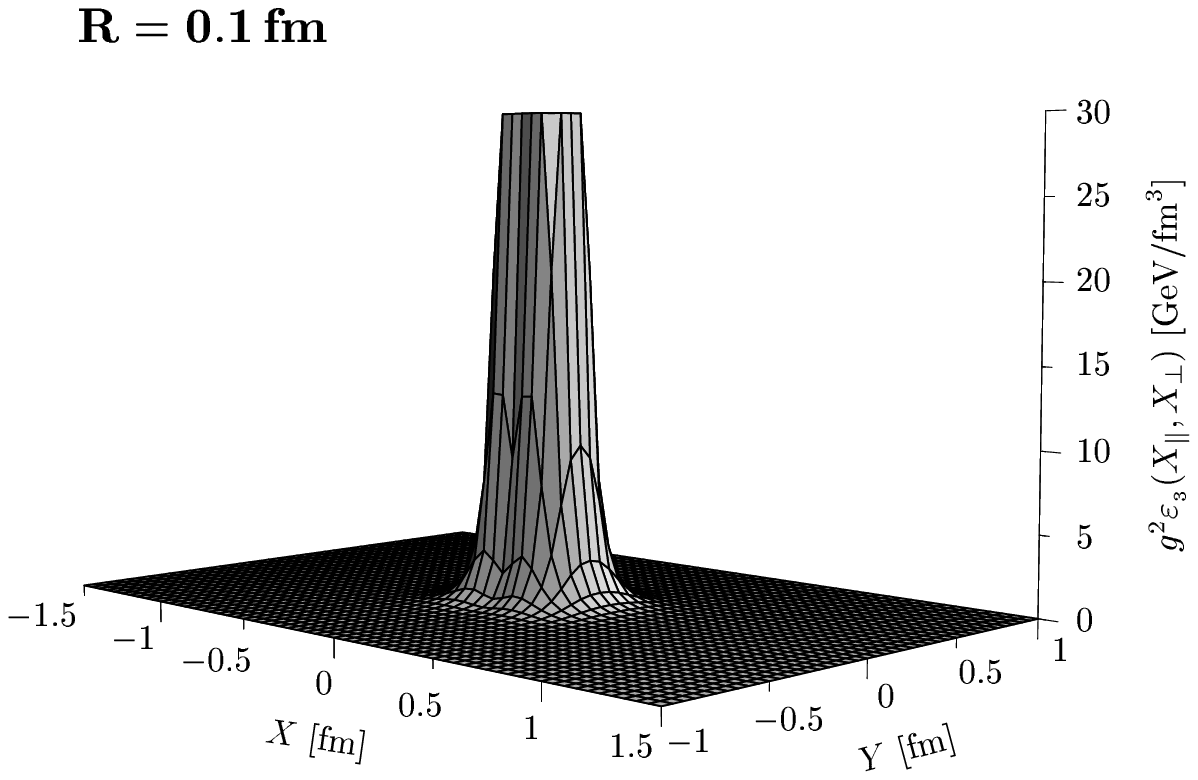,width=7.7cm}
\hspace*{-0.8cm}
\epsfig{figure=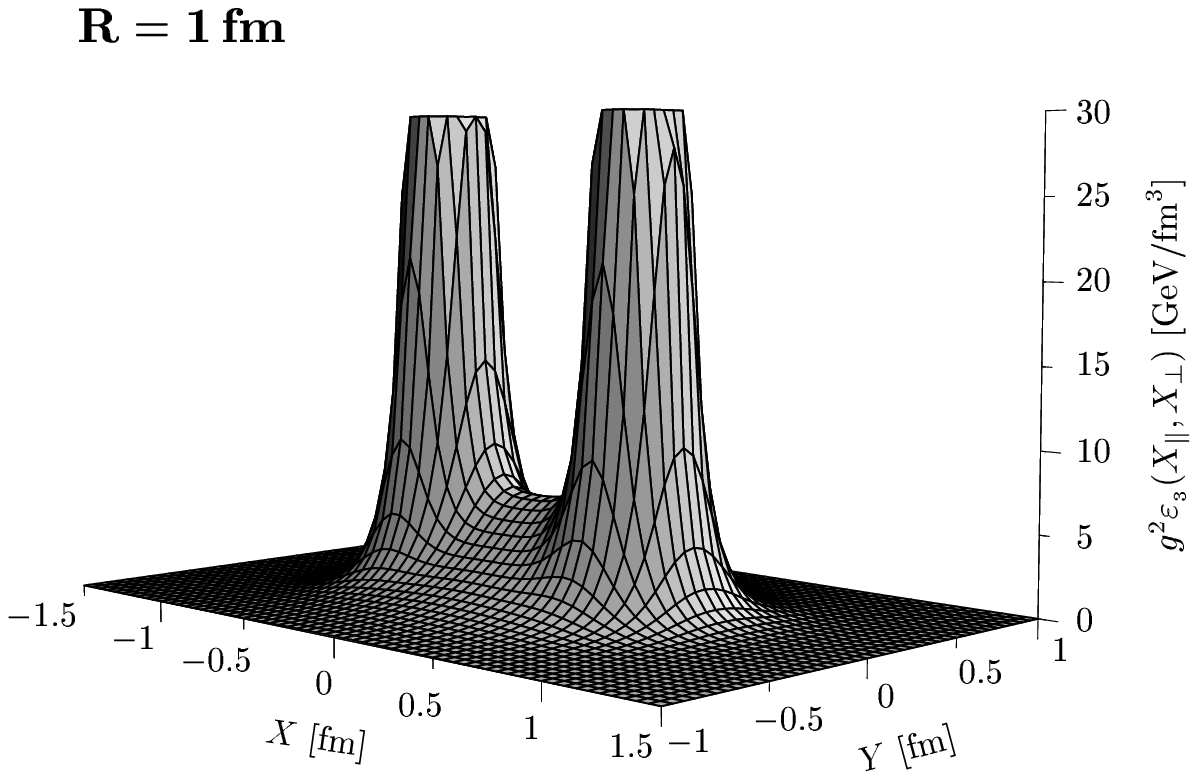,width=7.7cm}}
\centerline{
\epsfig{figure=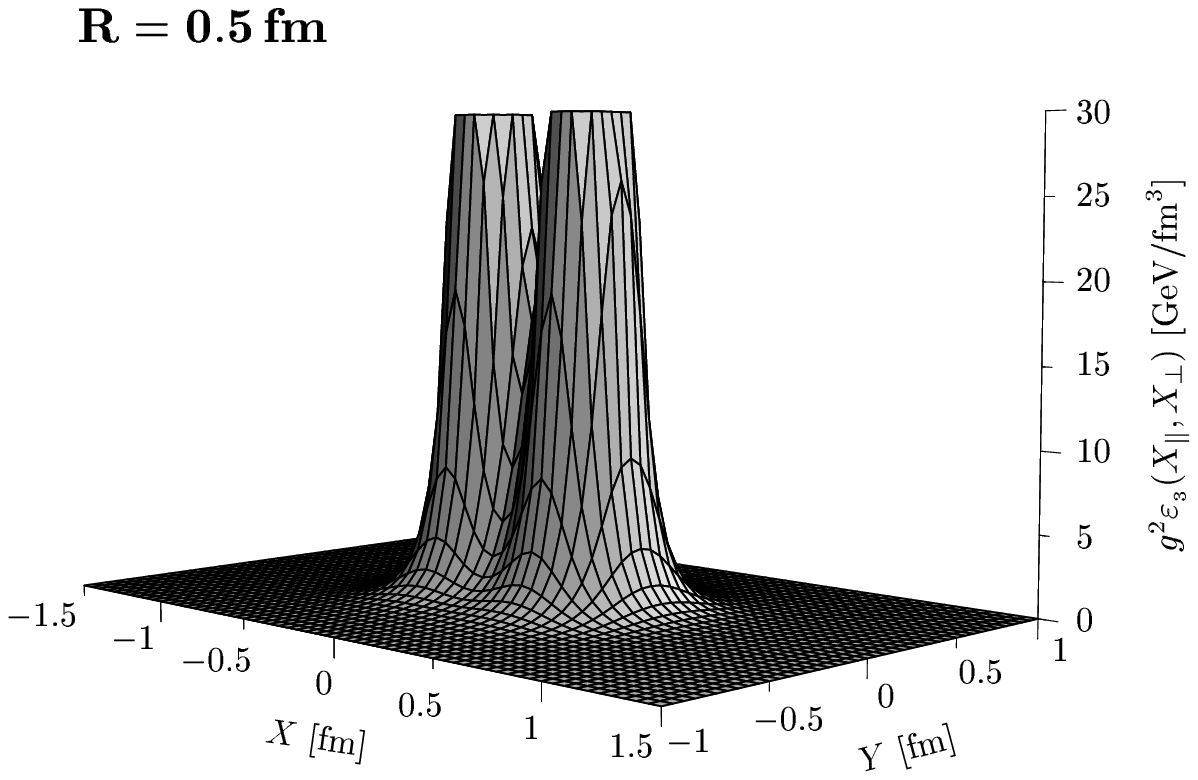,width=7.7cm}
\hspace*{-0.8cm}
\epsfig{figure=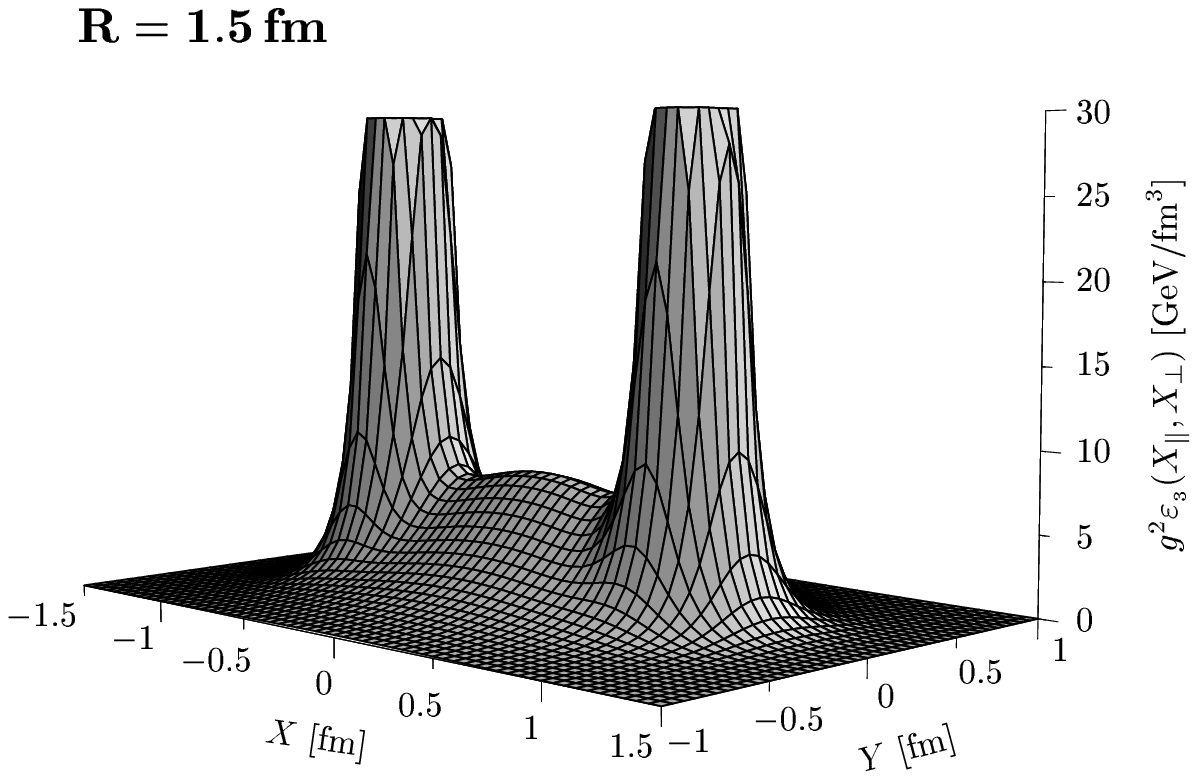,width=7.7cm}}
\caption{\small
  Energy density distributions $g^2\varepsilon_3(X_1,X_2\!=\!X_3)$
  generated by a color dipole in the fundamental $SU(3)$
  representation ($r\!=\!3$) for quark-antiquark separations of $R =
  0.1,\,0.5,\,1$ and $1.5\,\fm$. Flux-tube formation leads to the
  confining QCD string with increasing dipole size $R$.}
\label{Fig_3D_profiles}
\end{figure}
In Fig.~\ref{Fig_3D_profiles} the energy density distributions
$g^2\varepsilon_3(X_1,X_2\!=\!X_3)$ generated by a color dipole in the
fundamental $SU(3)$ representation ($r\!=\!3$) are shown for
quark-antiquark separations of $R = 0.1,\,0.5,\,1$ and $1.5\,\fm$.
With increasing dipole size $R$, one sees explicitly the formation of
the flux tube which represents the confining QCD string.

\begin{figure}[p]
\centerline{\epsfig{figure=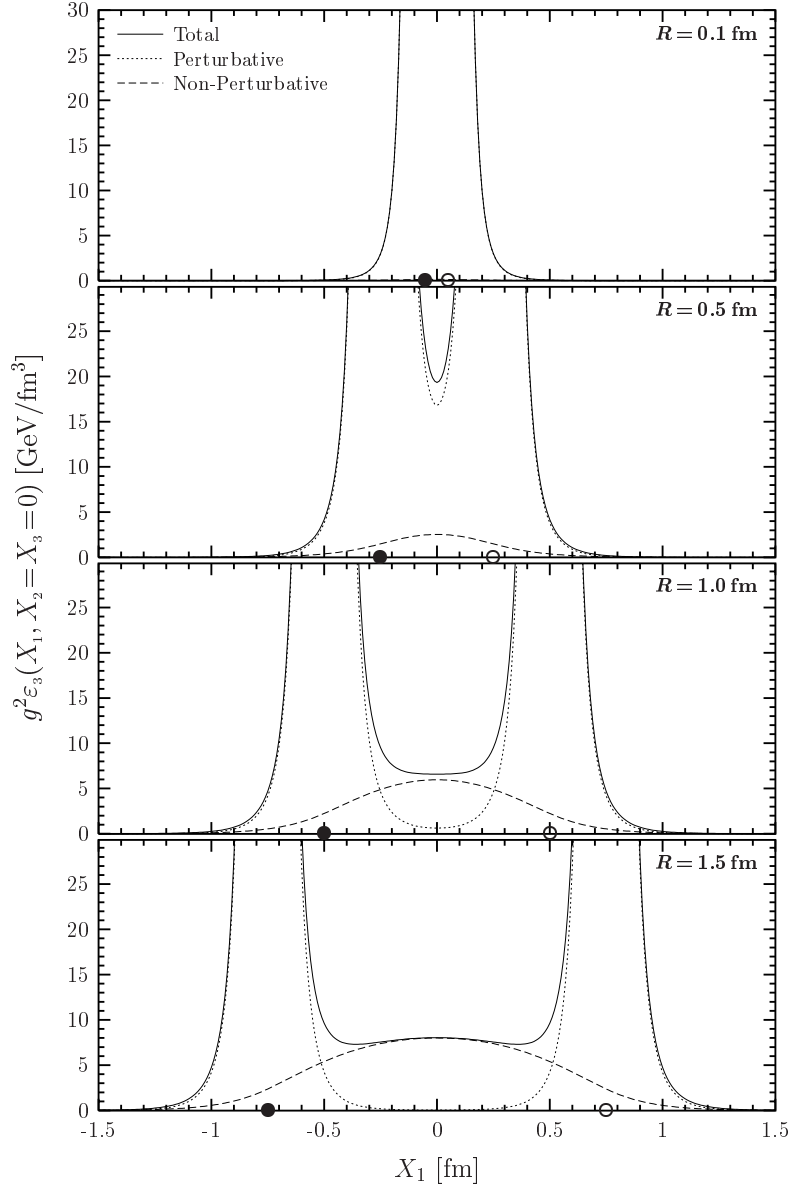,width=10cm}}
\caption{\small
  Longitudinal energy density profiles
  $g^2\varepsilon_3(X_1,X_2\!=\!X_3\!=\!0)$ generated by a color
  dipole in the fundamental $SU(3)$ representation ($r\!=\!3$) for
  quark-antiquark separations of $R = 0.1,\,0.5,\,1$ and $1.5\,\fm$.
  The dotted and dashed lines give the perturbative and
  non-perturbative contributions, respectively, and the solid lines
  the sum of both.  The open and filled circles indicate the quark and
  antiquark positions. For small dipoles, $R=0.1\,\fm$, perturbative
  physics dominates and non-perturbative correlations are negligible.
  For large dipoles, $R\gtsim 1\,\fm$, the formation the confining
  string (flux tube) can be seen which dominates the chromo-electric
  fields between the color sources.}
\label{Fig_L_profiles}
\end{figure}

\begin{figure}[p]
\centerline{\epsfig{figure=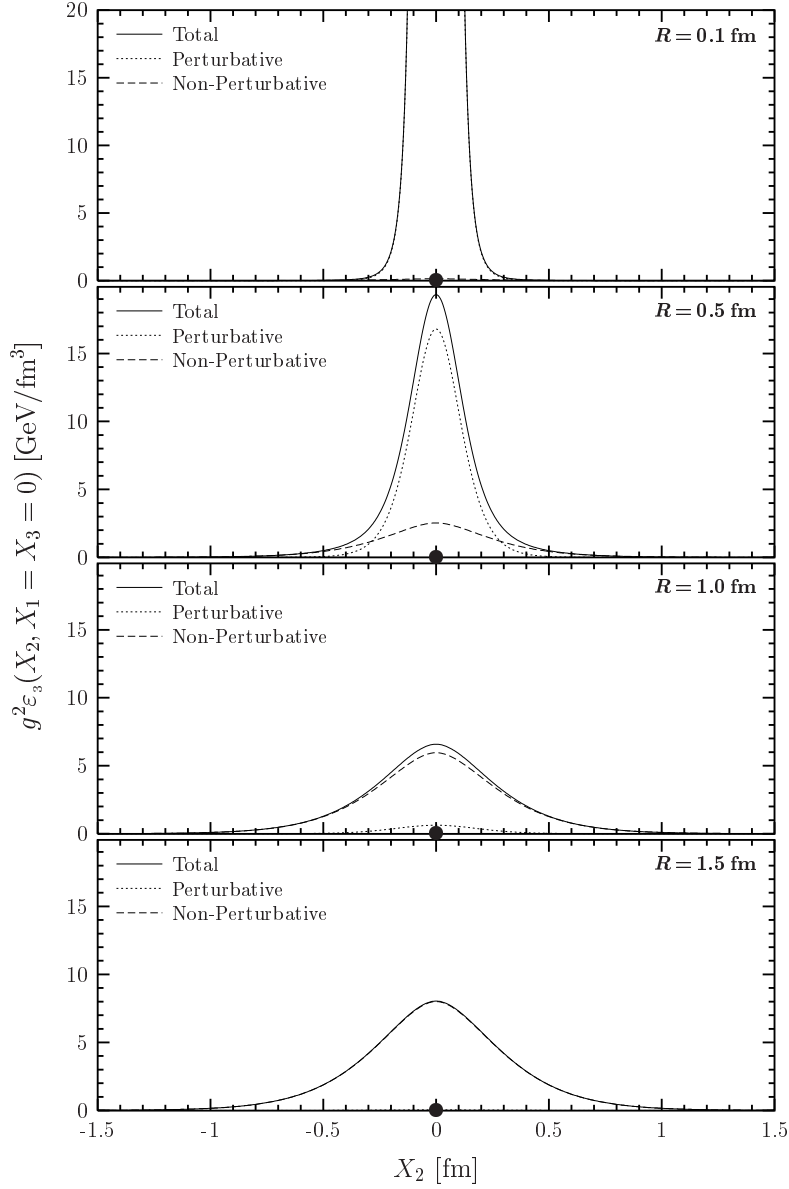,width=10cm}}
\caption{\small
  Transverse energy density profiles
  $g^2\varepsilon_3(X_2,X_1\!=\!X_3\!=\!0)$ generated by a color
  dipole in the fundamental $SU(3)$ representation ($r\!=\!3$) for
  quark-antiquark separations of $R = 0.1,\,0.5,\,1$ and $1.5\,\fm$.
  The dotted and dashed lines give the perturbative and
  non-perturbative contributions, respectively, and the solid lines
  the sum of both.  The filled circles indicate the positions of the
  color sources. For small dipoles, $R=0.1\,\fm$, perturbative physics
  dominates and non-perturbative correlations are negligible. For
  large dipoles, $R\gtsim 1\,\fm$, the formation the confining string
  (flux tube) can be seen which dominates the chromo-electric fields
  between the color sources.}
\label{Fig_T_profiles}
\end{figure}

The {\em longitudinal} and {\em transverse energy density profiles}
generated by a color dipole in the fundamental representation ($r=3$)
of $SU(N_c=3)$ are shown for quark-antiquark separations (dipole
sizes) of $R = 0.1,\,0.5,\,1$ and $1.5\,\fm$ in
Figs.~\ref{Fig_L_profiles} and~\ref{Fig_T_profiles}. The perturbative
and non-perturbative contributions are given in the dotted and dashed
lines, respectively, and the sum of both in the solid lines. The open
and filled circles indicate the quark and antiquark positions. As can
be seen from~(\ref{Eq_DeltaG2_formula})
and~(\ref{Eq_chromo-electromagnetic_fields}), we cannot compute the
energy density separately but only the product $g^2\varepsilon_r(X)$.
Nevertheless, a comparison of the total energy stored in
chromo-electric fields to the ground state energy of the color dipole
via low-energy theorems yields $g^2 = 10.2$ $(\equiv \alphaS=0.81)$
for the non-perturbative SVM component as shown in the next section.

In Figs.~\ref{Fig_L_profiles} and~\ref{Fig_T_profiles} the formation
of the confining string (flux tube) with increasing source separations
$R$ can again be seen explicitly: For small dipoles, $R=0.1\,\fm$,
perturbative physics dominates and non-perturbative correlations are
negligible. For large dipoles, $R\gtsim 1\,\fm$, the non-perturbative
correlations lead to formation of a narrow flux tube which dominates
the chromo-electric fields between the color sources.

\begin{figure}[t]
\centerline{\epsfig{figure=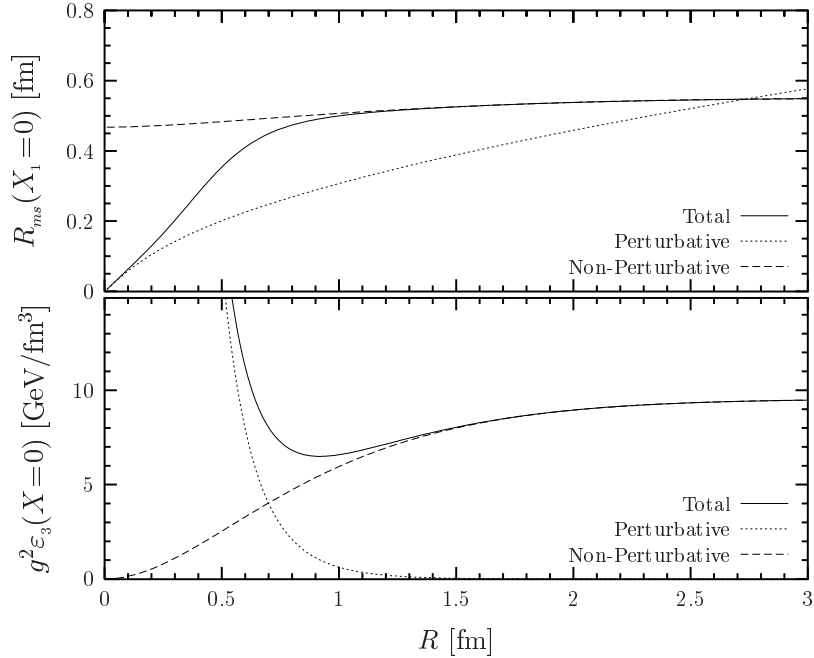,width=12cm}}
\caption{\small
  Root mean squared radius $R_{ms}$ of the flux tube and energy
  density in the center of a fundamental $SU(3)$ dipole
  $g^2\varepsilon_3(X=0)$ as a function of the dipole size $R$.
  Perturbative and non-perturbative contributions are given
  respectively in the dotted and dashed lines and the sum of both in
  the solid lines. For large $R$, both the width and height of the
  flux tube in the central region are governed completely by
  non-perturbative physics and saturate respectively at
  $R_{ms}^{R\to\infty}\approx 0.55\,\fm$ and
  $\varepsilon_3^{R\to\infty}(X=0)\approx 1\,\GeV/\fm^3$. The latter
  value is extracted with the result $g^2 = 10.2$ deduced from
  low-energy theorems in the next section.}
\label{Fig_R_ms_and_g2epsilon(0)}
\end{figure}

Figure~\ref{Fig_R_ms_and_g2epsilon(0)} 
shows the evolution of the transverse width (upper plot) and height
(lower plot) of the flux tube in the central region of the {\WW} loop
as a function of the dipole size $R$ where perturbative and
non-perturbative contributions are given in the dotted and dashed
lines, respectively, and the sum of both in the solid lines. The width
of the flux tube is best described by the root mean squared ($ms$)
radius
\be
        R_{ms}
        = \sqrt{\frac{\int dX_{\perp}\,X_{\perp}^3\,g^2\varepsilon_r(X_1=0,X_{\perp})}
        {\int dX_{\perp}\,X_{\perp}\,g^2\varepsilon_r(X_1=0,X_{\perp})}}
        \ ,
\label{Eq_R_ms}
\ee
which is universal for dipoles in all $SU(N_c)$ representations $r$ as
the Casimir factors divide out. The height of the flux tube is given
by the energy density in the center of the considered dipole,
$g^2\varepsilon_r(X=0)$. For large source separations, $R \gtsim
1\,\fm$, both the width and height of the flux tube in the central
region of the {\WW} loop are governed completely by non-perturbative
physics and saturate for a fundamental $SU(3)$ dipole
($r=\Fundamental=3$) at reasonable values of
\be
        R_{ms}^{R\to\infty}\approx 0.55\,\fm
        \quad \mbox{and} \quad 
        \varepsilon_3^{R\to\infty}(X=0)\approx 1\,\GeV/\fm^3
        \quad \mbox{with} \quad g^2 = 10.2
        \ .
\label{Eq_R_ms_and_g2epsilon(0)_saturation_values}
\ee

Note that the qualitative features of the non-perturbative SVM
component do not depend on the specific choice for the parameters,
surfaces, and correlation functions and have already been discussed
with the pyramid mantle choice of the surface and different
correlation functions in the first investigation of flux-tube
formation in the SVM~\cite{Rueter:1994cn}. The quantitative results,
however, are sensitive to the parameter values, the surface choice,
and the correlation functions and are presented above with the LLCM
parameters, the minimal surfaces, and the exponential correlation
function~\cite{Shoshi:2002rd}.

\section{Low-Energy Theorems}
\label{Sec_Low_Energy_Theorems}

Many low-energy theorems have been derived in continuum theory by
Novikov, Shifman, Vainshtein, and Zakharov~\cite{Novikov:xi+X} and in
lattice gauge theory by Michael~\cite{Michael:1986yi}. Here  we
consider the energy and action sum rules -- known in lattice QCD as
{\em Michael sum rules} -- that relate the energy and action stored
in the chromo-fields of a static color dipole to the corresponding
ground state energy~\cite{Wilson:1974sk,Brown:1979ya}
\be
        E_r(R) = 
        - \lim_{T \to \infty} \inv{T} 
        \ln \langle W_r[C] \rangle
        \ .
\label{Eq_Er(R)_def}        
\ee
In their original form~\cite{Michael:1986yi}, however, the Michael sum
rules are incomplete~\cite{Dosch:1995fz,Rothe:1995hu+X}. In
particular, significant contributions to the energy sum rule from the
trace anomaly of the energy-momentum tensor have been
found~\cite{Rothe:1995hu+X} that modify the naively expected relation
in line with the importance of the trace anomaly found for hadron
masses~\cite{Ji:1995sv}. Taking all these contributions into account,
the {\em energy} and {\em action sum rule} read
respectively~\cite{Rothe:1995hu+X,Michael:1995pv,Green:1996be}
\bea
        && 
        E_r(R) 
        = \int d^3X\,\varepsilon_r(X)
        - \inv{2}\frac{\beta(g)}{g}
        \int d^3X\,\actiondensity_r(X)
\label{Eq_energy_sum_rule}\\
        &&
        E_r(R) + R\,\frac{\partial E_r(R)}{\partial R}
        = - \frac{2\beta(g)}{g}
        \int d^3X\,\actiondensity_r(X)
\label{Eq_action_sum_rule}
\eea
where $\beta(g)=\mu \partial g/\partial\mu$ with the renormalization
scale $\mu$. Inserting~(\ref{Eq_action_sum_rule})
into~(\ref{Eq_energy_sum_rule}), we find the following relation
between the total energy stored in the chromo-fields
$E_r^{\mbox{\scriptsize tot}}(R)$ and the ground state energy $E_r(R)$
\be
        E_r^{\mbox{\scriptsize tot}}(R) 
        := \int d^3X\,\varepsilon_r(X)
        = \inv{4}\left(3\,E_r(R) - R\frac{\partial E_r(R)}{\partial R} \right)
        \ .
\label{Eq_Etot-Er(R)_relation}
\ee
The difference from the naive expectation that the full ground state
energy of the static color sources is stored in the chromo-fields is
due to the trace anomaly contribution~\cite{Rothe:1995hu+X} described
by the second term on the right-hand side (rhs)
of~(\ref{Eq_energy_sum_rule}).

With the low energy theorems~(\ref{Eq_action_sum_rule})
and~(\ref{Eq_Etot-Er(R)_relation}) the ratio of the integrated squared
chromo-magnetic to the integrated squared chromo-electric field
distributions can be derived
\be
        Q(R) := \frac{\int d^3X \vec{B}^2(X)}{\int d^3X \vec{E}^2(X)}
        = \frac{\left(2+6\,\beta(g)/g\right)\,E_r(R) 
        + \left(1-\beta(g)/g\right)\,R\,\frac{\partial E_r(R)}{\partial R}}
        {\left(2-6\,\beta(g)/g\right)\,E_r(R) 
        + \left(1+\beta(g)/g\right)\,R\,\frac{\partial E_r(R)}{\partial R}}
      \ ,
\label{Eq_Q_ratio_general}
\ee
which becomes for $E_r(R)=\sigma_r R+E_{\self}$ after subtraction of
the self-energy contributions, i.e.\ the linear potential
$V_r(R)=\sigma_r R$ with string tension $\sigma_r$ in the considered
representation $r$,
\be
        Q(R) \Big|_{V_r(R)=\sigma_r R}
        = \frac{2+\beta(g)/g}{2-\beta(g)/g}
      \ .
\label{Eq_Q_ratio_linear_potential}
\ee

In our model there are no contributions to the chromo-magnetic
fields~(\ref{Eq_B^2=0}) so that -- as already discussed in the
previous section -- the energy and action densities are identical and
completely determined by the squared chromo-electric
fields~(\ref{Eq_energy=action_density}). Since the non-perturbative
SVM component of our model describes the confining linear potential
for large source separations $R$, this allows us to determine
from~(\ref{Eq_Q_ratio_linear_potential}) immediately the value of the
$\beta$\,-\,function at the scale $\mu_{\nprt}$ at which the
non-perturbative component is working
\be
        \frac{\beta(g)}{g}\Big|_{\mu =\mu_{\nprt}} = -2
        \ .
\label{Eq_beta/g=-2}
\ee

Concentrating on the confining non-perturbative component ($\nprt c$)
we now use (\ref{Eq_Etot-Er(R)_relation}) to determine the value of
$\alphaS = g^2/(4\pi)$ at which the non-perturbative SVM component is
working. The rhs of~(\ref{Eq_Etot-Er(R)_relation}) is obtained
directly from the confining contribution to the static potential
$E_r^{\nprt c}(R)=V_r^{\nprt c}(R)$ given in~(\ref{Eq_Vr(R)_NP_c}).
The lhs of~(\ref{Eq_Etot-Er(R)_relation}), however, involves a
division by the {\em a priori} unknown value of $g^2$ after
integrating $g^2\varepsilon_r(X)$ for the chromo-electric field of the
confining non-perturbative component~(\ref{Eq_Chi_PW_np_c_14}). As
discussed in the previous section, we cannot compute the energy
density separately but only the product $g^2\varepsilon_r(X)$.
Adjusting the value of $g^2$ such that~(\ref{Eq_Etot-Er(R)_relation})
is exactly fulfilled for source separations of $R=1.5\,\fm$, we find
that the non-perturbative component is working at the scale
$\mu_{\nprt}$ at which
\be
        g^2(\mu_{\nprt}) = 10.2
        \quad \equiv \quad
        \alphaS(\mu_{\nprt}) = 0.81
        \ .
\label{Eq_alphaS=0.81}
\ee
As already mentioned in Sec.~\ref{Sec_QCD_Components}, we use this
value as a practical asymptotic limit for the simple one-loop
coupling~(\ref{Eq_g2(z_perp)}) used in our perturbative component.
Note that earlier SVM investigations along these lines have found a
smaller value of $\mbox{$\alphaS(\mu_{\nprt}) = 0.57$}$ with the
pyramid mantle choice for the
surface~\cite{Rueter:1994cn,Dosch:1995fz} but were incomplete since
only the contribution from the traceless part of the energy-momentum
tensor has been considered in the energy sum rule.

\begin{figure}[t]
\centerline{\epsfig{figure=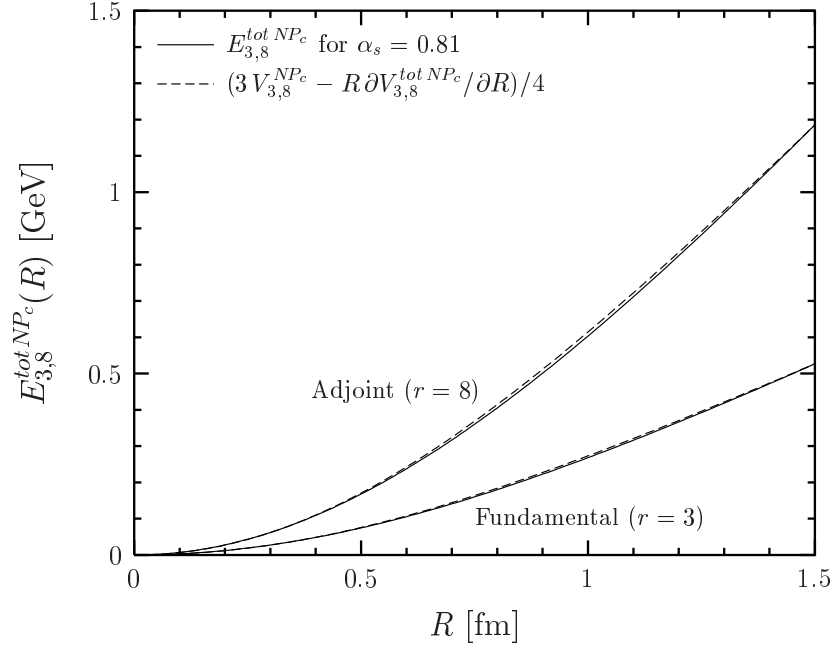,width=12cm}}
\caption{\small
  The total energy stored in the chromo-field distributions around a
  static color dipole of size $R$ in the fundamental ($r=3$) and
  adjoint ($r=8$) representation of $SU(3)$ from the confining
  non-perturbative SVM component, $E_{3,8}^{\mbox{\scriptsize
      tot}\,\nprt_c}(R)$, for $\alphaS = 0.81$ (solid lines) compared
  with the relation to the corresponding ground state energy (dashed
  lines) given by the low-energy
  theorem~(\ref{Eq_Etot-Er(R)_relation}). Good consistency is found
  even down to very small values of $R$.}
\label{Fig_NP_c_consistency}
\end{figure}

In Fig.~\ref{Fig_NP_c_consistency}
we show the total energy stored in the chromo-field distributions
around a static color dipole in the fundamental ($r=3$) and adjoint
($r=8$) representation of $SU(3)$ from the confining non-perturbative
SVM component, $E_{3,8}^{\mbox{\scriptsize tot}\,\nprt_c}(R)$, for
$\alphaS = 0.81$ (solid lines) as a function of the dipole size $R$.
Comparing this total energy, which appears on the lhs
of~(\ref{Eq_Etot-Er(R)_relation}), with the corresponding rhs
of~(\ref{Eq_Etot-Er(R)_relation}) (dashed lines), we find good
consistency even down to very small values of $R$. This is a
nontrivial and important result as it confirms the consistency of our
loop-loop correlation result -- needed to compute the chromo-electric
field -- with the result obtained for the VEV of one loop -- needed
to compute the static potential $V_r^{\nprt_c}(R)$. Moreover, it shows
that the minimal surfaces ensure the consistency of our
non-perturbative component. The good consistency found for the pyramid
mantle choice of the surface relies on the naively expected energy sum
rule~\cite{Rueter:1994cn,Dosch:1995fz} in which the contribution from
the traceless part of the energy-momentum tensor is not taken into
account.


%
\cleardoublepage
%
\chapter{Euclidean Approach to High-Energy Scattering}
\label{Sec_DD_Scattering}

In this chapter we present a Euclidean approach to high-energy
reactions of color dipoles in the eikonal approximation. We give a
short review of the functional integral approach to high-energy
scattering, which is the basis for the presented Euclidean approach
and for our investigations of hadronic high-energy reactions in the
following chapters. We generalize the analytic continuation introduced
by Meggiolaro~\cite{Meggiolaro:1996hf+X} from parton-parton scattering
to dipole-dipole scattering. This shows how one can access high-energy
reactions directly in lattice QCD. We apply this approach to compute
the scattering of dipoles in the fundamental and adjoint
representation of $SU(N_c)$ at high-energy in the Euclidean LLCM. The
result shows the consistency with the analytic continuation of the
gluon field strength correlator used in all earlier applications of
the SVM and LLCM to high-energy scattering. Finally, we comment on the
QCD van der Waals potential which appears in the limiting case of two
static color dipoles.

\section[Functional Integral Approach to High-Energy Scattering]
{\!\!\!\!\!Functional Integral Approach to High-Energy\! Scattering}
\label{Sec_Functional_Integral_Approach}

In {\em Minkowski space-time} high-energy reactions of color dipoles
in the eikonal approximation are considered -- as basis for
hadron-hadron, photon-hadron, and photon-photon reactions -- in the
functional integral approach to high-energy collisions developed
originally for parton-parton
scattering~\cite{Nachtmann:1991ua+X,Nachtmann:ed.kt} and then extended
to gauge-invariant dipole-dipole
scattering~\cite{Kramer:1990tr,Dosch:1994ym,Dosch:RioLecture}. The
corresponding $T$-matrix element for the elastic scattering of two
color dipoles at transverse momentum transfer ${\vec q}_{\!\perp}$ ($t
= -{\vec q}_{\!\perp}^{\,\,2}$) and c.m.\ energy squared~$s$ reads
\be
        T^M_{r_1 r_2}(s,t,z_1,\vec{r}_{1\perp},z_2,\vec{r}_{2\perp}) =
        2is \int \!\!d^2b_{\!\perp} 
        e^{i {\vec q}_{\!\perp} {\vec b}_{\!\perp}}
        \left[1-S^M_{r_1 r_2}(s,{\vec b}_{\!\perp},z_1,\vec{r}_{1\perp},z_2,\vec{r}_{2\perp})\right]
\label{Eq_T_DD_Minkowski}
\ee
with the $S$-matrix element ($M$ refers to Minkowski space-time)
\be
        S^M_{r_1 r_2}(s,{\vec b}_{\!\perp},z_1,\vec{r}_{1\perp},z_2,\vec{r}_{2\perp})
        = \lim_{T \rightarrow \infty}
        \frac{\langle W_{r_1}[C_1] W_{r_2}[C_2]\rangle_M}
        {\langle W_{r_1}[C_1]\rangle_M \langle W_{r_2}[C_2]\rangle_M}
        \ .
\label{Eq_S_DD_Minkowski}
\ee
The color dipoles are considered in the $SU(N_c)$ representation $r_i$
and have transverse size and orientation ${\vec r}_{i\perp}$. The
longitudinal momentum fraction carried by the quark of dipole $i$ is
$z_i$.  (Here and in the following we use several times the term quark
generically for color sources in arbitrary $SU(N_c)$ representations.)
The impact parameter between the dipoles is~\cite{Dosch:1997ss}
\be
        {\vec b}_{\!\perp} 
        \,=\, {\vec r}_{1q} + (1-z_1) {\vec r}_{1\perp} 
            - {\vec r}_{2q} - (1-z_2) {\vec r}_{2\perp} 
        \,=\, {\vec r}_{1\,cm} - {\vec r}_{2\,cm} 
        \ ,
\label{Eq_impact_vector}
\ee
where ${\vec r}_{iq}$ (${\vec r}_{i\qbar}$) is the transverse position
of the quark (antiquark), ${\vec r}_{i\perp} = {\vec r}_{i\qbar} -
{\vec r}_{iq}$, and ${\vec r}_{i\,cm} = z_i {\vec r}_{iq} +
(1-z_i){\vec r}_{i\qbar}$ is the center of light-cone momenta.
Figure~\ref{Fig_loop_loop_scattering_surfaces} illustrates the (a)
space-time and (b) transverse arrangement of the dipoles.
\befig[p!]
  \begin{center}
        \epsfig{file=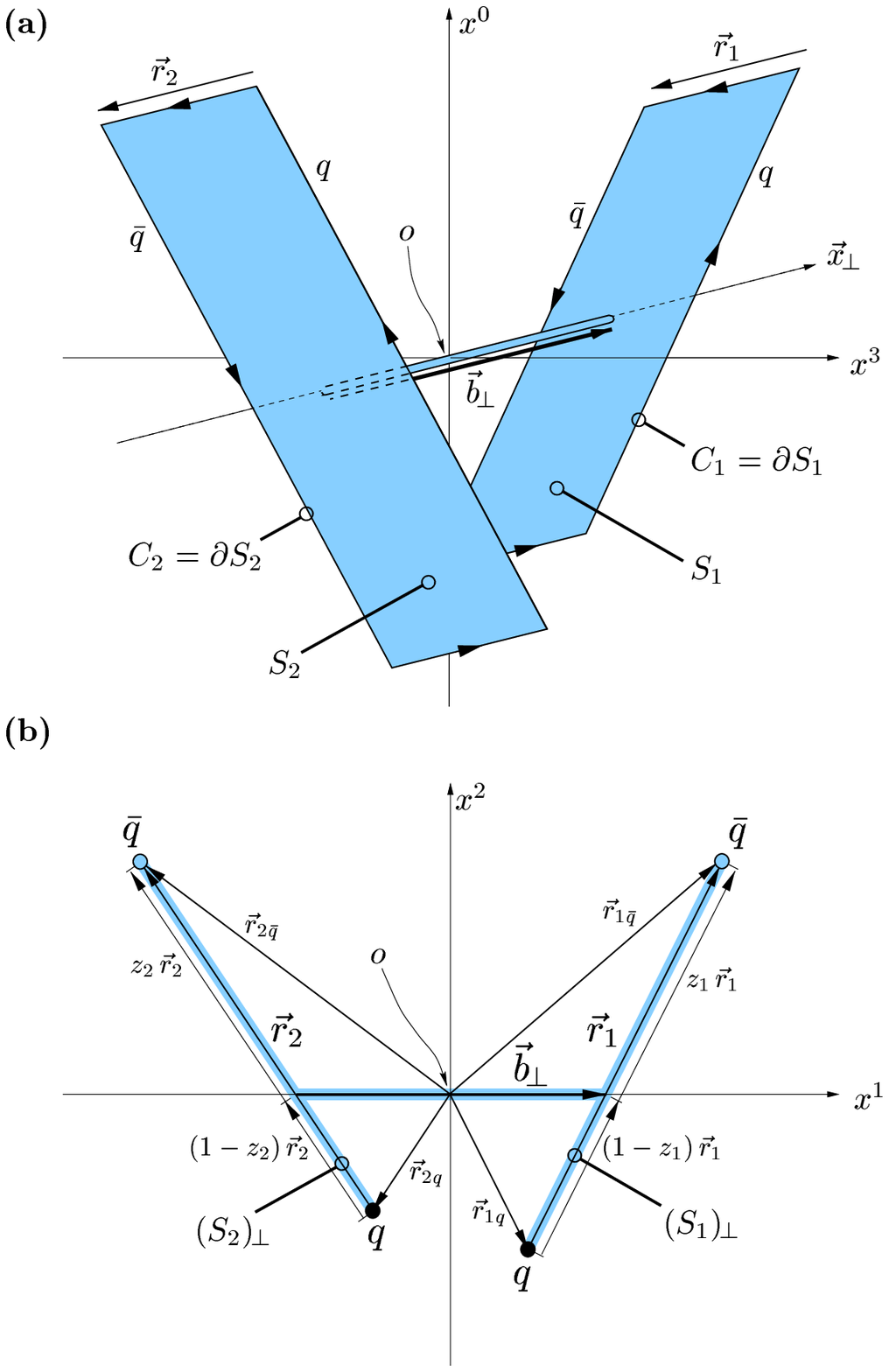,width=10.cm}
  \end{center}
\caption{\small High-energy dipole-dipole scattering in the eikonal
  approximation represented by Wegner-Wilson loops in the fundamental
  representation of $SU(N_c)$: (a) Space-time and (b) transverse
  arrangement of the Wegner-Wilson loops. The shaded areas represent
  the strings extending from the quark to the antiquark path in each
  color dipole.  The thin tube allows us to compare the gluon field
  strengths in surface $S_1$ with the gluon field strengths in surface
  $S_2$. The impact parameter $\vec{b}_{\perp}$ connects the centers
  of light-cone momenta of the dipoles.}
\label{Fig_loop_loop_scattering_surfaces}
\efig
The dipole trajectories $C_i$ defining the {\WW} loops
in~(\ref{Eq_S_DD_Minkowski}) are described as straight lines.  This is
a good approximation as long as the kinematical assumption behind the
eikonal approximation, $s \gg -t$, holds that allows us to neglect the
change of the dipole velocities $v_i = p_i/m$ in the scattering
process, where $p_i$ is the momentum and $m$ the mass of the
considered dipole.  Moreover, the paths $C_i$ are considered
light-like\footnote{In fact, exactly light-like trajectories ($\gamma
  \to \infty$) are considered in most applications of the functional
  integral approach to high-energy
  collisions~\cite{Kramer:1990tr,Dosch:1994ym,Dosch:RioLecture,Rueter:1996yb,Dosch:1997ss,Dosch:1998nw,Rueter:1998qy,Kulzinger:1999hw,Rueter:1998up,D'Alesio:1999sf,Berger:1999gu,Donnachie:2000kp,Donnachie:2001wt,Dosch:2001jg,Shoshi:2002in,Shoshi:2002ri,Shoshi:2002fq,Shoshi:2002mt}.
  A detailed investigation of the more general case of finite rapidity
  $\gamma$ can be found in~\cite{Kulzinger:2002iu}.} in line with the
high-energy limit, $m^2 \ll s \to \infty$. For the {\em hyperbolic
  angle} or {\em rapidity gap} between the dipole trajectories $\gamma
= (v_1 \cdot v_2)$ -- which is the central quantity in the analytic
continuation discussed below and also defined through $s =
4m^2\cosh^2(\gamma/2)$ -- the high-energy limit implies
\be
        \lim_{m^2\ll s\to\infty} \gamma \approx \ln(s/m^2) \to \infty 
        \ .
\label{Eq_rapidity_light-like_loops}
\ee
The QCD VEVs $\langle\ldots\rangle_M$ in the $S$-matrix
element~(\ref{Eq_S_DD_Minkowski}) represent {\em Minkowskian}
functional integrals~\cite{Nachtmann:ed.kt} in which -- as in the
Euclidean case discussed above -- the functional integration over the
fermion fields has already been carried out.

The $S$-matrix element $S^M_{DD}:=S^M_{\fundamental\fundamental}$ for
the scattering of light-like dipoles in the fundamental
$SU(N_c\!=\!3)$ representation ($r_1\!=\!r_2\!=\!\Fundamental\!=\!3$)
is the key to our unified description of hadron-hadron, photon-hadron,
and photon-photon reactions in the following chapters. With color
dipoles given by the quark and antiquark in the meson or photon or in
a simplified picture by a quark and diquark in the baryon, we describe
hadrons and photons as quark-antiquark or quark-diquark systems, i.e.\ 
fundamental $SU(3)$ dipoles, with size and orientation determined by
appropriate light-cone wave
functions~\cite{Dosch:1994ym,Dosch:RioLecture}.  Accordingly, the
$T$-matrix element for the reaction $ab \to cd$ factorizes into the
universal $S$-matrix element $S^M_{DD}$ and reaction-specific
light-cone wave functions $\psi_{a,b}$ and $\psi_{c,d}$ that describe
the ${\vec r}_i$ and $z_i$ distribution of the color
dipoles~\cite{Dosch:1994ym,Dosch:RioLecture,Nachtmann:ed.kt}
\bea
        \!\!\!\!\!\!
        &&\hspace{-1cm}
        T_{ab \rightarrow cd}(s,t) =
        2is \int \!\!d^2b_{\!\perp} 
        e^{i {\vec q}_{\!\perp} {\vec b}_{\!\perp}}
        \int \!\!dz_1 d^2r_1 \!\int \!\!dz_2 d^2r_2      
        \hphantom{\hspace*{5.cm}}   
\label{Eq_model_T_amplitude}\\ 
        & \!\! \times \!\! & 
        \psi_c^*(z_1,\vec{r}_{1\perp})\,\psi_d^*(z_2,\vec{r}_{2\perp})
        \left[1-S^M_{DD}(s,{\vec b}_{\!\perp},z_1,\vec{r}_{1\perp},z_2,\vec{r}_{2\perp})\right]
        \psi_a(z_1,\vec{r}_{1\perp})\,\psi_b(z_2,\vec{r}_{2\perp}) 
        \ . 
\nonumber
\eea
Concentrating in this work on reactions with $a = c$ and $b = d$, the
squared wave functions
$|\psi_1(z_1,\vec{r}_{1\perp})|^2:=\psi_c^*(z_1,\vec{r}_{1\perp})\,\psi_a(z_1,\vec{r}_{1\perp})$
and
$|\psi_2(z_2,\vec{r}_{2\perp})|^2:=\psi_d^*(z_2,\vec{r}_{2\perp})\,\psi_b(z_2,\vec{r}_{2\perp})$
are needed. We use for hadrons the phenomenological Gaussian wave
function~\cite{Dosch:2001jg,Wirbel:1985ji} and for photons the
perturbatively derived wave function with running quark masses
$m_f(Q^2)$ to account for the non-perturbative region of low photon
virtuality $Q^2$~\cite{Dosch:1998nw}. In Sec.~\ref{Sec_Wave_Functions}
we specify and discuss these wave functions explicitly. The scattering
of dipoles with fixed size $|\vec{r}_i|$ and fixed longitudinal quark
momentum fraction $z_i$ averaged over all orientations,
\be
        |\psi_{\!\mbox{\tiny\it D}_i}(z_i,\vec{r}_i)|^2 =
        \inv{2\pi|\vec{r}_{\!\mbox{\tiny\it D}_i}|}\,\delta
        (|\vec{r}_i|-|\vec{r}_{\!\mbox{\tiny\it D}_i}|)\,\delta
        (z_i-z_{\!\mbox{\tiny\it D}_i})
        \ ,
\label{Eq_dip_wf}
\ee
is considered in Sec.~\ref{Sec_S-Matrix_Unitarity} to show that
$S$-matrix unitarity constraints are respected in our model. For the
analytic continuation of high-energy scattering to Euclidean
space-time, we now return to the scattering of dipoles with fixed size
{\em and} orientation $\vec{r}_i$ and fixed longitudinal quark
momentum fraction $z_i$.

\section{Analytic Continuation of Dipole-Dipole Scattering}
\label{Sec_Analytic_Continuation_Dipoles}

The Euclidean approach to the described elastic scattering of dipoles
in the eikonal approximation is based on {\em Meggiolaro's analytic
  continuation} of the high-energy parton-parton scattering
amplitude~\cite{Meggiolaro:1996hf+X}. Meggiolaro's analytic
continuation has been derived in the functional integral approach to
high-energy collisions~\cite{Nachtmann:1991ua+X,Nachtmann:ed.kt} in
which parton-parton scattering is described in terms of {\WW} lines:
The Minkowskian amplitude, $g^M(\gamma,T,t)$, given by the expectation
value of two {\WW} lines, forming an hyperbolic angle $\gamma$ in
Minkowski space-time, and the Euclidean ``amplitude,''
$g^E(\Theta,T,t)$, given by the expectation value of two {\WW} lines,
forming an angle $\Theta \in [0,\pi]$ in Euclidean space-time, are
connected by the following analytic continuation in the angular
variables and the temporal extension $T$, which is needed as an IR
regulator in the case of {\WW} lines,
\bea
        g^E(\Theta,T,t) & = & g^M(\gamma\to i\Theta,T\to -iT,t) 
        \ ,
\label{Eq_gE=gM}\\
        g^M(\gamma,T,t) & = & g^E(\Theta\to -i\gamma, T\to iT,t)
        \ .
\label{Eq_gM=gE}
\eea
Generalizing this relation to {\em gauge-invariant} dipole-dipole
scattering described in terms of {\WW}
loops~\cite{Dosch:1994ym,Dosch:RioLecture,Nachtmann:ed.kt}, the IR
divergence known from the case of {\WW} lines vanishes and no finite
IR regulator $T$ is necessary. Thus, the Minkowskian $S$-matrix
element~(\ref{Eq_S_DD_Minkowski}), given by the expectation values of
two {\WW} loops, forming an hyperbolic angle $\gamma$ in Minkowski
space-time, can be computed from the Euclidean ``$S$-matrix element''
\be
        S^E_{r_1 r_2}(\Theta,{\vec b}_{\!\perp},z_1,\vec{r}_{1\perp},z_2,\vec{r}_{2\perp})
        = \lim_{T \rightarrow \infty}
        \frac{\langle W_{r_1}[C_1] W_{r_2}[C_2]\rangle_E}
        {\langle W_{r_1}[C_1]\rangle_E \langle W_{r_2}[C_2]\rangle_E}
\label{Eq_S_DD_Euclidean}
\ee
given by the expectation values of two {\WW} loops, forming an angle
$\Theta \in [0,\pi]$ in Euclidean space-time, via an analytic
continuation in the angular variable
\be
        S^M_{r_1 r_2}(\gamma\approx\ln[s/m^2],{\vec b}_{\!\perp},z_1,\vec{r}_{1\perp},z_2,\vec{r}_{2\perp})
        = S^E_{r_1 r_2}(\Theta\to -i\gamma,{\vec b}_{\!\perp},z_1,\vec{r}_{1\perp},z_2,\vec{r}_{2\perp})
        \ ,
\label{Eq_SM=SE(theta->-igamma)}
\ee
where $E$ indicates Euclidean space-time and the QCD VEVs
$\langle\ldots\rangle_E$ represent Euclidean functional integrals that
are equivalent to the ones denoted by $\langle\ldots\rangle_G$ in the
preceding sections, i.e.\ in which the functional integration over the
fermion fields has already been carried out.

\begin{figure}[t!]
\centerline{\epsfig{figure=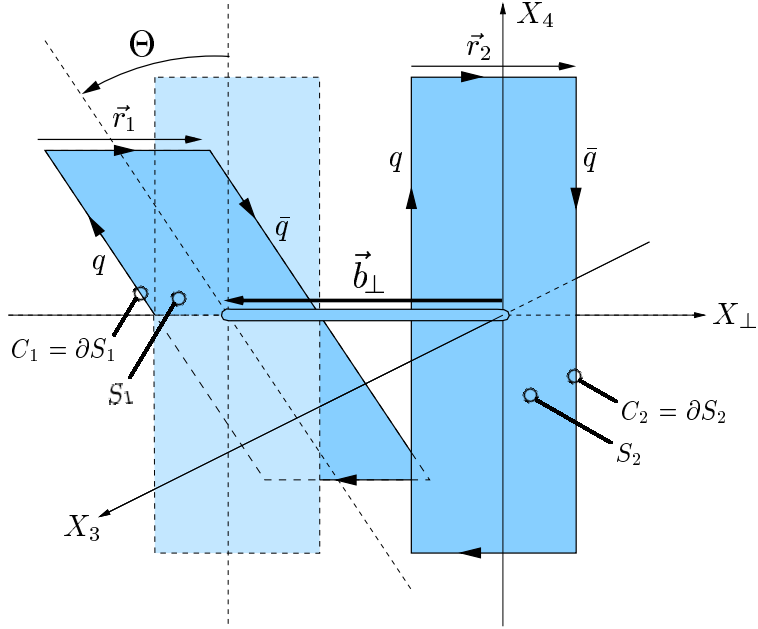,width=10.cm}}
\caption{\small 
  The loop-loop geometry necessary to compute $S^E_{r_1
    r_2}(\Theta\neq 0, \cdots)$ illustrated as a generalization of the
  geometry relevant for the computation of the van der Waals potential
  between two static dipoles ($\Theta=0$). While the potential between
  two static dipoles is computed from two loops along parallel
  ``temporal'' unit vectors, $t_1 = t_2 = (0,0,0,1)$, the Euclidean
  $S$-matrix element~(\ref{Eq_S_DD_Euclidean}) involves the tilting of
  one of the two loops, e.g.\ the tilting of $t_1$ by the angle
  $\Theta$ towards the $X_3$\,-\,axis, $t_1
  =(0,0,-\sin\Theta,\cos\Theta)$.}
\label{Fig_tilted_loops}
\end{figure}

The angle $\Theta$ is best illustrated in the relation of the
Euclidean $S$-matrix element~(\ref{Eq_S_DD_Euclidean}) to the van der
Waals potential between two static dipoles $V_{r_1 r_2}(\Theta=0,
\vec{b}, z_1, \vec{r}_1, z_2, \vec{r}_2)$, discussed in
Sec.~\ref{Sec_VDW_Potential},
\be
        S^E_{r_1 r_2}(\Theta,\vec{b}_{\!\perp},z_1,\vec{r}_{1\perp},z_2,\vec{r}_{2\perp})
        = \lim_{T \rightarrow \infty}
        \exp\!\left[-\,T\,V_{r_1 r_2}(\Theta,\vec{b}_{\!\perp},z_1,\vec{r}_{1\perp},z_2,\vec{r}_{2\perp})\right]
        \ .
\label{Eq_S_DD_<->_V_DD}
\ee
Figure~\ref{Fig_tilted_loops} shows the loop-loop geometry necessary
to compute $S^E_{r_1 r_2}(\Theta\neq 0, \cdots)$ and how it is
obtained by generalizing the geometry relevant for the computation of
the potential between two static dipoles ($\Theta=0$): While the
potential between two static dipoles is computed from two loops along
parallel ``temporal'' unit vectors, $t_1 = t_2 = (0,0,0,1)$, the
Euclidean $S$-matrix element~(\ref{Eq_S_DD_Euclidean}) involves the
tilting of one of the two loops, e.g.\ the tilting of $t_1$ by the
angle $\Theta$ towards the $X_3$\,-\,axis, $t_1 =
(0,0,-\sin\Theta,\cos\Theta)$. The ``temporal'' unit vectors $t_i$ are
also discussed in Appendix~\ref{Sec_Parameterizations} together with
another illustration of the tilting angle $\Theta$.

Since the Euclidean $S$-matrix element~(\ref{Eq_S_DD_Euclidean})
involves only configurations of {\WW} loops in Euclidean space-time
and {\em Euclidean} functional integrals, it can be computed directly
on a Euclidean lattice. First attempts in this direction have been
carried out but only very few signals could be extracted, while most
of the data was dominated by noise~\cite{DiGiacomo:2002PC}. Once
precise results are available, the analytic
continuation~(\ref{Eq_SM=SE(theta->-igamma)}) will allow us to access
hadronic high-energy reactions directly in lattice QCD, i.e.\ within a
non-perturbative description of QCD from first principles. More
generally, the presented gauge-invariant analytic
continuation~(\ref{Eq_SM=SE(theta->-igamma)}) makes any approach
limited to a Euclidean formulation of the theory applicable for
investigations of high-energy reactions.  Indeed, Meggiorlaro's
analytic continuation has already been used to access high-energy
scattering from the supergravity side of the AdS/CFT
correspondence~\cite{Janik:2000zk+X}, which requires a positive
definite metric in the definition of the minimal
surface~\cite{Rho:1999jm}, and to examine the effect of instantons on
high-energy scattering~\cite{Shuryak:2000df+X}.

\section[Dipole-Dipole Scattering in the Loop-Loop Correlation Model]
{Dipole-Dipole Scattering in the LLCM}
\label{Sec_DD_Scattering_LLCM}

Let us now perform the analytic continuation explicitly in our
Euclidean loop-loop correlation model. For the scattering of two color
dipoles in the {\em fundamental representation} of $SU(N_c)$, the
Euclidean $S$-matrix element becomes with the
VEVs~(\ref{Eq_final_result_<W[C]>})
and~(\ref{Eq_final_Euclidean_result_<W[C1]W[C2]>_fundamental})
\bea
        S^E_{DD}(\Theta,\vec{b}_{\!\perp},z_1,\vec{r}_{1\perp},z_2,\vec{r}_{2\perp})
         && \!\!\!\!\!\!
        :=\,\,S^E_{\fundamental\fundamental}
        (\Theta,\vec{b}_{\!\perp},z_1,\vec{r}_{1\perp},z_2,\vec{r}_{2\perp})
\nonumber\\
        && \hspace{-4cm} = \lim_{T \rightarrow \infty}
        \left(
        \frac{N_c\!+\!1}{2N_c}\exp\!\left[-\frac{N_c\!-\!1}{2 N_c}\chi_{S_1 S_2}\right]
        + \frac{N_c\!-\!1}{2N_c}\exp\!\left[ \frac{N_c\!+\!1}{2 N_c}\chi_{S_1 S_2}\right]
        \right)
        \ ,
\label{Eq_S_DD_1}
\eea
where $\chi_{S_i S_j}$ -- defined in~(\ref{Eq_chi_Si_Sj}) --
decomposes into a perturbative ($\pert$) and non-per\-tur\-ba\-tive
($\nprt$) component according to our decomposition of the gluon field
strength correlator~(\ref{Eq_F_decomposition}),
\be
        \chi_{S_1 S_2} 
        \,\,=\,\, 
        \chi_{S_1 S_1}^{\pert} 
        \,+\, \chi_{S_1 S_2}^{\nprt}
        \,\,=\,\, 
        \chi_{S_1 S_2}^{\pert} 
        \,+\, \left(\chi_{S_1 S_2}^{\nprt\,\,nc} 
          \,+\, \chi_{S_1 S_2}^{\nprt\,\,c}\right)
        \ .
\label{Eq_chi_decomposition}        
\ee
In the limit $T_1=T_2=T\to\infty$ and for $\Theta \in [0,\pi]$, the
components read
\be
        \chi_{S_1 S_2}^{\pert} 
        =\cot\Theta\,\,\chi^{\pert}
        \,\, , \quad
        \chi_{S_1 S_2}^{\nprt\,\,nc} 
        =\cot\Theta\,\,\chi^{\nprt\,\,nc}
        \,\, , \quad
        \chi_{S_i S_j}^{\nprt\,\,c} 
        =\cot\Theta\,\,\chi^{\nprt\,\,c}
\label{Eq_S_DD_p_npc_npnc_E}
\ee
with
\bea
        \!\!\!\!\!\!\!\!\!\!\!\!\!
        \chi^{\pert} &\!\!=\!\!& 
        \left[ 
        g^2 D^{\prime\,(2)}_{\pert}
        \left(|\vec{r}_{1q}-\vec{r}_{2\qbar}|\right)
        +g^2 D^{\prime\,(2)}_{\pert}
        \left(|\vec{r}_{1\qbar}-\vec{r}_{2q}|\right)
        \right.
\nonumber \\
        \!\!\!\!\!\!\!\!\!\!\!\!\!
        &&
        \left.
        -\,g^2 D^{\prime\,(2)}_{\pert}
        \left(|\vec{r}_{1q}-\vec{r}_{2q}|\right)
        -g^2 D^{\prime\,(2)}_{\pert}
        \left(|\vec{r}_{1\qbar}-\vec{r}_{2\qbar}|\right)
        \right]
\label{Eq_S_DD_chi_p_M}\\ 
        \!\!\!\!\!\!\!\!\!
        \chi^{\nprt\,\,nc} &\!\!=\!\!& 
        \frac{\pi^2 G_2 (1-\kappa)}{3(N_c^2-1)} 
        \left[ 
        D^{\prime\,(2)}_1
        \left(|\vec{r}_{1q}-\vec{r}_{2\qbar}|\right)
        +D^{\prime\,(2)}_1
        \left(|\vec{r}_{1\qbar}-\vec{r}_{2q}|\right)
         \right.
\nonumber \\
        \!\!\!\!\!\!\!\!\!\!\!\!\!
        &&
        \hphantom{-\frac{\pi^2 G_2 (1-\kappa)}{3(N_c^2-1)}}
        \left.
        -\,D^{\prime\,(2)}_1
        \left(|\vec{r}_{1q}-\vec{r}_{2q}|\right) 
        -D^{\prime\,(2)}_1
        \left(|\vec{r}_{1\qbar}-\vec{r}_{2\qbar}|\right)
       \right]
\label{Eq_S_DD_chi_np_nc_M}\\
        \!\!\!\!\!\!\!\!\!\!\!\!\!
        \chi^{\nprt\,\,c} &\!\!=\!\!& 
        \frac{\pi^2 G_2 \kappa}{3(N_c^2-1)}\,
        \left(\vec{r}_1\cdot\vec{r}_2\right)
        \int_0^1 \! dv_1 \int_0^1 \! dv_2 \,\, 
        D^{(2)}\left(|\vec{r}_{1q}\! +\! v_1\vec{r}_{1\perp} 
        \!-\! \vec{r}_{2q}\! -\! v_2\vec{r}_{2\perp}|\right)
\label{Eq_S_DD_chi_np_c_M}
\eea
as derived explicitly in Appendix~\ref{Sec_Chi_Computation} with the
minimal surfaces illustrated in Fig.~\ref{Fig_tilted_loops}. In
Eq.~(\ref{Eq_S_DD_chi_p_M}) the shorthand notation $g^2
D^{\prime\,(2)}_{\pert}(|\vec{Z_\perp}|) =
g^2(|\vec{Z_\perp}|)\,D^{\prime\,(2)}_{\pert}(|\vec{Z_\perp}|)$ is
used with $g^2(|\vec{Z_\perp}|)$ again understood as the running
coupling~(\ref{Eq_g2(z_perp)}). The transverse Euclidean correlation
functions
\be
        D_x^{(2)}(\vec{Z}^2)      
        := \int \frac{d^4K}{(2\pi)^2}\,e^{iKZ}\,
        \tilde{D}_x(K^2)\,\delta(K_3)\,\delta(K_4)
\label{Eq_D(2)x}
\ee
are obtained from the (massive) gluon
propagator~(\ref{Eq_massive_gluon_propagator}) and the exponential
correlation function~(\ref{Eq_SVM_correlation_functions})
\bea
        \!\!\!\!\!\!\!\!\!\!\!\!\!\!\!\!
        D^{\prime\,(2)}_{\pert}(\vec{Z}_{\!\perp}^2)
        & \!\!=\!\! &
        \inv{2\pi} K_0\left(m_G |\vec{Z}_{\!\perp}|\right)
\label{Eq_D'(2)p(z,mg)}\\ 
        \!\!\!\!\!\!\!\!\!\!\!\!\!\!\!\!
        D^{\prime\,(2)}_1(\vec{Z}_{\!\perp}^2)
        & \!\!=\!\! &
        \pi a^4  \Big(
        3 \!+\! 3\frac{|\vec{Z}_{\!\perp}|}{a} \!+\! \frac{|\vec{Z}_{\!\perp}|^2}{a^2} 
        \Big)
        \exp\!\Big(\!-\frac{|\vec{Z}_{\!\perp}|}{a}\Big)
\label{Eq_D'(2)np_nc(z,a)}\\
        \!\!\!\!\!\!\!\!\!\!\!\!\!\!\!\!
        D^{(2)}(\vec{Z}_{\!\perp}^2)      
        & \!\!=\!\! &
        2 \pi a^2 
        \Big(1\!+\!\frac{|\vec{Z}_{\!\perp}|}{a}\Big) 
        \exp\!\Big(\!-\frac{|\vec{Z}_{\!\perp}|}{a}\Big)
\label{Eq_D(2)np_c(z,a)}
\eea
With the full $\Theta$-dependence exposed
in~(\ref{Eq_S_DD_p_npc_npnc_E}), the analytic
continuation~(\ref{Eq_SM=SE(theta->-igamma)}) reads
\be
        \chi_{S_1 S_2} = \cot\Theta\,\,\chi
        \quad\underrightarrow{\,\,\Theta\to -i\gamma\,\,\,\,}\quad
        \cot(-i\gamma)\,\chi
        \quad\underrightarrow{\,\,s\to\infty\,\,\,\,}\quad
        i\chi
\label{Eq_analytic_continuation_of_chi}
\ee
and leads to the desired Minkowskian $S$-matrix element for elastic
dipole-dipole scattering ($DD$) in the high-energy limit in which the
dipoles move on the light-cone
\bea
        \lim_{s \rightarrow \infty} 
        S_{DD}^{M}(s,{\vec b}_{\!\perp},z_1,\vec{r}_{1\perp},z_2,\vec{r}_{2\perp}) 
        &&\!\!\!\!\!\!:=\,\, 
        \lim_{s \rightarrow \infty} 
        S^M_{\fundamental\fundamental}(s,{\vec b}_{\!\perp},z_1,\vec{r}_{1\perp},z_2,\vec{r}_{2\perp})
\nonumber\\
        && \hspace{-4cm} = 
        S^E_{DD}(\cot\Theta \to i,{\vec b}_{\!\perp},z_1,\vec{r}_{1\perp},z_2,\vec{r}_{2\perp})
\nonumber\\
        && \hspace{-4cm} = 
        \lim_{T \rightarrow \infty}
        \left(
        \frac{N_c\!+\!1}{2N_c}\exp\!\left[-i\frac{N_c\!-\!1}{2N_c}\chi\right]
        + \frac{N_c\!-\!1}{2N_c}\exp\!\left[i\frac{N_c\!+\!1}{2N_c}\chi\right]
        \right)
\label{Eq_S_DD_1_M}
\eea
where $\chi =\chi^{\pert}+\chi^{\nprt\,\,nc}+\chi^{\nprt\,\,c}$
with~(\ref{Eq_S_DD_chi_p_M}), (\ref{Eq_S_DD_chi_np_nc_M}),
and~(\ref{Eq_S_DD_chi_np_c_M}).

It is striking that exactly the same result has been obtained
in~\cite{Shoshi:2002in}\footnote{To see this identity, recall that
  $\langle W[C]\rangle = 1$ for light-like loops and consider
  in~\cite{Shoshi:2002in} the result~(2.30) for the loop-loop
  correlation function (2.3) together with the $\chi$-function (2.40)
  and its components given in (2.49), (2.54), and (2.57) with the
  transverse Minkowskian correlation functions (2.50), (2.55), and
  (2.58).} with the alternative analytic continuation introduced for
applications of the SVM to high-energy
reactions~\cite{Kramer:1990tr,Dosch:1994ym,Dosch:RioLecture}. In this
complementary approach the gauge-invariant bilocal gluon field
strength correlator is analytically continued from Euclidean to
Minkowskian space-time by the substitution $\delta_{\mu\rho}
\rightarrow - g_{\mu\rho}$ and the analytic continuation of the
Euclidean correlation functions to real time $D^E_x(Z^2) \rightarrow
D^M_x(z^2)$. In the subsequent steps, one finds $\langle W[C]\rangle_M
= 1$ due to the light-likeness of the loops and that the longitudinal
correlations can be integrated out $\langle
W_{r_1}[C_1]W_{r_2}[C_2]\rangle_M = f(s,{\vec b}_{\!\perp},\cdots)$.
One is left with exactly the Euclidean correlations in transverse
space that have been obtained above. This confirms the analytic
continuation used in the earlier LLCM investigations in Minkowski
space-time~\cite{Shoshi:2002in,Shoshi:2002ri,Shoshi:2002fq,Shoshi:2002mt}
and in all earlier SVM applications to high-energy
scattering~\cite{Kramer:1990tr,Dosch:1994ym,Dosch:RioLecture,Rueter:1996yb,Dosch:1997ss,Dosch:1998nw,Rueter:1998qy,Kulzinger:1999hw,Rueter:1998up,D'Alesio:1999sf,Berger:1999gu,Kulzinger:2002iu}.

In the limit of small $\chi$-functions, $|\chi^{\pert}| \ll 1$ and
$|\chi^{\nprt}| \ll 1$, (\ref{Eq_S_DD_1_M}) reduces to
\be
        \lim_{s \rightarrow \infty} 
        S_{DD}^{M}(s,{\vec b}_{\!\perp},z_1,\vec{r}_{1\perp},z_2,\vec{r}_{2\perp})
        \approx 1 + \frac{N_c^2-1}{8 N_c^2}\,\chi^2 
        = 1 + \frac{C_2(\Fundamental)}{4 N_c}\,\chi^2 
        \ .
\label{Eq_S_DD_M_small_chi}
\ee
The perturbative correlations, $(\chi^{\pert})^2$, describe the
well-known {\em two-gluon exchange}
contribution~\cite{Low:1975sv+X,Gunion:iy} to dipole-dipole
scattering, which is, of course, an important successful cross-check
of the presented Euclidean approach to high-energy scattering. The
non-perturbative correlations, $(\chi^{\nprt})^2$, describe the
corresponding non-perturbative two-point interactions that contain
contributions of the confining QCD strings to dipole-dipole
scattering.  We have analyzed these string contributions
systematically as manifestations of confinement in high-energy
scattering reactions and have indeed found a new characteristic
structure (different from the perturbative dipole factors) in momentum
space~\cite{Shoshi:2002fq}. This analysis has also shown explicitly
that the non-perturbative contribution governs -- as expected -- the
region of low transverse momenta $|\vec{k}_{\!\perp}|$.  Here, we
focus on the structure in space-time representation and refer the
reader for complementary insights to our momentum-space
analysis~\cite{Shoshi:2002fq}.

As evident from the $v_1$ and $v_2$ integrations
in~(\ref{Eq_S_DD_chi_np_c_M}) and
Fig.~\ref{Fig_loop_loop_scattering_surfaces}b, there are contributions
from the transverse projections of the minimal surfaces
$(S_{1,2})_{\!\perp}$ connecting the quark and antiquark in each of
the two dipoles. These are the contributions that we interpret as
manifestations of the strings confining the quark and antiquark in
each dipole. We thus understand the confining component
$\chi_{c}^{\nprt}$ as a {\em string-string interaction}.
Interestingly, we have found in dipole-hadron and dipole-photon
interactions that the strings confining the quark-antiquark pair in
the dipole can represented as an integral over stringless dipoles with
a given dipole number density. As already mentioned, this {\em
  decomposition of the confining string into dipoles} even allows us
to compute unintegrated gluon distributions of hadrons and photons and
thus gives new insights into the microscopic structure of the
non-perturbative SVM~\cite{Shoshi:2002fq}.

Both non-perturbative components, $\chi^{\nprt}_c$ and
$\chi^{\nprt}_{nc}$, show {\em color transparency} for small dipoles,
i.e.\ a dipole-dipole cross section with
$\sigma_{DD}(\vec{r}_1,\vec{r}_2) \propto |\vec{r}_1|^2|\vec{r}_2|^2$
for $|\vec{r}_{1,2}| \to 0$, as known for the perturbative
case~\cite{Nikolaev:1991ja}. This can be seen by
squaring~(\ref{Eq_S_DD_chi_np_nc_M}) and~(\ref{Eq_S_DD_chi_np_c_M}) to
obtain the leading terms in the $T$-matrix element for small dipoles;
see~(\ref{Eq_S_DD_M_small_chi}).

Due to the truncation of the cumulant expansion in the Gaussian
approximation, a considerable dependence of $\chi_{c}^{\nprt}$ on the
specific surface choice is observed. In fact, a different and more
complicated result for $\chi_{c}^{\nprt}$ was obtained with the
pyramid mantle choice for the surfaces $S_{1,2}$ in earlier
applications of the \SVM\ to high-energy
scattering~\cite{Dosch:1994ym,Dosch:RioLecture,Rueter:1996yb,Dosch:1997ss,Dosch:1998nw,Rueter:1998qy,Kulzinger:1999hw,Rueter:1998up,D'Alesio:1999sf,Berger:1999gu,Dosch:2001jg}.
However, we use minimal surfaces in line with our model applications
in Euclidean space-time discussed in the previous chapter.  Moreover,
the simplicity of the minimal surfaces allows us to give an analytic
expression for the leading term of the non-perturbative dipole-dipole
cross section~\cite{Shoshi:2002fq}.  Phenomenologically, in comparison
with pyramid mantles, the description of the slope parameter $B(s)$,
the differential elastic cross section $d\sigma^{el}/dt(s,t)$, and the
elastic cross section $\sigma^{el}(s)$ can be improved with minimal
surfaces as shown in Chap.~\ref{Sec_Comparison_Data}. In contrast to
the confining component $\chi_{c}^{\nprt}$, the non-confining
components, $\chi_{nc}^{\nprt}$ and $\chi^{\pert}$, depend only on the
transverse position between the quark and antiquark of the two dipoles
and are therefore independent of the surface choice.

With the insights from the small-$\chi$ limit, one sees clearly that
the full $S$-matrix element~(\ref{Eq_S_DD_1_M}) describes {\em
  multiple gluonic interactions}.  Indeed, the higher order terms in
the expansion of the exponential functions in~(\ref{Eq_S_DD_1_M}) are
crucial to respect $S$-matrix unitarity constraints in impact
parameter space~\cite{Berger:1999gu,Shoshi:2002in} as shown explicitly
in Chap.~\ref{Sec_Impact_Parameter}.

Concerning the energy dependence, the $S$-matrix
element~(\ref{Eq_S_DD_1_M}) leads to energy-independent cross sections
in contradiction to the experimental observation. Although
disappointing from the phenomenological point of view, this is not
surprising since our approach does not describe explicit gluon
radiation needed for a non-trivial energy dependence. However, based
on the $S$-matrix element~(\ref{Eq_S_DD_1_M}), a phenomenological
energy dependence can be constructed -- see
Sec.~\ref{Sec_Energy_Dependence} -- that allows a unified description
of high-energy hadron-hadron, photon-hadron, and photon-photon
reactions and an investigation of saturation effects in hadronic cross
sections manifesting $S$-matrix
unitarity~\cite{Shoshi:2002in,Shoshi:2002ri,Shoshi:2002mt}.  This, of
course, can only be an intermediate step. For a more fundamental
understanding of hadronic high-energy reactions in our model, gluon
radiation and quantum evolution have to be implemented explicitly.

Note that $\chi = \chi_{c}^{\nprt} + \chi_{nc}^{\nprt} + \chi^{\pert}$
is a real-valued function. Since, in addition, the wave functions
$|\psi_i(z_i,\vec{r}_{i\perp})|^2$ used in this work -- see
Sec.~\ref{Sec_Wave_Functions} -- are invariant under the replacement
$(\vec{r}_{i\perp} \rightarrow -\vec{r}_{i\perp}, z_i \rightarrow
1-z_i)$, the $T$-matrix element~(\ref{Eq_model_T_amplitude}) with
$S^M_{DD}$ given in~(\ref{Eq_S_DD_1_M}) becomes purely imaginary and
reads for $N_c=3$
\bea
        \!\!\!\!\!\!\!\!\!\!\!\!\!
        T(s,t) 
        & = & 2is \int \!\!d^2b_{\!\perp} 
                e^{i {\vec q}_{\!\perp} {\vec b}_{\!\perp}}
                \int \!\!dz_1 d^2r_1 \!
                \int \!\!dz_2 d^2r_2 \,\,
                |\psi_1(z_1,\vec{r}_{1\perp})|^2   \,\,
                |\psi_2(z_2,\vec{r}_{2\perp})|^2       
\label{Eq_model_purely_imaginary_T_amplitude}\\    
        && \!\!\!\!\!\!\!\!\!\!
        \times 
        \left[1-\frac{2}{3} 
        \cos\!\left(\frac{1}{3}
        \chi({\vec b}_{\!\perp},z_1,\vec{r}_{1\perp},z_2,\vec{r}_{2\perp})\!\right)
        - \frac{1}{3}
        \cos\!\left(\frac{2}{3}
        \chi({\vec b}_{\!\perp},z_1,\vec{r}_{1\perp},z_2,\vec{r}_{2\perp})\!\right)
        \right].
\nonumber
\eea
The real part averages out in the integration over ${\vec r}_i$ and
$z_i$ because of
\be
        \chi(\vec{b}_{\!\perp},1-z_1,-\vec{r}_{1\perp},z_2,\vec{r}_{2\perp})
        = - \chi(\vec{b}_{\!\perp},z_1,\vec{r}_{1\perp},z_2,\vec{r}_{2\perp})
        \ ,
\label{Eq_odd_eikonal_function}
\ee
which can be seen directly
from~(\ref{Eq_S_DD_chi_p_M}),(\ref{Eq_S_DD_chi_np_nc_M}), and
(\ref{Eq_S_DD_chi_np_c_M}) as $(\vec{r}_{1\perp} \rightarrow
-\vec{r}_{1\perp}, z_1 \rightarrow 1-z_1)$ implies $\vec{r}_{1q}
\rightarrow \vec{r}_{1\qbar}$. In physical terms, $(\vec{r}_{i\perp}
\rightarrow -\vec{r}_{i\perp}, z_i \rightarrow 1-z_i)$ corresponds to
{\em charge conjugation}, i.e.\ the replacement of each parton with
its antiparton and the associated reversal of the loop direction.
Consequently, the
$T$-matrix~(\ref{Eq_model_purely_imaginary_T_amplitude}) describes
only charge conjugation $C = +1$ exchange. Since in our quenched
approximation purely gluonic interactions are modeled,
(\ref{Eq_model_purely_imaginary_T_amplitude}) describes only
pomeron\footnote{Odderon $C = -1$ exchange is excluded in our model.
  It would survive in the following cases: (a) Wave functions are used
  that are not invariant under the transformation $(\vec{r}_i
  \rightarrow -\vec{r}_i, z_i \rightarrow 1-z_i)$. (b) The proton is
  described as a system of three quarks with finite separations
  modeled by three loops with one common light-like line. (c) The
  Gaussian approximation that enforces the truncation of the cumulant
  expansion is relaxed and additional higher cumulants are taken into
  account.}  but not reggeon exchange.

Although the scattering of two color dipoles in the fundamental
representation of $SU(N_c)$ is the most relevant case, we can derive
immediately also the Minkowskian $S$-matrix element for the scattering
of a fundamental ($D$) and an adjoint dipole (``glueball''
$\glueball$) in the Euclidean LLCM.
Using~(\ref{Eq_final_Euclidean_result_<Wf[C1]Wa[C2]>}) and proceeding
otherwise as above, we find in the high-energy limit
\bea
        && \!\!\!\!\!\!\!\!
        \lim_{s \rightarrow \infty} 
        S^M_{D\,\glueball}(s, \vec{b}, z_1, \vec{r}_{1\perp}, z_2, \vec{r}_{2\perp}) 
        \,\,:=\,\, \lim_{s \rightarrow \infty} 
        S^M_{\fundamental\,\adjoint}(\Theta, \vec{b}, z_1, \vec{r}_{1\perp}, z_2, \vec{r}_{2\perp}) 
\label{Eq_S_fa_final_result}\\
        && \!\!\!\!\!\!\!\!
        = \lim_{T \rightarrow \infty}
        \Bigg(\,\inv{N_c^2\!-\!1}\,\exp\!\Big[i\,\frac{N_c}{2}\,\chi\Big]
        +\frac{N_c\!+\!2}{2(N_c\!+\!1)}\exp\!\Big[\!-\,i\,\inv{2}\,\chi\Big]
        +\frac{N_c\!-\!2}{2(N_c\!-\!1)}\exp\!\Big[i\,\inv{2}\,\chi\Big]
        \Bigg)
        \ .
\nonumber
\eea
where $\chi =\chi^{\pert}+\chi^{\nprt\,nc}+\chi^{\nprt\,c}$
with~(\ref{Eq_S_DD_chi_p_M}), (\ref{Eq_S_DD_chi_np_nc_M}),
and~(\ref{Eq_S_DD_chi_np_c_M}).

\section{Comments on the QCD van der Waals Potential}
\label{Sec_VDW_Potential}

Finally, we would like to comment on the {\em QCD van der Waals
  interaction} between two color dipoles, which is -- as mentioned
together with~(\ref{Eq_S_DD_<->_V_DD}) -- related to the Euclidean
$S$-matrix element in the limiting case of $\Theta=0$: The QCD van der
Waals potential between two static dipoles reads in terms of {\WW}
loops~\cite{Appelquist:1978rt,Bhanot:1979af}
\be
        V_{r_1 r_2}(\Theta=0, \vec{b}, z_1=1/2, \vec{r}_1, z_2=1/2, \vec{r}_2) =
        - \lim_{T \rightarrow \infty} \frac{1}{T} 
        \ln \frac{\langle W_{r_1}[C_1] W_{r_2}[C_2] \rangle}
        {\langle W_{r_1}[C_1] \rangle \langle W_{r_2}[C_2] \rangle}
        \ .
\label{Eq_V_DD}        
\ee
In this limit ($\Theta=0$) intermediate octet states and their limited
lifetime become important as is well known from perturbative
computations of the QCD van der Waals potential between two static
color dipoles~\cite{Appelquist:1978rt,Bhanot:1979af,Peskin:1979va+X}:
Working with static dipoles, i.e.\ infinitely heavy color sources,
there is an energy degeneracy between the intermediate octet states
and the initial (final) singlet states that leads for perturbative
two-gluon exchange to a linear divergence in $T$ as $T\to\infty$. This
IR divergence can be lifted by introducing manually an energy gap
between the singlet ground state and the excited octet state and thus
a limit on the lifetime of the intermediate octet
state~\cite{Appelquist:1978rt,Bhanot:1979af,Peskin:1979va+X}.

In the perturbative limit of $g^2\to 0$ and $T$ large but finite,
i.e.\ $\chi^{\pert} \ll 1$, the perturbative component of our model
describes the two-gluon exchange contribution to the van der Waals
potential which is plagued by the discussed IR divergence resulting
from the static limit. In the more general case of $g^2$ finite and
$T\to\infty$, which does not exclude non-perturbative physics, one
cannot use the small-$\chi$ limit and multiple gluonic interactions
become important. Here our perturbative component describes multiple
gluon exchanges that reduce to an effective one-gluon exchange
contribution to the van der Waals potential whose interaction range
($\propto 1/m_G$) contradicts the common expectations. Indeed, it is
also in contradiction to our results for the glueball mass
$M_{\glueball}$ which determines the interaction range ($\propto
1/M_{\glueball}$) between two color dipoles for large dipole
separations. As already mentioned in Sec.~\ref{Sec_QCD_Components}, we
find for the perturbative component, $M_{\glueball}^{\pert} = 2 m_G$,
i.e.\ half of the interaction range of one-gluon exchange, by
computing the exponential decay of the correlation of two small
quadratic loops $P^{\alpha \beta}_{r_i}$ for large Euclidean times
$\tau\to\infty$
\be
         M_{\glueball} :=
        - \lim_{\tau \rightarrow \infty} \frac{1}{\tau} 
        \ln \frac{\langle P_{r_1}^{\alpha \beta}(0) P_{r_2}^{\alpha \beta}(\tau)\rangle}
        {\langle P_{r_1}^{\alpha\beta}(0) \rangle 
         \langle P_{r_2}^{\alpha\beta}(\tau) \rangle}
        \ .
\label{Eq_Glueball_mass}
\ee
Note that we find for the non-perturbative component,
$M_{\glueball}^{\nprt} = 2/a$, which is smaller than
$M_{\glueball}^{\pert} = 2 m_G$ with the LLCM parameters and thus
governs the long range correlations in the LLCM.

Thus, for a meaningful investigation of the QCD van der Waals forces
within our model, one has to go beyond the static limit in order to
describe the limited lifetime of the intermediate octet states
appropriately. This we postpone for future work since the focus in
this work is on high-energy scattering where the gluons are always
exchanged within a short time interval due to the light-likeness of
the scattered particles and the finite correlation lengths.
Nevertheless, going beyond the static limit in the dipole-dipole
potential means going beyond the eikonal approximation in high-energy
scattering and it is, of course, of utmost importance to see how such
generalizations alter our results.


%
\cleardoublepage
%
\chapter{Hadronic Wave Functions and Universal Energy Dependence}
\label{Sec_High-Energy_Scattering}

In this chapter hadron and photon wave functions are provided and a
universal energy dependence is constructed. Together with the
$T$-matrix element~(\ref{Eq_model_purely_imaginary_T_amplitude}),
these ingredients are crucial for our unified description of
hadron-hadron, photon-hadron, and photon-photon reactions in the
following chapters. The model parameters adjusted in fits to
experimental data are summarized at the end of this chapter.

\section{Hadron and Photon Wave Functions}
\label{Sec_Wave_Functions}

The light-cone wave functions $\psi_i(z_i,\vec{r}_i)$ provide the
distribution of transverse size and orientation ${\vec r}_{i}$ and
longitudinal quark momentum fraction $z_i$ to the light-like
Wegner-Wilson loops $W[C_i]$ that represent the scattering
color dipoles. In this way, they specify the projectiles as mesons,
baryons described as quark-diquark systems, or photons that fluctuate
into a quark-antiquark pair before the interaction.

\subsection*{The Hadron Wave Function}

In this work mesons and baryons are assumed to have a quark-antiquark
and quark-diquark valence structure, respectively. As quark-diquark
systems are equivalent to quark-antiquark systems~\cite{Dosch:1989hu},
this allows us to model not only mesons but also baryons as color
dipoles represented by Wegner-Wilson loops. To characterize mesons and
baryons, we use the phenomenological Gaussian Wirbel-Stech-Bauer
ansatz~\cite{Wirbel:1985ji}
\be
        \psi_h(z_i,\vec{r}_i) 
        = \sqrt{\frac{z_i(1-z_i)}{2 \pi S_h^2 N_h}}\, 
        e^{-(z_i-\inv{2})^2 / (4 \Delta z_h^2)}\,  
        e^{-|\vec{r}_i|^2 / (4 S_h^2)} 
        \ ,
\label{Eq_hadron_wave_function}
\ee
where the hadron wave function normalization to unity
\be
        \int \!\!dz_i d^2r_i \ |\psi_i(z_i,\vec{r}_i)|^2 = 1  
        \ ,
\label{Eq_hadron_wave_function_normalization}
\ee
requires the normalization constant
\be
        N_h = \int_0^1 dz_i \ z_i(1-z_i) \ e^{-(z_i-\inv{2})^2 / (2
        \Delta z_h^2)} 
        \ .
\label{Eq_N_h}
\ee
The different hadrons considered -- protons, pions, and kaons -- are
specified by $\Delta z_h$ and $S_h$ providing the width for the
distributions of the longitudinal momentum fraction carried by the
quark $z_i$ and transverse spatial extension $|\vec{r}_i|$,
respectively. In this work the extension parameter $S_h$ is a fit
parameter that should resemble approximately the electromagnetic
radius of the corresponding hadron~\cite{Dosch:2001jg}, while $\Delta
z_h = w/(\sqrt{2}\,m_h)$~\cite{Wirbel:1985ji} is fixed by the hadron
mass $m_h$ and the value $w = 0.35 - 0.5\,\GeV$ extracted from
experimental data. We find for (anti-)protons $\Delta z_p = 0.3$ and
$S_p = 0.86\,\fm$, for pions $\Delta z_{\pi} = 2$ and $S_{\pi} =
0.607\,\fm$, and for kaons $\Delta z_{K} = 0.57$ and $S_{K} =
0.55\,\fm$ which are the values used in the main text. For convenience
they are summarized in Table~\ref{Tab_Hadron_Parameters}.
\begin{table}
\caption{\small Hadron Parameters} 
\vspace{0.3cm}
\centering      
\begin{tabular}{|l|l|l|}\hline
Hadron        & $\Delta z_h$    & $S_h\;[\fm]$  \\ [1ex] \hline\hline       
$p, \bar{p}$    & $0.3$           & $0.86$  \\ \hline
$\pi^{\pm}$     & $2$             & $0.607$ \\ \hline
$K^{\pm}$      & $0.57$          & $0.55$ \\ \hline
\end{tabular}
\label{Tab_Hadron_Parameters}
\end{table}

Concerning the quark-diquark structure of the baryons, the more
conventional three-quark structure of a baryon would complicate the
model significantly but would lead to similar predictions once the
model parameters are readjusted~\cite{Dosch:1994ym}. In fact, there
are also physical arguments that favor the quark-diquark structure of
the baryon such as the $\delta I = 1/2$ enhancement in
semi-leptonic decays of baryons~\cite{Dosch:1989hu} and the strong
attraction in the scalar diquark channel in the instanton
vacuum~\cite{Schafer:1994ra}.

\subsection*{The Photon Wave Function}   

The photon wave function $\psi_{\gamma}(z_i,\vec{r}_i,Q^2)$ describes
the fluctuation of a photon with virtuality $Q^2$ into a
quark-antiquark pair with longitudinal quark momentum fraction $z_i$
and spatial transverse size and orientation $\vec{r}_i$. The
computation of the corresponding transition amplitude $\langle
q\qbar(z_i,\vec{r}_i)|\gamma^*(Q^2)\rangle$ can be performed
conveniently in light-cone perturbation theory~\cite{Bjorken:1971ah+X}
and leads to the following squared wave functions for transverse $(T)$
and longitudinally $(L)$ polarized photons~\cite{Nikolaev:1991ja}
\bea
\!\!\!\!\!\!\!\!\!\!\!\!\!|\psi_{\gamma_T^*}(z_i,\vec{r}_i,Q^2)|^2\! 
        &\!\!=\!\!&\!\frac{3\,\alphaEM}{2\,\pi^2} \sum_f e_f^2
                \left\{ 
                  \left[ z_i^2 + (1-z_i)^2\right]\,\epsilon_f^2\,K_1^2(\epsilon_f\,|\vec{r}_i|) 
                  + m_f^2\,K_0^2(\epsilon_f\,|\vec{r}_i|) 
                \right\}
        \label{Eq_photon_wave_function_T_squared} \\
\!\!\!\!\!\!\!\!\!\!\!\!\!|\psi_{\gamma_L^*}(z_i,\vec{r}_i,Q^2)|^2\! 
        &\!\!=\!\!&\!\frac{3\,\alphaEM}{2\,\pi^2} \sum_f e_f^2
                \left\{ 4\,Q^2\,z_i^2(1-z_i)^2\,K_0^2(\epsilon_f\,|\vec{r}_i|) \right\},
        \label{Eq_photon_wave_function_L_squared}
\eea
where $\alphaEM$ is the fine-structure constant, $e_f$ is the electric
charge of the quark with flavor $f$, and $K_0$ and $K_1$ are the modified
Bessel functions (McDonald functions). In the above expressions,
\be
        \epsilon_f^2 = z_i(1-z_i)\,Q^2 + m_f^2
\label{Eq_photon_extension_parameter}
\ee
controlls the transverse size(-distribution) of the emerging dipole,
$|\vec{r}_i| \propto 1/ \epsilon_f$, that depends on the quark flavor
through the current quark mass $m_f$.

For small $Q^2$, the perturbatively derived wave functions,
(\ref{Eq_photon_wave_function_T_squared}) and
(\ref{Eq_photon_wave_function_L_squared}), are not appropriate since
the resulting color dipoles of size $|\vec{r}_i| \propto 1/m_f \gg
1\,\fm$ should encounter non-perturbative effects such as confinement
and chiral symmetry breaking. To take these effects into account the
vector meson dominance (VMD) model~\cite{Bauer:1978iq} is usually
used. However, the transition from the ``partonic'' behavior at large
$Q^2$ to the ``hadronic'' one at small $Q^2$ can be modeled as well
by introducing $Q^2$-dependent quark masses, $m_f = m_f(Q^2)$, that
interpolate between the current quarks at large $Q^2$ and the
constituent quarks at small $Q^2$~\cite{Dosch:1998nw}. Following this
approach, we use~(\ref{Eq_photon_wave_function_T_squared})
and~(\ref{Eq_photon_wave_function_L_squared}) also in the low-$Q^2$
region but with the running quark masses
\bea
        m_{u,d}(Q^2) 
        &=& 0.178\,\GeV\,(1-\frac{Q^2}{Q^2_{u,d}})\,\Theta(Q^2_{u,d}-Q^2) 
        \ , 
        \label{Eq_m_ud_(Q^2)}\\
        m_s(Q^2) 
        &=& 0.121\,\GeV + 0.129\,\GeV\,(1-\frac{Q^2}{Q^2_s})\,\Theta(Q^2_s-Q^2) 
        \label{Eq_m_s_(Q^2)}
        \ ,
\eea
and the fixed charm quark mass
\be
        m_c = 1.25\,\GeV
        \ ,
\ee
where the parameters $Q^2_{u,d} = 1.05\,\GeV^2$ and $Q^2_s =
1.6\,\GeV^2$ are taken directly from~\cite{Dosch:1998nw} while we
reduced the values for the constituent quark masses $m_f(Q^2 = 0)$
of~\cite{Dosch:1998nw} by about 20\%. The smaller constituent quark
masses are necessary to reproduce the total cross sections for
$\gamma^* p$ and $\gamma^* \gamma^*$ reactions at low $Q^2$.  Similar
running quark masses are obtained in a QCD-motivated model of
spontaneous chiral symmetry breaking in the instanton
vacuum~\cite{Petrov:1998kf} that improve the description of $\gamma^*
p$ scattering at low $Q^2$~\cite{Martin:1999bh+X}.

\section{Universal Energy Dependence}
\label{Sec_Energy_Dependence}

Until now the $T$-matrix
element~(\ref{Eq_model_purely_imaginary_T_amplitude}) leads to energy
independent total cross sections in contradiction to the experimental
observation. In this section we introduce the energy dependence in a
phenomenological way inspired by other successful models.

Most models for high-energy scattering are constructed to describe
either hadron-hadron or photon-hadron reactions.  For example,
Kopeliovich et al.~\cite{Kopeliovich:2001pc} as well as Berger and
Nachtmann~\cite{Berger:1999gu} focus on hadron-hadron scattering. In
contrast, Golec-Biernat and W\"usthoff~\cite{Golec-Biernat:1999js+X}
and Forshaw, Kerley, and Shaw~\cite{Forshaw:1999uf} concentrate on
photon-proton reactions. A model that describes the energy dependence
in both hadron-hadron and photon-hadron reactions up to large photon
virtualities is the two-pomeron model of Donnachie and
Landshoff~\cite{Donnachie:1998gm+X}. Based on Regge
theory~\cite{Donnachie:en}, they find a soft pomeron trajectory with
intercept $ 1 + \epsilon_{\mathrm{soft}} \approx 1.08$ that governs the weak
energy dependence of hadron-hadron or $\gamma^{*} p$ reactions with
low $Q^2$~\cite{Donnachie:1992ny} and a hard pomeron trajectory with
intercept $1 + \epsilon_{\mathrm{hard}} \approx 1.4$ that governs the strong
energy dependence of $\gamma^*p$ reactions with high
$Q^2$~\cite{Donnachie:1998gm+X}. Similarly, we aim at a simultaneous
description of hadron-hadron, photon-proton, and photon-photon
reactions involving real and virtual photons as well.

In line with other two-component (soft $+$ hard)
models~\cite{Donnachie:1998gm+X,Forshaw:1999uf,Rueter:1998up,D'Alesio:1999sf,Donnachie:2001wt}
and the different hadronization mechanisms in soft and hard
collisions, our physical ansatz demands that the perturbative and
non-perturbative contributions do not interfere. Therefore, we modify
the cosine-summation in~(\ref{Eq_model_purely_imaginary_T_amplitude})
allowing only even numbers of soft and hard correlations, $\left
  (\chi^{\nprt} \right)^{2n} \left ( \chi^{\pert} \right)^{2m}$ with
$n,m \in I\!\!N$.  Interference terms with odd numbers of soft and
hard correlations are subtracted by the replacement
\be
        \cos\left[ c\chi \right] =  
        \cos\left[c\left( \chi^{\nprt} + \chi^{\pert} \right)\right] 
        \rightarrow 
        \cos\left[c\chi^{\nprt}\right]\cos\left[c\chi^{\pert}\right] 
        \ ,
\label{Eq_interference_term_subtraction}
\ee
where $c = 1/3$ or $2/3$. This prescription leads to the following
factorization of soft and hard physics in the $T$-matrix
element,
\bea
        && \!\!\!\!\!\!\!\!\!\!
        T(s,t) 
        = 2is \int \!\!d^2b_{\!\perp} 
        e^{i {\vec q}_{\!\perp} {\vec b}_{\!\perp}}
        \int \!\!dz_1 d^2r_1 \!\int \!\!dz_2 d^2r_2\,\,
        |\psi_1(z_1,\vec{r}_1)|^2 \,\, 
        |\psi_2(z_2,\vec{r}_2)|^2       
        \nonumber \\    
        &&\!\!\!\!\!\!\!\!\!\!\!\! 
        \times \left[ 1 - \frac{2}{3} 
        \cos\!\left(\!\frac{1}{3}\chi^{\nprt}\!\right)
        \cos\!\left(\!\frac{1}{3}\chi^{\pert}\!\right)         
        - \frac{1}{3}
        \cos\!\left(\!\frac{2}{3}\chi^{\nprt}\!\right)
        \cos\!\left(\!\frac{2}{3}\chi^{\pert}\!\right)
        \right] \ . 
\label{Eq_model_purely_imaginary_T_amplitude_almost_final_result}
\eea
In the limit of small $\chi$-functions, $\chi^{\nprt}
\ll 1$ and $\chi^{\pert} \ll 1$, one gets
\bea
        T(s,t) 
        & = & 2is \!\int \!\!d^2b_{\!\perp} 
        e^{i {\vec q}_{\!\perp} {\vec b}_{\!\perp}}
        \!\int \!\!dz_1 d^2r_1 \!\int \!\!dz_2 d^2r_2\,\,
        |\psi_1(z_1,\vec{r}_1)|^2 \,\,
        |\psi_2(z_2,\vec{r}_2)|^2       
        \nonumber \\  
        && \times \frac{1}{9}\left[
        \left(\chi^{\nprt}\right)^2
        +\left(\chi^{\pert}\right)^2
        \right]
\label{Eq_model_purely_imaginary_T_amplitude_small_chi_limit}
\eea
so that the $T$-matrix element becomes a sum of a perturbative and a
non-perturbative component. As already discussed in
Sec.~\ref{Sec_DD_Scattering_LLCM}, the perturbative component
$(\chi^{\pert})^2$ coincides with the well-known perturbative {\em
  two-gluon exchange}~\cite{Low:1975sv+X,Gunion:iy} and the
non-perturbative component $(\chi^{\nprt})^2$ represents the
corresponding non-perturbative gluonic interaction on the
``two-gluon-exchange'' level~\cite{Shoshi:2002fq}.

As the two-component structure of
(\ref{Eq_model_purely_imaginary_T_amplitude_small_chi_limit}) reminds
of the two-pomeron model of Donnachie and
Landshoff~\cite{Donnachie:1998gm+X}, we adopt the powerlike energy
increase and ascribe a weak energy dependence to the non-perturbative
component $\chi^{\nprt}$ and a strong one to the perturbative component
$\chi^{\pert}$
\bea
        \left(\chi^{\nprt}\right)^2 \quad & \rightarrow & \quad 
        \left(\chi^{\nprt}(s)\right)^2 := \left(\chi^{\nprt}\right)^2 
        \left(\frac{s}{s_0}
        \frac{\vec{r}_1^{\,2}\,\vec{r}_2^{\,2}}{R_0^4}\right)^{\epsilon^{\nprt}}
        \nonumber \\
        \left(\chi^{\pert}\right)^2 \quad & \rightarrow & \quad 
        \left(\chi^{\pert}(s)\right)^2 := \left(\chi^{\pert}\right)^2
        \left(\frac{s}{s_0} 
        \frac{\vec{r}_1^{\,2}\,\vec{r}_2^{\,2}}{R_0^4}\right)^{\epsilon^{\pert}}
\label{Eq_energy_dependence}
\eea
with the scaling factor $s_0 R_0^4$. The powerlike energy dependence
with the exponents $0\approx \epsilon^{\nprt} < \epsilon^{\pert} < 1$
guarantees Regge type behavior at moderately high energies, where the
small-$\chi$
limit~(\ref{Eq_model_purely_imaginary_T_amplitude_small_chi_limit}) is
appropriate. In~(\ref{Eq_energy_dependence}) the energy variable $s$
is scaled by the factor $\vec{r}_1^{\,2}\,\vec{r}_2^{\,2}$ that allows
to rewrite the energy dependence in photon-hadron scattering in terms
of the appropriate Bjorken scaling variable $x$
\be
        s\,\vec{r}_1^{\,2} \propto \frac{s}{Q^2} = \inv{x}
        \ ,
\label{Eq_x_Bj_<->_s}
\ee
where $|\vec{r}_1|$ is the transverse extension of the $q\qbar $
dipole in the photon. A similar factor has been used before in the
dipole model of Forshaw, Kerley, and Shaw~\cite{Forshaw:1999uf} and
also in the model of Donnachie and Dosch~\cite{Donnachie:2001wt} in
order to respect the scaling properties observed in the structure
function of the proton.\footnote{In the model of Donnachie and
  Dosch~\cite{Donnachie:2001wt}, $s\,|\vec{r}_1|\,|\vec{r}_2|$ is used
  as the energy variable if both dipoles are small, which is in
  accordance with the choice of the typical BFKL energy scale but
  leads to discontinuities in the dipole-dipole cross section. In
  order to avoid such discontinuities, we use the energy
  variable~(\ref{Eq_energy_dependence}) also for the scattering of two
  small dipoles.} In the dipole-proton cross section of Golec-Biernat
and W\"usthoff~\cite{Golec-Biernat:1999js+X}, Bjorken $x$ is used
directly as energy variable which is important for the success of the
model. In fact, also in our model, the
$\vec{r}_1^{\,2}\,\vec{r}_2^{\,2}$ factor improves the description of
$\gamma^{*} p$ reactions at large $Q^2$.

The powerlike Regge type energy dependence introduced
in~(\ref{Eq_energy_dependence}) is, of course, not mandatory but
allows successful fits and can also be derived in other theoretical
frameworks: A powerlike energy dependence is found for hard
photon-proton reactions from the BFKL equation~\cite{BFKL} and for
hadronic reactions by Kopeliovich et al.~\cite{Kopeliovich:2001pc}.
However, these approaches need unitarization since their powerlike
energy dependence will ultimately violate $S$-matrix unitarity at
asymptotic energies. In our model we use the following $T$-matrix
element for investigations in the remaining chapters
\bea
        &&\!\!\!\!\!\!\!\!\!\!
        T(s,t)  
        =  2is \int \!\!d^2b_{\!\perp} 
        e^{i {\vec q}_{\!\perp} {\vec b}_{\!\perp}}
        \int \!\!dz_1 d^2r_1 \!\int \!\!dz_2 d^2r_2\,\, 
        |\psi_1(z_1,\vec{r}_1)|^2 \,\, 
        |\psi_2(z_2,\vec{r}_2)|^2       
\label{Eq_model_purely_imaginary_T_amplitude_final_result}\\    
        &&\!\!\!\!\!\!\!\!\!\!\!\! 
        \times \left[1 - \frac{2}{3} 
        \cos\!\left(\!\frac{1}{3}\chi^{\nprt}(s)\!\right)
        \cos\!\left(\!\frac{1}{3}\chi^{\pert}(s)\!\right)         
        - \frac{1}{3}
        \cos\!\left(\!\frac{2}{3}\chi^{\nprt}(s)\!\right)
        \cos\!\left(\!\frac{2}{3}\chi^{\pert}(s)\!\right)
        \right] 
        \ , \nonumber
\eea
where the cosine functions ensure the unitarity condition in impact
parameter space as shown in Chap.~\ref{Sec_Impact_Parameter}. Indeed,
the multiple gluonic interactions associated with the higher order
terms in the expansion of the cosine functions are important for the
saturation effects observed within our model at ultra-high energies.

Having ascribed the energy dependence to the $\chi$-function, the
energy behavior of hadron-hadron, photon-hadron, and photon-photon
scattering results exclusively from the {\em universal} dipole-dipole
scattering kernel.

\section{Model Parameters for High-Energy Scattering}
\label{Sec_Model_Parameters}

Lattice QCD simulations provide important information and constraints
on the model parameters as discussed in Chap.~\ref{Sec_The_Model}. The
fine tuning of the parameters was, however, directly performed on
high-energy scattering data for hadron-hadron, photon-hadron, and
photon-photon reactions~\cite{Shoshi:2002in} where an error ($\chi^2$)
minimization was not fea\-si\-ble because of the non-trivial
multi-dimensional integrals in the $T$-matrix
element~(\ref{Eq_model_purely_imaginary_T_amplitude_final_result}).

The parameters $a$, $\kappa$, $G_2$, $m_G$, $M^2$, $s_0R^4_0$,
$\epsilon^{\nprt}$ and $\epsilon^{\pert}$ determine the dipole-dipole
scattering amplitude and are universal for all reactions considered.
In addition, there are reaction-dependent parameters in the wave
functions which are given in Sec.~\ref{Sec_Wave_Functions}.

The non-perturbative component involves the correlation length $a$,
the gluon condensate $G_2$, and the parameter $\kappa$ indicating the
non-Abelian character of the correlator. With the simple exponential
correlation function~(\ref{Eq_SVM_correlation_functions}), we obtain
the values given in~(\ref{Eq_MSV_scattering_fit_parameter_results})
that have already been applied in Chap.~\ref{Sec_Static_Sources}
\be
        a =  0.302\,\fm, \quad 
        \kappa = 0.74, \quad 
        G_2 = 0.074\,\GeV^4
        \ .
\nonumber
\ee

The perturbative component involves the gluon mass $m_G$ as IR
regulator (or inverse ``perturbative correlation length'') and the
parameter $M^2$ that freezes the running
coupling~(\ref{Eq_g2(z_perp)}) for large distance scales. The
following values are used in the following investigations
\be
        m_G =  m_{\rho} = 0.77\,\GeV 
        \quad \mbox{and }\quad 
        M^2 = 1.04\,\GeV^2
        \ .
\label{Eq_PGE_scattering_fit_parameter_results}
\ee

The energy dependence of the model is associated with the energy
exponents $\epsilon^{\nprt}$ and $\epsilon^{\pert}$, and the  scaling
parameter $s_0R^4_0$ 
\be
        \epsilon^{\nprt} = 0.125, \quad 
        \epsilon^{\pert} = 0.73, \quad \mbox{and} \quad 
        s_0 R_0^4 = (\,47\,\GeV\,\fm^2\,)^2
        \ .
\label{Eq_energy_dependence_scattering_fit_parameter_results}
\ee
In comparison to the energy exponents of Donnachie and
Landshoff~\cite{Donnachie:1992ny,Donnachie:1998gm+X}, $\epsilon_{\mathrm{soft}}
\approx 0.08$ and $\epsilon_{\mathrm{hard}} \approx 0.4$, our exponents are
significantly larger. However, the cosine functions in our $T$-matrix
element~(\ref{Eq_model_purely_imaginary_T_amplitude_final_result})
reduce the large exponents so that the energy dependence of the cross
sections agrees with the experimental data as illustrated in
Chap.~\ref{Sec_Comparison_Data}.


%
\cleardoublepage
%
\chapter{Impact Parameter Profiles and Gluon Saturation}
\label{Sec_Impact_Parameter}

In this chapter $S$-matrix unitarity constraints are considered in our
model. On the basis of the impact parameter dependence of the
scattering amplitude, saturation effects can be exposed that manifest
the $S$-matrix unitarity. For each impact parameter the energy at
which the unitarity limit becomes important can be determined. The
results are used to discuss gluon saturation and to localize
saturation effects in experimental observables.

The impact parameter dependence of the scattering amplitude is given
by $\impactT(s,|\vec{b}_{\!\perp}|)$,
\be
        T(s,t=-{\vec q}_{\!\perp}^{\,\,2}) \;=\;
        4s\!\int \!\!d^2b_{\!\perp}\,
        e^{i {\vec q}_{\!\perp} {\vec b}_{\!\perp}}\,
        \impactT(s,|\vec{b}_{\!\perp}|)
        \ ,
\label{Eq_Fourier_transformed_T-matrix_element}
\ee
and in particular by the {\em profile function}
\be
        J(s,|\vec{b}_{\!\perp}|) 
        = 2\,\im\impactT(s,|\vec{b}_{\!\perp}|)
        \ ,
\label{Eq_profile_function_def}
\ee 
which describes the {\em blackness} or {\em opacity} of the
interacting particles as a function of the impact parameter $|{\vec
  b}_{\!\perp}|$ and the c.m.\ energy $\sqrt{s}$. In fact, the profile
function~(\ref{Eq_profile_function_def}) determines all observables if
the $T$-matrix is -- as in our model -- purely imaginary.

\section[$S$-Matrix Unitarity Constraints]
{\boldmath$S$-Matrix Unitarity Constraints}
\label{Sec_S-Matrix_Unitarity}

The $S$-matrix unitarity, $SS^{\dagger} = S^{\dagger}S = \Identity$,
leads directly to the {\em unitarity condition} in impact parameter
space\footnote{Integrating
  (\ref{Eq_unitarity_condition}) over the impact parameter space and
  multiplying by a factor of $4$ one obtains the relation
  $\sigma^{tot}(s) = \sigma^{el}(s) + \sigma^{inel}(s)$.}~\cite{Amaldi:1976gr,Castaldi:1983ft}
\be
        \im\impactT(s,|\vec{b}_{\!\perp}|)
        = |\impactT(s,|\vec{b}_{\!\perp}|)|^2 + G_{inel}(s,|\vec{b}_{\!\perp}|)
        \ ,
\label{Eq_unitarity_condition}
\ee 
where $G_{inel}(s,|\vec{b}_{\!\perp}|) \ge 0$ is the inelastic overlap
function~\cite{VanHove:1964rp}. This
unitarity condition imposes an absolute limit on the profile function
\be
        0 \;\;\leq\;\;
        2\,|\impactT(s,|\vec{b}_{\!\perp}|)|^2
        \;\;\leq\;\; 
        J(s,|\vec{b}_{\!\perp}|) 
        \;\;\leq\;\; 2
\label{Eq_absolute_unitarity_limit}
\ee 
and the inelastic overlap function, $G_{inel}(s,|\vec{b}_{\!\perp}|)
\le 1/4$.  At high energies, the elastic amplitude is expected to be
purely imaginary. Consequently, the solution
of~(\ref{Eq_unitarity_condition}) reads
\be
        J(s,|\vec{b}_{\!\perp}|) = 1 \pm \sqrt{1-4\,G_{inel}(s,|\vec{b}_{\!\perp}|)}
\label{Eq_solution_unitarity_condition}
\ee
and leads with the minus sign corresponding to the physical situation
to the {\em reduced unitarity bound}
\be
        0 \;\;\leq\;\;
        J(s,|\vec{b}_{\!\perp}|) 
        \;\;\leq\;\; 1
        \ .
\label{Eq_reduced_unitarity_bound}
\ee 
Reaching the {\em black disc limit} or {\em maximum opacity} at a
certain impact parameter $|\vec{b}_{\!\perp}|$,
$J(s,|\vec{b}_{\!\perp}|) = 1$, corresponds to maximal inelastic
absorption $G_{inel}(s,|\vec{b}_{\!\perp}|) = 1/4$ and equal elastic
and inelastic contributions to the total cross section at that impact
parameter.

In our model every reaction is reduced to dipole-dipole scattering
with well defined dipole sizes $|{\vec r}_i|$ and longitudinal quark
momentum fractions $z_i$ as discussed in
Sec.~\ref{Sec_Functional_Integral_Approach}. Thus, the most general
test of our model with respect to the unitarity constraints is
performed with the profile function
\be
        J_{DD}(s,|\vec{b}_{\!\perp}|,z_1,|\vec{r}_1|,z_2,|\vec{r}_2|)  
        = \int \frac{d\phi_1}{2\pi}  \int \frac{d\phi_2}{2\pi} 
        \left[1 - S_{DD}(s,\vec{b}_{\!\perp},z_1,{\vec r}_1,z_2,{\vec r}_2)\right]
        \ ,
\label{Eq_DD_profile_function}
\ee
where $\phi_i$ describes the dipole orientation, i.e.\ the angle
between ${\vec r}_i$ and $\vec{b}_{\!\perp}$, and $S_{DD}$ describes
{\em elastic dipole-dipole scattering}
\be
        S_{DD}
        = \frac{2}{3} 
        \cos\!\left(\frac{1}{3}\chi^{\nprt}(s)\right)
        \cos\!\left(\frac{1}{3}\chi^{\pert}(s)\right)         
        + \frac{1}{3}
        \cos\!\left( \frac{2}{3}\chi^{\nprt}(s)\right)
        \cos\!\left( \frac{2}{3}\chi^{\pert}(s)\right)
\label{Eq_S_DD_final_result}
\ee
with the purely real-valued eikonal functions $\chi^{\nprt}(s)$ and
$\chi^{\pert}(s)$ defined in~(\ref{Eq_energy_dependence}). Because of
$|S_{DD}| \leq 1$, a consequence of the cosine functions
in~(\ref{Eq_S_DD_final_result}) describing multiple gluonic
interactions, $J_{DD}$ respects the absolute
limit~(\ref{Eq_absolute_unitarity_limit}). Thus, the elastic
dipole-dipole scattering respects the unitarity
condition~(\ref{Eq_unitarity_condition}).  At high energies, the
arguments of the cosine functions in $S_{DD}$ become so large that
these cosines average to zero in the integration over the dipole
orientations. This leads to the black disc limit $J_{DD}^{\mathrm{max}} = 1$
reached at high energies first for small impact parameters.

\section{The Profile Function for Proton-Proton Scattering}
\label{Sec_PP_Profile_Function}

The profile function for proton-proton scattering
\be
        J_{pp}(s,|\vec{b}_{\!\perp}|)  = 
        \int \!\!dz_1 d^2r_1 \!\int \!\!dz_2 d^2r_2      
        |\psi_p(z_1,\vec{r}_1)|^2 |\psi_p(z_2,\vec{r}_2)|^2
        \left[1-S_{DD}(s,\vec{b}_{\!\perp},z_1,{\vec r}_1,z_2,{\vec r}_2)\right]
\label{Eq_model_pp_profile_function}
\ee
is obtained from~(\ref{Eq_DD_profile_function}) by weighting the
dipole sizes $|{\vec r}_i|$ and longitudinal quark momentum fractions
$z_i$ with the proton wave function $|\psi_p(z_i,\vec{r}_i)|^2$ from
Sec.~\ref{Sec_Wave_Functions}.

Using the model parameters from Sec.~\ref{Sec_Model_Parameters}, we
obtain the profile function $J_{pp}(s,|\vec{b}_{\!\perp}|)$ shown in
Fig.~\ref{Fig_J_pp(b,s)} for c.m.\ energies from $\sqrt{s} = 10\,\GeV$
to $10^8\,\GeV$.
\befig[t!]  
\centerline{\epsfig{figure=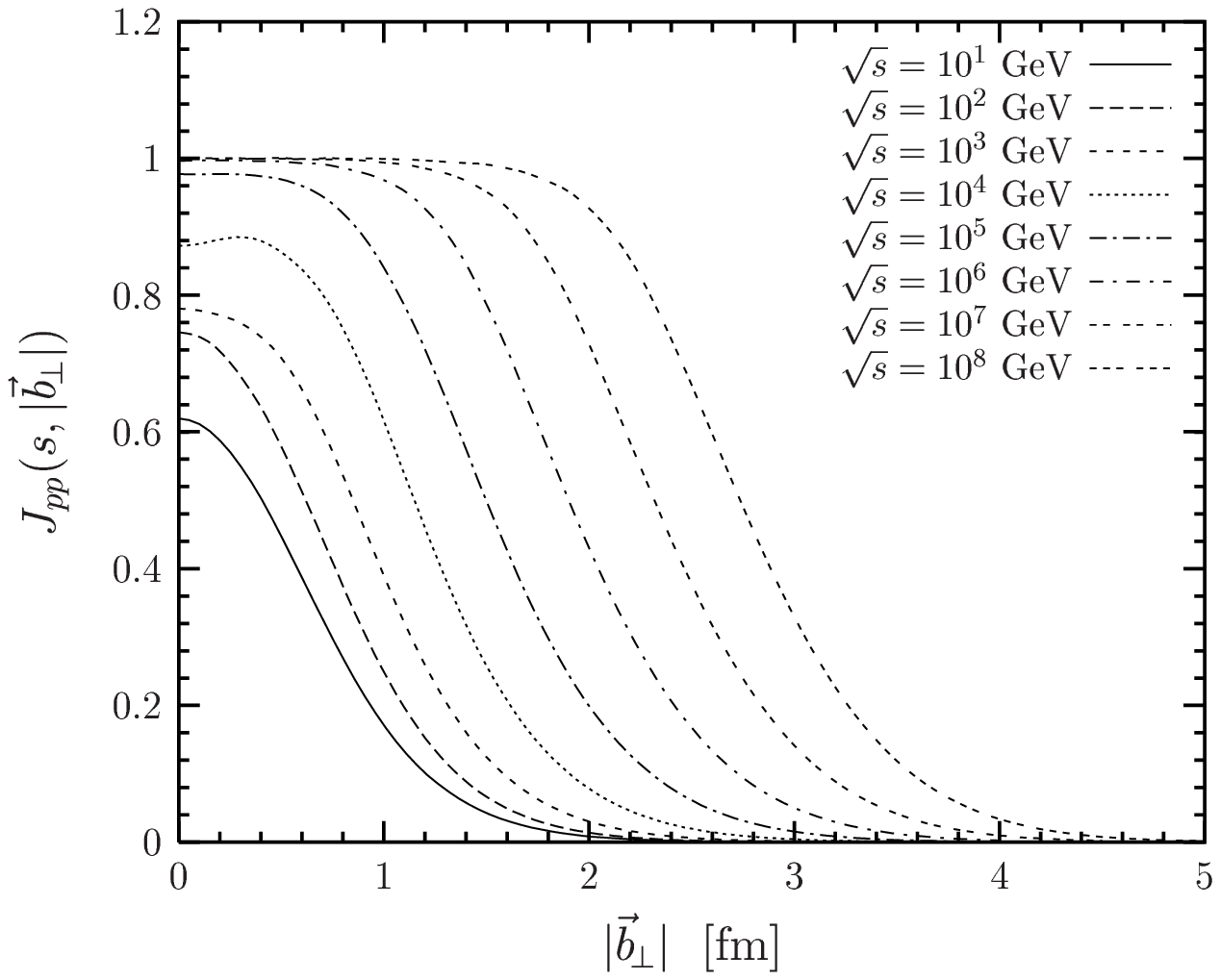,width=11.cm}}
\protect\caption{\small 
  The profile function for proton-proton scattering
  $J_{pp}(s,|\vec{b}_{\!\perp}|)$ as a function of the impact
  parameter $|\vec{b}_{\!\perp}|$ for c.m.\ energies from $\sqrt{s} =
  10\,\GeV$ to $10^8\,\GeV$. The unitarity
  limit~(\ref{Eq_absolute_unitarity_limit}) corresponds to
  $J_{pp}(s,|\vec{b}_{\!\perp}|) = 2$ and the black disc
  limit~(\ref{Eq_reduced_unitarity_bound}) to
  $J_{pp}(s,|\vec{b}_{\!\perp}|) = 1$.
}
\label{Fig_J_pp(b,s)}
\efig
Up to $\sqrt{s} \approx 100\,\GeV$, the profile has approximately a
Gaussian shape. Above $\sqrt{s}=1\,\TeV$, it develops into a broader
and higher profile until the black disc limit is reached for $\sqrt{s}
\approx 10^6\,\GeV$ and $|\vec{b}_{\!\perp}|=0$.  At this point, the
cosine functions in $S_{DD}$ average to zero
\be
        \int \!\!dz_1 d^2r_1 \!\int \!\!dz_2 d^2r_2  
        |\psi_p(z_1,{\vec r}_1)|^2|\psi_p(z_2,{\vec r}_2)|^2
        S_{DD}(\sqrt{s}\gtsim10^6\,\GeV,|\vec{b}_{\!\perp}|=0,\dots) 
        \approx 0
\label{Eq_pp_black_disc_limit_explained}
\ee
so that the proton wave function normalization determines the maximum
opacity
\be
        J_{pp}^{\mathrm{max}}
        =\int \!\!dz_1 d^2r_1 \!\int \!\!dz_2 d^2r_2\,        
        |\psi_p(z_1,{\vec r}_1)|^2\,|\psi_p(z_2,{\vec r}_2)|^2
        = 1
        \ .
\label{Eq_pp_black_disc_limit}
\ee
Once the maximum opacity is reached at a certain impact parameter, the
height of the profile function saturates at that $|\vec{b}_{\!\perp}|$
while the width of the profile function extends towards larger impact
parameters with increasing energy. Thus, multiple gluonic interactions
important to respect the $S$-matrix unitarity
constraint~(\ref{Eq_unitarity_condition}) lead to saturation for
$\sqrt{s} \gtsim 10^6\,\GeV$.

The above behavior of the profile function illustrates the evolution
of the proton with increasing c.m.\ energy. The proton is gray and of
small transverse size at small $\sqrt{s}$ but becomes blacker and more
transversally extended with increasing $\sqrt{s}$ until it reaches the
black disc limit in its center at $\sqrt{s} \approx 10^6\,\GeV$.
Beyond this energy, the proton cannot become blacker in its central
region but in its periphery with continuing transverse growth.
Furthermore, the proton boundary stays diffusive as claimed also
in~\cite{Frankfurt:2001nt+X}.

According to our model the black disc limit will not be reached at
LHC. Our prediction of $\sqrt{s} \approx 10^6\,\GeV = 10^3\,\TeV$ for
the onset of the black disc limit in proton-proton collisions is about
two orders of magnitude beyond the LHC energy $\sqrt{s} = 14\,\TeV$.
This is in contrast to, e.g.~\cite{Desgrolard:1999pr}, where the value
predicted for the onset of the black disc limit is $\sqrt{s} =
2\,\TeV$, i.e.\ small enough to be reached at LHC. However, note that
our profile function $J_{pp}(s,|\vec{b}_{\!\perp}|)$ yields good
agreement with experimental data for cross sections up to the highest
energies as shown in Chap.~\ref{Sec_Comparison_Data}.

For hadron-hadron reactions in general, the wave function
normalization of the hadrons determines the maximum opacity analogous
to~(\ref{Eq_pp_black_disc_limit}) and the transverse hadron size the
c.m.\ energy at which it is reached. Consequently, the maximum opacity
obtained for $\pi p$ and $K p$ scattering is identical to the one for
$pp$ scattering due to the
normalization~(\ref{Eq_hadron_wave_function_normalization}). The
smaller size of pions and kaons in comparison to protons, however,
demands slightly higher c.m.\ energies to reach this maximum opacity.
Such size effects become more pronounced in longitudinal photon-proton
scattering, where the size of the dipole emerging from the photon can
be controlled by the photon virtuality.

\section{The Profile Function for Photon-Proton Scattering}
\label{Sec_GP_Profile_Function}

The profile function for a longitudinal photon $\gamma_L^*$ scattering
off a proton $p$
\bea
        J_{\gamma_L^* p}(s,|\vec{b}_{\!\perp}|,Q^2)  
        & = & 
        \int \!\!dz_1 d^2r_1 \!\int \!\!dz_2 d^2r_2\,
        |\psi_{\gamma_L^*}(z_1,\vec{r}_1,Q^2)|^2 \,
        |\psi_p(z_2,\vec{r}_2)|^2
        \nonumber \\
        && 
        \times
        \left[1-S_{DD}(\vec{b}_{\!\perp},s,z_1,{\vec r}_1,z_2,{\vec r}_2)\right]
\label{Eq_model_gp_profile_function}
\eea
is calculated with the longitudinal photon wave function
$|\psi_{\gamma_L^*}(z_i,\vec{r}_i,Q^2)|^2$ given
in~(\ref{Eq_photon_wave_function_L_squared}). In this way, the profile
function~(\ref{Eq_model_gp_profile_function}) is ideally suited for
the investigation of dipole size effects since the photon virtuality
$Q^2$ determines the transverse size of the dipole into which the
photon fluctuates before it interacts with the proton.

Figure~\ref{Fig_J_gp_(b,s,Q^2)} shows the $|\vec{b}_{\!\perp}|$
dependence of the profile function
$J_{\gamma_L^*p}(s,|\vec{b}_{\!\perp}|,Q^2)$ divided by
$\alphaEM/{\pi}$ at a photon virtuality of $Q^2 = 1\,\GeV^2$ for c.m.\ 
energies $\sqrt{s}$ from $10\,\GeV$ to $10^9\,\GeV$, where $\alphaEM$
is the fine-structure constant.
\befig[t] \centerline{\epsfig{figure=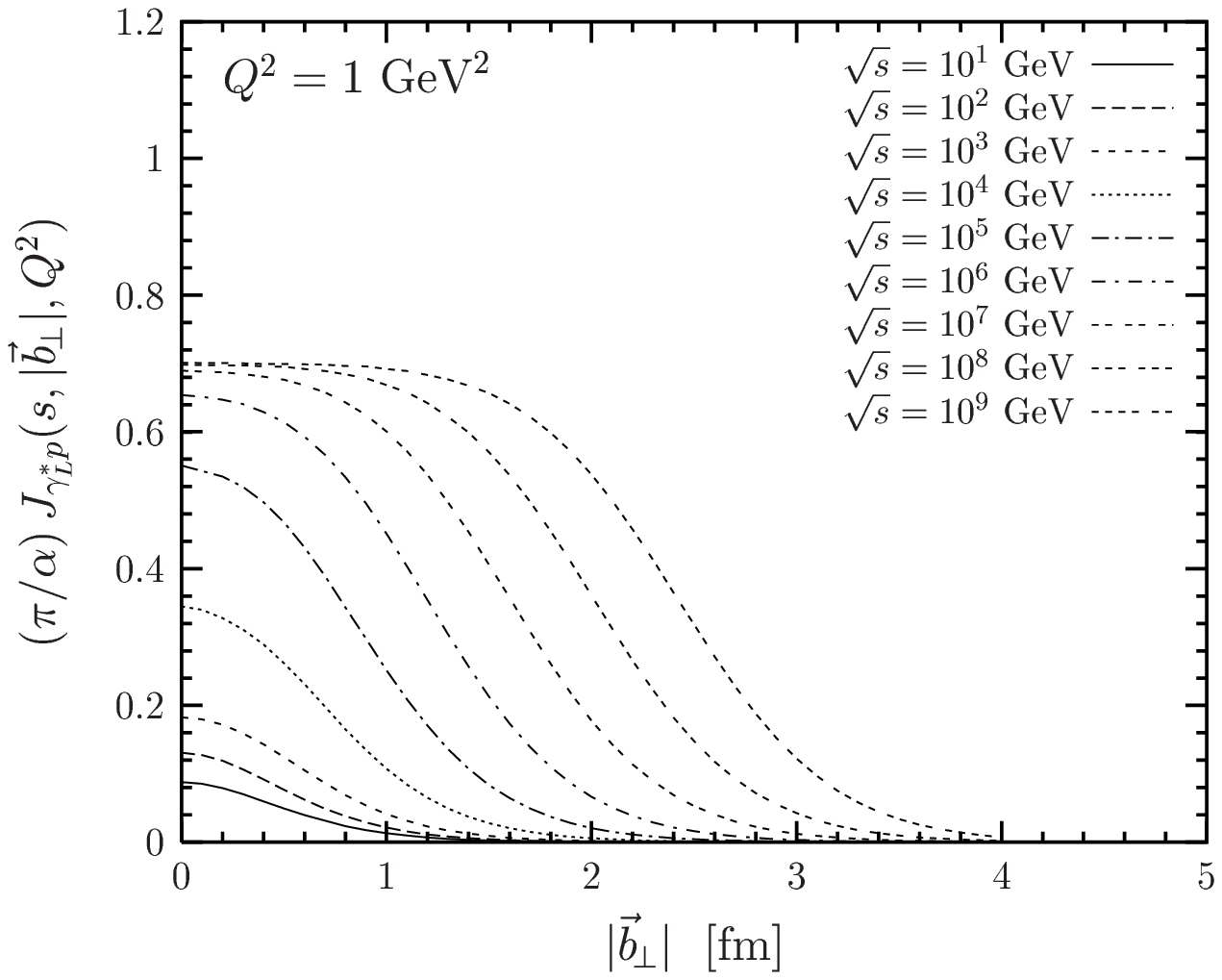,width=11cm}}
\protect\caption{\small 
  The profile function for a longitudinal photon scattering off a
  proton $J_{\gamma_L^* p}(s,|\vec{b}_{\!\perp}|,Q^2)$ divided by
  $\alphaEM/{\pi}$ as a function of the impact parameter
  $|\vec{b}_{\!\perp}|$ at a photon virtuality of $Q^2 = 1\,\GeV^2$
  and c.m.\ energies from $\sqrt{s} = 10\,\GeV$ to $10^9\,\GeV$. The
  value of the black disc limit is
  $J_{\gamma_L^*p}^{\mathrm{max}}(Q^2=1\,\GeV^2) = 0.00164$\ .
}
\label{Fig_J_gp_(b,s,Q^2)}
\end{figure}
One clearly sees that the qualitative behavior of this rescaled
profile function is similar to the one for proton-proton scattering.
However, the black disc limit induced by the underlying dipole-dipole
scattering depends on the photon virtuality $Q^2$ and is given by the
normalization of the longitudinal photon wave function
\bea
        J_{\gamma_L^* p}^{\mathrm{max}}(Q^2)  
        = \int \!\!dz d^2r |\psi_{\gamma_L^*}(z,\vec{r},Q^2)|^2
\label{Eq_gp_black_disc_limit}
\eea
since the proton wave function is normalized to one.

The photon virtuality $Q^2$ does not only determine the absolute value
of the black disc limit but also the c.m.\ energy at which it is
reached. This is illustrated in Fig.~\ref{Fig_J_gp_(b=0,s,Q^2)}, where
the $\sqrt{s}$ dependence of $J_{\gamma_L^*
  p}(s,|\vec{b}_{\!\perp}|=0,Q^2)$ divided by $\alphaEM/{\pi}$ is
presented for $Q^2 = 1,\,10,\,\mbox{and}\,100\,\GeV^2$.
\befig[t]
\centerline{\epsfig{figure=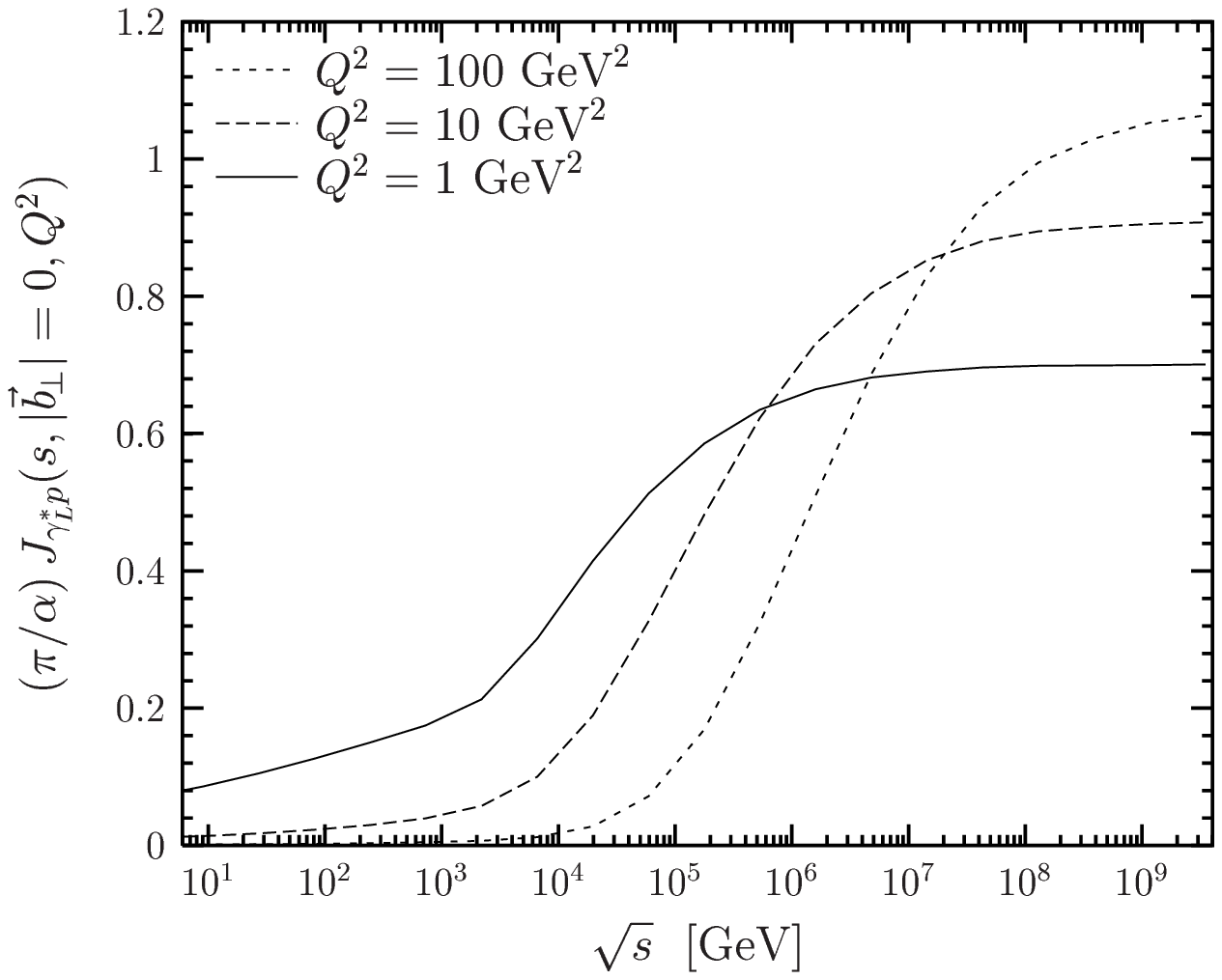,width=10.5cm}}
\caption{\small 
  The profile function for a longitudinal photon scattering off a
  proton $J_{\gamma_L^*p}(s,|\vec{b}_{\!\perp}|,Q^2)$ divided by
  $\alphaEM/{\pi}$ as a function of the c.m.\ energy $\sqrt{s}$ at
  zero impact parameter ($|\vec{b}_{\!\perp}|=0$) for photon
  virtualities of $Q^2 = 1,\,10,\,\mbox{and}\,100\,\GeV^2$.}
\label{Fig_J_gp_(b=0,s,Q^2)}
\end{figure}
With increasing resolution $Q^2$, i.e.\ decreasing dipole sizes,
$|\vec{r}_{\gamma_L^*}|^2 \propto 1/Q^2$, the absolute value of the
black disc limit grows and higher energies are needed to reach this
limit.\footnote{Note that the Bjorken $x$ at which the black disc
  limit is reached decreases with increasing photon virtuality $Q^2$.
  See also Fig.~\ref{Fig_xg(x,Q^2,b=0)_vs_x}.} The growth of the
absolute value of the black disc limit is simply due to the
normalization of the longitudinal photon wave function while the need
for higher energies to reach this limit is due to the decreasing
interaction strength with decreasing dipole size. The latter explains
also why the energies needed to reach the black disc limit in $\pi p$
and $K p$ scattering are higher than in $pp$ scattering. Comparing
$\gamma_L^*p$ scattering at $Q^2=1\,\GeV^2$ with $pp$ scattering
quantitatively, the black disc limit
$J_{\gamma_L^*p}^{\mathrm{max}}(Q^2=1\,\GeV^2) = 0.00164$ is about three orders
of magnitude smaller because of the photon wave function normalization
($\propto \alphaEM/{\pi}$). At $|\vec{b}_{\!\perp}|=0$ it is reached
at an energy of $\sqrt{s} \approx 10^8\,\GeV$, which is about two
orders of magnitude higher because of the smaller dipoles involved.

The way in which the profile function $J_{\gamma_L^*
  p}(s,|\vec{b}_{\!\perp}|,Q^2)$ approaches the black disc limit at
high energies depends on the shape of the proton and longitudinal
photon wave function at small dipole sizes $|\vec{r}_{1,2}|$. At high
energies, dipoles of typical sizes $0 \leq |\vec{r}_{1,2}| \leq
R_0\,(s_0/s)^{1/4}$ give the main contribution to
$S_{\gamma_L^*p} = 1 - J_{\gamma_L^*p}$ because
of~(\ref{Eq_energy_dependence}) and the fact that the contribution of
the large dipole sizes averages to zero upon integration over the
dipole orientations,
cf.~also~(\ref{Eq_pp_black_disc_limit_explained}). Since
$S_{\gamma_L^*p}$ is a measure of the proton transmittance, this means
that only small dipoles can penetrate the proton at high energies.
Increasing the energy further, even these small dipoles are absorbed
and the black disc limit is reached. However, the dependence of the
profile function on the short distance behavior of normalizable wave functions is
weak which can be understood as follows. Because of color
transparency, the eikonal functions $\chi^{\nprt}(s)$ and
$\chi^{\pert}(s)$ are small for small dipole sizes $0 \leq
|\vec{r}_{1,2}| \leq R_0\,(s_0/s)^{1/4}$ at large
$\sqrt{s}$. Consequently, $S_{DD} \approx 1$ and
\bea
        \!\!\!\!\!\!\!\!
        && 
        \!\!\!\!\!\!\!\!\!\!\!\!\!\!\!\!
        J_{\gamma_L^* p}(s,|\vec{b}_{\!\perp}|,Q^2)  
\nonumber \\
        \!\!\!\!\!\!\!\!
        &&
        \!\!\!\!\!\!\!\!\!\!\!\!\!\!\!\!
         \approx 
        J_{\gamma_L^* p}^{\mathrm{max}}(Q^2) - 4\pi^2\!\!
        \int\limits_0^1 \!\!dz_1 \!\!
        \int\limits_0^{r_c(s)}\!\!\!dr_1 r_1 
        |\psi_{\gamma_L^*}(z_1,r_1,Q^2)|^2 
        \int\limits_0^1 \!\!dz_2 \!\!
        \int\limits_0^{r_c(s)}\!\!\!dr_2 r_2 
        |\psi_p(z_2,r_2)|^2
\label{Eq_model_gp_profile_function_wavefunction_independence}
\eea
where $r_c(s) \approx R_0\,(s_0/s)^{1/4}$. Clearly, the
linear behavior from the phase space factors $r_{1,2}$ dominates over
the $r_{1,2}$-dependence of normalizable wave functions.\footnote{For our
  choice of the wave functions
  in~(\ref{Eq_model_gp_profile_function_wavefunction_independence}),
  one sees very explicitly that the specific Gaussian behavior of
  $|\psi_p(z_2,r_2)|^2$ and the logarithmic short distance behavior of
  $|\psi_{\gamma_L^*}(z_1,r_1,Q^2)|^2$ is dominated by the phase space
  factors $r_{1,2}$.} More generally, for any profile function
involving normalizable wave functions, the way in which the black disc
limit is approached depends only weakly on the short distance behavior
of the wave functions.

\section{A Scenario for Gluon Saturation}
\label{Sec_Gluon_Saturation}

In this section we discuss saturation of the {\em impact parameter
  dependent gluon distribution} of the proton
$xG(x,Q^2,|\vec{b}_{\!\perp}|)$.  Using a leading twist,
next-to-leading order QCD relation between $xG(x,Q^2)$ and the
longitudinal structure function $F_L(x,Q^2)$, we relate
$xG(x,Q^2,|\vec{b}_{\!\perp}|)$ to the profile function $J_{\gamma_L^*
  p}(s=Q^2/x,|\vec{b}_{\!\perp}|,Q^2)$ and find low-$x$ saturation of
$xG(x,Q^2,|\vec{b}_{\!\perp}|)$ as a manifestation of $S$-matrix
unitarity. The resulting $xG(x,Q^2,|\vec{b}_{\!\perp}|)$ is, of
course, only an estimate since our profile function contains also
higher twist contributions. Furthermore, in the considered low-$x$
region, the leading twist, next-to-leading order QCD formula may be
inadequate as higher twist contributions~\cite{Martin:1998kk+X} and
higher order QCD corrections~\cite{Gribov:1983tu,Mueller:1986wy} are
expected to become important. Nevertheless, still assuming a close
relation between $F_L(x,Q^2)$ and $xG(x,Q^2)$ at low $x$, we think
that our approach provides some insight into the gluon distribution as
a function of the impact parameter and into its saturation.

The {\em gluon distribution}\ of the proton $~xG(x,Q^2)~$ is defined
as follows: $xG(x,Q^2)dx$ gives the momentum fraction of the proton
which is carried by the gluons in the interval $[x, x+dx]$ as seen by
probes of virtuality $Q^2$. The {\em impact parameter dependent gluon
  distribution} $xG(x,Q^2,|\vec{b}_{\!\perp}|)$ is the gluon
distribution $xG(x,Q^2)$ at a given impact parameter
$|\vec{b}_{\!\perp}|$ so that
\be
        xG(x,Q^2) = \int
        \!\!d^2b_{\!\perp}\,xG(x,Q^2,|\vec{b}_{\!\perp}|) \ .
\label{Eq_def_xg(x,Q^2)}
\ee

In leading twist, next-to-leading order QCD, the gluon distribution
$xG(x,Q^2)$ is related to the structure functions $F_L(x,Q^2)$ and
$F_2(x,Q^2)$ of the proton~\cite{Martin:1988vw}
\be
        F_L(x, Q^2) 
        = \frac{\alphaS}{\pi}\!
        \left[
        \frac{4}{3}\int_x^1 \!
        \frac{dy}{y}\!\left(\frac{x}{y}\right)^{\!\!2} \!F_2(y,Q^2)
        + 2 \sum_f e_f^2\!\int_x^1 \!
        \frac{dy}{y}\!\left(\frac{x}{y}\right)^{\!\!2} \!\!
        \left(\!1-\frac{x}{y}\right) yG(y,Q^2)
        \right]
\label{Eq_FL_QCD_prediction}
\ee
where $\sum_f e_f^2$ is a flavor sum over the quark charges squared.
For four active flavors and $x \ltsim 10^{-3}$,
(\ref{Eq_FL_QCD_prediction}) can be approximated as
follows~\cite{Cooper-Sarkar:1988ds+X}
\be
        xG(x,Q^2) 
        \approx \frac{3}{5}\,5.8\, 
        \left[ 
        \frac{3\pi}{4\alphaS}\, F_L(0.417 x, Q^2)
        - \inv{1.97}\, F_2(0.75 x, Q^2) 
        \right] 
        \ .
\label{Eq_xg(x,Q^2)_approximation}
\ee
For typical $\Lambda_{QCD} = 100-300\,\MeV$ and $Q^2 = 50 -
100\,\GeV^2$, the coefficient of $F_L$ in
(\ref{Eq_xg(x,Q^2)_approximation}), $3\pi/(4\alphaS) = {\cal{O}}(10)$,
is large compared to the one of $F_2$. Taking into account also the
values of $F_2$ and $F_L$, the gluon distribution is mainly determined
by the longitudinal structure function for $x \ltsim 10^{-3}$ in this
$Q^2$ region. The longitudinal structure function can be expressed in
terms of the profile function for longitudinal photon-proton
scattering using the optical theorem (cf.~(\ref{Eq_optical_theorem}))
\be
        F_L(x,Q^2) 
        = \frac{Q^2}{4\,\pi^2\,\alphaEM}\,
        \sigma^{tot}_{\gamma^*_L p}(x,Q^2) 
        = \frac{Q^2}{4\,\pi^2\,\alphaEM}\, 
        2\!\int \!\!d^2b_{\!\perp}\,
        J_{\gamma_L^*p}(x,|\vec{b}_{\!\perp}|,Q^2) 
        \ ,
\label{fl}
\ee
where the $s$-dependence of the profile function is rewritten in terms
of the Bjorken scaling variable $x = Q^2/s$. Thus, neglecting the
$F_2$ term in~(\ref{Eq_xg(x,Q^2)_approximation}), the gluon
distribution reduces to
\be
        xG(x,Q^2) 
        \approx
        1.305\,\frac{Q^2}{\pi^2 \alphaS}\,\frac{\pi}{\alphaEM}
        \int \!\!d^2b_{\!\perp}\,
        J_{\gamma_L^*p}(0.417 x,|\vec{b}_{\!\perp}|,Q^2)
        \ . 
\label{Eq_xg(x,Q^2)-J_gLp(x,b,Q^2)_connection}
\ee
Comparing (\ref{Eq_def_xg(x,Q^2)}) with
(\ref{Eq_xg(x,Q^2)-J_gLp(x,b,Q^2)_connection}), it seems natural to
relate the integrand of (\ref{Eq_xg(x,Q^2)-J_gLp(x,b,Q^2)_connection})
to the impact parameter dependent gluon distribution
\be
        xG(x,Q^2,|\vec{b}_{\!\perp}|) 
        \approx
        1.305\,\frac{Q^2}{\pi^2 \alphaS}\,\frac{\pi}{\alphaEM}\,
        J_{\gamma_L^*p}(0.417 x,|\vec{b}_{\!\perp}|,Q^2)
        \ .
\label{Eq_xg(x,Q^2,b)-J_gLp(x,b,Q^2)_relation}
\ee

The black disc limit of the profile function for longitudinal
photon-proton scattering~(\ref{Eq_gp_black_disc_limit}) imposes
accordingly an upper bound on $xG(x,Q^2,|\vec{b}_{\!\perp}|)$
\be
        xG(x,Q^2,|\vec{b}_{\!\perp}|)\ \leq \ 
        xG^{\mathrm{max}}(Q^2)
        \approx 
        1.305\,\frac{Q^2}{\pi^2 \alphaS}\,\frac{\pi}{\alphaEM}\,
        J_{\gamma^*_L p}^{\mathrm{max}}(Q^2)
        \ ,
\label{Eq_low_x_saturation} 
\ee
which is the low-$x$ saturation value of the gluon distribution
$xG(x,Q^2,|\vec{b}_{\!\perp}|)$ in our approach. With $\pi
J_{\gamma^*_L p}^{\mathrm{max}}(Q^2)/\alphaEM \approx 1$ as shown in
Fig.~\ref{Fig_J_gp_(b=0,s,Q^2)}, a compact approximation
of~(\ref{Eq_low_x_saturation}) is obtained
\be
        xG(x,Q^2,|\vec{b}_{\!\perp}|)\ \leq \ 
        xG^{\mathrm{max}}(Q^2)
        \approx 
        \frac{Q^2}{\pi^2 \alphaS}
        \ ,
\label{Eq_low_x_saturation_approximation} 
\ee
which is consistent with the results
in~\cite{Mueller:1986wy,Mueller:1999wm,Iancu:2001md} and indicates
strong color field strengths $G^a_{\mu \nu} \sim 1/ \sqrt{\alphaS}$ as
well.

According to our
relations~(\ref{Eq_xg(x,Q^2,b)-J_gLp(x,b,Q^2)_relation})
and~(\ref{Eq_low_x_saturation}), the {\em blackness} described by the
profile function is a measure for the gluon distribution and the {\em
  black disc limit} corresponds to the maximum gluon distribution
reached at the impact parameter under consideration. In accordance
with the behavior of the profile function $J_{\gamma_L^*p}$, see
Fig.~\ref{Fig_J_gp_(b,s,Q^2)}, the gluon distribution
$xG(x,Q^2,|\vec{b}_{\!\perp}|)$ decreases with increasing impact
parameter for given values of $x$ and $Q^2$. Consequently, the gluon
density has its maximum in the geometrical center of the proton, i.e.\ 
at zero impact parameter, and decreases towards the periphery.  With
decreasing $x$ at given $Q^2$, the gluon distribution
$xG(x,Q^2,|\vec{b}_{\!\perp}|)$ increases and extends towards larger
impact parameters just as the profile function $J_{\gamma_L^*p}$ for
increasing $s$.  The saturation of the gluon distribution
$xG(x,Q^2,|\vec{b}_{\!\perp}|)$ sets in first in the center of the
proton ($|\vec{b}_{\!\perp}|=0$) at very small Bjorken $x$.

In Fig.~\ref{Fig_xg(x,Q^2,b=0)_vs_x} the gluon distribution
$xG(x,Q^2,|\vec{b}_{\!\perp}|=0)$ is shown as a function of $x$ for
$Q^2 = 1,\,10,\,\mbox{and}\,100\,\GeV^2$, where
relation~(\ref{Eq_xg(x,Q^2,b)-J_gLp(x,b,Q^2)_relation}) has been used
also for low photon virtualities.
\begin{figure}[t]
  \centerline{\epsfig{figure=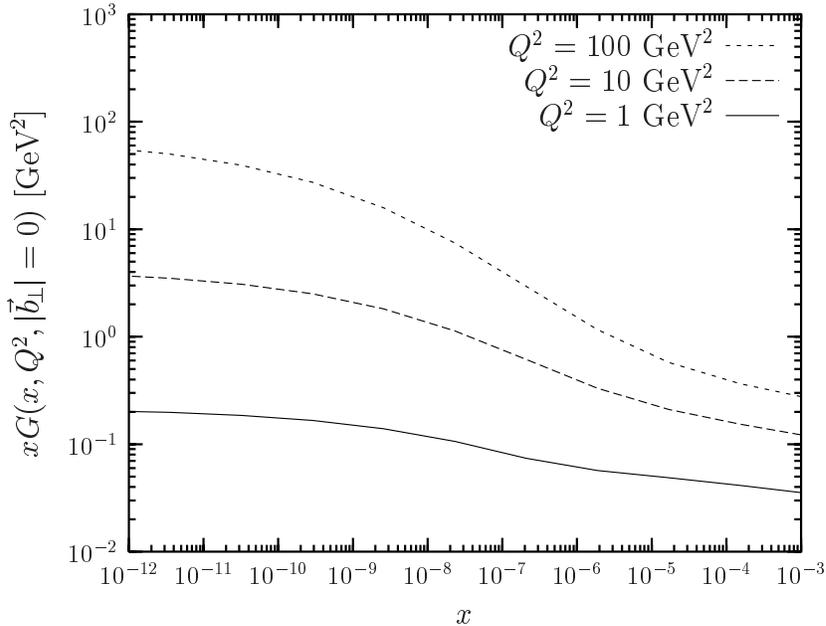,width=11cm}}
  \caption{\small 
    The gluon distribution of the proton at zero impact parameter
    $xG(x,Q^2,|\vec{b}_{\!\perp}|=0)$ as a function of $x$ for $Q^2 =
    1,\,10,\,\mbox{and}\,100\,\GeV^2$. The results are obtained within
    approximation~(\ref{Eq_xg(x,Q^2,b)-J_gLp(x,b,Q^2)_relation}).}
\label{Fig_xg(x,Q^2,b=0)_vs_x}
\end{figure}
Evidently, the gluon distribution $xG(x,Q^2,|\vec{b}_{\!\perp}|=0)$
saturates at very low values of $x \ltsim 10^{-10}$ for $Q^2 \gtsim
1\,\GeV^2$. The photon virtuality $Q^2$ determines the saturation
value~(\ref{Eq_low_x_saturation}) and the Bjorken $x$ at which it is
reached; cf.\ also Fig.~\ref{Fig_J_gp_(b,s,Q^2)}.  For larger $Q^2$,
the low-$x$ saturation value is larger and is reached at smaller
values of $x$, as claimed also in \cite{Gotsman:2001ku}.  Moreover, the
growth of $xG(x,Q^2,|\vec{b}_{\!\perp}|=0)$ with decreasing $x$ becomes
stronger with increasing $Q^2$. This results from the stronger energy
increase of the perturbative component, $\epsilon^{\pert} = 0.73$, that
becomes more important with decreasing dipole size.

According to our approach, the onset of the
$xG(x,Q^2,|\vec{b}_{\!\perp}|)$-saturation appears for $Q^2 \gtsim
1\,\GeV^2$ at $x \ltsim 10^{-10}$, which is far below the $x$-region
accessible at HERA ($x \gtsim 10^{-6}$). Even for THERA ($x\gtsim
10^{-7}$), gluon saturation is not predicted for $Q^2 \gtsim 1
\,\GeV^2$. However, since the HERA data can be described by models
with and without saturation embedded~\cite{Gotsman:2001ku}, the
present situation is not conclusive.\footnote{So far, the most
  striking hint for saturation in the present HERA data at $x\approx
  10^{-4}$ and $Q^2 < 2\,\GeV^2$ has been the turnover of
  $dF_2(x,Q^2)/d\ln(Q^2)$ towards small $x$ in the Caldwell
  plot~\cite{Abramowicz:1999ii}, which is still a controversial issue
  due to the correlation of $Q^2$ and $x$ values.}

Note that the $S$-matrix unitarity
condition~(\ref{Eq_unitarity_condition}) together
with~(\ref{Eq_xg(x,Q^2,b)-J_gLp(x,b,Q^2)_relation}) requires the
saturation of the impact parameter dependent gluon distribution
$xG(x,Q^2,|\vec{b}_{\!\perp}|)$ but not the saturation of the
integrated gluon distribution $xG(x,Q^2)$. Due to multiple gluonic
interactions in our model, this requirement is fulfilled, as can be
seen from Fig.~\ref{Fig_J_gp_(b,s,Q^2)} and
relation~(\ref{Eq_xg(x,Q^2,b)-J_gLp(x,b,Q^2)_relation}). Indeed,
approximating the gluon distribution $xG(x,Q^2,|\vec{b}_{\!\perp}|)$
in the saturation regime of very low $x$ by a step-function
\be
        xG(x,Q^2,|\vec{b}_{\!\perp}|) 
        \approx xG^{\mathrm{max}}(Q^2)\,
        \Theta(\,R(x,Q^2)-|\vec{b}_{\!\perp}|\,)
        \ ,
\label{Eq_J_gp_(x,b,Q^2)_Theta-approximation}
\ee
where $R(x,Q^2)$ denotes the full width at half maximum of the profile
function, one obtains with~(\ref{Eq_def_xg(x,Q^2)}),
(\ref{Eq_low_x_saturation}) and
(\ref{Eq_low_x_saturation_approximation}) the integrated gluon
distribution
\be
        xG(x,Q^2) 
        \;\approx\;
        1.305\,\frac{Q^2\,R^2(x,Q^2)}{\pi \alphaS}\,
        \frac{\pi}{\alphaEM}\,
        J_{\gamma^*_L p}^{\mathrm{max}}(Q^2)
        \;\approx\;
        \frac{Q^2\,R^2(x,Q^2)}{\pi\alphaS}
        \ ,
\label{Eq_xg(x,Q^2)_saturation_regime}
\ee
which does not saturate because of the increase of the effective
proton radius $R(x,Q^2)$ with decreasing $x$. Nevertheless, although
$xG(x,Q^2)$ does not saturate, the saturation of
$xG(x,Q^2,|\vec{b}_{\!\perp}|)$ leads to a slow-down in its growth
towards small $x$.\footnote{This is analogous to the rise of the total
  $pp$ cross section with growing c.m.\ energy that slows down as the
  corresponding profile function $J_{pp}(s,|\vec{b}_{\!\perp}|)$
  reaches its black disc limit as shown in
  Sec.~\ref{Sec_Total_Cross_Sections}.} Interestingly, our
result~(\ref{Eq_xg(x,Q^2)_saturation_regime}) coincides with the
result of Mueller and Qiu~\cite{Mueller:1986wy}.

Finally, it must be emphasized that the low-$x$ saturation of
$xG(x,Q^2,|\vec{b}_{\!\perp}|)$, required in our approach by the
$S$-matrix unitarity, is realized by {\em multiple gluonic
  interactions}. In other approaches that describe the evolution of
the gluon distribution with varying $x$ and $Q^2$, {\em gluon
  recombination} leads to gluon
saturation~\cite{Gribov:1983tu,Mueller:1986wy,McLerran:1994ni+X,Jalilian-Marian:1999dw+X,Iancu:2001hn+X},
which is reached when the probability of a gluon splitting up into two
is equal to the probability of two gluons fusing into one. A more
phenomenological understanding of saturation is attempted
in~\cite{Golec-Biernat:1999js+X,Capella:2001hq+X}.


%
\cleardoublepage
%
\chapter{Comparison with Experimental Data}
\label{Sec_Comparison_Data}

In this chapter we present the phenomenological performance of our
model. We compute total, differential, and elastic cross sections,
structure functions, and diffractive slopes for hadron-hadron,
photon-proton, and photon-photon scattering, compare the results with
experimental data including cosmic ray data, and provide predictions
for future experiments. Having studied the saturation of the impact
parameter profiles, we show here how this manifestation of unitarity
translates into the quantities mentioned above and how it could become
observable.

Using the
$T$-matrix~(\ref{Eq_model_purely_imaginary_T_amplitude_final_result})
with the wave functions and parameters from
Secs.~\ref{Sec_Wave_Functions} and~\ref{Sec_Model_Parameters}, we
compute the {\em pomeron} contribution to $pp$, $p\pbar$,
$\pi^{\pm}p$, $K^{\pm}p$, $\gamma^{*} p$, and $\gamma \gamma$
reactions in terms of the universal dipole-dipole scattering amplitude
$S_{DD}$. This allows one to compare reactions induced by hadrons and
photons in a systematic way. In fact, it is our aim to provide a
unified description of all these reactions and to show in this way
that the pomeron contribution to the above reactions is universal and
can be traced back to the dipole-dipole scattering amplitude $S_{DD}$.

Our model describes pomeron ($C=+1$ gluon exchange) but neither
odderon ($C=-1$ gluon exchange) nor reggeon exchange (quark-antiquark
exchange) as discussed in Sec.~\ref{Sec_DD_Scattering_LLCM}. Only in the
computation of the hadronic total cross sections the reggeon
contribution is added~\cite{Donnachie:1992ny,Donnachie:2000kp}. This
improves the agreement with the data for $\sqrt{s} \ltsim 100\,\GeV$
and describes exactly the differences between $ab$ and $\bar{a}b$
reactions.

The fine tuning of the model and wave function parameters was
performed on the data shown below. The resulting parameter set given
in Secs.~\ref{Sec_Wave_Functions} and~\ref{Sec_Model_Parameters}
is used throughout this chapter.

\section{Total Cross Sections}
\label{Sec_Total_Cross_Sections}

The total cross section for the high-energy reaction $ab \to X$ is
related via the {\em optical theorem} to the imaginary part of the
forward elastic scattering amplitude and can also be expressed in
terms of the profile function~(\ref{Eq_profile_function_def})
\be
        \sigma^{tot}_{ab}(s) 
        \;=\; \inv{s}\,\im\,T(s, t=0) 
        \;=\; 2 \int \!d^2b_{\!\perp}\,J_{ab}(s,|\vec{b}_{\!\perp}|)
        \ ,  
\label{Eq_optical_theorem}
\ee
where $a$ and $b$ label the initial particles whose masses were
neglected as they are small in comparison to the c.m.\ energy
$\sqrt{s}$. 

We compute the pomeron contribution to the total cross section,
$\sigma^{tot, \Pomeron}_{ab}(s)$, from the
$T$-matrix~(\ref{Eq_model_purely_imaginary_T_amplitude_final_result}),
as explained above, and add only here a reggeon contribution of the
form~\cite{Donnachie:1992ny,Donnachie:2000kp}
\be
        \sigma^{tot, \Reggeon}_{ab}(s)
        = X_{ab}\, \left(  \frac{s}{1\,\GeV^2} \right)^{-0.4525} 
        \ ,
\label{Eq_DL_reggeon_contribution}
\ee
where $X_{ab}$ depends on the reaction considered: $X_{pp} =
56.08\,\mb$, $X_{p\pbar} = 98.39\,\mb$, $X_{\pi^+p} = 27.56\,\mb$,
$X_{\pi^-p} = 36.02\,\mb$, $X_{K^+p} = 8.15\,\mb$, $X_{K^-p} =
26.36\,\mb$, $X_{\gamma p} = 0.129\,\mb$, and $X_{\gamma \gamma} =
605\,\nb$. Accordingly, we obtain the total cross section
\be
        \sigma^{tot}_{ab}(s)
        = \sigma^{tot, \Pomeron}_{ab}(s) 
        + \sigma^{tot, \Reggeon}_{ab}(s)
\label{Eq_total_cross_section_final_result}
\ee
for $pp$, $p\pbar$, $\pi^{\pm}p$, $K^{\pm}p$, $\gamma p$, and $\gamma
\gamma$ scattering.

\begin{figure}[p]
\setlength{\unitlength}{1.cm}
\begin{center}
\epsfig{file=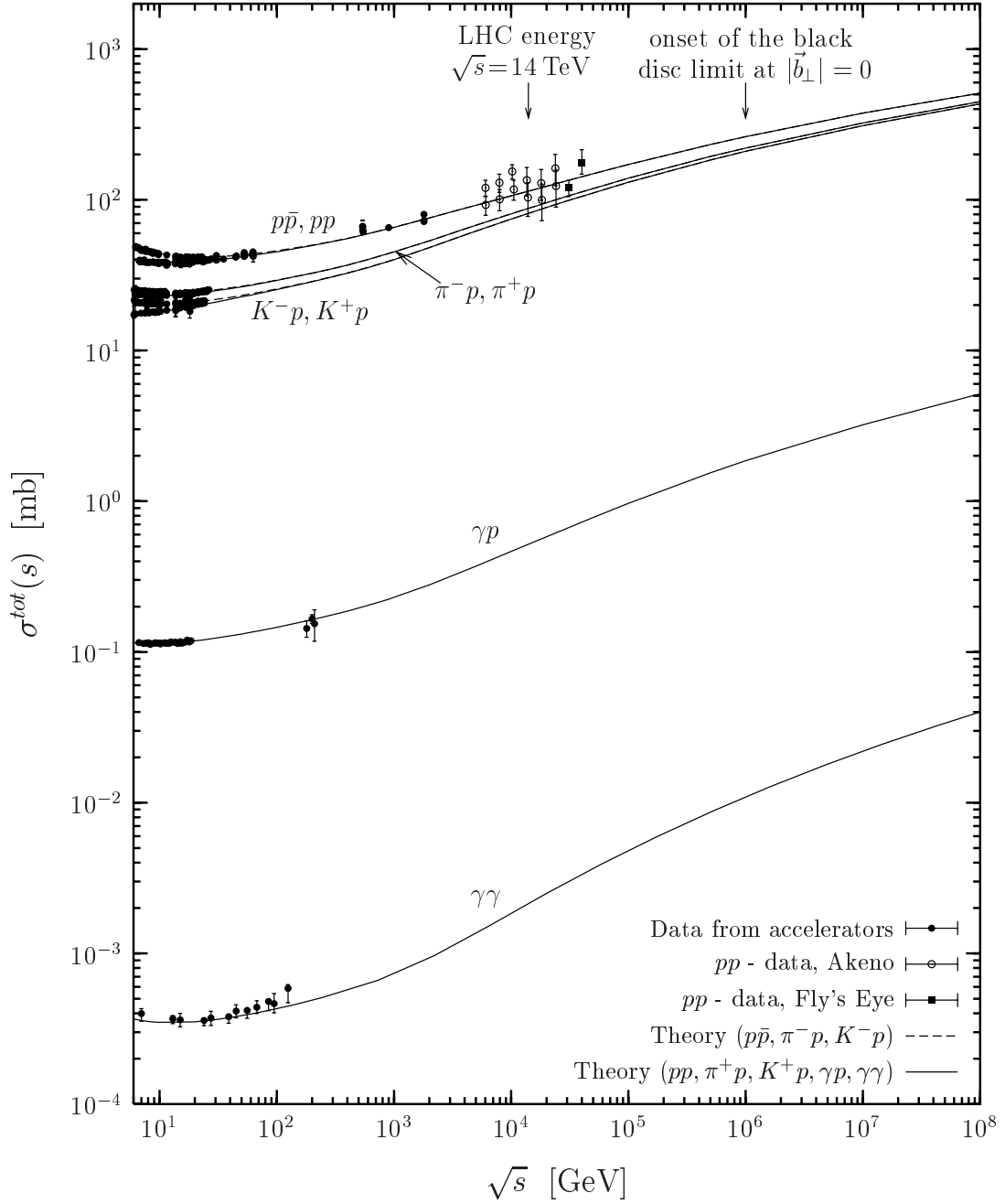,width=14.5cm}
\end{center}
\caption{ \small 
  The total cross section $\sigma^{tot}$ as a function of the c.m.\ 
  energy $\sqrt{s}$ for $pp$, $p\pbar$, $\pi^{\pm}p$, $K^{\pm}p$,
  $\gamma p$, and $\gamma \gamma$ scattering.  The solid lines
  represent the model results for $pp$, $\pi^+p$, $K^+p$, $\gamma p$
  and $\gamma \gamma$ scattering and the dashed lines the ones for
  $p\pbar$, $\pi^-p$, and $K^-p$ scattering. The $pp$, $p\pbar$,
  $\pi^{\pm}p$, $K^{\pm}p$, $\gamma p$~\cite{Hagiwara:fs} and $\gamma
  \gamma$ data~\cite{Abbiendi:2000sz+X} taken at accelerators are
  indicated by the closed circles while the closed squares (Fly's eye
  data)~\cite{Baltrusaitis:1984ka+X} and the open circles (Akeno
  data)~\cite{Honda:1993kv+X} indicate cosmic ray data.  The arrows at
  the top point to the LHC energy, $\sqrt{s} = 14\,\TeV$, and to the
  onset of the black disc limit in $pp$ ($p\pbar$) reactions,
  $\sqrt{s} \approx 10^6\,\GeV$.}
\label{Fig_sigma_tot}
\end{figure}

The good agreement of the computed total cross sections with the
experimental data is shown in Fig.~\ref{Fig_sigma_tot}.
Here the solid lines represent the theoretical results for $pp$,
$\pi^+p$, $K^+p$, $\gamma p$, and $\gamma \gamma$ scattering and the
dashed lines the ones for $p\pbar$, $\pi^-p$, and $K^-p$ scattering.
The $pp$, $p\pbar$, $\pi^{\pm}p$, $K^{\pm}p$, $\gamma
p$~\cite{Hagiwara:fs} and $\gamma \gamma$
data~\cite{Abbiendi:2000sz+X} taken at accelerators are indicated by
the closed circles while the closed squares (Fly's eye
data)~\cite{Baltrusaitis:1984ka+X} and the open circles (Akeno
data)~\cite{Honda:1993kv+X} indicate cosmic ray data. Only real
photons are considered which are, of course, transverse polarized.

The prediction for the total $pp$ cross section at LHC ($\sqrt{s} =
14\,\TeV$) is $\sigma^{tot}_{pp} = 114.2\,\mb$ in good agreement with
cosmic ray data. Compared with other works, our LHC prediction is
close to the one of Block et al.~\cite{Block:1999hu},
$\sigma^{tot}_{pp} = 108 \pm 3.4\,\mb$, but considerably larger than
the one of Donnachie and Landshoff~\cite{Donnachie:1992ny},
$\sigma^{tot}_{pp} = 101.5\,\mb$.

The differences between $ab$ and $\bar{a}b$ reactions for $\sqrt{s}
\ltsim 100\,\GeV$ result solely from the different reggeon
contributions which die out rapidly as the energy increases. The
pomeron contribution to $ab$ and $\bar{a}b$ reactions is, in
contrast, identical and increases as the energy increases. It thus
governs the total cross sections for $\sqrt{s} \gtsim 100\,\GeV$ where
the results for $ab$ and $\bar{a}b$ reactions coincide.

The differences between $pp$ ($p\pbar$), $\pi^{\pm}p$, and $K^{\pm}p$
scattering result from the different transverse extension parameters,
$S_p = 0.86\,\fm > S_{\pi} = 0.607\,\fm > S_{K} = 0.55\,\fm$, cf.\ 
Sec.~\ref{Sec_Wave_Functions}.  Since a smaller transverse extension
parameter favors smaller dipoles, the total cross section becomes
smaller and the short distance physics described by the perturbative
component becomes more important which leads to a stronger energy
growth due to $\epsilon^{\pert} = 0.73 > \epsilon^{\nprt} = 0.125$. In
fact, the ratios $\sigma^{tot}_{pp}/\sigma^{tot}_{\pi p}$ and
$\sigma^{tot}_{pp}/\sigma^{tot}_{Kp}$ converge slowly towards unity
with increasing energy as can already be seen in
Fig.~\ref{Fig_sigma_tot}.

For real photons, the transverse size is governed by the constituent
quark masses $m_f(Q^2=0)$, cf.\ Sec.~\ref{Sec_Wave_Functions},
where the light quarks have the strongest effect, i.e.\ 
$\sigma^{tot}_{\gamma p} \propto 1/m_{u,d}^2$ and
$\sigma^{tot}_{\gamma \gamma} \propto 1/m_{u,d}^4$. Furthermore, in
comparison with hadron-hadron scattering, there is the additional
suppression factor of $\alphaEM$ for $\gamma p$ and $\alphaEM^2$ for
$\gamma \gamma$ scattering coming from the photon-dipole transition.
In the $\gamma \gamma$ reaction, also the box diagram
contributes~\cite{Budnev:1975zs,Donnachie:2000kp} but is neglected
since its contribution to the total cross section is less than
1\%~\cite{Donnachie:2001wt}.

It is worthwhile mentioning that total cross sections for $pp$
($p\pbar$), $\pi^{\pm}p$, and $K^{\pm}p$ scattering do not depend on
the longitudinal quark momentum distribution in the hadrons since the
underlying dipole-dipole cross section is independent of the
longitudinal quark momentum fraction $z_i$ for $t = 0$. We have shown
this analytically on the two-gluon-exchange level
in~\cite{Shoshi:2002fq}.

Saturation effects as a manifestation of the $S$-matrix unitarity can
be seen in Fig.~\ref{Fig_sigma_tot}. Having investigated the profile
function for $pp$ ($p\pbar$) scattering, we know that this profile
function becomes higher and broader with increasing energy until it
saturates the black disc limit first for zero impact parameter
($|\vec{b}_{\!\perp}|=0$) at $\sqrt{s} \approx 10^6\,\GeV$.
Beyond this energy, the profile function cannot become higher but
expands towards larger values of $|\vec{b}_{\!\perp}|$. Consequently,
the total cross section~(\ref{Eq_optical_theorem}) increases no longer
due to the growing blackness at the center but only due to the
transverse expansion of the hadrons. This tames the growth of the
total cross section as can be seen for the total $pp$ cross section
beyond $\sqrt{s} \approx 10^6\,\GeV$ in Fig.~\ref{Fig_sigma_tot}.

At energies beyond the onset of the black disc limit at zero impact
parameter, the profile function can be approximated by
\be 
      J_{ab}^{\mathrm{approx}}(s,|\vec{b}_{\!\perp}|) =
      N_a\,N_b\,\Theta\left(R(s)-|\vec{b}_{\!\perp}|\right) \,
\label{Eq_J_ab_asymptotic_energies}
\ee
where $N_{a,b}$ denotes the normalization of the wave functions of the
scattered particles and $R(s)$ the full width at half maximum of the
exact profile function $J_{ab}(s,|\vec{b}_{\!\perp}|)$ that reflects
the effective radii of the interacting particles. Thus, the energy
dependence of the total cross section~(\ref{Eq_optical_theorem}) is
driven exclusively by the increase of the transverse extension of the
particles $R(s)$
\be
        \sigma^{tot}_{ab}(s) = 2 \pi N_a N_b R(s)^2
        \ ,
\label{Eq_sigma_tot_ab_asymptotic_energies}
\ee
which is known as {\em geometrical
  scaling}~\cite{Amaldi:1980kd,Castaldi:1983ft}. The growth of $R(s)$
can at most be logarithmic for $\sqrt{s} \to \infty$ because of the
Froissart bound~\cite{Froissart:1961ux}. In fact, a transition from a
power-like to an $\ln^2$-increase of the total $pp$ cross section
seems to set in at about $\sqrt{s} \approx 10^6\,\GeV$ as visible in
Fig.~\ref{Fig_sigma_tot}. Moreover, since the hadronic cross sections
join for $\sqrt{s} \to \infty$, $R(s)$ becomes independent of the
hadron-hadron reaction considered at asymptotic energies as long as
$N_{a,b}=1$. Also for photons of different virtuality $Q_1^2$ and
$Q_2^2$ one can check that the ratio of the total cross sections
$\sigma^{tot}_{\gamma^* p}(Q_1^2)/\sigma^{tot}_{\gamma^* p}(Q_2^2)$
converges to unity at asymptotic energies in agreement with the
conclusion in~\cite{Schildknecht:2001qe}.

\section{The Proton Structure Function}
\label{Sec_Structure_Functions}

The total cross section for the scattering of a transverse ($T$) and
longitudinally ($L$) polarized photon off the proton,
$\sigma_{\gamma^*_{T\!,L}p}^{tot}(x,Q^2)$, at photon virtuality $Q^2$
and c.m.\ energy\footnote{Here $\sqrt{s}$ refers to the c.m.\ energy
  in the $\gamma^* p$ system.} squared $s=Q^2/x$ is equivalent to the
{\em structure functions} of the proton
\be
        F_{T,L}(x,Q^2) 
        = \frac{Q^2}{4\pi^2\alphaEM} 
        \sigma_{\gamma^*_{T\!,L}p}^{tot}(x,Q^2)
\label{Eq_FTL}
\ee
and
\be
        F_2(x,Q^2) = F_{T}(x,Q^2) + F_{L}(x,Q^2)
        \ .
\label{Eq_F2}
\ee

Reactions induced by virtual photons are particularly interesting
because the transverse separation of the quark-antiquark pair that
emerges from the virtual photon decreases as the photon virtuality
increases (cf.\ Sec.~\ref{Sec_Wave_Functions})
\be
        |\vec{r}_\gamma| \approx \frac{2}{\sqrt{Q^2+4m_{u,d}^2}} 
        \ ,
\label{pts}
\ee
where $m_{u,d}$ is a mass of the order of the constituent $u$-quark
mass. With increasing virtuality, one probes therefore smaller
transverse distance scales of the proton.

\begin{figure}[p]
  \centerline{\psfig{figure=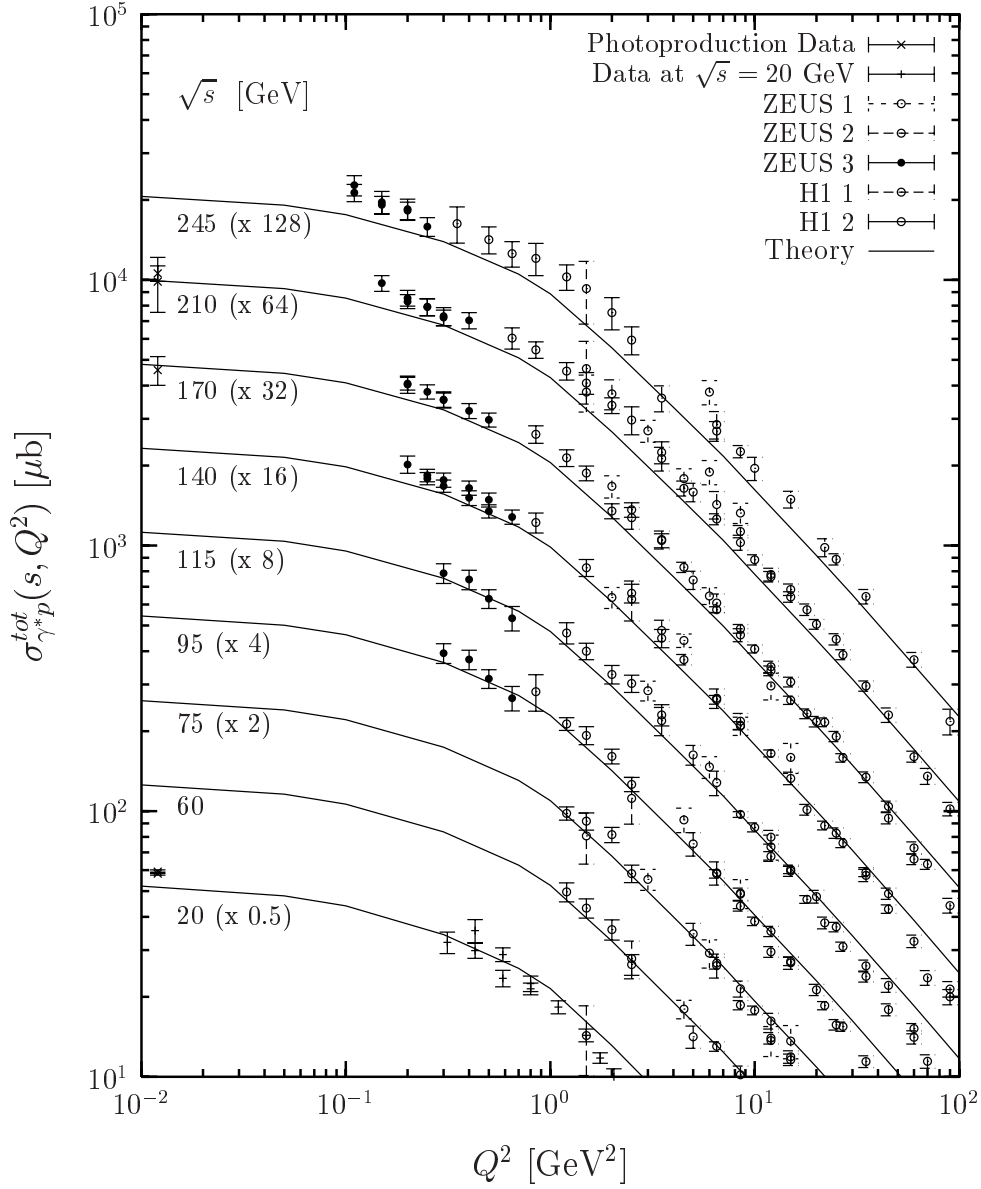,
      width=13cm}}\bigskip 
\protect\caption{\small 
  The total $\gamma^* p$ cross section
  $\sigma^{tot}_{\gamma^*p}(s,Q^2)$ as a function of the photon
  virtuality $Q^2$ for c.m.\ energies from $\sqrt{s} = 20\,\GeV$ to
  $245\,\GeV$, where the model results (solid lines) and the
  experimental data at different $\sqrt{s}$ values are scaled with the
  indicated factors. The low energy data at $\sqrt{s} = 20\,\GeV$ are
  from~\cite{Benvenuti:1989rh+X}, the data at higher energies from the
  H1~\cite{Aid:1996au,Adloff:1997mf} and ZEUS
  collaboration~\cite{Derrick:1996ef+X,Breitweg:1997hz}. The
  photoproduction ($Q^2=0$) data from~\cite{Caldwell:1978yb+X} are
  displayed at $Q^2 = 0.012\,\GeV^2$.
}
\label{Fig_sigma_tot_gp_vs_Q^2}
\end{figure}

In Fig.~\ref{Fig_sigma_tot_gp_vs_Q^2} the $Q^2$-dependence of the
total $\gamma^* p$ cross section
\be
        \sigma^{tot}_{\gamma^*p}(s,Q^2)
        = \sigma^{tot}_{\gamma_T^*p}(s,Q^2)
        + \sigma^{tot}_{\gamma_L^*p}(s,Q^2)
\label{Eq_sigma_tot_gp_=_sigma_T_+_sigma_L}
\ee
is presented, where the model results (solid lines) are compared with
the experimental data for c.m.\ energies from $\sqrt{s} = 20\,\GeV$ up
to $245\,\GeV$. Note the indicated scaling factors at different
$\sqrt{s}$ values. The low-energy data at $\sqrt{s} = 20\,\GeV$ are
from~\cite{Benvenuti:1989rh+X} while the data at higher energies have
been measured at HERA by the H1~\cite{Aid:1996au,Adloff:1997mf} and ZEUS
collaboration~\cite{Derrick:1996ef+X,Breitweg:1997hz}. At $Q^2 = 0.012\,\GeV^2$, also
the photoproduction ($Q^2=0$) data from~\cite{Caldwell:1978yb+X} are
displayed.

The model results are in reasonable agreement with the experimental
data in the window shown in Fig.~\ref{Fig_sigma_tot_gp_vs_Q^2}. The
total $\gamma^* p$ cross section levels off towards small values of
$Q^2$ as soon as the photon size $|\vec{r}_\gamma|$, i.e\ the
resolution scale, becomes comparable to the proton size. Our model
reproduces this behavior by using the perturbative photon wave
functions with $Q^2$-dependent quark masses, $m_f(Q^2)$, that
interpolate between the current (large $Q^2$) and the constituent
(small $Q^2$) quark masses as explained in detail in
Sec.~\ref{Sec_Wave_Functions}. The decrease of $\sigma^{tot}_{\gamma^*
  p}$ with increasing $Q^2$ results from decreasing dipole sizes and
the fact that small dipoles do not interact as strongly as large
dipoles.

\begin{figure}[h]
\centerline{\psfig{figure=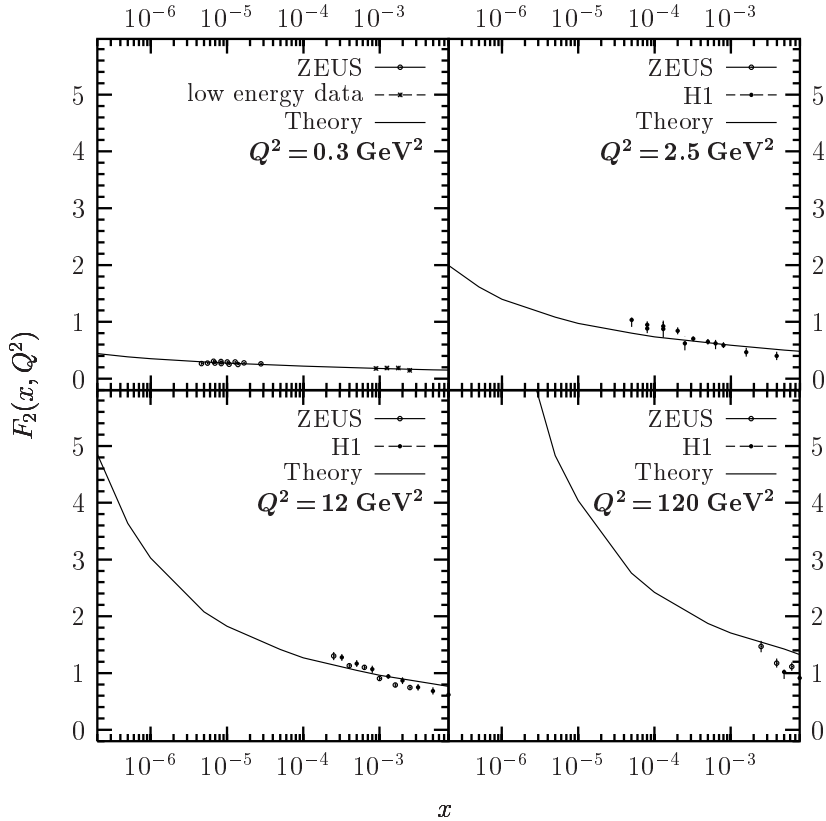, width=11.cm}}
\caption{\small 
  The $x$-dependence of the computed proton structure function
  $F_2(x,Q^2)$ (solid line) for $Q^2 = 0.3,\,2.5,\,12,$ and
  $120\,\GeV^2$ in comparison to the data measured by the
  H1~\cite{Abt:1993cb+X} and ZEUS~\cite{Derrick:1993ft+X} detector,
  and the low energy data at $\sqrt{s} = 20\,\GeV$
  from~\cite{Benvenuti:1989rh+X}.}
\label{F2_p}
\end{figure}

The $x$-dependence of the computed proton structure function
$F_2(x,Q^2)$ is shown in Fig.~\ref{F2_p} for $Q^2 = 0.3,\,2.5,\,12,$
and $120\,\GeV^2$ in comparison to the data measured by the
H1~\cite{Abt:1993cb+X} and ZEUS~\cite{Derrick:1993ft+X} detector.
Within our model, the increase of $F_2(x,Q^2)$ towards small Bjorken
$x$ becomes stronger with increasing $Q^2$ in agreement with the trend
in the HERA data. This behavior results from the fast energy growth of
the perturbative component that becomes more important with increasing
$Q^2$ due to the smaller dipole sizes involved.

As can be seen in Fig.~\ref{F2_p}, the data show a faster increase
with decreasing $x$ than the model outside the low-$Q^2$ region. This
results from the weak energy boost of the non-perturbative component
that dominates $F_2(x,Q^2)$ in our model. In fact, even for large
$Q^2$ the non-perturbative contribution overwhelms the perturbative
one, which explains also the overestimation of the data for $x \gtsim
10^{-3}$.
 
This problem is typical for the \SVM\ model applied to the scattering
of a small size dipole off a proton. In an earlier application by
R\"uter~\cite{Rueter:1998up}, an additional cut-off was introduced to
switch from the non-perturbative to the perturbative contribution as
soon as one of the dipoles is smaller than $r_{cut} = 0.16\,\fm$. This
yields a better agreement with the data also for large $Q^2$ but leads
to a discontinuous dipole-proton cross section. In the model of
Donnachie and Dosch~\cite{Donnachie:2001wt}, a similar \SVM-based
component is used also for dipoles smaller than $R_c = 0.22\,\fm$ with
a strong energy boost instead of a perturbative component.
Furthermore, their \SVM-based component is tamed for large $Q^2$ by an
additional $\alphaS(Q^2)$ factor.

We did not follow these lines in order to keep a continuous,
$Q^2$-independent dipole-proton cross section and, therefore, cannot
improve the agreement with the $F_2(x,Q^2)$ data without losing
quality in the description of hadronic observables. Since our
non-perturbative component relies on lattice QCD, we are more
confident in describing non-perturbative physics and thus put more
emphasis on the hadronic observables. Admittedly, our model misses
details of the proton structure that become visible with increasing
$Q^2$. However, most other existing models provide neither the profile
functions nor a simultaneous description of hadronic and
$\gamma^*$-induced processes.

\section[The Slope $B$ of Elastic Forward Scattering]
{The Slope \boldmath$B$ of Elastic Forward Scattering}
\label{Sec_Slope_B}

The {\em local slope} of elastic scattering $B(s,t)$ is defined as
\be
        B(s,t) := 
        \frac{d}{dt} \left( \ln \left[ \frac{d\sigma^{el}}{dt}(s,t) \right] \right)
\label{Eq_elastic_local_slope}
\ee
and characterizes the diffractive peak of the differential elastic
cross section $d\sigma^{el}/dt(s,t)$ discussed below. Here we
concentrate on the slope for elastic forward ($t=0$) scattering also
called {\em slope parameter}
\be
        B(s) 
        := B(s,t=0) 
        = \inv{2} 
        \frac{\int\!d^2b_{\!\perp}\,|\vec{b}_{\!\perp}|^2\,J(s,|\vec{b}_{\!\perp}|)}
        {\int\!d^2b_{\!\perp}\,J(s,|\vec{b}_{\!\perp}|)}
        = \inv{2} \langle b^2 \rangle \ ,
\label{Eq_elastic_forward_slope}
\ee
which measures the root mean squared interaction radius $\langle b^2
\rangle$ of the scattered particles, and does not depend on the
opacity.

We compute the slope parameter with the profile function from the
$T$-matrix~(\ref{Eq_model_purely_imaginary_T_amplitude_final_result})
and neglect the reggeon contributions, which are relevant only at
small c.m.\ energies, so that the same result is obtained for $ab$ and
$\bar{a}b$ scattering.

\begin{figure}[t] 
\centerline{\epsfig{figure=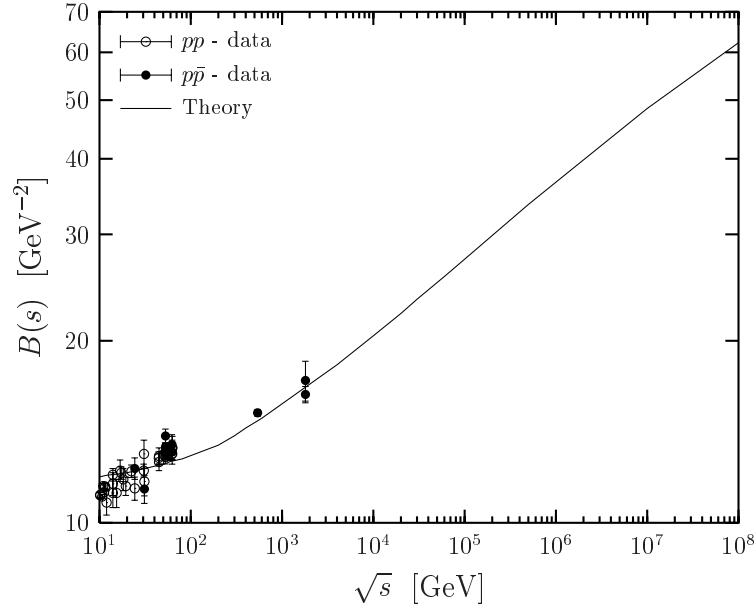,width=10.cm}}
\caption{\small
  The elastic slope parameter $B(s)$ as a function of the c.m.\ energy
  $\sqrt{s}$ for $pp$ and $p\pbar$ forward ($t=0$) scattering. The
  solid line represents the model result that is compared to the data
  for $pp$ (open circles) and $p\pbar$ (closed circles) reactions
  from~\cite{Amaldi:1971kt+X,Bozzo:1984ri,Amos:1989at}.}
\label{Fig_B_pp}
\end{figure}

In Fig.~\ref{Fig_B_pp} the resulting slope parameter $B(s)$ is shown
as a function of $\sqrt{s}$ for $pp$ and $p\pbar$ scattering (solid
line) and compared with the $pp$ (open circles) and $p\pbar$ (closed
circles) data from~\cite{Amaldi:1971kt+X,Bozzo:1984ri,Amos:1989at}.
As expected from the opacity independence of the slope parameter
(\ref{Eq_elastic_forward_slope}), saturation effects as seen in the
total cross sections do not occur. Indeed, one observes an approximate
$B(s) \propto R^2(s) \propto \ln^2(\sqrt{s}/\sqrt{s_0})$ growth for
$\sqrt{s} \gtsim 10^4\,\GeV$.  This behavior agrees, of course, with
the transverse expansion of $J_{pp}(s,|\vec{b}_{\!\perp}|)$ for
increasing $\sqrt{s}$ shown in Fig.~\ref{Fig_J_pp(b,s)}. Analogous
results are obtained also for $\pi p$ and $Kp$ scattering.

For the good agreement of our model with the data, the finite width of
the longitudinal quark momentum distribution in the hadrons, i.e.\ 
$\Delta z_p,\,\Delta z_{\pi},\,\mbox{and}\,\Delta z_{K}\neq 0$
in~(\ref{Eq_hadron_wave_function}), is important as the numerator in
(\ref{Eq_elastic_forward_slope}) depends on this width. In fact,
$B(s)$ comes out more than 10\% smaller with $\Delta z_p,\,\Delta
z_{\pi},\,\mbox{and}\,\Delta z_{K}= 0$. Furthermore, a strong growth
of the perturbative component, $\epsilon^{\pert} = 0.73$, is important
to achieve the increase of $B(s)$ for $\sqrt{s} \gtsim 500\,\GeV$
indicated by the data.

It must be emphasized that only the simultaneous fit of the total
cross section and the slope parameter provides the correct shape of
the profile function $J(s,|\vec{b}_{\!\perp}|)$. This shape leads then
automatically to a good description of the differential elastic cross
section $d\sigma^{el}/dt(s,t)$ as demonstrated below. Astonishingly,
only few phenomenological models provide a satisfactory description of
both quantities~\cite{Block:1999hu,Kopeliovich:2001pc}. In the
approach of~\cite{Berger:1999gu}, for example, the total cross section
is described correctly while the slope parameter exceeds the data by
more than 20\% already at $\sqrt{s} = 23.5\,\GeV$ and thus indicates
deficiencies in the form of $J(s,|\vec{b}_{\!\perp}|)$.

\section{The Differential Elastic Cross Section}
\label{Sec_Diff_El_Cross_Section}

The {\em differential elastic cross section} obtained from the squared
absolute value of the $T$-matrix element
\be
        \frac{d\sigma^{el}}{dt}(s,t) 
        = \inv{16 \pi s^2}|T(s,t)|^2
\label{Eq_dsigma_el_dt}
\ee
can be expressed for our purely imaginary
$T$-matrix~(\ref{Eq_model_purely_imaginary_T_amplitude_final_result})
in terms of the profile function
\be
        \frac{d\sigma^{el}}{dt}(s,t) 
        = \inv{4\pi} \left[ 
        \int \!\!d^2b_{\!\perp}\,
        e^{i {\vec q}_{\!\perp} {\vec b}_{\!\perp}}\,
        J(s,|\vec{b}_{\!\perp}|)
        \right ]^2
        \ .
\label{Eq_dsigma_el_dt_model}
\ee
and is, thus, very sensitive to the transverse extension {\em and}
opacity of the scattered particles.
Equation~(\ref{Eq_dsigma_el_dt_model}) shows the analogy to optical
diffraction, where $J(s,|\vec{b}_{\!\perp}|)$ describes the
distribution of an absorber that causes the diffraction pattern
observed for incident plane waves.

\begin{figure}[p]
  \centerline{\psfig{figure=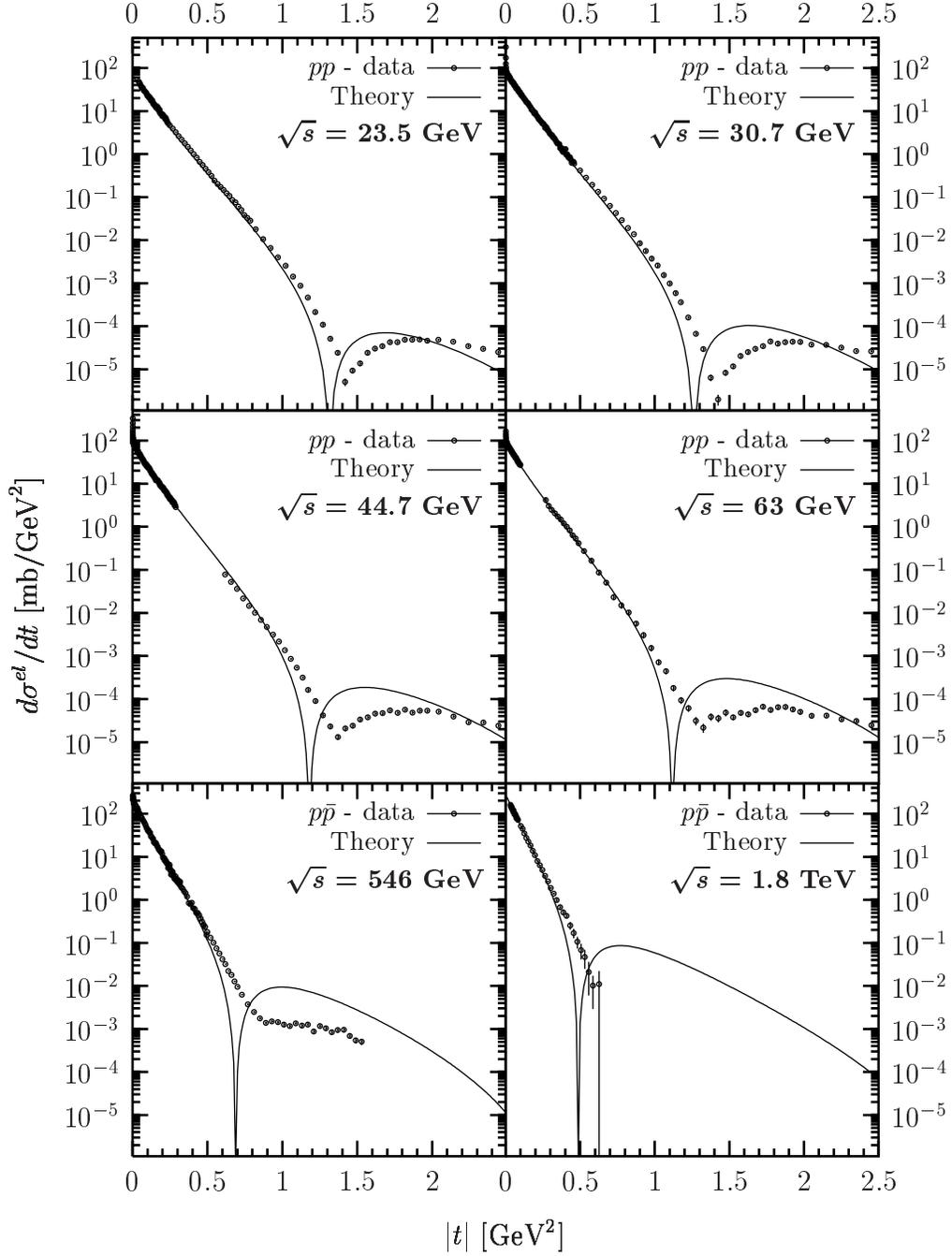,width=13.5cm}} 
\protect\caption{ \small 
  The differential elastic cross section for $pp$ and $p\pbar$
  scattering as a function of $|t|$. The result of our model,
  indicated by the solid line, is compared for $\sqrt{s} =
  23.5,\,30.7,\,44.7,\,\mbox{and}\,63\,\GeV$ with the CERN ISR $pp$
  data~\cite{Amaldi:1980kd}, for $\sqrt{s} = 546\,\GeV$ with the CERN
  $Sp{\pbar}S$ data~\cite{Bozzo:1984ri}, and for $\sqrt{s} =
  1.8\,\TeV$ with the Fermilab Tevatron $p\pbar$
  data~\cite{Amos:1989at,Amos:1990jh}, all indicated by the open
  circles with error bars.
}
\label{Fig_dsigma_el_dt_pp}
\end{figure}

\begin{figure}[tb]
  \centerline{\psfig{figure=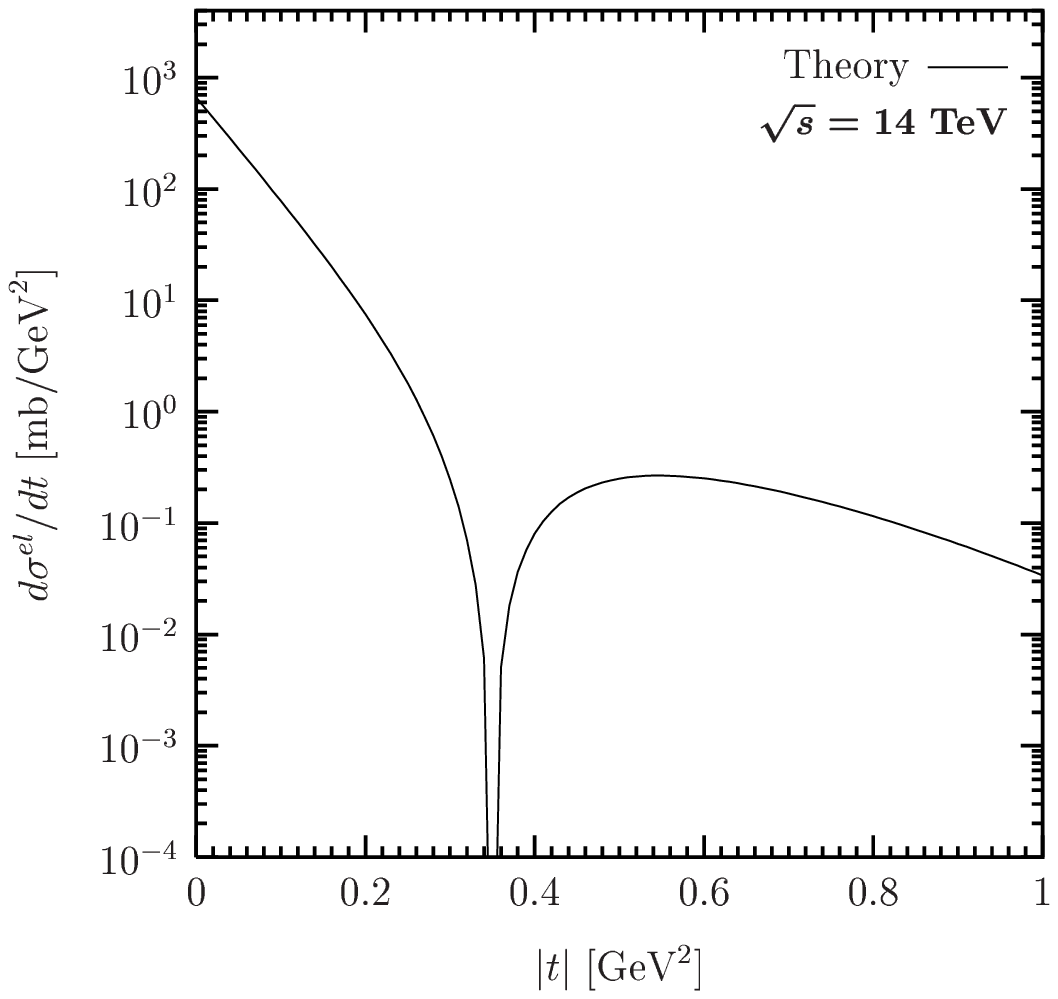,width=9.cm}}
  \protect\caption{ \small 
    The prediction of our model for the $pp$ differential elastic
    cross section at LHC ($\sqrt{s} = 14\,\TeV$) as a
    function of momentum transfer $|t|$ up to $1\,\GeV^2$.
}
\label{Fig_dsigma_el_dt_pp_LHC}
\end{figure}

In Fig.~\ref{Fig_dsigma_el_dt_pp} the differential elastic cross
section computed for $pp$ and $p\pbar$ scattering (solid line) is
shown as a function of $|t|=\vec{q}^{\,2}_{\!\perp}$ at $\sqrt{s} =
23.5,\,30.7,\,44.7,\,63,\,546$, and $1800\,\GeV$ and compared with
experimental data (open circles), where the $pp$ data at $\sqrt{s} =
23.5,\,30.7,\,44.7,\,\mbox{and}\,63\,\GeV$ were measured at the CERN
ISR~\cite{Amaldi:1980kd}, the $p\pbar$ data at $\sqrt{s} = 546\,\GeV$
at the CERN $Sp{\pbar}S$~\cite{Bozzo:1984ri}, and the $p\pbar$ data at
$\sqrt{s} = 1.8\,\TeV$ at the Fermilab
Tevatron~\cite{Amos:1989at,Amos:1990jh}. The prediction of our model
for the $pp$ differential elastic cross section at the CERN LHC,
$\sqrt{s} = 14\,\TeV$, is given in Fig.~\ref{Fig_dsigma_el_dt_pp_LHC}.

For all energies, the model reproduces the experimentally observed
diffraction pattern, i.e\ the characteristic {\em diffraction peak} at
small $|t|$ and the {\em dip} structure at medium $|t|$. As the energy
increases, also the {\em shrinking of the diffraction peak} is
described which reflects the rise of the slope parameter $B(s,t=0)$
already discussed above. The shrinking of the diffraction peak comes
along with a dip structure that moves towards smaller values of
$|t|$ as the energy increases. This motion of the dip is also
described approximately.

The dip in the theoretical curves reflects a change of sign in the
$T$-matrix
element~(\ref{Eq_model_purely_imaginary_T_amplitude_final_result}). As
the latter is purely imaginary, it is not surprising that there are
deviations from the data in the dip region. Here  the real part is
expected to be important~\cite{Amos:1990jh} which is in the small
$|t|$ region negligible in comparison to the imaginary part.

The difference between the $pp$ and $p\pbar$ data, a deep dip for $pp$
but only a bump or shoulder for $p\pbar$ reactions, requires a $C = -
1$ contribution. Besides the reggeon contribution at small
energies,\footnote{Zooming in on the result for $\sqrt{s} =
  23.5\,\GeV$, one finds an underestimation of the data for all $|t|$
  before the dip, which is correct as it leaves room for the reggeon
  contribution being non-negligible at small energies.} one expects
here an additional perturbative $C=-1$ contribution such as
three-gluon exchange~\cite{Fukugita:1979fe,Donnachie:1984hf+X} or an
odderon~\cite{Lukaszuk:1973nt+X,Rueter:1999gj,Dosch:2002ai}. In fact,
allowing a finite size diquark in the (anti-)proton an odderon appears
that supports the dip in $pp$ but leads to the shoulder in $p\pbar$
reactions~\cite{Dosch:2002ai}.

For the differential elastic cross section at the LHC energy,
$\sqrt{s} = 14\,\TeV$, the above findings suggest an accurate
prediction in the small-$|t|$ region but a dip at a position smaller
than the predicted value at $|t| \approx 0.35\,\GeV^2$. Our confidence
in the validity of the model at small $|t|$ is supported additionally
by the total cross section that fixes $d\sigma^{el}/dt(s,t=0)$ and is
in agreement with the cosmic ray data shown in
Fig.~\ref{Fig_sigma_tot}. Concerning our prediction for the dip
position, it is close to the value $|t| \approx 0.41\,\GeV^2$
of~\cite{Block:1999hu} but significantly below the value $|t| \approx
0.55\,\GeV^2$ of~\cite{Berger:1999gu}. Beyond the dip position, the
height of the computed shoulder is always above the data and, thus,
very likely to exceed also the LHC data. In comparison with other
works, the height of our shoulder is similar to the one
of~\cite{Block:1999hu} but almost one order of magnitude above the one
of~\cite{Berger:1999gu}.

Considering Figs.~\ref{Fig_dsigma_el_dt_pp}
and~\ref{Fig_dsigma_el_dt_pp_LHC} more quantitatively in the
small-$|t|$ region, one can use the well known parametrization of the
differential elastic cross section in terms of the slope parameter
$B(s)$ and the {\em curvature} $C(s)$
\be
        d\sigma^{el}/dt(s,t) 
        = d\sigma^{el}/dt(s,t=0)\,\exp\left[B(s)t+C(s)t^2\right]  
        \ .
\label{Eq_dsigma_el_dt_exp_parameterization}
\ee
Using $B(s)$ from the preceding section and assuming for the moment
$C(s) = 0$, one achieves a good description at small momentum
transfers and energies, which is consistent with the approximate
Gaussian shape of $J_{pp}(s,|\vec{b}_{\!\perp}|)$ at small energies
shown in Fig.~\ref{Fig_J_pp(b,s)}. The dip, of course, is generated by
the deviation from the Gaussian shape at small impact parameters.
According to (\ref{Eq_dsigma_el_dt_exp_parameterization}), the
shrinking of the diffraction peak with increasing energy simply
reflects the increasing interaction radius described by $B(s)$.

For small energies $\sqrt{s}$, our model reproduces the experimentally
observed change in the slope at $|t| \approx
0.25\,\GeV^2$~\cite{Barbiellini:1972ua+X} that is characterized by a
positive curvature. For LHC, we find clearly a negative value for the
curvature in agreement with~\cite{Block:1999hu} but in contrast
to~\cite{Berger:1999gu}. The change of sign in the curvature reflects
the transition of $J(s,|\vec{b}_{\!\perp}|)$ from the approximate
Gaussian shape at low energies to the approximate step-function
shape~(\ref{Eq_J_ab_asymptotic_energies}) at high energies.

Important for the good agreement with the data is the longitudinal
quark momentum distribution in the proton. Besides the slope
parameter, which characterizes the diffraction peak, also the dip
position is very sensitive to the distribution width $\Delta z_p$,
i.e.\ with $\Delta z_p= 0$ the dip position appears at more than 10\%
lower values of $|t|$. In the earlier \SVM\ 
approach~\cite{Berger:1999gu}, the reproduction of the correct dip
position was possible without the $z$-dependence of the hadronic wave
functions but a deviation from the data in the low-$|t|$ region had to
be accepted. In this low-$|t|$ region, we achieved a definite
improvement with the new correlation
functions~(\ref{Eq_SVM_correlation_functions}) and the minimal
surfaces used in our model.
\begin{figure}[htb]
  \centerline{\psfig{figure=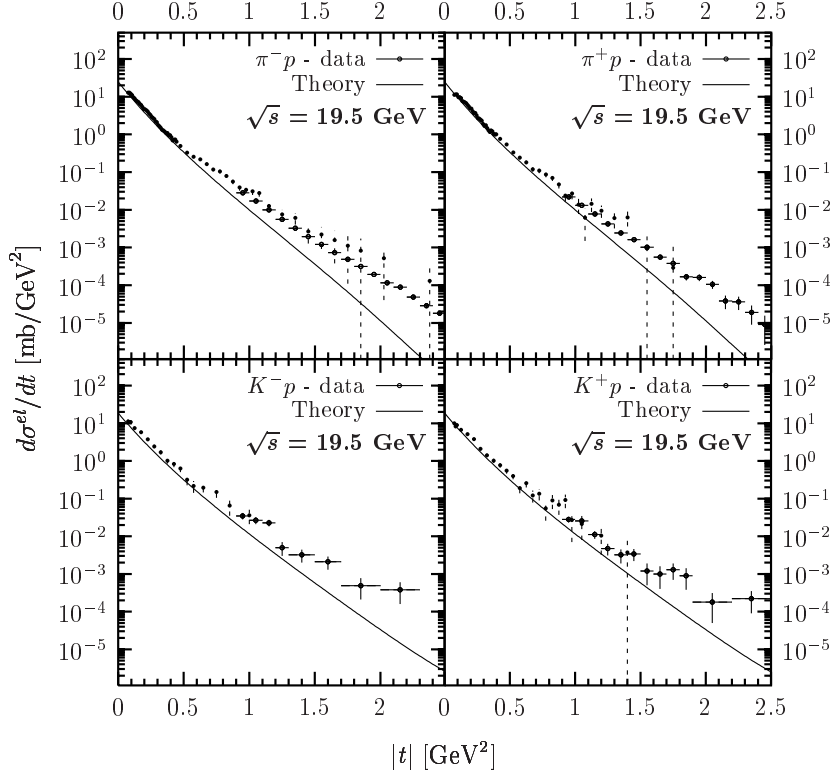,width=11.cm}} 
  \protect\caption{ \small 
    The differential elastic cross section $d\sigma^{el}/dt(s,t)$ as a
    function of $|t|$ for $\pi^{\pm}p$ and $K^{\pm}p$ reactions at a
    c.m.\ energy of $\sqrt{s} = 19.5\,\GeV$. The model results (solid
    line) are compared with the data (closed circles with error bars)
    from~\cite{Akerlof:1976gk+X}.}
\label{Fig_dsigdt_kp_pip}
\end{figure} 

The differential elastic cross section computed for $\pi^{\pm}p$ and
$K^{\pm}p$ reactions has the same behavior as the one for $pp$
($p\pbar$) reactions. The only difference comes from the different
$z$-distribution widths, $\Delta z_{\pi}$ and $\Delta z_{K}$, and the
smaller extension parameters, $S_{\pi}$ and $S_{K}$, which shift the
dip position to higher values of $|t|$. This is illustrated in
Fig.~\ref{Fig_dsigdt_kp_pip}, where the model results (solid line) for
the $\pi^{\pm}p$ and $K^{\pm}p$ differential elastic cross section are
shown at $\sqrt{s} = 19.5\,\GeV$ as a function of $|t|$ in comparison
to experimental data (closed circles) from~\cite{Akerlof:1976gk+X}.
The deviations from the data towards large $|t|$ leave room for
odderon and reggeon contributions. Indeed, with a finite diquark size
in the proton, an odderon occurs that improves the description of the
data at large values of $|t|$~\cite{Berger:PhDthesis:1999}.

\section[The Elastic Cross Section $\sigma^{el}$, $\sigma^{el}/ \sigma^{tot}$ and $\sigma^{tot}/B$]{The Elastic Cross Section \boldmath$\sigma^{el}$, \boldmath$\sigma^{el}/ \sigma^{tot}$ and \boldmath$\sigma^{tot}/B$}

The {\em elastic cross section} obtained by integrating the
differential elastic cross section
\be
        \sigma^{el}(s) 
        = \int_0^{-\infty}\!dt\,\frac{d\sigma^{el}}{dt}(s,t) 
        = \int_0^{-\infty}\!dt\,\inv{16 \pi s^2}|T(s,t)|^2
\label{Eq_total_elastic_cross_section}
\ee
reduces for our purely imaginary
$T$-matrix~(\ref{Eq_model_purely_imaginary_T_amplitude_final_result})
to
\be
        \sigma^{el}(s) 
        = \int \!\!d^2b_{\!\perp}\,|J(s,|\vec{b}_{\!\perp}|)|^2 
        \ .
\label{Eq_total_elastic_cross_section_J}
\ee
Due to the squaring, it exhibits the saturation of
$J(s,|\vec{b}_{\!\perp}|)$ at the black disc limit more clearly than
$\sigma^{tot}(s)$. Even more transparent is the saturation in the
following ratios given here for a purely imaginary $T$-matrix
\bea
        \frac{\sigma^{el}}{\sigma^{tot}}(s) 
        & = & 
        \frac
        {\int\!d^2b_{\!\perp}\,|J(s,|\vec{b}_{\!\perp}|)|^2}
        {2\int\!d^2b_{\!\perp}\,J(s,|\vec{b}_{\!\perp}|)}
        \ ,
\label{Eq_sigma_el/sigma_tot} \\
        \frac{\sigma^{tot}}{B}(s) 
        & = & 
        \frac
        {\left(2\int\!d^2b_{\!\perp}\,J(s,|\vec{b}_{\!\perp}|)\right)^2}
        {\int\!d^2b_{\!\perp}\,|\vec{b}_{\!\perp}|^2\,J(s,|\vec{b}_{\!\perp}|)} 
\label{Eq_sigma_tot/B}
        \ ,
\eea
which are directly sensitive to the opacity of the particles. This
sensitivity can be illustrated within the approximation 
\be
        T(s,t) = i\, s\, \sigma^{tot}(s)\, \exp[B(s) t/2]
\label{Eq_T_matrix_exp_parameterization}
\ee
that leads to the differential cross
section~(\ref{Eq_dsigma_el_dt_exp_parameterization}) with $C(s) = 0$,
i.e.\ an exponential decrease over $|t|$ with a slope $B(s)$. As the
purely imaginary $T$-matrix
element~(\ref{Eq_T_matrix_exp_parameterization}) is equivalent to
\be
        J(s,|\vec{b}_{\!\perp}|)
        =(\sigma^{tot}/4\pi B)\,\exp[-|\vec{b}_{\!\perp}|^2/2B] 
        =(4\sigma^{el}/\sigma^{tot})\,\exp[-|\vec{b}_{\!\perp}|^2/2B]
        \ ,
\label{Eq_J(b,s)_exp_parameterization}
\ee
one finds that the above ratios are a direct measure for the opacity
at zero impact parameter
\be
        J(s,|\vec{b}_{\!\perp}|=0) 
        = (\sigma^{tot}/4\pi B)
        = (4\sigma^{el}/\sigma^{tot}) 
        \ .
\label{Eq_J(b=0,s)_exp_parameterization}
\ee
For a general purely imaginary $T$-matrix, $T(s,t) =
i\,s\,\sigma^{tot}\,g(|t|)$ with an arbitrary real-valued function
$g(|t|)$, $J(s,|\vec{b}_{\!\perp}|=0)$ is given by
$\sigma^{el}/\sigma^{tot}$ times a pure number which depends on the
shape of $g(|t|)$.

\begin{figure}[p]
  \centerline{\psfig{figure=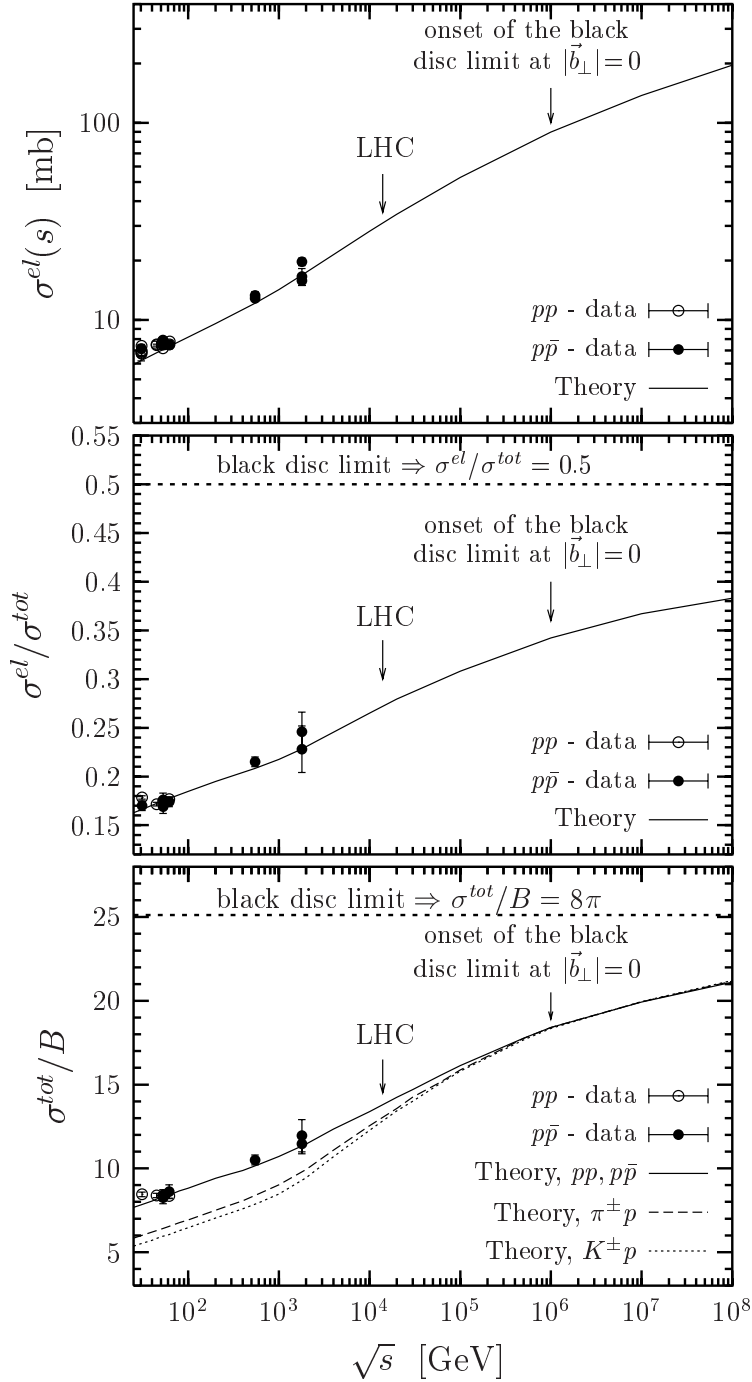, width=10.cm}}
  \protect\caption{ \small The elastic cross section $\sigma^{el}$ and
    the ratios $\sigma^{el}/ \sigma^{tot}$ and $\sigma^{tot}/B$ as a
    function of the c.m.\ energy $\sqrt{s}$. The model results for
    $pp$ ($p\pbar$), $\pi p$, and $K p$ scattering are represented by
    the solid, dashed and dotted lines, respectively. The experimental
    data for the $pp$ and $p\pbar$ reactions are indicated by the open
    and closed circles, respectively. The data for the elastic cross
    section are taken from~\cite{Hagiwara:fs} and the data for
    $\sigma^{tot}$ and $B$ from the references given in previous
    sections.}
\label{Fig_sigtot_el_and_ratios}
\end{figure}

We compute the elastic cross section $\sigma^{el}$ and the ratios
$\sigma^{el}/ \sigma^{tot}$ and $\sigma^{tot}/B$ in our model without
taking into account reggeons. In Fig.~\ref{Fig_sigtot_el_and_ratios}
the results for $pp$ and $p\pbar$ reactions (solid lines) are compared
with the experimental data (open and closed circles). The data for the
elastic cross section are taken from~\cite{Hagiwara:fs} and the data
for $\sigma^{tot}$ and $B$ from the references given in previous
sections.
For $pp$ ($p\pbar$) scattering, we indicate explicitly the prediction
for LHC at $\sqrt{s}=14\,\TeV$ and the onset of the black disc limit
at $\sqrt{s} = 10^6\,\GeV$. The model results for $\pi p$ and $K p$
reactions are presented by the dashed and dotted line, respectively.
For the ratios, the asymptotic limits are indicated: Since the maximum
opacity or black disc limit governs the $\sqrt{s} \to \infty$
behavior, $\sigma^{el}/\sigma^{tot}$ ($\sigma^{tot}/B$) cannot exceed
$0.5$ ($8\pi$) in hadron-hadron scattering.

In the investigation of $pp$ ($p\pbar$) scattering, our theoretical
curves successfully confront the experimental data for the elastic
cross section and both ratios. At low energies, the data are
underestimated as reggeon contributions are not taken into account.
Again, the agreement is comparable to the one achieved
in~\cite{Block:1999hu} and better than in the approach
of~\cite{Berger:1999gu}, where $\sigma^{el}$ comes out too small due
to an underestimation of $d\sigma^{el}/dt$ in the low-$|t|$ region.

Concerning the energy dependence, $\sigma^{el}$ shows a similar
behavior as $\sigma^{tot}$ but with a more pronounced flattening
around $\sqrt{s} \gtsim 10^6\,\GeV$. This flattening is even stronger
for the ratios, drawn on a linear scale, and reflects very clearly the
onset of the black disc limit. As expected from the simple
approximation~(\ref{Eq_J(b=0,s)_exp_parameterization}),
$\sigma^{el}/\sigma^{tot}$ and $\sigma^{tot}/B$ show a similar
functional dependence on $\sqrt{s}$. At the highest energy shown,
$\sqrt{s} = 10^8\,\GeV$, both ratios are still below the indicated
asymptotic limits, which reflects that $J(s,|\vec{b}_{\!\perp}|)$
still deviates from the step-function
shape~(\ref{Eq_J_ab_asymptotic_energies}). The ratios
$\sigma^{el}/\sigma^{tot}$ and $\sigma^{tot}/B$ reach their upper
limits $0.5$ and $8\pi$, respectively, at asymptotic energies,
$\sqrt{s} \to \infty$, where the hadrons become infinitely large,
completely black discs.

Comparing the $pp$ ($p\pbar$) results with the ones for $\pi p$ and
$Kp$, one finds that the results for $\sigma^{tot}/B$ converge at high
energies as shown in Fig.~\ref{Fig_sigtot_el_and_ratios}. This follows
from the identical normalizations of the hadron wave functions that
lead to an identical black disc limit for hadron-hadron reactions.


%
\cleardoublepage
%
\chapter{Conclusion}
\label{Sec_Conclusion}

We have developed a loop-loop correlation model (LLCM)
\cite{Shoshi:2002rd,Shoshi:2002in} in which the QCD vacuum is
described by perturbative gluon exchange and the non-perturbative
stochastic vacuum model (SVM) \cite{Dosch:1987sk+X}.  This combination
leads to a static quark-antiquark potential with color Coulomb
behavior for small and confining linear rise for large source
separations in good agreement with lattice QCD results. We have
computed in the LLCM the vacuum expectation value of one Wegner-Wilson
loop, $\langle W_{r}[C] \rangle$, and the correlation of two
Wegner-Wilson loops, $\langle W_{r_1}[C_1] W_{r_2}[C_2] \rangle$, for
arbitrary loop geometries and general representations $r_{(i)}$ of
$SU(N_c)$.  Specifying the loop geometries, these results allow us to
compute the static quark-antiquark potential, the glueball mass, the
chromo-field distributions of static color dipoles, the QCD van der
Waals potential between two static color dipoles, and the $S$-matrix
element for high-energy dipole-dipole scattering. Together with a
phenomenological universal energy dependence and hadron and photon
wave functions, the latter provides the basis for a unified
description of hadron-hadron, photon-hadron, and photon-photon
reactions within the functional integral approach to high-energy
scattering.

We have applied the LLCM to compute the chromo-electric fields
generated by a static color dipole in the fundamental and adjoint
representation of $SU(N_c)$. The formation of a confining color flux
tube is described by the non-perturbative SVM
correlations~\cite{Rueter:1994cn} and the color Coulomb field is
obtained from perturbative gluon exchange. We have found Casimir
scaling for both the perturbative and non-perturbative contributions
to the chromo-electric fields in agreement with recent lattice
data~\cite{Bali:2000un}. String breaking is neither described for
sources in the fundamental representation nor for sources in the
adjoint representation which indicates that in our approach not only
dynamical fermions (quenched approximation) are missing but also some
gluon dynamics. Transverse and longitudinal energy density profiles
have been provided: For small dipoles, $R=0.1\,\fm$, perturbative
physics dominates and non-perturbative correlations are negligible.
For large dipoles, $R\gtsim 1\,\fm$, the non-perturbative confining
string dominates the chromo-electric fields between the color sources.
The transition from perturbative to string behavior takes place at
source separations of about $0.5\,\fm$ in agreement with the recent
results of L\"uscher and Weisz~\cite{Luscher:2002qv}.  The root mean
squared radius $R_{ms}$ of the confining string and the energy density
in the center of a fundamental $SU(3)$ dipole $\varepsilon_3(X=0)$ are
governed completely by non-perturbative physics for large $R$ and
saturate as $R$ increases at $R_{ms}^{R\to\infty}\approx 0.55\,\fm$
and $\varepsilon_3^{R\to\infty}(X=0)\approx 1\,\GeV/\fm^3$.

We have presented the low-energy
theorems~\cite{Rothe:1995hu+X,Michael:1995pv,Green:1996be}, known in
lattice QCD as Michael sum rules~\cite{Michael:1986yi}, in their
complete form in continuum theory taking into account the
contributions found in~\cite{Dosch:1995fz,Rothe:1995hu+X} that are
missing in the original formulation~\cite{Michael:1986yi}. We have
used the complete theorems to compare the energy and action stored in
the confining string with the confining part of the static
quark-antiquark potential. The comparison shows consistency of the
model results and indicates that the non-perturbative SVM component is
working at the renormalization scale at which $\beta(g)/g=-2$ and
$\alphaS = 0.81$.  Earlier SVM investigations along these lines have
found a different value of $\alphaS = 0.57$ with the pyramid mantle
choice for the surface~\cite{Rueter:1994cn,Dosch:1995fz} but were
incomplete since only the contribution from the traceless part of the
energy-momentum tensor has been considered in the energy sum.

A Euclidean approach to high-energy dipole-dipole scattering has been
established by generalizing Meggiolaro's analytic
continuation~\cite{Meggiolaro:1996hf+X} from parton-parton scattering
to gauge-invariant dipole-dipole scattering. The generalized analytic
continuation allows us to derive $S$-matrix elements for high-energy
reactions from configurations of {\WW} loops in Euclidean space-time
with Euclidean functional integrals. It thus shows how one can access
high-energy reactions directly in lattice QCD. First attempts in this
direction have already been carried out but only very few signals
could be extracted, while most of the data was dominated by
noise~\cite{DiGiacomo:2002PC}. We have applied this approach to
compute in the Euclidean LLCM the scattering of dipoles at
high-energy. The result derived in the Minkowskian version of the
LLCM~\cite{Shoshi:2002in} has exactly been recovered including the
well-known two-gluon exchange contribution to dipole-dipole
scattering~\cite{Low:1975sv+X,Gunion:iy}. This confirms the analytic
continuation of the gluon field strength correlator used in all
earlier applications of the SVM to high-energy
scattering~\cite{Kramer:1990tr,Dosch:1994ym,Dosch:RioLecture,Rueter:1996yb,Dosch:1997ss,Dosch:1998nw,Rueter:1998qy,Kulzinger:1999hw,Rueter:1998up,D'Alesio:1999sf,Berger:1999gu,Donnachie:2000kp,Donnachie:2001wt,Dosch:2001jg,Kulzinger:2002iu}.

The $S$-matrix element obtained in our approach allows us to
investigate manifestations of the confining QCD string in high-energy
reactions of photons and hadrons~\cite{Shoshi:2002fq} but leads to
energy-independent cross sections in contradiction to the experimental
observation~\cite{Shoshi:2002in}. The missing energy dependence is
disappointing but not surprising since our approach does not describe
explicit gluon radiation needed for a non-trivial energy dependence.

We have introduced a phenomenological energy dependence into the
$S$-matrix element $S_{DD}$ that allows a unified description of
hadron-hadron, photon-hadron, and photon-photon reactions and respects
$S$-matrix unitarity constraints in impact parameter
space~\cite{Shoshi:2002in,Shoshi:2002ri,Shoshi:2002mt}. Motivated by
the two-pomeron model of Donnachie and
Landshoff~\cite{Donnachie:1998gm+X}, we have ascribed to the
non-perturbative and to the perturbative component a weak and a strong
energy dependence, respectively.  The constructed $T$-matrix element
shows Regge behavior at moderately high energies and describes
multiple gluonic interactions important to respect unitarity
constraints in impact parameter space at ultra-high energies. However,
for a more fundamental understanding of hadronic high-energy reactions
in our model, one faces the highly ambitious task to implement gluon
radiation and quantum evolution explicitly.

The model parameters have been adjusted to reproduce a wealth of
experimental data (including cosmic ray data) for total, differential,
and elastic cross sections, structure functions, and slope parameters
over a large range of c.m.\ energies.  The model parameters that
allowed a good fit to high-energy scattering data are in good
agreement with complementary investigations: The parameters of the
non-perturbative component -- the correlation length $a$, the
non-Abelian strength $\kappa$, and the gluon condensate $G_2$ -- are
constrained by lattice QCD investigations, by the string tension
$\sigma$ of a static quark-antiquark pair, and by the SVZ gluon
condensate $G_2$ essential in QCD sum rule investigations. For the
energy dependence, the exponents of the Donnachie-Landshoff
two-pomeron fit, $\epsilon_{\mathrm{soft}}$ and
$\epsilon_{\mathrm{hard}}$, have been used as an orientation for our
energy exponents $\epsilon^{\nprt}$ and $\epsilon^{\pert}$. Besides
these parameters describing dipole-dipole scattering, the
reaction-dependent parameters in the light-cone wave functions are
also consistent with other approaches: In the hadron wave functions,
the transverse extension parameters $S_h$ are in good agreement with
the corresponding electromagnetic radii~\cite{Dosch:2001jg} and the
width of the longitudinal quark momentum distributions $\Delta z_h$
has been computed from~\cite{Wirbel:1985ji}. In the photon wave
function, the running quark masses, which coincide with the current
quark masses for large $Q^2$ and the constituent quark masses for
small $Q^2$~\cite{Dosch:1998nw}, have been chosen in agreement with
sum rule derivations.

Having adjusted the model parameters, we have studied $S$-matrix
unitarity limits of the scattering amplitudes in impact parameter
space. On the basis of dipole-dipole scattering, we have shown
explicitly that our model respects unitarity constraints in impact
parameter space and, in particular, the black disc limit at ultra-high
energies. The profile functions have been calculated for proton-proton
and longitudinal photon-proton scattering. They show very clearly that
the interacting particles become blacker and larger with increasing
c.m.\ energy. At ultra-high energies, the opacity saturates at the
black disc limit while the transverse expansion of the scattered
particles continues. Increasing the photon virtuality $Q^2$ in
longitudinal photon-proton scattering, the maximum opacity increases
and also the energy at which it is reached for zero impact parameter.

We have related the impact parameter dependent gluon distribution
$xG(x,Q^2,|\vec{b}_{\perp}|)$ to the profile function for longitudinal
photon-proton scattering and found low-$x$ saturation of
$xG(x,Q^2,|\vec{b}_{\perp}|)$ as a manifestation of $S$-matrix
unitarity. In accordance with the profile function,
$xG(x,Q^2,|\vec{b}_{\perp}|)$ decreases from the center towards the
periphery of the proton.  With increasing photon virtuality $Q^2$, the
increase of $xG(x,Q^2,|\vec{b}_{\perp}|=0)$ becomes stronger towards
small $x$ and the saturation value of $xG(x,Q^2,|\vec{b}_{\perp}|=0)$
increases but is reached at decreasing values of $x$. However, the
integrated gluon distribution $xG(x,Q^2)$ does not saturate because of
the growth of the proton radius with decreasing $x$ observed in our
approach.  Similar results are obtained in complementary
approaches~\cite{Mueller:1986wy,Mueller:1999wm,Iancu:2001md}.

More model dependent are the specific energies at which these
saturation effects set in. The profile function saturates the black
disc limit at zero impact parameter for $\sqrt{s} \gtsim 10^6\,\GeV$
in proton-proton scattering and for $\sqrt{s} \gtsim 10^7\,\GeV$ in
longitudinal photon-proton scattering with $Q^2 \gtsim 1\,\GeV^2$. In
both reactions, the wave function normalization determines the maximum
opacity. The saturation of $xG(x,Q^2,|\vec{b}_{\perp}|)$ occurs in our
approach for $Q^2\gtsim 1\,\GeV^2$ at values of $x \ltsim 10^{-10}$,
i.e.\ far below the HERA and THERA range.

For proton-proton scattering, we have found that the rise of the total
and elastic cross section becomes weaker for $\sqrt{s} \gtsim
10^6\,\GeV$ due to the onset of the black disc limit at
$|\vec{b}_{\perp}|=0$ in the profile function. This saturation of the
profile function becomes even more apparent in the ratios
$\sigma^{el}/\sigma^{tot}$ and $\sigma^{tot}/B$ which are a measure of
the proton opacity. In contrast, no saturation effect has been
observed in the slope parameter $B(s)$ which is a measure for the
transverse expansion of the proton. Considering the differential
elastic cross section $d\sigma^{el}/dt$, the model has described the
diffraction pattern and also the shrinkage of the diffraction peak
with increasing energy in good agreement with experimental data at
small momentum transfers $|t|$. Around the dip region, where a real
part is expected to be important, deviations from the data have
reflected that our $T$-matrix is purely imaginary. Our predictions for
proton-proton scattering at LHC ($\sqrt{s} = 14\,\TeV$) are a total
cross section of $\sigma^{tot}_{pp} = 114.2\,\mb$ in good agreement
with cosmic ray data and a differential elastic cross section
$d\sigma^{el}/dt$ with a slope parameter of $B = 21.26\,\GeV^{-2}$, a
negative curvature $C<0$, and a dip at $|t| \approx 0.35\,\GeV^2$.

For pion-proton and kaon-proton scattering, results analogous to
proton-proton scattering have been obtained but with a slightly
stronger rise observed in the total cross section. This has been
traced back to the smaller size of pions and kaons in comparison to
protons, $S_p = 0.86\,\fm > S_{\pi} = 0.607\,\fm > S_{K} = 0.55\,\fm$,
and the perturbative component becoming increasingly important with
decreasing dipole sizes involved. Furthermore, a weak convergence of
the ratios $\sigma^{tot}_{pp}/\sigma^{tot}_{\pi p}$ and
$\sigma^{tot}_{pp}/\sigma^{tot}_{Kp}$ towards unity is predicted as
the energy increases. The smaller size of the pion and kaon is
reflected in the differential elastic cross sections $d\sigma^{el}/dt$
by the dip shifted towards larger values of $|t|$.

For photon-proton and photon-photon reactions, an even stronger rise
of the total cross section has been observed with increasing energy.
As illustrated in the proton structure function $F_2(x,Q^2)$, this
rise becomes steeper with increasing photon virtuality $Q^2$. Again,
we have traced back the strong energy boost to the growing importance
of the perturbative component with decreasing dipole size.  Besides
some deviations from the experimental data with increasing $Q^2$, our
model has described $\sigma^{tot}_{\gamma^{*} p}(s,Q^2)$ successfully
in the low-$Q^2$ region where the running quark masses become
constituent quark masses.

Finally, let us emphasize that the presented Euclidean approach to
high-energy scattering is independent of our specific loop-loop
correlation model. It indeed makes any method limited to a Euclidean
formulation of the theory applicable for investigations of high-energy
reactions. Here encouraging new results have been obtained with
instantons~\cite{Shuryak:2000df+X} and within the AdS/CFT
correspondence~\cite{Janik:2000zk+X} and it will be interesting to see
precise results from the lattice. A promising complementary Euclidean
approach has been proposed in~\cite{Hebecker:1999pb+X} where the
structure functions of deep inelastic scattering at small Bjorken $x$
are related to an effective Euclidean field theory. Here one hopes
that the limit $x\to 0$ corresponds to critical behavior in the
effective theory. The aim is again to provide a framework in which
structure functions can be calculated from first principles using
genuine non-perturbative methods such as lattice computations. In
another recent attempt, the energy dependence of the proton structure
function has been related successfully to critical properties of an
effective near light-cone Hamiltonian in a non-perturbative lattice
approach~\cite{Pirner:2001pv+X}. It will be interesting to see further
developments along these lines aiming at an understanding of hadronic
high-energy scattering from the QCD Lagrangian.


%
%
\begin{appendix}
\cleardoublepage
%
\chapter{Loop and Minimal Surface Parametrizations}
\label{Sec_Parameterizations}

\begin{figure}[h]
\centerline{\epsfig{figure=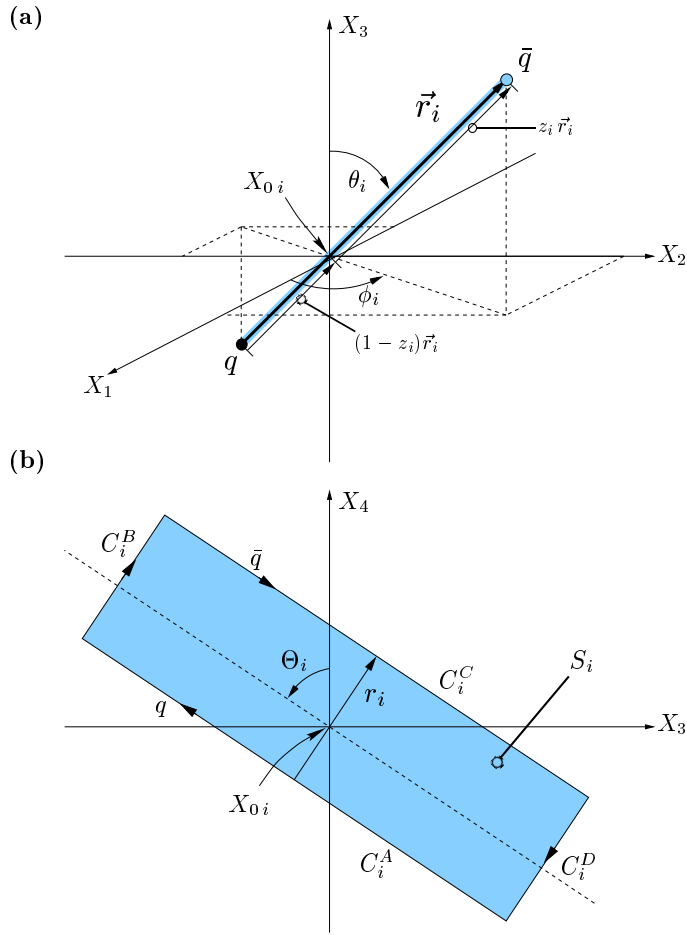,width=9.cm}}\hskip 0.5cm
\caption{\small
  (a) Spatial arrangement of a color dipole and (b) its world-line in
  Euclidean ``longitudinal'' space given by the rectangular {\em loop}
  $C_i$ that defines the {\em minimal surface} $S_i$ with $\partial
  S_i = C_i$. The minimal surface is represented by the shaded area.
  In our model, it is interpreted as the world-sheet of the QCD string
  that confines the quark and antiquark in the dipole.}
\label{Fig_OneLoop_MinimalSurface}
\end{figure}

A rectangular {\em loop} $C_i$ with ``spatial'' extension $R_i$ and
``temporal'' extension $T_i$ placed in four-dimensional Euclidean
space --- as shown in Fig.~\ref{Fig_OneLoop_MinimalSurface} --- has
the following parameter representation
\be
        C_i 
        \,\,=\,\, 
        C_i^A \,\cup\, C_i^B \,\cup\, C_i^C \,\cup\, C_i^D
\label{Eq_Ci_parameterization}
\ee
with
\bea
        C_i^A \,\,=\,\,  
        \Big\{ 
        X_i^A(u_i) 
        & = & 
        X_{\!0\,i} - (1-z_i)\,r_i + u_i\,t_i,\quad\hphantom{v_i\,r_i}
        u_i \in [-T_i,T_i]
        \Big\} 
\label{Eq_Ci^A_parameterization}\\
        C_i^B \,\,=\,\, 
        \Big\{ 
        X_i^B(v_i) 
        & = & 
        X_{\!0\,i} - (1-z_i)\,r_i + v_i\,r_i + T_i\,t_i,\;\quad
        v_i \in [0,1]
        \Big\} 
\label{Eq_Ci^B_parameterization}\\
        C_i^C \,\,=\,\, 
        \Big\{ 
        X_i^C(u_i) 
        & = & 
        X_{\!0\,i} + z_i\,r_i + u_i\,t_i,\quad\hphantom{(1-z_i)\,r_i}
        u_i \in [T_i,-T_i]
        \Big\} 
\label{Eq_Ci^C_parameterization}\\
        C_i^D \,\,=\,\,
        \Big\{ 
        X_i^D(v_i) 
        & = & 
        X_{\!0\,i} - (1-z_i)\,r_i + v_i\,r_i + T_i\,t_i,\;\quad
        v_i \in [1,0]
        \Big\} 
\label{Eq_Ci^D_parameterization}
\eea
where
\be
        r_i
        := \left( \barray{c} 
        R_i\,\sin\theta_i\,\cos\phi_i \\ 
        R_i\,\sin\theta_i\,\sin\phi_i \\ 
        R_i\,\cos\theta_i\,\cos\Theta_i \\ 
        R_i\,\cos\theta_i\,\sin\Theta_i
        \earray \right)
        \quad \mbox{and} \quad
        t_i
        := \left( \!\! \barray{c} 
        0 \\ 
        0 \\
        -\sin\Theta_i \\
        \hphantom{-}\cos\Theta_i
        \earray \right)
        \ .
\label{Eq_ri_ti_four_vectors}
\ee
The ``center'' of the loop $C_i$ is given by $X_{\!0\,i}$. The
parameters $z_i$, $R_i$, $\theta_i$, $\phi_i$, and $\Theta_i$ are
defined in Fig.~\ref{Fig_OneLoop_MinimalSurface} that illustrates (a)
the spatial arrangement of a color dipole and (b) its world-line $C_i$
in Euclidean ``longitudinal'' space. The tilting angle $\Theta_i\neq
0$ is the central quantity in the analytic continuation presented in
Chap.~\ref{Sec_DD_Scattering}.  Moreover, $\Theta_1 =\pi/2$ together
with $\Theta_2 = 0$ allows us to compute conveniently the
chromo-magnetic field distributions in
Appendix~\ref{Sec_Chi_Computation}.

The {\em minimal surface} $S_i$ is the planar surface bounded by the
loop $C_i=\partial S_i$ given in~(\ref{Eq_Ci_parameterization}). It
can be parametrized as follows
\be
        S_i =  
        \Big\{ 
        X_i(u_i,v_i) 
        = X_{\!0\,i} - (1-z_i)\,r_i + v_i\,r_i + u_i\,t_i,
        \;u_i \in [-T_i,T_i], \;v_i \in [0,1] 
        \Big\}
\label{Eq_Si_parameterization}
\ee
with $r_i$ and $t_i$ given in~(\ref{Eq_ri_ti_four_vectors}). The
corresponding infinitesimal surface element reads
\be
        d\sigma_{\mu\nu}(X_i)
        = \Bigg( 
        \frac{\partial X_{\!i\mu}}{\partial u_i} 
        \frac{\partial X_{\!i\nu}}{\partial v_i}
        - \frac{\partial X_{\!i\mu}}{\partial v_i} 
        \frac{\partial X_{\!i\nu}}{\partial u_i} 
        \Bigg)\,du_i\,dv_i
        = \Bigg( 
        t_{i\mu} r_{i\nu} - r_{i\mu} t_{i\nu} 
        \Bigg)\,du_i\,dv_i
        \ .
\label{Eq_Si_surface_element}
\ee

\bigskip

%

  
\cleardoublepage
%
\chapter[$\chi$-Computations with Minimal Surfaces]
{\boldmath$\chi$-Computations with Minimal Surfaces}
\label{Sec_Chi_Computation}

The quantities considered in the main text are computed from the VEV
of one loop $\langle W[C] \rangle$ and the loop-loop correlation
function $\langle W[C_1] W[C_2] \rangle$. Using the Gaussian
approximation in the gluon field strengths, both are expressed in
terms of $\chi_{S_i S_j}$-functions (\ref{Eq_chi_SS})
and~(\ref{Eq_chi_Si_Sj}) as shown in Secs.~\ref{Sec_<W[C]>}
and~\ref{Sec_<W[C_1]W[C_2]>}. These $\chi$-functions are central
quantities since here the ansatz of the gauge-invariant bilocal gluon
field strength correlator and the surface choice enter the model. In
this Appendix these functions are computed explicitly for minimal
surfaces~(\ref{Eq_Si_parameterization}) and the
$F_{\mu\nu\rho\sigma}$\,-\,ansatz given in~(\ref{Eq_F_decomposition}),
(\ref{Eq_PGE_Ansatz_F}), and~(\ref{Eq_MSV_Ansatz_F}). Note that the
contributions from the infinitesimally thin tube --- which allows us
to compare the gluon field strengths in surface $S_1$ with the gluon
field strength in surface $S_2$ --- cancel mutually.

Depending on the geometries and the relative arrangement of the loops,
the $\chi$-functions determine the physical quantities investigated
within the LLCM such as the static $q\qbar$
potential~(\ref{Eq_Vr(R)_Gaussian_approximation}), the chromo-field
distributions of a
color dipole~(\ref{Eq_chromo_fields_F_final_result}), and the
$S$-matrix element for elastic dipole-dipole
scattering~(\ref{Eq_S_DD_1}).

We compute separately the three components $\chi_{S_1 S_2}^{\pert}$,
$\chi_{S_1 S_2}^{\nprt\,nc}$, and $\chi_{S_1 S_2}^{\nprt\,c}$ for
general loop arrangements from which the considered quantities are
obtained as special cases. Without loss of generality, the center of
the loop $C_2$ is placed at the origin of the coordinate system,
$X_{\!0\,2}=(0,0,0,0)$. Moreover, $C_2$ is kept untilted, $\Theta_2 =
0$, and $\Theta := \Theta_1$ is used to simplify notation. We limit
our general computation to loops with $r_{1,2} =
(\vec{r}_{1,2\perp},0,0) \,\,\equiv\,\, \theta_{1,2} = \pi/2$ and
transverse ``impact parameters'' $b = X_{\!0\,1} - X_{\!0\,2} =
X_{\!0\,1} = (b_1, b_2, 0, 0) = (\vec{b}_{\!\perp},0,0)$ which allows
to compute all of the considered quantities.

\subsection*{\boldmath $\chi_{S_1 S_2}^{\nprt\,c}$-Computation}

Starting with the definition
\bea
        \chi_{S_1 S_2}^{\nprt\,c} 
        & := & \frac{\pi^2}{4} 
        \int_{S_1} \!\! d\sigma_{\mu\nu}(X_1) 
        \int_{S_2} \!\! d\sigma_{\rho\sigma}(X_2)\,
        F_{\mu\nu\rho\sigma}^{\nprt\,c}(Z = X_1 - X_2)
\label{Eq_chi_S1S2_NP_c_def_E}\\ 
        & = & 
        \frac{\pi^2 G_2 \kappa}{12(N_c^2-1)}
        \int_{S_1} \!\! d\sigma_{\mu\nu}(X_1) 
        \int_{S_2} \!\! d\sigma_{\rho\sigma}(X_2)
        \left(\delta_{\mu\rho}\delta_{\nu\sigma}
         -\delta_{\mu\sigma}\delta_{\nu\rho}\right) D(Z^2)
        \ ,
\nonumber
\eea
one exploits the anti-symmetry of the surface elements,
$d\sigma_{\mu\nu} = - d\sigma_{\nu\mu}$, and applies the surface
parametrization~(\ref{Eq_Si_parameterization}) with the corresponding
surface elements~(\ref{Eq_Si_surface_element}) to obtain
\be
        \chi_{S_1 S_2}^{\nprt\,c} = 
        \cos\Theta\,
        \frac{\pi^2 G_2 \kappa}{3(N_c^2-1)}\,
        (r_1\cdot r_2)
        \int_0^1 \!\! dv_1 \int_0^1 \!\! dv_2
        \int_{-T_1}^{T_1} \!\! du_1 
        \int_{-T_2}^{T_2} \!\! du_2 \,D(Z^2)
\label{Eq_chi_S1S2_NP_c_T<infty_result_E}
\ee
with
\be
        Z = X_{1} - X_{2}
        = \left( \barray{l}
   \!\!\hphantom{-\,}
        \vec{b}_{\!\perp} - (1-z_1)\,\vec{r}_{1\perp} + v_1\,\vec{r}_{1\perp} 
        + (1-z_2)\,\vec{r}_{2\perp} - v_2\,\vec{r}_{2\perp} \\
   \!\!
        -\,u_1 \sin\Theta \\
   \!\!\hphantom{-\,}
        u_1 \cos\Theta - u_2
        \earray \right)
        \ ,
\label{Eq_Z_S1_S2_E}
\ee
where the identities $t_1\cdot r_2 = r_1 \cdot t_2= 0$ and $t_1 \cdot
t_2 = \cos\Theta$, evident from~(\ref{Eq_ri_ti_four_vectors}) with the
mentioned specification of the loop geometries, have been used. In the
limit $T_2 \rightarrow \infty$, the $u_2$ integration can be performed
\bea
        && \lim_{T_2\to\infty}\int_{-T_2}^{T_2}\!\!du_2\,D(Z^2)
        \,\, = \,\, 
        \int \frac{d^4K}{(2\pi)^4}\,\tilde{D}(K^2) 
                \lim_{T_2\to\infty}\int_{-T_2}^{T_2}\!\!du_2\,e^{iKZ}
        \nonumber \\
        && = \int \frac{d^4K}{(2\pi)^4}\,\tilde{D}(K^2)\, 
                2\pi\,\delta(K_4)\,
                \exp\!\left[
                i\vec{K}_{\!\perp}\vec{Z}_{\!\perp} + iK_3 u_1\sin\Theta + iK_4 u_1\cos\Theta
                \right]
        \nonumber \\
        && = \int \frac{d^3K}{(2\pi)^3}\,\tilde{D}^{(3)}(\vec{K}^2)\,
                \exp\!\left[
                i\vec{K}_{\!\perp}\vec{Z}_{\!\perp} + iK_3 u_1\sin\Theta
                \right]
        \,\, = \,\, 
        D^{(3)}(\vec{Z}^2)
        \ ,
\label{Eq_chi_S1S2_NP_c_u2_integration}
\eea
which leads to
\be
        \lim_{T_2 \to \infty}
        \chi_{S_1 S_2}^{\nprt\,c} = 
        \cos\Theta\,
        \frac{\pi^2 G_2 \kappa}{3(N_c^2-1)}\,
        (\vec{r}_{1\perp}\cdot\vec{r}_{2\perp})
        \int_0^1 \!\! dv_1 \int_0^1 \!\! dv_2
        \int_{-T_1}^{T_1} \!\! du_1 \,
        D^{(3)}(\vec{Z}^2)
        \ .
\label{Eq_chi_S1S2_NP_c_T2->infty_result_E}
\ee
Taking in addition the limit $T_{1} \rightarrow \infty$, the $u_1$
integration can be performed as well
\be
        \lim_{T_1 \to \infty} 
        \int_{-T_1}^{T_1} \!\! du_1 
        e^{iK_3 u_1\sin\Theta}
        = \left\{\begin{array}{lc}
        2\pi\delta(K_3\sin\Theta)
        = \frac{2\pi\delta(K_3)}{|\sin\Theta|}
        & \quad \mbox{for} \quad \sin\Theta \neq 0 \\
        \lim_{T_1\to\infty}\,2\,T_1
        & \quad \mbox{for} \quad \sin\Theta = 0 \\
        \end{array}\right.
        \ .
\label{Eq_chi_S1S2_NP_c_u1_integration}
\ee
With $T_1 = T_2 = T/2 \to \infty$, one obtains for $\sin\Theta\neq 0$
\be
        \lim_{T\to\infty}
        \chi_{S_1 S_2}^{\nprt\,c} = 
        \frac{\cos\Theta}{|\sin\Theta|}\,
        \frac{\pi^2 G_2 \kappa}{3(N_c^2-1)}\,
        (\vec{r}_{1\perp}\cdot \vec{r}_{2\perp})
        \int_0^1 \!\! dv_1 \int_0^1 \!\! dv_2\,
        D^{(2)}(\vec{Z}_{\!\perp}^2)
\label{Eq_chi_S1S2_NP_c_T->infty_S_E}
\ee
and for $\sin\Theta = 0$ 
\be
        \lim_{T\to\infty}
        \chi_{S_1 S_2}^{\nprt\,c} = 
        \lim_{T\to\infty}\,T\,
        \cos\Theta\,
        \frac{\pi^2 G_2 \kappa}{3(N_c^2-1)}\,
        (\vec{r}_{1\perp}\cdot \vec{r}_{2\perp})
        \int_0^1 \!\! dv_1 \int_0^1 \!\! dv_2\,
        D^{(3)}(\vec{Z}^2)
        \ .
\label{Eq_chi_S1S2_NP_c_T->infty_V_E}
\ee
Evidently,~(\ref{Eq_chi_S1S2_NP_c_T->infty_S_E}) is the result given
in~(\ref{Eq_S_DD_p_npc_npnc_E}) and~(\ref{Eq_S_DD_chi_np_c_M}) that
describes the confining contribution to the dipole-dipole scattering
matrix element $S_{DD}$.

From~(\ref{Eq_chi_S1S2_NP_c_T->infty_V_E}), one obtains the confining
contribution to the static color dipole potential for $S_1 = S_2 = S$
which implies $T_1 = T_2 = T/2$, $\Theta = 0$, $z_1 = z_2$, $r_1 = r_2
= r$, and $\vec{r}_{1\perp}\cdot \vec{r}_{2\perp} = r^2 = R^2$ so that
\bea
        \lim_{T\to\infty}
        \chi_{SS}^{\nprt\,c} 
        & = &
        \lim_{T\to\infty}\,T\,
        \frac{\pi^2 G_2 \kappa}{3(N_c^2-1)}\,
        R^2
        \int_0^1 \!\! dv_1 \int_0^1 \!\! dv_2\,
        D^{(3)}(\vec{Z}^2=(v_1-v_2)^2R^2)
\nonumber\\
        & = &
        \lim_{T\to\infty}\,T\,
        \frac{2 \pi^2 G_2 \kappa}{3(N_c^2-1)}\,
        R^2
        \int_0^1 \!\! d\rho\,
        (1-\rho)\,
        D^{(3)}(\rho^2 R^2)
        \ ,
\label{Eq_chi_SS_NP_c_T->infty_V_E}
\eea
which leads directly to~(\ref{Eq_Vr(R)_NP_c}).
 
From~(\ref{Eq_chi_S1S2_NP_c_T2->infty_result_E}) the confining
contribution to the chromo-field distributions $\Delta G_{\alpha
  \beta}^2(X)$ can be computed conveniently.
Equation~(\ref{Eq_chi_S1S2_NP_c_T2->infty_result_E}) reads for $S_1 =
S_P$, $T_1 = R_P/2$ and $R_1 = R_P$, and $S_2 = S_W$, $T_2 = T/2$ and
$R_2 = R$
\be
        \lim_{T \to \infty}
        \chi_{S_P S_W}^{\nprt\,c} =
        \cos \Theta\, 
        \frac{\pi^2 G_2 \kappa}{3(N_c^2-1)}\,
        (\vec{r}_{1\perp}\cdot\vec{r}_{2\perp})
        \int_0^1 \!\! dv_1 \int_0^1 \!\! dv_2
        \int_{-R_P/2}^{R_P/2} \!\! du_1 \,
        D^{(3)}(\vec{Z}^2)
\label{Eq_chi_SPSW_NP_c_T2->infty_result_E}
\ee
with
\be
        \vec{Z} = \vec{X}_{1} - \vec{X}_{2}
        = \left( \barray{l}
   \!\!\hphantom{-\,}
        \vec{b}_{\!\perp} - (1-z_1)\,\vec{r}_{1\perp} + v_1\,\vec{r}_{1\perp} 
        + (1-z_2)\,\vec{r}_{2\perp} - v_2\,\vec{r}_{2\perp} \\
   \!\!
        -\,u_1 \sin\Theta 
        \earray \right)
        \ .
\label{Eq_vec3_Z_S1_S2_E}
\ee
The confining non-perturbative contribution to the chromo-magnetic
fields vanishes as it is obtained for plaquettes with $\Theta =
\pi/2$. The corresponding contribution to the chromo-electric fields
can be computed with $\Theta=0$ as follows: Due to $R_1 = R_p\to 0$, the $u_1$ and
$v_1$ integrations in~(\ref{Eq_chi_SPSW_NP_c_T2->infty_result_E}) can
be performed with the mean value theorem. Keeping only terms up to
$\Order(R_p^2)$, the confining non-perturbative contribution to the
chromo-field distributions $\Delta G_{\alpha \beta}^2(X)$ is obtained
as given in~(\ref{Eq_Chi_PW_np_c_14}).

\subsection*{\boldmath$\chi_{S_1 S_2}^{\nprt\,nc}$-Computation}

We start again with the definition
\bea
        \chi_{S_1 S_2}^{\nprt\,nc}
        & := & \frac{\pi^2}{4} 
        \int_{S_1} \!\! d\sigma_{\mu\nu}(X_1) 
        \int_{S_2} \!\! d\sigma_{\rho\sigma}(X_2)\,
        F_{\mu\nu\rho\sigma}^{\nprt\,nc}(Z = X_1 - X_2)
\nonumber \\ 
        & = & 
        \frac{\pi^2 G_2 (1\!-\!\kappa)}{12(N_c^2-1)} \,
        \int_{S_1} \!\! d\sigma_{\mu\nu}(X_1) 
        \int_{S_2} \!\! d\sigma_{\rho\sigma}(X_2)
\label{Eq_chi_S1S2_NP_nc_def_E}\\ 
        && \times
        \inv{2}\Bigl[
        \frac{\partial}{\partial Z_\nu}
        \left(Z_\sigma \delta_{\mu\rho}
          -Z_\rho \delta_{\mu\sigma}\right)
        +\frac{\partial}{\partial Z_\mu}
        \left(Z_\rho \delta_{\nu\sigma}
          -Z_\sigma \delta_{\nu\rho}\right)\Bigr]\,
        D_1(Z^2)
\nonumber
\eea
and use the anti-symmetry of both surface elements to obtain
\bea
        \!\!\!\!\!\!\!\!\!\!\!\!\!\!\!\!\!\!
        \chi_{S_1 S_2}^{\nprt\,nc}\!\!
        & = & 
        \frac{\pi^2 G_2 (1\!-\!\kappa)}{6(N_c^2-1)} \,
        \int_{S_1} \!\! d\sigma_{\mu\nu}(X_1) 
        \int_{S_2} \!\! d\sigma_{\rho\sigma}(X_2)\,
        \frac{\partial}{\partial Z_\nu}\,Z_\sigma\,
        \delta_{\mu\rho}\,D_1(Z^2)
\label{Eq_chi_S1S2_NP_nc_1_E}\\ 
        & = & 
        \frac{\pi^2 G_2 (1\!-\!\kappa)}{3(N_c^2-1)}
        \!\int_{S_1} \!\! d\sigma_{\mu\nu}(X_1) 
        \!\int_{S_2} \!\! d\sigma_{\rho\sigma}(X_2)\,
        \frac{\partial}{\partial Z_\nu}\,
        \frac{\partial}{\partial Z_\sigma}\,
        \delta_{\mu\rho}\,D_1^{\prime}(Z^2)
\label{Eq_chi_S1S2_NP_nc_2_E}\\
        & = & 
        \!\!\!\!
        -\,\frac{\pi^2 G_2 (1\!-\!\kappa)}{3(N_c^2-1)}
        \!\!\int_{S_1} \!\!\! d\sigma_{\mu\nu}(X_1)
        \frac{\partial}{\partial X_{1\nu}}
        \!  \int_{S_2} \!\!\! d\sigma_{\rho\sigma}(X_2)
        \frac{\partial}{\partial X_{2\sigma}}\,
        \delta_{\mu\rho}\,D_1^{\prime}(Z^2)
\label{Eq_chi_S1S2_NP_nc_3_E}
\eea
with
\be
        D_1^{\prime}(Z^2)
        = \int\!\!\frac{d^4K}{(2\pi)^4}\,e^{iKZ}\,
        \tilde{D}_1^{\prime}(K^2)
        = \int\!\!\frac{d^4K}{(2\pi)^4}\,e^{iKZ}\,
        \frac{d}{dK^2}\,\tilde{D}_1(K^2)
        \ .
\label{Eq_D1'(Z^2)_def_E}
\ee
As evident from~(\ref{Eq_chi_S1S2_NP_nc_3_E}), Stokes' theorem can be
used to transform each of the surface integrals in $\chi_{S_1
  S_2}^{\nprt\,nc}$ into a line integral
\bea
        \chi_{S_1 S_2}^{\nprt\,nc}\!\!
        & = & 
        -\,\frac{\pi^2 G_2 (1\!-\!\kappa)}{3(N_c^2-1)}
        \int_{S_1} \!\!\! d\sigma_{\mu\nu}(X_1)
        \frac{\partial}{\partial Z_{\nu}}
        \!  \oint_{C_2} \!\!\! dZ_{\rho}(X_2)\,
        \delta_{\mu\rho}\,D_1^{\prime}(Z^2)
\label{Eq_chi_S1C2_NP_nc_D1'_E}\\
         & = & 
        -\,\frac{\pi^2 G_2 (1\!-\!\kappa)}{6(N_c^2-1)}
        \int_{S_1} \!\!\! d\sigma_{\mu\nu}(X_1)
        \!  \oint_{C_2} \!\!\! dZ_{\rho}(X_2)\,
        \delta_{\mu\rho}\,Z_{\nu}\,D_1(Z^2)
\label{Eq_chi_S1C2_NP_nc_D1_E}\\
        & = & 
        -\,\frac{\pi^2 G_2 (1\!-\!\kappa)}{3(N_c^2-1)}\,
        \oint_{C_1} \!\!\! dZ_{\mu}(X_1)
        \oint_{C_2} \!\!\! dZ_{\rho}(X_2)\,
        \delta_{\mu\rho}\,D_1^{\prime}(Z^2)
\label{Eq_chi_C1C2_NP_nc_D1'_E}
        \ .
\eea
With the line parametrizations of $C_1$ and $C_2$ given
in~(\ref{Eq_Ci_parameterization}) and the specification of the loop
geometries mentioned at the beginning of this appendix,
(\ref{Eq_chi_C1C2_NP_nc_D1'_E}) becomes
\bea
        && 
        \!\!\!\!\!\!\!\!
        \chi_{S_1 S_2}^{\nprt\,nc}
        = -\,\frac{\pi^2 G_2 (1-\kappa)}{3(N_c^2-1)}
\label{Eq_chi_C1C2_NP_nc_D1_1_E}\\
        &&
        \!\!\!\!\!\!\!\!
        \times
        \Bigg\{
        \cos\Theta\!\!
        \int_{-T_1}^{T_1} \!\!\! du_1 \!\!
        \int_{-T_2}^{T_2} \!\!\! du_2
        \Big[
        D_1^{\prime}(Z_{AA}^2) \!-\! D_1^{\prime}(Z_{AC}^2) 
        \!-\! D_1^{\prime}(Z_{CA}^2) \!+\! D_1^{\prime}(Z_{CC}^2) 
        \Big]
\nonumber\\
        \!\!\!\!\!\!\!\!
        &&\!\!\!\!\!\!
        +\,(\vec{r}_{1\perp}\cdot\vec{r}_{2\perp})\!\!
        \int_0^1 \!\!\! dv_1 \!\!
        \int_0^1 \!\!\! dv_2
        \Big[
        D_1^{\prime}(Z_{BB}^2) \!-\! D_1^{\prime}(Z_{BD}^2) 
        \!-\!D_1^{\prime}(Z_{DB}^2) \!+\! D_1^{\prime}(Z_{DD}^2) 
        \Big]
        \Bigg\}
\nonumber
\eea
where the following shorthand notation is used
\be
        Z_{XY} := X_{1}^{X} - X_{2}^{Y}
        \quad \mbox{with} \quad
        X_{2}^X \in C_2^X
        \quad \mbox{and} \quad
        X_{2}^Y \in C_2^Y
        \ .
\label{Eq_Z_C1X_C2Y_E}
\ee
In the limit $R_{1,2} \ll T_{1,2} \to \infty$, the term
$\propto\!(\vec{r}_{1\perp}\cdot\vec{r}_{2\perp})$ on the rhs
of~(\ref{Eq_chi_C1C2_NP_nc_D1_1_E}) can be neglected and, thus,
(\ref{Eq_chi_C1C2_NP_nc_D1_1_E}) reduces to
\bea
        &&
        \lim_{T_1\to\infty\atop T_2\to\infty}
        \chi_{S_1 S_2}^{\nprt\,nc}
        =
        -\,\cos\Theta\,
        \frac{\pi^2 G_2 (1-\kappa)}{3(N_c^2-1)}\,
        \lim_{T_1\to\infty}\int_{-T_1}^{T_1} \!\! du_1
        \lim_{T_2\to\infty}\int_{-T_2}^{T_2} \!\! du_2
\label{Eq_chi_C1C2_NP_nc_D1_T->infty_E}\\
        &&\hspace{2.5cm}
        \times
        \Big[
        D_1^{\prime}(Z_{AA}^2) - D_1^{\prime}(Z_{AC}^2) 
        - D_1^{\prime}(Z_{CA}^2) + D_1^{\prime}(Z_{CC}^2) 
        \Big] 
        \ .
\nonumber
\eea
Here  the integrations over $u_1$ and $u_2$ can be performed
analytically proceeding analogously
to~(\ref{Eq_chi_S1S2_NP_c_u2_integration})
and~(\ref{Eq_chi_S1S2_NP_c_u1_integration}). With $T_1 = T_2 =
T/2\to\infty$, one obtains for $\sin\Theta \neq 0$
\bea
        &&\!\!\!\!\!\!
        \lim_{T\to\infty}
        \chi_{S_1 S_2}^{\nprt\,nc}
        = 
        -\,\frac{\cos\Theta}{|\sin\Theta|}\,
        \frac{\pi^2 G_2 (1-\kappa)}{3(N_c^2-1)}\,
\label{Eq_chi_C1C2_NP_nc_T->infty_S_E}\\
        && \!\!\!\!\!\!\hspace{0.5cm}
        \times
        \Big[
        D_1^{\prime\,(2)}(\vec{Z}_{AA\perp}^2) 
        - D_1^{\prime\,(2)}(\vec{Z}_{AC\perp}^2) 
        - D_1^{\prime\,(2)}(\vec{Z}_{CA\perp}^2) 
        +  D_1^{\prime\,(2)}(\vec{Z}_{CC\perp}^2) 
        \Big] 
\nonumber
\eea
and for $\sin\Theta = 0$ 
\bea
        &&\!\!\!\!\!\!
        \lim_{T\to\infty}
        \chi_{S_1 S_2}^{\nprt\,nc}
        = 
        -\,\lim_{T\to\infty}T\,
        \cos\Theta\,
        \frac{\pi^2 G_2 (1-\kappa)}{3(N_c^2-1)}\,
\label{Eq_chi_C1C2_NP_nc_T->infty_V_E}\\
        && \!\!\!\!\!\!\hspace{0.5cm}
        \times
        \Big[
        D_1^{\prime\,(3)}(\vec{Z}_{AA}^2) 
        - D_1^{\prime\,(3)}(\vec{Z}_{AC}^2) 
        - D_1^{\prime\,(3)}(\vec{Z}_{CA}^2) 
        +  D_1^{\prime\,(3)}(\vec{Z}_{CC}^2) 
        \Big]
        \ . 
\nonumber
\eea
With the following identities 
\be
        \vec{Z}_{AA\perp}  =  \vec{r}_{1q}-\vec{r}_{2q}\ ,\,\,\,\,
        \vec{Z}_{AC\perp}  =  \vec{r}_{1q}-\vec{r}_{2\qbar}\ ,\,\,\,\,
        \vec{Z}_{CA\perp}  =  \vec{r}_{1\qbar}-\vec{r}_{2q}\ ,\,\,\,\,
        \vec{Z}_{CC\perp}  =  \vec{r}_{1\qbar}-\vec{r}_{2\qbar}\ ,\,\,\,\,
\label{Eq_Z's_E}
\ee
one sees immediately that~(\ref{Eq_chi_C1C2_NP_nc_T->infty_S_E}) is
the result given in~(\ref{Eq_S_DD_p_npc_npnc_E})
and~(\ref{Eq_S_DD_chi_np_nc_M}) that describes the non-confining
non-perturbative contribution to the dipole-dipole scattering matrix
element $S_{DD}$.

From~(\ref{Eq_chi_C1C2_NP_nc_T->infty_V_E}) one obtains the
non-confining contribution to the static potential for $S_1 = S_2 =
S$, i.e., $T_1 = T_2 = T/2$, $\Theta = 0$, $r_1 = r_2 = r$,
\bea
        &&\!\!\!\!\!\!
        \lim_{T\to\infty}
        \chi_{S S}^{\nprt\,nc}
        = 
        -\,\lim_{T\to\infty}T\,\,
        \frac{\pi^2 G_2 (1-\kappa)}{3(N_c^2-1)}\,\,
\label{Eq_chi_SS_NP_nc_T->infty_V_E}\\
        && \!\!\!\!\!\!\hspace{0.5cm}
        \times
        \Big[
        D_1^{\prime\,(3)}(\vec{Z}_{AA}^2) - D_1^{\prime\,(3)}(Z_{AC}^2) 
        - D_1^{\prime\,(3)}(\vec{Z}_{CA}^2) +  D_1^{\prime\,(3)}(\vec{Z}_{CC}^2) 
        \Big]
        \ ,
\nonumber
\eea
which contributes to the self-energy of the color sources with
\bea
        \lim_{T\to\infty}
        \chi_{S S\,\mbox{\scriptsize self}}^{\nprt\,nc}
        &=& 
        -\,\lim_{T\to\infty}T\,\,
        \frac{\pi^2 G_2 (1-\kappa)}{3(N_c^2-1)}\,\,
        \Big[
        D_1^{\prime\,(3)}(\vec{Z}_{AA}^2) +  D_1^{\prime\,(3)}(\vec{Z}_{CC}^2) 
        \Big]
\nonumber\\
        &=& 
        -\,\lim_{T\to\infty}T\,\,
        \frac{2\pi^2 G_2 (1-\kappa)}{3(N_c^2-1)}\,\,
                D_1^{\prime\,(3)}(\vec{Z}_{AA}^2)
\label{Eq_chi_SS_NP_nc_T->infty_self_E}
\eea
and to the potential energy between the color sources with
\bea
        \lim_{T\to\infty}
        \chi_{S S\,\mbox{\scriptsize pot}}^{\nprt\,nc}
        &=& 
        \lim_{T\to\infty}T\,\,
        \frac{\pi^2 G_2 (1-\kappa)}{6(N_c^2-1)}\,\,
        \Big[
        D_1^{\prime\,(3)}(\vec{Z}_{AC}^2) +  D_1^{\prime\,(3)}(\vec{Z}_{CA}^2) 
        \Big]
\nonumber\\
        &=& 
        \lim_{T\to\infty}T\,\,
        \frac{\pi^2 G_2 (1-\kappa)}{3(N_c^2-1)}\,\,
        D_1^{\prime\,(3)}(\vec{Z}_{AC}^2)
        \ .
\label{Eq_chi_SS_NP_nc_T->infty_pot_E}
\eea
The latter gives the non-confining contribution to the static
potential~(\ref{Eq_Vr(R)_NP_nc}).

From~(\ref{Eq_chi_S1C2_NP_nc_D1_E}) the non-confining non-perturbative
contribution to the chromo-electric fields ($\Delta G_{\alpha
  \beta}^2(X)$ with $\alpha\beta=i4=4i$) can be computed most
conveniently with zero plaquette tilting angle $\Theta = 0$. The
corresponding contribution to the chromo-magnetic fields ($\Delta
G_{\alpha \beta}^2(X)$ with $\alpha\beta=ij=ji$) is obtained for
plaquette tilting angle $\Theta = \pi/2$ and thus vanishes which can
be seen most directly from the surface
integrals~(\ref{Eq_chi_S1S2_NP_nc_1_E}). Now, we set $\Theta = 0 $ to
compute the contribution to the chromo-electric fields: Using the
surface $S_1 = S_P$ and loop $C_2 = \partial S_W$ parametrizations,
(\ref{Eq_Si_parameterization}) and~(\ref{Eq_Ci_parameterization}),
with our specification of the loop geometries, one obtains
from~(\ref{Eq_chi_S1C2_NP_nc_D1_E})
\bea
        \chi_{S_P S_W}^{\nprt\,nc}\!\!
        & = & 
        -\,\frac{\pi^2 G_2 (1\!-\!\kappa)}{3(N_c^2-1)}
        \int_{-R_P/2}^{R_P/2} \!\! du_1 
        \int_0^1 \!\! dv_1 
\label{Eq_chi_SPCW_NP_nc_D1_1_E}\\
        &&
        \times
        \left\{
        \int_{-T/2}^{T/2} \!\! du_2\,
        \left[
        (\vec{r}_{1\perp} \cdot \vec{Z}_{1A\perp})\,D_1(Z_{1A}^2)
        - (\vec{r}_{1\perp} \cdot \vec{Z}_{1C\perp})\,D_1(Z_{1C}^2)
        \right]
        \right.
\nonumber\\
        &&\hphantom{\times\,}
        \left.
        -\,
        (\vec{r}_{1\perp} \cdot \vec{r}_{2\perp})
        \int_0^1 \!\! dv_2\,
        \left[
        (\vec{r}_{1\perp} \cdot \vec{Z}_{1B\perp})\,D_1(Z_{1B}^2)
        - (\vec{r}_{1\perp} \cdot \vec{Z}_{1D\perp})\,D_1(Z_{1D}^2)
        \right]
        \right\}
\nonumber
\eea
with $T_1 = R_P/2$, $R_1 = R_P$, $T_2 = T/2$, $R_2 = R$, and the shorthand notation
\be
        Z_{1X} := X_{1} - X_{2}^{X}
        \quad \mbox{with} \quad
        X_1 \in S_1 = S_P
        \quad \mbox{and} \quad
        X_{2}^{X} \in C_2^X = \partial S_W^X
        \ .
\label{Eq_Z_SP_CWX_E}
\ee
In the limit $R \ll T \to \infty$, the term
$\propto\!(\vec{r}_{1\perp}\cdot\vec{r}_{2\perp})$ on the rhs
of~(\ref{Eq_chi_SPCW_NP_nc_D1_1_E}) can be neglected
\bea
        \lim_{T \to \infty}
        \chi_{S_P S_W}^{\nprt\,nc}\!\!
        & = & 
        -\,\frac{\pi^2 G_2 (1\!-\!\kappa)}{3(N_c^2-1)}
        \int_{-R_P/2}^{R_P/2} \!\! du_1
        \int_0^1 \!\! dv_1 
        \lim_{T\to \infty}
        \int_{-T/2}^{T/2} \!\! du_2\,
\label{Eq_chi_SPCW_NP_nc_D1_T2->infty_E}\\
        &&
        \times
        \left[
        (\vec{r}_{1\perp} \cdot \vec{Z}_{1A\perp})\,D_1(Z_{1A}^2)
        - (\vec{r}_{1\perp} \cdot \vec{Z}_{1C\perp})\,D_1(Z_{1C}^2)
        \right]
        \ .
\nonumber
\eea
With an infinitesimal plaquette used to measure the chromo-electric
field, $R_1 = R_p\to 0$, the mean value theorem can be used to perform
the $u_1$ and $v_1$ integrations
in~(\ref{Eq_chi_SPCW_NP_nc_D1_T2->infty_E}). Keeping only terms up to
$\Order(R_p^2)$, this leads directly to the non-confining
non-per\-tur\-ba\-tive contribution to the chromo-field distributions
$\Delta G_{\alpha \beta}^2(X)$ as given in~(\ref{Eq_Chi_PW_np_nc_14})
and~(\ref{Eq_Chi_PW_np_nc_24}).

\subsection*{\boldmath$\chi^{\pert}$-Computation}

Comparing the definition of the perturbative component
\bea
        \chi_{S_1 S_2}^{\pert} 
        & := & \frac{\pi^2}{4} 
        \int_{S_1} \!\! d\sigma_{\mu\nu}(X_1) 
        \int_{S_2} \!\! d\sigma_{\rho\sigma}(X_2)\,
        F_{\mu\nu\rho\sigma}^{\pert}(Z = X_1 - X_2)
\nonumber \\ 
        & = & 
        \frac{g^2}{4} \,
        \int_{S_1} \!\! d\sigma_{\mu\nu}(X_1) 
        \int_{S_2} \!\! d\sigma_{\rho\sigma}(X_2)
\label{Eq_chi_S1S2_P_def_E}\\ 
        && \times
        \inv{2}\Bigl[
        \frac{\partial}{\partial Z_\nu}
        \left(Z_\sigma \delta_{\mu\rho}
          -Z_\rho \delta_{\mu\sigma}\right)
        +\frac{\partial}{\partial Z_\mu}
        \left(Z_\rho \delta_{\nu\sigma}
          -Z_\sigma \delta_{\nu\rho}\right)\Bigr]\,
        D_{\pert}(Z^2)
\nonumber
\eea
with the one of the non-confining non-per\-tur\-ba\-tive component
$\chi_{S_1 S_2}^{\nprt\,nc}$ given in~(\ref{Eq_chi_S1S2_NP_nc_def_E}),
one finds an identical structure. Thus, accounting for the different
prefactors and the different correlation function, the results for
$\chi_{S_1 S_2}^{\pert}$ can be read off directly from the results for
$\chi_{S_1 S_2}^{\nprt\,nc}$ given above:

With $T_1 = T_2 = T/2\to\infty$ and our specification of the loop
geometries, one obtains the result for $\sin\Theta \neq 0$
from~(\ref{Eq_chi_C1C2_NP_nc_T->infty_S_E})
\bea
        &&\!\!\!\!\!\!
        \lim_{T\to\infty}
        \chi_{S_1 S_2}^{\pert}
        = 
        -\,
        \frac{\cos\Theta}{|\sin\Theta|}\,
        g^2\,
\label{Eq_chi_C1C2_P_T->infty_S_E}\\
        && \!\!\!\!\!\!\hspace{0.5cm}
        \times
        \Big[
        D_{\pert}^{\prime\,(2)}(\vec{Z}_{AA\perp}^2) - D_{\pert}^{\prime\,(2)}(Z_{AC\perp}^2) 
        - D_{\pert}^{\prime\,(2)}(\vec{Z}_{CA\perp}^2) +  D_{\pert}^{\prime\,(2)}(\vec{Z}_{CC\perp}^2) 
        \Big] 
\nonumber
\eea
and the result for $\sin\Theta = 0$
from~(\ref{Eq_chi_C1C2_NP_nc_T->infty_V_E})
\bea
        &&\!\!\!\!\!\!
        \lim_{T\to\infty}
        \chi_{S_1 S_2}^{\pert}
        = 
        -\,
        \lim_{T\to\infty}T\,
        \cos\Theta\,
        g^2\,        
\label{Eq_chi_C1C2_P_T->infty_V_E}\\
        && \!\!\!\!\!\!\hspace{0.5cm}
        \times
        \Big[
        D_{\pert}^{\prime\,(3)}(\vec{Z}_{AA}^2) - D_{\pert}^{\prime\,(3)}(Z_{AC}^2) 
        - D_{\pert}^{\prime\,(3)}(\vec{Z}_{CA}^2) +  D_{\pert}^{\prime\,(3)}(\vec{Z}_{CC}^2) 
        \Big]
        \ , 
\nonumber
\eea
where $Z_{XY}$ is defined in~(\ref{Eq_Z_C1X_C2Y_E}) and $Z_{XY\perp}$
is given explicitly in~(\ref{Eq_Z's_E}). Evidently,
(\ref{Eq_chi_C1C2_P_T->infty_S_E}) is the final result given
in~(\ref{Eq_S_DD_p_npc_npnc_E}) and~(\ref{Eq_S_DD_chi_p_M}) that
describes the perturbative contribution the dipole-dipole scattering
matrix element $S_{DD}$.

The perturbative contribution to the static potential is obtained from the expression corresponding to~(\ref{Eq_chi_SS_NP_nc_T->infty_V_E}),
\bea
        &&\!\!\!\!\!\!
        \lim_{T\to\infty}
        \chi_{S S}^{\pert}
        = 
        -\,\lim_{T\to\infty}T\,\,
        g^2\,\,
\label{Eq_chi_SS_P_T->infty_V_E}\\
        && \!\!\!\!\!\!\hspace{0.5cm}
        \times
        \Big[
        D_\pert^{\prime\,(3)}(\vec{Z}_{AA}^2) - D_\pert^{\prime\,(3)}(Z_{AC}^2) 
        - D_\pert^{\prime\,(3)}(\vec{Z}_{CA}^2) +  D_\pert^{\prime\,(3)}(\vec{Z}_{CC}^2) 
        \Big]
        \ ,
\nonumber
\eea
which contributes to the self-energy of the color sources with
\bea
        \lim_{T\to\infty}
        \chi_{S S\,\mbox{\scriptsize self}}^{\pert}
        &=& 
        -\,\lim_{T\to\infty}T\,\,
        g^2\,\,
        \Big[
        D_\pert^{\prime\,(3)}(\vec{Z}_{AA}^2) +  D_\pert^{\prime\,(3)}(\vec{Z}_{CC}^2) 
        \Big]
\nonumber\\
        &=& 
        -\,\lim_{T\to\infty}T\,\,
        2\,g^2\,\,
                D_\pert^{\prime\,(3)}(\vec{Z}_{AA}^2)
\label{Eq_chi_SS_P_T->infty_self_E}
\eea
and to the potential energy between the color sources with
\bea
        \lim_{T\to\infty}
        \chi_{S S\,\mbox{\scriptsize pot}}^{\pert}
        &=& 
        -\,\lim_{T\to\infty}T\,\,
        g^2\,\,
        \Big[
        D_\pert^{\prime\,(3)}(\vec{Z}_{AC}^2) +  D_\pert^{\prime\,(3)}(\vec{Z}_{CA}^2) 
        \Big]
\nonumber\\
        &=& 
        -\,\lim_{T\to\infty}T\,\,
        2\,g^2\,\,
        D_\pert^{\prime\,(3)}(\vec{Z}_{AC}^2)
        \ .
\label{Eq_chi_SS_P_T->infty_pot_E}
\eea
The latter gives the perturbative contribution to the static
potential~(\ref{Eq_Vr(R)_color-Yukawa}).

The perturbative contribution to the chromo-magnetic fields ($\Delta
G_{\alpha \beta}^2(X)$ with $\alpha\beta=ij=ji$) vanishes while the
one to the chromo-electric fields ($\Delta G_{\alpha \beta}^2(X)$ with
$\alpha\beta=i4=4i$) for which a plaquette with $\Theta = 0$ is
needed, is obtained from the expression corresponding
to~(\ref{Eq_chi_SPCW_NP_nc_D1_T2->infty_E}),
\bea
        \lim_{T \to \infty}
        \chi_{S_P S_W}^{\pert}\!\!
        & = & 
        -\,g^2
        \int_{-R_P/2}^{R_P/2} \!\! du_1
        \int_0^1 \!\! dv_1 
        \lim_{T\to \infty}
        \int_{-T/2}^{T/2} \!\! du_2\,
\label{Eq_chi_SPCW_P_T2->infty_E}\\
        &&
        \times
        \left[
        (\vec{r}_{1\perp} \cdot \vec{Z}_{1A\perp})\,D_\pert(Z_{1A}^2)
        - (\vec{r}_{1\perp} \cdot \vec{Z}_{1C\perp})\,D_\pert(Z_{1C}^2)
        \right]
\nonumber
\eea
with $Z_{1X}$ as defined in~(\ref{Eq_Z_SP_CWX_E}). To perform the
$u_1$ and $v_1$ integrations in~(\ref{Eq_chi_SPCW_P_T2->infty_E}),
again the mean value theorem can be used since the plaquette has
infinitesimally small extensions, $R_1 = R_p\to 0$. Keeping only terms
up to $\Order(R_p^2)$, this leads directly to the perturbative
contribution to the chromo-field distribution $\Delta G_{\alpha
  \beta}^2(X)$ as given in~(\ref{Eq_Chi_PW_p_14})
and~(\ref{Eq_Chi_PW_p_24}).


%
\end{appendix}
%
%
%
\cleardoublepage
  
%
%
\pagestyle{empty}
\cleardoublepage
%
%
\begin{Large}
\begin{center}
{\bf ACKNOWLEDGMENTS}
\end{center}
\end{Large}

I am most grateful to Prof.~H.~G.~Dosch for the friendly supervision
of this thesis, for his strong support, and for the many interesting
and intensive discussions. I had many invaluable opportunities to
learn non-perturbative methods in QCD from one of the experts. I also
enjoyed very much his lectures on ``The Physics of Music.''

Next I want to thank Prof.~H.~J.~Pirner for the close and fruitful
collaboration, for his interest in my work, and for the very many
creative and instructive ideas. I also appreciated very much his
suggestions on the preliminary versions of this work. It has been a
pleasure to attend his interesting lectures and seminars.

I would like to thank Arif Shoshi, with whom I collaborated most
closely during large parts of this investigation, for countless
discussions and for the vivid exchange of ideas that helped enormously
to master the model in all its details.

I thank Prof.~B.~Povh for his friendly interest and readiness to
referee this thesis.

I want to express my sincere gratitude to Prof.~O.~Nachtmann for his
continuous willingness to clarify and help in subtle issues, for
inviting me to participate regularly in the pleasant Wednesday lunch
of his group, and for the letter of recommendation.

I am thankful to Prof.~J.~H\"ufner for many suggestions that have
helped to find the right perspective in many cases and for his
hospitality in the theoretical nuclear physics group. With his
invitation, I enjoyed very often the pleasant Friday afternoon tea.

I would like to thank Dr.~Carlo Ewerz for his proofreading, the many
suggestions, and for bringing birdtrack notation to my attention.
Dr.~Hilmar Forkel deserves a special thank for his magnificent lecture
on ``Instantons in QCD'' and many illuminating discussions. I am also
grateful to Dr.~Matthias Jamin for helping me to better understand
renormalization in QCD. Moreover, I thank Dr.~Peter John for advice in
computational issues, Dr.~Eduard Thommes and his crew for
administrative support, and Prof.~W.~Wetzel for providing an extremely
stable and reliable computer environment. For careful readings of the
manuskipt I would like to express my gratitude to Sonja Bartsch and
Felix Schwab.

For helpful introductions to the stochastic vacuum model (SVM), useful
computer code, and notes on analytic SVM computations, I thank
Dr.~Edgar Berger, Dr.~Gerhard Kulzinger, Dr.~Timo Paulus, and
Dr.~Steffen Weinstock. Volker Schatz is thanked for help in
mathematical problems.

I enjoyed many stimulating discussions related to this thesis for
which I thank Dr.~Nora Brambilla, Prof.~A.~Di Giacomo, Dr.~Alberto
Polleri, Prof.~E.~Meggiolaro, Prof.~A.~H.~Mueller, Dr.~J\"org
Raufeisen, Prof.~I.~Stamatescu, and Dr.~Antonio Vairo.

It is a great pleasure to work at the Institute for Theoretical
Physics in Heidelberg. I would like to thank the many friendly,
open-minded, and helpful members for making this institute a unique
place with an extraordinarily pleasant atmosphere. I enjoyed very much
to be part of the crew under the roof. I thank my colleagues Lala
Adueva (``party party''), Dr.~Tobias Baier, Juliane Behrend, Dr.~Eike
Bick (thanks for the nice gatherings and the excellent menues),
Dr.~Michael Doran (``daradada dadada daaaahhh''), Dr.~Markus
Eidem\"uller, Bj\"orn Feuerbacher, Dietrich Foethke, Gero von
Gersdorff, J\"org J\"ackel, Bj\"orn O.\ Lange, Christian M.\ M\"uller,
Markus M.\ M\"uller, Filipe Paccetti, Tassilo Ott, Tania Robens,
Gregor Sch\"afer, Xaver Schlagberger, Kai Schwenzer (sorry for the
``Badegumpen'' trip), Jan Schwindt and Claus Zahlten for many cheerful
conversations in and outside of physics. I am particularly grateful to
Joachim Holk for the delicious tee and the many kind Fanta Pina de
Coco offers during our night sessions. It was always pleasant not to
be alone with the cat at two o' clock in the morning. My sincere
thanks goes to Mrs.~G.~Rumpf for countless friendly conversations and
the valuable permanent catering service. Felix Schwab deserves a
special thank not only for the Bruce tickets but also for being the
best drummer in the institute.

My sincere thanks to the Graduiertenkolleg ``Physical Systems with
many Degrees of Freedom'' for supporting me since January 2000 and for
providing an ideal framework to organize our own workshops and
schools. In particular, I want to thank Prof.~F.~Wegner for his trust
in our organization of the autumn school ``Topology and Geometry in
Physics.'' I am extremely grateful to Prof.~J.-W.~van Holten,
Prof.~F.~Lenz, Prof.~T.~Sch\"ucker, Prof.~M.~Shifman, and
Prof.~J.~Zinn-Justin for making this autumn school a very special
event and for writing such highly pedagogical lecture notes for the
proceedings.

Now I come to the world outside of the physics. I have enjoyed very
much the sports program at the University of Heidelberg and the many
nice conversations with the friendly participants. I particularly
thank Matthias Wolf for the excellent TAEBO specials that helped a lot
in relaxing from physics. Moreover, I would like to thank Janine
Fritz, Angela Klein, Verena Schmidt-Steffens, Edgitha Stork, Zhe Xu
for their loyal friendship during all these years.

In a very special way I want to thank Natascha Kunert for her love,
all the wonderful moments, and her patience during the many weekends I
spent at the institute. I am also indebted to her family for the
continuous warm-hearted friendship and hospitality.

Finally, I am extremely thankful to my mother and my father for their
care and love. It is always a great pleasure to be at home.


%
%
\end{document}